\documentclass[traditabstract]{aa} 
\usepackage{graphicx}

\usepackage{lscape}
\usepackage{amssymb}
\usepackage{amsmath}

\usepackage{txfonts}

\usepackage{booktabs}
\usepackage{subfigure}
\usepackage{placeins}
\usepackage{verbatim}
\usepackage{arydshln}
\usepackage{adjustbox}
\usepackage[toc,page]{appendix} 
\usepackage{tabularx}
\usepackage{lscape}
\usepackage{textcomp}
\usepackage{gensymb}
\usepackage{hyperref}
\usepackage{multicol}
\usepackage[symbol]{footmisc}
\usepackage{tikz}
\newcommand*\circled[1]{\tikz[baseline=(char.base)]{
            \node[shape=circle,draw,inner sep=2pt] (char) {#1};}}
\usepackage{caption}

\usepackage{natbib}

\usepackage{color}

\newcommand{\sersic}{S\'{e}rsic}
\defcitealias{Schmidt2022}{S22}
\begin{document}

   \title{TDCOSMO. X. Automated modeling of nine strongly lensed quasars and comparison between lens-modeling software}

  \titlerunning{Automated modeling of nine strongly lensed quasars}

   \author{S. Ertl\inst{1}\inst{,2}
                  \and
                        S. Schuldt\inst{1}\inst{,2}
                        \and
                        S. H. Suyu\inst{1}\inst{,2}\inst{,3}
                        \and
                        T. Schmidt\inst{4}
                        \and
                        T. Treu\inst{4}
                        \and
                        S. Birrer\inst{5}\inst{,6}\inst{,7}
                        \and
                        A. J. Shajib\inst{8,4}
                        \and
                        D. Sluse\inst{9}
                        }

        \institute{Max-Planck-Institut f{\"u}r Astrophysik, Karl-Schwarzschild Stra{\ss}e 1, 85748 Garching, Germany\\
             e-mail: \href{mailto:ertlseb@mpa-garching.mpg.de}{\tt ertlseb@mpa-garching.mpg.de}
         \and
             Technical University of Munich, TUM School of Natural Sciences, Department of Physics,  James-Franck-Stra{\ss}e 1, 85748 Garching, Germany
          \and
             Academia Sinica Institute of Astronomy and Astrophysics (ASIAA), 11F of ASMAB, No.1, Section 4, Roosevelt Road, Taipei 10617, Taiwan
        \and
            Physics and Astronomy Department, University of California, Los Angeles, CA 90095, USA
        \and
            Kavli Institute for Particle Astrophysics and Cosmology and Department of Physics, Stanford University, Stanford, CA 94305, USA
        \and
            SLAC National Accelerator Laboratory, Menlo Park, CA, 94025, USA
        \and
            Department of Physics and Astronomy, Stony Brook University, Stony Brook, NY 11794, USA
        \and
            Department of Astronomy \& Astrophysics, University of Chicago, Chicago, IL 60637, USA
        \and
            STAR Institute, Quartier Agora - All\'{e}e du six Ao\^{u}t, 19c B-4000 Liege, Belgium
             }

   \date{Received --; accepted --}

 
  \abstract{
  When strong gravitational lenses are to be used as an astrophysical or cosmological probe, models of their mass distributions are often needed. We present a new, time-efficient automation code for the uniform modeling of strongly lensed quasars with \texttt{GLEE}, a lens-modeling software for multiband data. By using the observed positions of the lensed quasars and the spatially extended surface brightness distribution of the host galaxy of the lensed quasar, we obtain a model of the mass distribution of the lens galaxy. We applied this uniform modeling pipeline to a sample of nine strongly lensed quasars for which images were obtained with the Wide Field Camera 3 of the Hubble Space Telescope. The models show well-reconstructed light components and a good alignment between mass and light centroids in most cases. We find that the automated modeling code significantly reduces the input time during the modeling process for the user. The time for preparing the required input files is reduced by a factor of 3 from $\sim$3 hours to $\text{about }$one hour. The active input time during the modeling process for the user is reduced by a factor of 10 from $\sim$10 hours to $\text{about one}$ hour per lens system. This automated uniform modeling pipeline can efficiently produce uniform models of extensive lens-system samples that can be used for further cosmological analysis. A blind test that compared our results with those of an  independent automated modeling pipeline based on the modeling software \texttt{Lenstronomy} revealed important lessons. Quantities such as Einstein radius, astrometry, mass flattening, and position angle are generally robustly determined. Other quantities, such as the radial slope of the mass density profile and predicted time delays, depend crucially on the quality of the data and on the accuracy with which the point spread function is reconstructed.  Better data and/or a more detailed analysis are necessary to elevate our automated models to cosmography grade. Nevertheless, our pipeline enables the quick selection of lenses for follow-up and further modeling, which significantly speeds up the construction of  cosmography-grade models. This important step forward will help us to take advantage of the increase in the number of lenses that is expected in the coming decade, which is an increase of several orders of magnitude.}


   \maketitle

\section{Introduction}
\label{sec:intro}

Gravitational lensing describes the effect of the gravitational potential of a massive object on the light of a background source. In the case of strong lensing (SL), the light is deflected, such that multiple images of the source are observed. 

Because SL is sensitive to both luminous and dark matter (DM), it is a powerful tool for probing the distribution of the total galaxy mass and thus for giving insight into the DM distribution of galaxies \citep[e.g.,][]{Dye2004, Barnabe2012,Sonnenfeld2015, Schuldt2019, Shajib2021}. Mass clumps in the lensing galaxy affect the image magnification, and substructures can be detected by analyzing flux-ratio anomalies of point-like sources (e.g., \citealt{Dalal2001}; \citealt{Moustakas2002}; \citealt{Nierenberg2014}; \citealt{Nierenberg2017}; \citealt{Hsueh2020}; \citealt{Nierenberg2020}; \citealt{Gilman2020}) or distortions in the arcs formed by the lensing of spatially extended background galaxies
(e.g., \citealt{Koopmans2005}; \citealt{Vegetti2010}; \citealt{Vegetti2012}). Clumps with a mass lower than $10^8$ solar masses can be detected in this way. 
We can also analyze the structural parameters of early-type lens galaxies, such as the average slope of the total mass density profile or the DM fraction, with models (e.g., \citealt{Gavazzi2007}; \citealt{Auger2010}; \citealt{Lagattuta2010}; \citealt{Barnabe2011}; \citealt{Shu2015}).

In addition, the magnifying effect of SL can be used to observe the earliest galaxies and quasars in the Universe (e.g., \citealt{Kneib2004}; \citealt{Bradley2008}; \citealt{Coe2013}; \citealt{Salmon2020}). Modeling these high-redshift objects enables us to study their rotation curves and masses, for example (e.g., \citealt{Jones2010}; \citealt{Chirivi2020}; \citealt{Rizzo2020}). This helps us to understand galaxy evolution and structure formation in the early Universe. Despite major progress in the last years, this research field has many open questions.

Another important application of SL is the measurement of cosmological parameters such as the Hubble constant $H_0$ (e.g., \citealt{Suyu.2017}; \citealt{Wong2019}; \citealt{Shajib2019b}; \citealt{Chen2020}; \citealt{Millon2020}; \citealt{Birrer2020}). This parameter, which gives the current expansion rate of the Universe, is of great interest because early-universe measurements from the cosmic microwave background (CMB) give a value of 67.36 $\pm$ 0.54  $\rm{\, km\, s^{-1}\, Mpc^{-1}}$ (\citealt{Planck2019}), whereas local measurements from the SH0ES project (\citealt{Riess2021}) based on the cosmic distance ladder using Cepheids and type Ia supernovae (SNe) obtain a higher value of 73.04 $\pm$ 1.04  $\rm{\, km\, s^{-1}\, Mpc^{-1}}$. This 5$\sigma$ disagreement is commonly known as the Hubble tension and might be an indication of physics beyond a flat $\Lambda$CDM. The  flat $\Lambda$CDM currently is the standard cosmological model. SL can also be used to probe alternative cosmological models \citep[e.g.,][]{Jullo2010, Oguri2012,Cao2015, KrishnanEtal21}. To resolve the Hubble tension, independent methods are necessary, including megamasers (e.g., \citealt{Pesce2020}), standard sirens (e.g., \citealt{Abbott2017}), surface brightness fluctuations (e.g., \citealt{Khetan2020}), expanding photospheres of supernovae (e.g., \citealt{Schmidt1994}), and gravitational lensing.

To measure $H_0$ with SL, a strongly lensed variable background source such as a quasar or an SN is required. Because of the different path length and the gravitational potential differences on the way of the photons, the characteristic brightness fluctuations reach the observer at different times. We can measure this time delay in the light curves by monitoring the multiple images and comparing these flux variations with each other \citep[e.g.,][]{Millon2020b, Millon2020a}.

In addition to an accurate measurement of the time delays, the lens potential is required to determine $H_0$, which is obtained by modeling the observed lensing system. Major progress has been made in this field in recent decades, in part because of the development of advanced modeling software such as GLEE (\citealt{Suyu.2010}; \citealt{Suyu.2012}) and \texttt{Lenstronomy} (\citealt{Birrer2015}; \citealt{Birrer2018LT}). Modeling software like these allow us to constrain the model parameters of the lens and the source, such as their position and radial mass profile. A downside of this approach is that we need at least several weeks to model a lens system with spatially extended sources because it is time consuming to model the parameter space, for instance, with Markov chain Monte Carlo (MCMC) methods. Moreover, the modeling process itself is interactive. In addition, the input files required by \texttt{GLEE} (e.g., point spread function (PSF) and error map) have to be obtained manually from the available data in some cases. 

With the increasing number of wide-field imaging surveys, including the Hyper Suprime-Cam (HSC; \citealt{Aihara2018}), the Dark Energy Survey (DES; \citealt{DarkEnergySurveyCollaboration.2016}), the upcoming  Euclid (\citealt{Laureijs2011, Scaramella+2021}), and the Rubin Observatory Legacy Survey of Space and Time (LSST; \citealt{Ivezic.2019}),  the number of known strong-lensing systems is growing rapidly. Even though the number of detected lenses per square degree lies between 0.5 and 10 (\citealt{Li.2020}) depending on the image depth, the number of lens candidates will increase strongly as a result of the large survey areas. For example, the LSST will cover a total of $\sim$20,000 deg$^2$ in multiple bandpasses over its planned ten-year run time (\citealt{Tyson.2002}, \citealt{Lochner+2021}). We expect new images of billions of galaxies, about 100,000 of which are strong-lensing systems (\citealt{Collett.2015}). To be able to keep up with the increasing sample size, we need techniques that automate the lens-modeling process. 
By automating the creation of the input files, the calculation of the starting values of lens parameters and image positions, and the optimization and sampling of the parameter space, we can save many hours of user interaction during the modeling process.
Several projects in recent years have already successfully demonstrated the automation of lens modeling, especially when initial estimates of the lens mass distribution were to be obtained \citep[e.g.,][]{Shajib2019,Nightingale2021}. Techniques are also being developed based on machine learning \citep[e.g.,][]{Hezaveh+2017, PerreaultLevasseur+2017, Park+2021, Schuldt+2021, Schuldt2022}, although the modeling of strongly lensed quasars via neural networks has yet to be tested on real instead of mock data. Another method is the massive parallelization of graphics processing units \citep{Gu2022} to speed up the lens modeling. All these approaches cannot yet replace the detailed manual lens-by-lens analysis that is necessary to produce models that are accurate enough for time-delay cosmography \citep[e.g.,][]{Suyu+2010b,Suyu+2013,Birrer2018,Wong2019,Shajib2019b,Rusu2020}, but rather serve as a first step toward this and provide important information for follow-up observations.

In this paper, we present a new automated procedure for modeling of strongly lensed quasars that utilizes the lens-modeling software \texttt{GLEE} and efficiently employs the user time. The modeling is uniform (i.e., has the same assumptions) for all systems. We developed a pipeline written in Python that works with minimum user input. Providing only high-resolution imaging data and marking important regions in the field, the user can model in a nearly fully automated way with \texttt{GLEE.}  The automated modeling can be performed with single-band and multiband data. We focus on lensed quasars to model the lens mass distribution with an isothermal or power-law profile and the source brightness distribution on a grid of pixels. To test our code, we uniformly modeled a sample of nine strongly lensed quasars observed with the Hubble Space Telescope (HST). Furthermore, we predict the time delays and Fermat potential differences between the multiple images of the quasar. We emphasize that our approach differs from that of papers aimed at deriving cosmological parameters from individual lenses, because we prioritize speed over accuracy. Owing to the adopted shortcuts (e.g., we do not iteratively correct the PSF and usually model only the images in filters with visible arcs from the host galaxy) and the uneven quality of the data, we do not in general expect our results to reach cosmography grade. The output of this pipeline is aimed as a first step toward cosmography-grade models, and it is meant to assist in prioritizing follow-up monitoring and further modeling. 

\citet[][hereafter \citetalias{Schmidt2022}]{Schmidt2022} have also developed an automated modeling pipeline using \texttt{Lenstronomy} and applied it to a sample of 30 lensed quasars.  Our sample of 9 quasars is a subset of the systems in \citetalias{Schmidt2022}, allowing us a direct comparison of the  results between the two modeling pipelines as a blind test because the lead modelers of the two teams obtained their lens models independently and only compared the results after the models were completed. We note that our approach and that of \citetalias{Schmidt2022} are similar, but not identical. Important differences include the modeling of the PSF, which is known to be crucially important \citep{Shajib2022}, and the strategy adopted to cope with insufficient data quality (\citetalias{Schmidt2022} imposed informative priors, but we do not). Therefore, we do not expect the two efforts to agree within the formal uncertainties, which are only a representation of the statistical error and not of the systematics, which are notoriously dominant. This should therefore be kept in mind when interpreting the comparison results. Nevertheless, as we show below, important lessons can be gleaned from it. 

The outline of the paper is as follows: in Sec.~\ref{sec:modeling} we describe the steps in modeling a strong-lensing system with our automation code.  We briefly describe each system and the observational data in Sec.~\ref{sec:observations}. The modeling results are presented in Sec.~\ref{sec:results}. The performance of the automated modeling compared to manual modeling 
is discussed in Sec.~\ref{sec:discuss}, together with a comparison with the models obtained with \texttt{Lenstronomy} of the same systems (\citetalias{Schmidt2022}). In Sec.~\ref{sec:conclude} we summarize our results.

Throughout the paper, we adopt a flat $\Lambda$CDM cosmology with $H_0=70 \rm{\, km\, s^{-1}\, Mpc^{-1}}$ and $\Omega_{\rm M} = 1 - \Omega_{\Lambda}=0.3$. Parameter estimates are given by the median of its one-dimensional marginalized posterior probability density function, and the quoted uncertainties show the 16th$^{\rm }$ and 84th$^{\rm }$ percentiles (corresponding to a 68\% credible interval).

\section{Automated modeling process}
\label{sec:modeling}
\begin{figure}[h]
        \subfigure{\includegraphics[width=0.48\textwidth]{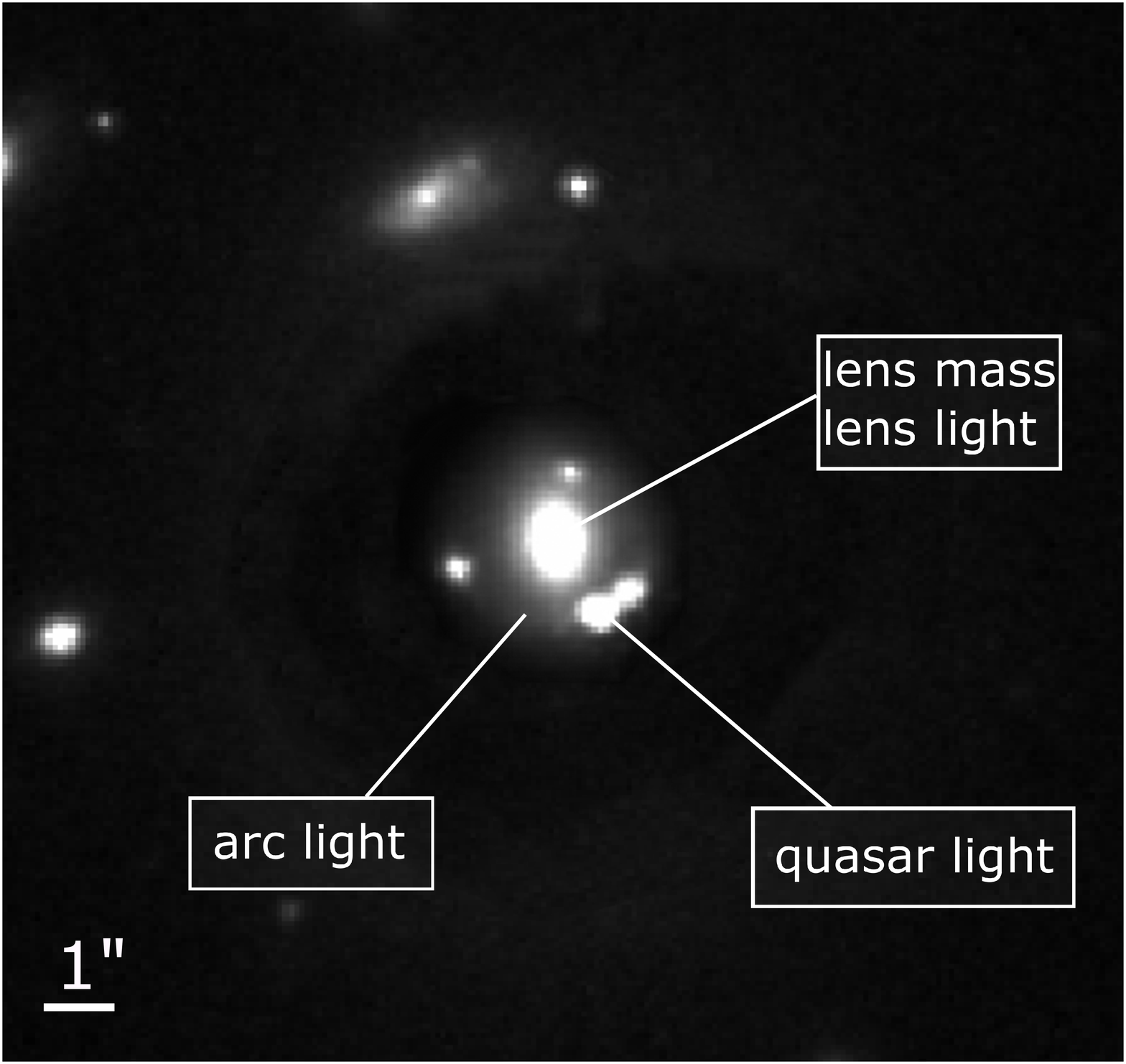}}
        \hfill
        \subfigure{\includegraphics[width=0.48\textwidth]{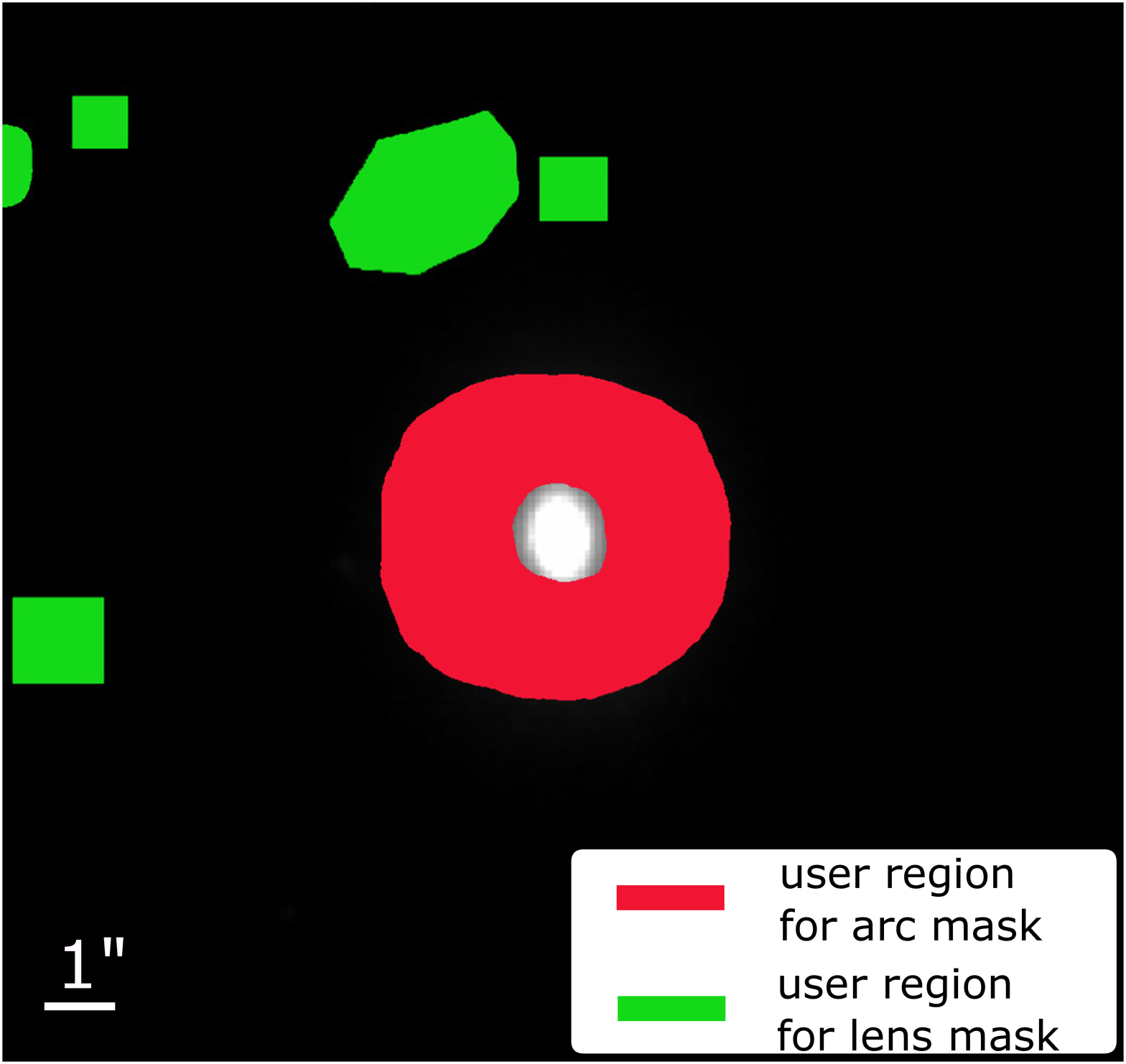}}
        \caption{Lensing system and regions for mask creation. \textit{Top}: Image cutout of J2100$-$4452 to illustrate its different components which are modeled with our automation code. \textit{Bottom}: Image cutout of J2100$-$4452 with regions provided by the user to create the masks. The region required for creating the arc mask, which covers the light of the lensed quasar and arcs, is plotted in red. For the lens mask, the user provides the region(s) of the surrounding luminous objects (green) that contain the pixels that are ignored by the model.\label{fig:sketch}}
\end{figure}

In this section, we describe the steps of modeling a strongly lensed quasar with \texttt{GLEE} in detail and how they are conducted in our automation code. We adopt different mass and light profiles, and use the Bayesian framework of MCMC sampling and simulated annealing \citep{Kirkpatrick1983} provided by \texttt{GLEE}. The goal is to model the lens mass distribution first using the source and image positions, and then by reconstructing the surface brightness distribution of the lens and background source galaxies.
Fig.~\ref{fig:sketch} shows the strongly lensed quasar J2100$-$4452 as an example. The figure shows the lens galaxy in the image center, four multiple images of the background quasar, and the arc, which is the lensed light of the quasar host galaxy. The code models the mass distribution of the lens galaxy and the light components of the system (i.e., lens, satellite galaxies of the lens, lensed quasar, and arc light). The modeled constituents of the lensing system are labeled in Fig.~\ref{fig:sketch}.

Because the modeling is typically first performed in a single filter and further bands may be included after the model has stabilized, we present in Sec.~\ref{sec:singleband} an automation code dedicated for the single-band modeling. We then introduce the automation code for the multiband modeling in Sec.~\ref{sec:multiband} and finally predict the time delays between the multiple images of strongly lensed quasars in Sec.~\ref{sec:timedelay}.

\subsection{Single-band modeling}
\label{sec:singleband}
\begin{figure}[h]
        \centering
        \includegraphics[scale=0.5]{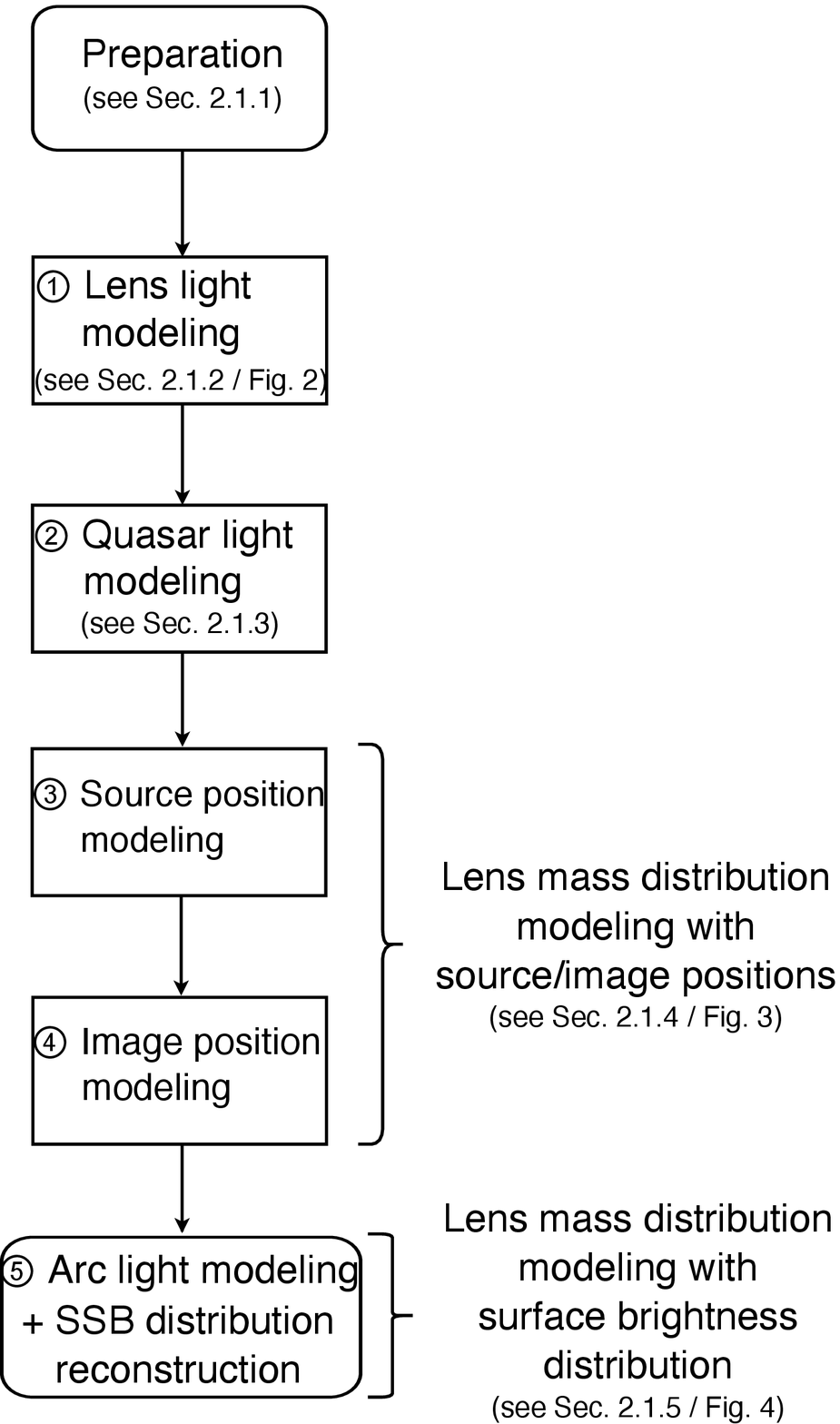}
        \caption{Overview flowchart for single-band modeling of a strongly lensed quasar system with our new automated code. The modeling consists of two main steps: the lens mass distribution modeling with source or image positions, and the reconstruction of the source surface brightness (SSB; which is the unlensed light distribution of the quasar host galaxy). Each step is divided into smaller modeling steps. The details are described in flowcharts for each step and in the corresponding subsection, as referenced in the flowchart.}
        \label{fig:flowchart_overview}
\end{figure}

In this section, we discuss the automated modeling of a lensing system with single-band data. We describe the semi-automated preparation of the required input files and each modeling step and their substeps in the following subsections. An overview of the basic modeling procedure is shown as a flowchart in Fig.~\ref{fig:flowchart_overview} and summarized in Table~\ref{tab:Modelingtable_single}. In particular, Table~\ref{tab:Modelingtable_single} shows the varying and fixed parameters in each modeling step.

\begin{table*}
    \caption{Modeling steps in the single-band modeling code.}
        \begin{tabular}{lp{5.8cm}lp{2.2cm}ccccccc}\toprule
                Step & Description & Section & Criteria & \multicolumn{2}{c}{Lens mass parameters} & Lens light & \multicolumn{4}{c}{Source light}
                \\\cmidrule(lr){5-6}\cmidrule(lr){8-9}
                & & & & profiles  & ext. shear &  & quasar    & host \\\midrule\midrule \vspace{5px}
                \circled{1} & modeling of the lens light & \ref{sec:lenslight} & $\circlearrowleft_2 (\chi^2_{\rm red}\leq1)$ & $\times$  & $\times$ & $\checkmark$ & $\times$ & $\times$ \\ \vspace{5px} 
                \circled{2} & modeling of the quasar light & \ref{sec:quasarlight} & $\circlearrowleft_2 (\Delta \textrm{logP}\leq5)$ & $\times$  & $\times$ & $\bigcirc$ & $\checkmark$ & $\times$ \\ \vspace{5px}
                \circled{3} & source position modeling & \ref{sec:srcimgpos} & $\circlearrowleft_1 (\chi^2<5)$ & $\checkmark$ & $\times$ & $\bigcirc$  & $\bigcirc$  & $\times$ \\ \vspace{5px}
                \circled{4} & image position modeling with external shear & \ref{sec:srcimgpos} & \raggedright $\circlearrowleft_2 (\chi^2<5,$ & $\checkmark$ & $\checkmark$ & $\bigcirc$  & $\bigcirc$ & $\times$ \\ \vspace{5px}
                &&&$\Delta$ \textrm{logP}$\leq$5)&&&&\\\vspace{5px}
                \circled{5} a & modeling of the arc light + SSB distribution reconstruction & \ref{sec:arclight} & $\circlearrowleft_2 (\Delta \textrm{logP}\leq2)$ & $\checkmark$ & $\checkmark$ & $\bigcirc$ & $\checkmark_{\rm A}$ & $\checkmark$\\
                \circled{5} b & modeling of the arc light + SSB distribution reconstruction & \ref{sec:arclight} & $\circlearrowleft_2 (\Delta \textrm{logP}\leq2)$ & $\checkmark$ & $\checkmark$ & $\checkmark$ & $\bigcirc$ & $\checkmark$ \\\bottomrule
        \end{tabular}
        \ \\ \ \\
        \textit{Symbols:}\\
        \begin{tabular}{ll}
                $\circlearrowleft_1$: optimizing cycle with simulated annealing & $\checkmark$: parameters vary\\
                $\circlearrowleft_2$: optimizing and sampling cycle with simulated annealing and MCMC \ \ \ \ & $\bigcirc$: parameters fixed \\
                $\checkmark_{\rm A}$: only the amplitude varies & $\times$: parameters not included
        \end{tabular}
    \ \\ \ \\
        \caption*{\textbf{Notes: }Each step is described briefly in the second column with the corresponding subsection in the third column. The modeling technique and stopping criteria are given in the fourth column. The cycle runs until the stopping criteria in the parentheses are met. The remaining columns show the parameters that vary, are fixed, or are not included.}
        \label{tab:Modelingtable_single}
\end{table*}

\subsubsection{Preparation}
\label{sec:preparation}
To model a strong-lensing system with our newly developed automated modeling codes, the user needs to provide information about the position and approximate light structure of the lens, the quasar images, and the satellites. In addition, several input files (see Table~\ref{tab:inputfiles}) are needed. Each input file is created in separate codes that are described in the following. 

\begin{table}[h]
        \caption[Sec.~\ref{sec:preparation}: Overview of input files]{Overview and description of required input files.}               \begin{tabular}{l|p{6.4cm}}
                        Input file & Description \\ \midrule
            Image cutout & Cutout of the lensing system \\
                        Error map & 
                        $1\sigma$ uncertainty of pixel intensity in the cutout, accounting for background and Poisson noise\\ 
                        PSF &  
                    Image of point-spread function\\  
                        Arc mask & Region of arc light (red area in Fig.~\ref{fig:sketch})\\
                        Lens mask &  Region for masking neighboring objects (green area in Fig.~\ref{fig:sketch})
                \end{tabular}
        \label{tab:inputfiles}
\end{table}

\paragraph{Science image cutout}

We start by creating a cutout of the lensing system from the preprocessed reduced data such that it includes the immediate environment ($\lesssim10''$) that is to be taken into account during the modeling.  In the automated pipeline, the user creates a DS9 (\citealt{Joye.2003}) region file containing the cutout area in the field and the positions of the lens, satellite galaxies, and the quasar images. Optionally, the background flux is subtracted by calculating the average flux of an empty patch in the field. The modeling code then creates the cutout, which is used for the subsequent modeling process, and calculates starting values for the lens mass and the light parameters and image positions from the region information. 
\paragraph{Error map}

The error map accounts for both background noise and Poisson noise. The background noise $\sigma_{\rm background}$ is approximated as a constant that is set to be the standard deviation from a small patch in the science image without astrophysical sources. The Poisson noise $\sigma_{\rm poisson}$ is computed from the weighted image (exposure time map) and the science image. For images in units of counts per second, we calculate
\begin{equation}
        \sigma^2_{\rm poisson, \it i} = \Big|\frac{d_i}{t_i\times G}\Big|,
\end{equation}
with  $d_i$ the observed intensity\footnote{Using the observed image instead of the model intensity may induce biases at low signal-to-noise ratios \citep{Horne1986}.}, $t_i$ the exposure time for pixel $i$, and $G$ the gain of the charge-coupled device (CCD) for the specific filter. In the case of data units of counts, we have
\begin{equation}
        \sigma^2_{\rm poisson, \it i} = \Big|\frac{d_i}{G}\Big|.
\end{equation}
We add $\sigma_{\rm poisson}$ in quadrature to the background noise level only if $\sigma_{\rm poisson} > 2\times\sigma_{\rm background}$ to avoid overestimating the noise in regions without flux from astrophysical objects. We check this condition for each pixel $i$ to obtain the final error map:\\
\begin{align}
    \sigma^2_{\rm tot, \it i} &=\begin{cases}
        \sigma^2_{\rm background, \it i} + \sigma^2_{\rm poisson} & \text{if $\sigma_{\rm poisson} > 2\times\sigma_{\rm background}$;}\\
        \sigma^2_{\rm background, \it i}                     & \text{otherwise.}
    \end{cases}
    \label{eq:sigmatotal}
\end{align}

\paragraph{Point spread function}
If the PSF is not directly provided with the science image, the PSF can be created with our PSF code. For this, the user marks several nonsaturated stars in the field. The field spans $\sim$ 2.5 $\times$ 2.5 arc minutes in the case of the HST WFC3 data we used. The PSF code then automatically creates cutouts of these stars that are centered at the brightest pixel. Since the star is not necessarily located at exactly the center of the brightest pixel, the PSF code interpolates and shifts the cutout within fractions of a pixel to center it in the cutout. In particular, the profile is shifted such that the brightness of the pixels to the left and right of the brightest pixel is within a margin of 10\%, and similarly for the pixels above and below the brightest pixel. This causes the central 3$\times$3 region to be roughly symmetric. To obtain the PSF, the recentered stars are normalized and then stacked by calculating the median per pixel among the cutouts, and they are renormalized such that the sum of all pixel values is 1. In addition, the automation code requires two PSFs, one for modeling the light of the lensed quasar, and one for the extended source. The first PSF includes the diffraction spikes as that PSF image is used directly as the spatial profile for the quasar images. The second PSF is cropped at the first diffraction minimum of the Airy disk because it is used for the convolution of the parameterized light distribution and arcs, and the computation time increases significantly with increasing size of the PSF. The cropped PSF is also used to model the lens light.  Subsampling of these two PSFs is required when the pixel sizes are large relative to the full width at half maximum of the PSF in order to avoid numerical inaccuracies caused by an undersampled PSF.\footnote{This is the case for the HST-IR F160W filter that is used.  In practice, we subsample the PSF when the pixel size is greater than 0.05\arcsec.}
The PSF code subsamples the PSF by a custom factor $n$ such that its pixel scale is $\frac{1}{n}$ of the original pixel size. The two subsampled PSFs are renormalized by the PSF code. 
\paragraph{Masks}
\texttt{GLEE} requires two masks for the modeling. The so-called arc mask contains the region of the lensed quasar and its host (red regions in Fig.~\ref{fig:sketch}). It is used during the modeling of the lens light to exclude quasar and host light from the modeling, and it later gives the pixel region that is used to reconstruct the distribution of the source surface brightness (SSB). The lens mask specifies the pixels of the light of surrounding luminous objects such as neighboring stars (green areas in Fig~\ref{fig:sketch}). \texttt{GLEE} sequentially fits the pixels of the lens light, excluding regions of the arc mask and neighboring objects specified by the lens mask, as we describe in the following subsections. The masks are created by the user, who marks the corresponding regions, saves the DS9 region file of the marked areas, and runs a mask-generation code on these DS9 region files.

\subsubsection{Modeling the lens light}
\label{sec:lenslight}
The first modeling step is to reconstruct the light of the lens galaxy and the multiple quasar images in order to subtract their contribution and reveal the underlying arc light more prominently. We model the lens light with the S\'{e}rsic profile (\citealt{Sersic.1963}), which is parameterized as
\begin{equation}
        I_{\rm S}(x,y) = A_{\rm S}\exp\Bigg[-k\Bigg\{\Bigg(\frac{\sqrt{(x-x_{\rm S})^2+\left( \frac{y-y_{\rm S} }{q_\text{S}} \right) ^2}}{r_{\rm eff}}\Bigg)^{\frac{1}{n_{\rm s}}}-1\Bigg\}\Bigg].
\end{equation}
\noindent
An overview of the parameters in the S\'{e}rsic profile and their priors that were used in the automated modeling is provided in Table \ref{tab:Sersicparam}. The constant $k$ normalizes $r_{\rm eff}$ such that $r_{\rm eff}$ is the half-light radius in the direction of the semi-major axis.

\begin{table*}[h]
    \caption{Prior ranges of the \sersic \, parameters for the lens galaxy light.}
                \begin{tabular}{l|l|l|p{5.5cm}|p{5.5cm}}
                        Component & Parameter & Description & Prior & Prior range / value \\ \midrule
                        & $x_{\rm S}\ ['']$   & $x$-centroid & Gaussian, flat prior in SSB reconstruction & centered on starting value, Gaussian $\sigma=0.3$   \\ \rule{0pt}{2ex}   
                        & $y_{\rm S}\ ['']$   & $y$-centroid  & Gaussian, flat prior in SSB reconstruction & centered on starting value, Gaussian $\sigma=0.3$ \\ \rule{0pt}{2ex}   
                        & $q_{\rm S}$         & axis ratio & flat & single-band: [0.3, 1], multi-band: [0.1, 1] \\ \rule{0pt}{2ex} 
                        S\'{e}rsic & $\phi_{\rm S}$ [$^\circ$]   & position angle & flat & [0, 360]\\ \rule{0pt}{2ex}
                        & $A_{\rm S}$  & amplitude & flat & [0.01, 10] \\  \rule{0pt}{2ex}
                        & $r_{\rm eff}\ ['']$         & effective radius & flat & [0.01, 10]\\ \rule{0pt}{2ex}
                        & $n_{\rm S}$      & S\'{e}rsic index & flat & single-band: [0.5, 6], multi-band: [0.4,9]\\
                \end{tabular}
    \ \\
        \caption*{\textbf{Notes: }The S\'{e}rsic parameters are varied during lens light modeling and the modeling of the arc light with SSB reconstruction. The position angle is measured counterclockwise from the positive $x$-axis (east of north).}
        \label{tab:Sersicparam}
\end{table*}

The Gaussian prior on the centroid coordinates is used to stabilize the optimization during lens light modeling, and it is later changed to a flat prior when the code models the light of the arc (Sec.~\ref{sec:arclight}). The prior range on the axis ratio $q_{\rm S}$ and \sersic\ index $n_{\rm S}$ are relaxed during multi-band modeling as more constraints are available when additional filters are added. As described in Sec.~\ref{sec:preparation}, we do not subsample the PSF for images with pixels smaller than 0.05\arcsec\, and use the cropped and subsampled PSF for images with a pixel scale greater than 0.05\arcsec. 

\begin{figure*}
        \centering
        \includegraphics[scale=0.42]{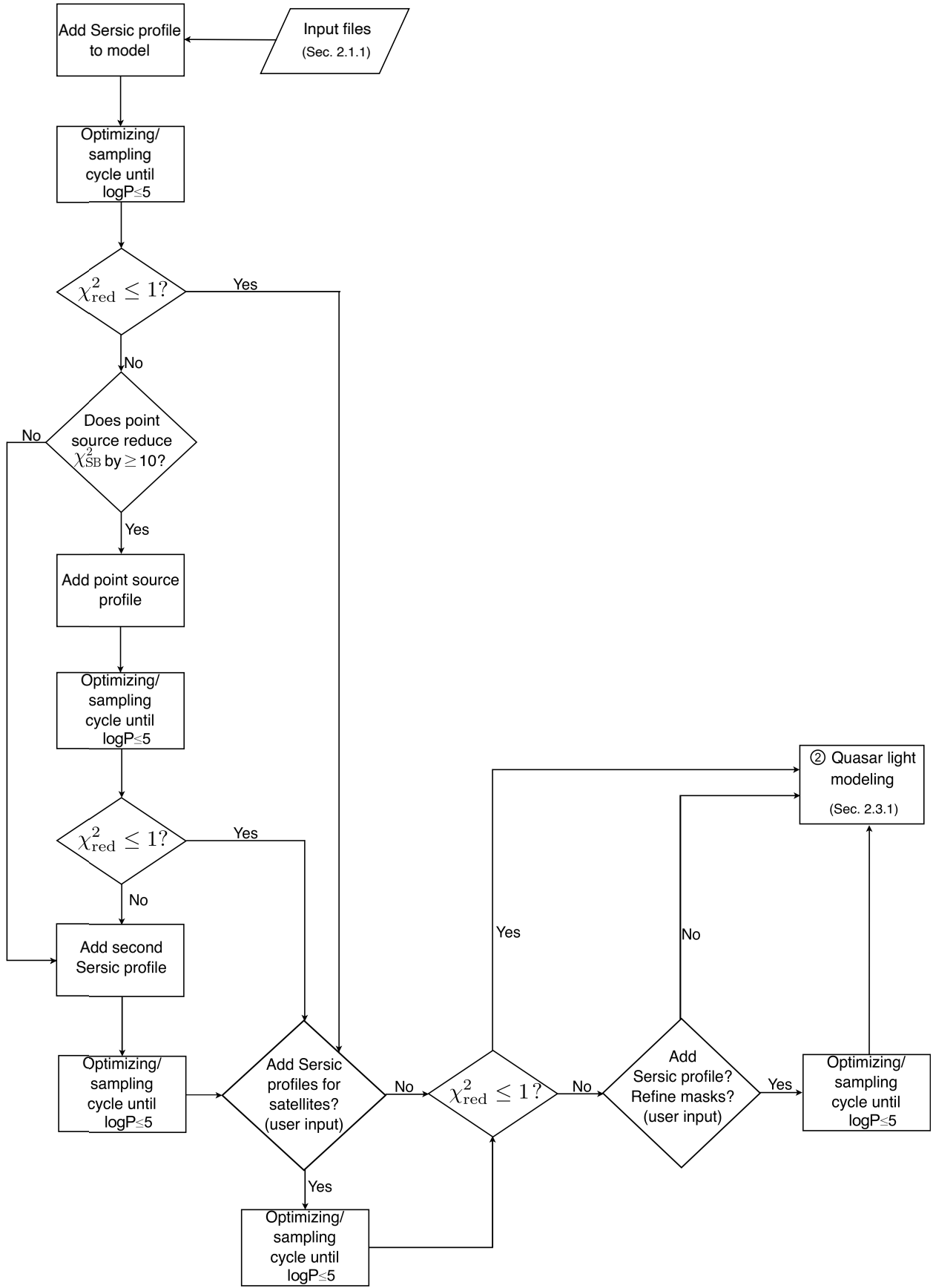}
        \caption[Sec.~\ref{sec:lenslight}: Flowchart for lens light modeling]{Flowchart showing the decision tree for the lens light modeling. The modeling code obtains the best configuration of S\'{e}rsic and point-source profiles to fit the primary lens light. The quality check is made with a $\chi^2_{\rm red}$ approach and a criterion for stability in the posterior values. In addition, the code also has the option of modeling the light of satellite lens galaxies.}
        \label{fig:flowchart_lenslight}
\end{figure*}

The parameters are fit to the observed intensity value $I_i^{\rm obs}$ of pixel $i$ by minimizing the $\chi^2_{\rm SB}$ of the surface brightness (SB)
\begin{equation}
        \chi^2_{\rm SB} = \sum_{i=1}^{N_{\rm p}}\frac{|I_i^{\rm obs}-\rm PSF\otimes \it 
        I_{{\rm S},i} |^2}{\sigma^2_{\rm tot,\it i}},
        \label{extsourcechi2}
\end{equation}
where $N_{\rm p}$ is the number of pixels in the cutout that are not masked, $\sigma^2_{\rm tot,\it i}$ is the total noise of pixel $i$ from Eq.~(\ref{eq:sigmatotal}), and the crossed circle represents the convolution of the PSF with the predicted S\'{e}rsic intensity 
$I_{{\rm S},i}$ 
of pixel $i$.
The modeling code runs cycles that consist of one simulated annealing in the beginning and then one MCMC chain using a sampling covariance matrix obtained from the previous chain, which is updated after every cycle. To obtain the uncertainties and optimized parameter values, we sample the parameter space with an MCMC method. Each chain samples 100,000 points, and the code removes the burn-in phase, which we conservatively set to be the first 50\% of the points in the chain. The modeling runs until two criteria are met. The first criterion is
\begin{equation}
        \chi^2_{\rm red} = \frac{\chi^2_{\rm SB}}{\textrm{DOF}} \leq 1,
        \label{lenschi2red}
\end{equation}
with the degrees of freedom (DOF), which is the number of pixels in the cutout minus the number of masked pixels minus the number of free parameters.
The number of masked pixels is the sum of all pixels inside the arc mask and of nearby objects masked with the lens mask. 
The second criterion is approximate convergence of the chain, that is,
\begin{equation}
        \Delta \textrm{logP} \leq 5.
        \label{lenslogP}
\end{equation}
This is the difference in the log-likelihood between the median of the first 2000 points in the chain after burn-in is removed and the median of the last 2000 points.
We visualize the lens light modeling steps with a flowchart in Figure \ref{fig:flowchart_lenslight}.

If the chain of the lens light model with one S\'{e}rsic profile (approximately) converges (i.e., fulfills Eq.~\ref{lenslogP}) without reaching the $\chi^2_{\rm red}$ threshold in Eq.~(\ref{lenschi2red}), the code checks whether a point source at the center of the lens improves the model given the possibility of an active galactic nucleus (AGN) within the lens galaxy. The light distribution of a point source is represented by the PSF. For this, the code checks how different point sources in an amplitude range between 0.01 and 100 change the $\chi^2_{\rm SB}$ of the model. If $\chi^2_{\rm SB}$ is lowered by more than 10, a point-source component with the corresponding amplitude is added to the model. This is motivated by the Bayesian information criterion (BIC; following \citealt{Rusu.2019}): the point source adds one additional parameter (the amplitude). A $\chi^2_{\rm SB}$ decrease by 10 ensures that the BIC does not increase. The model is subsequently optimized and sampled until the criterion in Eq.~(\ref{lenslogP}) is fulfilled.

When the criterion defined with Eq.~(\ref{lenslogP}) is met, the code checks the criterion of Eq.~(\ref{lenschi2red}). If the latter is fulfilled, the code stops the lens light modeling. When satellites are present in the cutout field, the code continues to model the light of the satellites. If the criterion of Eq.~(\ref{lenschi2red}) is not fulfilled, one additional S\'{e}rsic profile with the same centroid is added to the model for the primary lens light. The code runs optimizing and sampling cycles until the criterion defined by Eq.~(\ref{lenslogP}) is met.
We model the light of the satellites with S\'{e}rsic profiles when the chain of the model for the primary lens light has converged according to Eq.~(\ref{lenslogP}). The user is queried to replace the masks such that the satellites are no longer masked. The model is then optimized and sampled until the criterion in Eq.~(\ref{lenslogP}) is fulfilled. If this converged model of the primary lens light and of the satellite galaxy light still does not fulfill Eq.~(\ref{lenschi2red}), the user is asked to either add a third S\'{e}rsic profile, refine the masks, or directly continue with the next modeling step. In the first two cases, the code again runs an optimizing/sampling cycle until the $\Delta \textrm{logP}$ criterion (Eq.~\ref{lenslogP}) is met. 

\subsubsection{Modeling the quasar light}
\label{sec:quasarlight}
After the lens light modeling procedure (see Sec.~\ref{sec:lenslight} and Fig.~\ref{fig:flowchart_lenslight}), that is, when we have obtained a good fit for the lens and the satellite light, the code continues by modeling the light of the lensed quasar. Now we mask only the luminous objects surrounding the lens system, that is, we remove the arc mask such that the model now includes the lensed quasars and arcs. For each quasar image, we add a point-like light component that is represented by the PSF and is initially centered on the positions given by the user. The PSF used here was subsampled for images with pixels $>$0.05\arcsec and extended $\sim$4\arcsec$\times$4\arcsec\ to include the diffraction spikes. We only allowed the quasar positions and amplitudes to vary; the \sersic \, lens light parameters were fixed. To infer the best-fit parameters for the quasar images, we again used simulated annealing and an MCMC method with a sampling covariance matrix obtained from an earlier MCMC chain. We stopped the quasar light modeling when approximate convergence of the chain (Eq.~\ref{lenslogP}) was achieved.

\subsubsection{Modeling the lens mass distribution with source or image positions}
\label{sec:srcimgpos}

\begin{figure}
        \centering
        \includegraphics[scale=0.3]{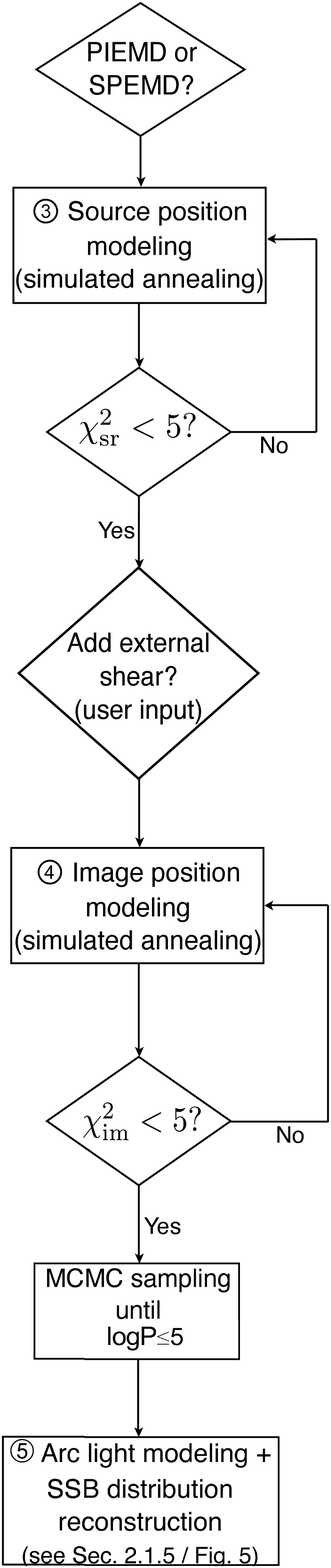}
        \caption[Sec.~\ref{sec:srcimgpos}: Flowchart for lens mass distribution modeling with source and image positions]{Flowchart showing the decision tree for modeling the lens mass distribution with source or image positions. After the user specifies a lens mass profile (PIEMD or SPEMD), the code models the lens mass distribution first by using the modeled source position, and then by using the positions of the multiple quasar images with the option to include an external shear component to the model. Both steps use simulated annealing until the respective $\chi^2$ is below 5. The image position modeling is completed after an approximately converged MCMC chain is achieved. After this, the code continues the lens mass modeling with the arc light modeling and reconstruction of the SSB distribution.}
        \label{fig:flowchart_lensmass}
\end{figure}

With the quasar image positions obtained from the quasar light modeling (see Sec.~\ref{sec:quasarlight}), we can model the mass distribution of the lens galaxy using the mapped source positions and subsequently the image positions.  The modeling code queries the user for the choice of the lens mass profile. Our automated modeling code supports both a pseudo-isothermal elliptical mass distribution (PIEMD; \citealt{Kassiola.1993}) and a singular power-law elliptical mass distribution (SPEMD; \citealt{Barkana.1998}). The steps conducted in the modeling of the lens mass distribution with source and image positions are visualized in Fig.~\ref{fig:flowchart_lensmass}.

The dimensionless surface mass density (or convergence) of the SPEMD profile is given by 
\begin{equation}
\label{eq:kappa_spemd}
        \kappa_{\rm SPEMD}(x, y) = (1-\tilde{\gamma}_{\rm PL})\left[\frac{\theta_{\rm E}}{\sqrt{q_{\rm m}(x-x_{\rm m})^2+\frac{(y-y_{\rm m})^2}{q_{\rm m}}}}\right]^{2\tilde{\gamma}_{\rm PL}},
\end{equation}
where $(x_{\rm m}, y_{\rm m})$ is the lens mass centroid, $q_{\rm m}$ is the axis ratio of the elliptical mass distribution, $r_{\rm c}$ is the core radius, which we set to $1\times10^{-4}$, and $\tilde{\gamma}_{\rm PL}=(\gamma_{\rm PL}-1)/2$ is the radial profile slope (where $\gamma_{\rm PL}$ is the power-law slope of the three-dimensional mass density $\rho(r) \propto r^{-\gamma_{\rm PL}}$).

The convergence $\kappa$ is related to the lens potential $\psi(\vec{\theta})$ via $\nabla^2 \psi(\vec{\theta}) = 2\kappa$. The scaled deflection angle for computing the deflection of light rays is the gradient of the lens potential: $\alpha(\vec{\theta}) = \nabla \psi(\vec{\theta})$.

An overview of the mass parameters and their priors is shown in Table~\ref{tab:lensmassparam}. The isothermal PIEMD profile is a special case of the SPEMD profile, where $\tilde{\gamma}_{\rm PL}$=0.5 holds (although with slightly different parameterizations for the $\kappa$) in Eq. (\ref{eq:kappa_spemd}). In case of the SPEMD profile, our uniform modeling pipeline first keeps the power-law index fixed to the isothermal case $\tilde{\gamma}_{\rm PL}$=0.5. Only later, when the code models the light of the arc (Sec.~\ref{sec:arclight}), do we allow $\tilde{\gamma}_{\rm PL}$ to vary between 0.3 and 0.7. 

\begin{table*}[h]
\caption{Lens mass parameters and their priors.}
                \begin{tabular}{l|l|p{3.5cm}|p{5.2cm}|p{4.7cm}}
                        Component & Parameter & Description & Prior & Prior range / value \\ \midrule
                        & $x_{\rm m}\ ['']$   & $x$-centroid & Gaussian; flat prior in SSB reconstruction & centered on starting value, Gaussian $\sigma=0.3$ \\ \rule{0pt}{2ex}
                        primary & $y_{\rm m}\ ['']$   & $y$-centroid  & Gaussian; flat prior in SSB reconstruction & centered on starting value, Gaussian $\sigma=0.3$  \\ \rule{0pt}{2ex}
                        deflector & $q_{\rm m}$         & axis ratio & flat & [0.3, 1]\\ \rule{0pt}{2ex}
                        (PIEMD/& $\phi_{\rm m}$ [$^\circ$]   & position angle & flat & [0, 360]\\ \rule{0pt}{2ex}
                        SPEMD)& $\theta_{\rm E}\ ['']$  & Einstein radius & flat & [0.5, 3]\\ \rule{0pt}{2ex}
                        & $r_{\rm c}\ ['']$         & core radius & exact & $10^{-4}$\\\rule{0pt}{2ex}
                        & $\tilde{\gamma}_{\rm PL}$      & power-law index (SPEMD)& exact; flat in SSB modeling & 0.5; [0.3, 0.7]\\ \midrule
                        & $x_{\rm m, sat}\ ['']$   & $x$-centroid & exact & fixed to satellite light centroid   \\ \rule{0pt}{2ex}
                        satellite & $y_{\rm m, sat}\ ['']$   & $y$-centroid  & exact & fixed to satellite light centroid  \\ \rule{0pt}{2ex}
                        deflector(s) & $q_{\rm m, sat}$         & axis ratio & exact & 1 \\ \rule{0pt}{2ex}
                        (SIS) & $\phi_{\rm m, sat}$ [$^\circ$]   & position angle & exact & 0\\ \rule{0pt}{2ex}
                        & $\theta_{\rm E, sat}\ ['']$  & Einstein radius & flat & [0.01, 3]\\ \rule{0pt}{2ex}
                        & $r_{\rm c, sat}\ ['']$         & core radius & exact & $10^{-4}$\\\rule{0pt}{2ex}
                        & $\tilde{\gamma}_{\rm sat}$      & power-law index & exact & 0.5\\ \midrule
                        external& $\gamma_{\rm ext}$ & magnitude & flat & [0, 0.6]\\ \rule{0pt}{2ex}
                        shear& $\phi_{\rm ext}$ [$^\circ$] & position angle & flat & [0, 360]
                \end{tabular}
                \ \\
        \caption*{\textbf{Notes: } The position angles $\phi_\text{m}$, $\phi_\text{m,sat}$, and $\phi_{\rm ext}$ are measured counterclockwise from the positive $x$-axis (east of north). In case of the centroid coordinates, the modeling code starts with a Gaussian prior centered on the starting value obtained from the user input, which is removed after lens light modeling. The power-law index $\tilde{\gamma}_{\rm PL}$ was fixed to 0.5 (isothermal) in the beginning and was changed to a flat prior during arc light modeling and SSB reconstruction.}
        \label{tab:lensmassparam}
\end{table*}

The lens mass parameters are first constrained by modeling a source position that reproduces the observed image positions of the quasars. First, the position of the source is obtained by using the lens equation
\begin{equation}
        \vec{\beta}_i=\vec{\theta}_i-\vec{\alpha}_i(\vec{\theta}_i), \ \ \ i=1,...,N_{\rm im},
\end{equation}
where the image positions $\vec{\theta}_i = (x_{{\rm QSO},i}, y_{{\rm QSO},i})$ are mapped back to the source plane with the scaled deflection angle $\vec{\alpha}_i(\vec{\theta}_i)$. $N_{\rm im}$ is the image multiplicity of the system.
The lens mass parameters $\vec{\eta}$ are then varied to minimize the quantity
\begin{equation}
        \chi^2_{\rm sr} = \sum_{i=1}^{N_{\rm im}}\frac{|\vec{\beta}_i(\vec{\eta}, \vec{\theta}_i)-\vec{\beta}^{\rm mod}(\vec{\eta}, \vec{\theta}_i)|^2}{(\frac{\sigma_i}{\sqrt{\mu_i}})^2},
\end{equation}
with $\vec{\beta}_i$ the mapped source position for image $i$, $\vec{\beta}^{\rm mod}$ the modeled source position (i.e., the mean of the mapped source positions, weighted by the magnification), $\sigma_i$ the positional uncertainty on the image plane, and $\mu_i$ the magnification of image $i$. We adopted 0.004\arcsec as the positional uncertainty, which accounts for both the astrometric uncertainties from PSF fitting in Sec.~\ref{sec:quasarlight} and for astrometric perturbations of typically several milliarcseconds (mas) due to mass substructures in the lens galaxy (e.g., \citealt{Chen2006}). For HST images where the PSF is well sampled, the centroid position of the PSF can be measured with 2-4 mas for galaxy-scale lenses (e.g., \citealt{Lehar1999}; \citealt{Sluse2012}), which is substantially more precise than the pixel size; in Sec.~\ref{sec:astrometry} we show that our astrometric uncertainties from PSF fitting reach an accuracy of 2 mas. In order to minimize $\chi^2_{\rm sr}$, we used simulated annealing to optimize the parameters until $\chi^2_{\rm sr}<5$.

Based on the model obtained by minimizing the $\chi^2_{\rm sr}$, which is based on the mapped source positions, the parameter values are further optimized by using the image positions as constraints. When a satellite galaxy is located close to the lensing system, we can model its mass distribution with a singular isothermal sphere (SIS) profile. The satellite mass centroid is fixed to the satellite light centroid obtained in Sec.~\ref{sec:lenslight}.
In addition, the user can choose to add an external shear profile to the model after the code has constrained the lens mass parameters with the modeled source position. This accounts for the tidal gravitational field of objects surrounding the lensing system.  The shear magnitude is calculated by $\gamma_{\rm ext} = \sqrt{\gamma_{\rm ext,1}^2 + \gamma_{\rm ext,2}^2}$, with $\gamma_{\rm ext,1}$ and $\gamma_{\rm ext,2}$ the components of the shear matrix.  The parameters $\vec{\eta}$ are varied to minimize
\begin{equation}
        \chi^2_{\rm im} = \sum_{i=1}^{N_{\rm im}}\frac{|\vec{\theta}_i^{\rm obs}-\vec{\theta}_i^{\rm pred}(\vec{\eta}, \vec{\beta}^{\rm mod})|^2}{\sigma_i^2},
\end{equation}
with $\vec{\theta}_i^{\rm obs}$ and $\vec{\theta}_i^{\rm pred}$ the observed and the predicted position for image $i$, respectively. Again, the code optimizes with simulated annealing until $\chi^2_{\rm im}<5$. It then samples with MCMC chains until approximate convergence ($\Delta$\textrm{logP}$\leq$5).\\

\subsubsection{Modeling the lens mass distribution with surface brightness distribution}
\label{sec:arclight}
\begin{figure}
        \centering
        \includegraphics[scale=0.6]{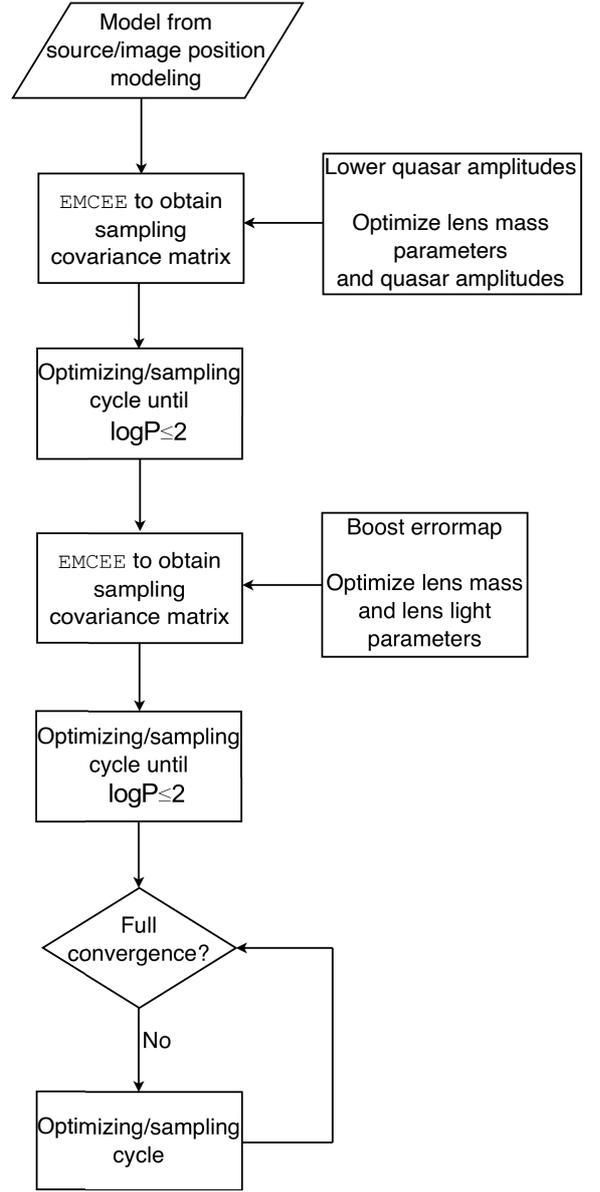}
        \caption[Sec.~\ref{sec:arclight}: Flowchart for arc light modeling and reconstruction of the SSB distribution]{Flowchart for arc light modeling and SSB reconstruction. In the first stage, the quasar amplitudes are lowered, and in the second stage, the error map is boosted in the bright quasar regions in order to model the light of the arc. The lens mass parameters are allowed to vary in this modeling step. In a final step, the code runs optimizing and sampling cycles until full convergence of the chain is achieved.}
        \label{fig:flowchart_arclight}
\end{figure}

When the lens mass distribution is modeled by using the multiple quasar image positions, we proceed to constrain the lens mass parameters by modeling the SB distribution of the unlensed source and the light of the arc, which is the lensed light of the quasar host galaxy.  We illustrate the following modeling steps with a flowchart in Fig.~\ref{fig:flowchart_arclight}.

Our automated modeling pipeline uses \texttt{GLEE} to reconstruct the most probable SSB distribution on a pixelated grid through a linear inversion of the pixel intensities in the arc mask (\citealt{Suyu.2006}). The optimizing or sampling of the posterior probability distribution includes both the $\chi^2_{\rm SB}$ term in Eq.~(\ref{extsourcechi2}) and a prior term that acts on the source pixels with a quadratic regularizing function. The SSB distribution is then mapped back to the image plane to reconstruct the arc light. We use the properties of the arc light to further constrain the lens mass parameters. The inferred quasar light intensity in the image cutout (see Sec.~\ref{sec:quasarlight}) is a superposition of the real quasar amplitude and a roughly constant contribution from the lensed host light. To reconstruct the light of the host galaxy, we lower the parameter value of the quasar amplitude. The new amplitudes are determined such that the condition 
\begin{equation}
        I_{\rm circ,avrg} = 3 \times I_{\rm arc,avrg}
        \label{qsoampcrit}
\end{equation}
is fulfilled, where $I_{\rm circ,avrg}$ is the average intensity in a circle with a radius of 4 pixels around the center of the quasar after the quasar and lens light is removed, and $I_{\rm arc,avrg}$ is the average intensity of a region provided by the user that contains a small patch of arc light. The factor of 3 is chosen empirically such that the flux is continuous across the arc. Our automation code then adjusts each quasar amplitude such that the criterion in Eq.~(\ref{qsoampcrit}) is met. For the optimizing or sampling process, we first allow the lens mass parameters, now also including the power-law index $\tilde{\gamma}_{\rm PL}$ of the SPEMD profile, and the quasar amplitudes to vary while fixing all the parameters of the lens or satellite light profiles and the quasar centroids. From this point on, we remove the Gaussian prior of the lens galaxy centroids in the lens mass and S\'{e}rsic profiles.

To sample the parameter space efficiently, we use \texttt{EMCEE}, an ensemble sampler based on MCMC chains that is highly parallelizable \citep{ForemanMackey.2013}. With \texttt{EMCEE}, we obtain a sampling covariance matrix for the parameter space, which makes the subsequent sampling with Metropolis-Hastings MCMC more efficient. We consider this modeling part successful when the Metropolis-Hastings chain has approximately converged, now with a stricter requirement:
\begin{equation}
        \Delta \textrm{logP} \leq 2.
        \label{eq:logP2}
\end{equation}

The quasar images are much brighter than the light of the arc, and residuals of quasar light fitting with imperfect PSF can often overwhelm the signal in the arcs. Therefore, we allow for larger uncertainties in the quasar image areas by boosting the error map in these regions. The user marks the areas dominated by quasar light, and the code will boost all the pixels of the error map within this region such that their normalized residuals are within $\pm1\sigma$. The error map can also be boosted in the satellite galaxy region if the user chooses to do so. This can be necessary if the satellite galaxy is very close to the arc of the lensing system.
In the next modeling step, now including the boosted error map, we fix all quasar light parameters and allow the lens mass and the lens light parameters to vary. If the uncertainties in the satellite region are not boosted, the satellite light parameters are varied as well. We again use \texttt{EMCEE} to obtain a sampling covariance matrix and optimize or sample with simulated annealing or an MCMC until approximate convergence of the chain (Eq.~\ref{eq:logP2}) is achieved. In a last step, the code checks the model for full convergence by analyzing the discrete power spectrum of the chain (following \citealt{Dunkley.2005}).
If full convergence is not yet achieved, the code runs optimizing and sampling cycles with an increased chain length by now sampling 200,000 points. When full convergence is achieved, the code stops and the modeling is completed.

\subsection{Multiband modeling}
\label{sec:multiband}
\begin{figure}[h]
        \centering
        \includegraphics[scale=0.55]{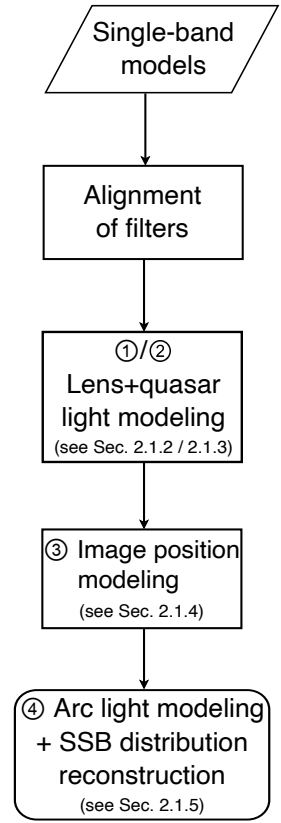}
        \caption{Overview flowchart for multiband modeling of a lensed quasar system with the multiband automated code. Details are described in flowcharts for each step and in the corresponding subsections, as referenced in the flowchart.}
        \label{fig:flowchart_multiband}
\end{figure}

In addition to the single-band automation code described in Sec.~\ref{sec:singleband}, we developed a code to simultaneously model systems with strongly lensed quasars using multiband data. The modeling procedure in most parts is analogous to the single-band case and is described in Table \ref{tab:Modelingtable_multi}. We present an overview flowchart in Fig.~\ref{fig:flowchart_multiband}.

When we obtain a single-band lens mass model in one band (the reference band) and a lens and quasar light model in the other bands with the pipeline described in Sec.~\ref{sec:singleband}, we proceed to jointly model the lens and quasar light in all the considered bands. We chose the F160W band as our reference band unless otherwise specified because the arc light is most prominent in this wavelength. 
We use the modeled quasar positions of the individual bands to align the multiple filters. The lens mass distribution model in the reference filter serves as a starting model for the multiband modeling.  

In the first step, we pair the reference filter with one of the additional filters and model the lens and quasar light together with MCMC chains. We fix the amplitudes of the lens light components in the reference filter, and force the lens light amplitudes in the other filter to follow the same ratio of light amplitudes as in the reference filter. In addition, we link the remaining lens light parameters (centroid, axis ratio, position angle, effective radius, and S\'{e}rsic index) between the filters. The quasar light centroids are linked as well, while the quasar amplitudes vary independently. After a MCMC chain is approximately converged with $\Delta \textrm{logP} \leq 5$, we fix the lens light amplitudes of this filter, and add the next filter, which is modeled together with the reference filter in the same way as the filter before. We also allow variation in the quasar light parameters of the previous filters. When approximate convergence ($\Delta \textrm{logP} \leq 5$) is achieved for all pairs, we model the lens and quasar light parameters of all filters simultaneously and now allow all lens light amplitudes to vary independently while keeping the remaining lens light parameters and quasar light centroids linked.

Like in the single-band modeling, the modeled quasar positions are then used for image position modeling (see Sec.~\ref{sec:srcimgpos}) to constrain the lens mass parameters. In the last step, the light of the arc and the SSB distribution are reconstructed analogously to the single-band modeling case described in Sec.~\ref{sec:arclight}. Finally, we run additional MCMC chains to achieve full convergence of our lens model parameters.

\begin{table*}
\caption{Modeling steps in the multiband modeling code.}
        \begin{tabular}{lp{6.4cm}p{1cm}lccccccc}\toprule
                Step & Description & Section & Criteria & \multicolumn{2}{c}{Lens mass parameters} & Lens light & \multicolumn{4}{c}{Source light}
                \\\cmidrule(lr){5-6}\cmidrule(lr){8-9}
                & & & & profiles  & ext. shear &  & quasar    & host \\\midrule\midrule \vspace{5px}
                \circled{1} & pair each band with reference filter and model lens + quasar light together & \ref{sec:lenslight}/ \ref{sec:quasarlight}, \ref{sec:multiband} & $\circlearrowleft_2 (\chi^2_{\rm red}\leq1)$ & $\bigcirc$ & $\bigcirc$ & $\checkmark$ & $\checkmark$ & $\times$ \\  \vspace{5px} 
                \circled{2} & modeling of lens + quasar light with all bands at once & \ref{sec:lenslight}/ \ref{sec:quasarlight}, \ref{sec:multiband}  &  $\circlearrowleft_2 (\Delta \textrm{logP}\leq5)$ & $\bigcirc$ & $\bigcirc$ & $\checkmark$ & $\checkmark$ & $\times$ \\  \vspace{5px} 
                \circled{3} & image position modeling with external shear & \ref{sec:srcimgpos}, \ref{sec:multiband} & \begin{tabular}[t]{@{}l@{}}$\circlearrowleft_1 (\chi^2<5,$ \\ $\Delta \textrm{logP}\leq5)$\end{tabular} & $\checkmark$ & $\checkmark$ & $\bigcirc$  & $\bigcirc$ & $\times$ \\  \vspace{5px} 
                \circled{4} a & modeling of arc light + SSB distribution reconstruction & \ref{sec:arclight}, \ref{sec:multiband} &  $\circlearrowleft_2 (\Delta \textrm{logP}\leq2)$ & $\checkmark$ & $\checkmark$ & $\bigcirc$ & $\checkmark_{\rm A}$ & $\checkmark$\\  \vspace{5px} 
                \circled{4} b & modeling of arc light + SSB distribution reconstruction  & \ref{sec:arclight}, \ref{sec:multiband} &  $\circlearrowleft_2 (\Delta \textrm{logP}\leq2)$ & $\checkmark$ & $\checkmark$ & $\checkmark$ & $\bigcirc$ & $\checkmark$ \\\bottomrule
        \end{tabular}
        \ \\ \ \\
        \textit{Symbols:}\\
        \begin{tabular}{ll}
                $\circlearrowleft_1$: optimizing cycle with simulated annealing & $\checkmark$: parameters vary\\
                $\circlearrowleft_2$: optimizing and sampling cycle with simulated annealing and MCMC \ \ \ \ \ & $\bigcirc$: parameters fixed \\
                $\checkmark_{A}$: only the amplitude varies & $\times$: parameters not included
        \end{tabular}
    \ \\
        \caption*{\textbf{Notes: }Each step is described briefly in the second column. The modeling technique and stopping criteria are given in the third column with the symbols described below the table. The cycle runs until the stopping criteria in the parentheses are met. The remaining columns show the parameters that vary, are fixed, or are not included.}
        \label{tab:Modelingtable_multi}
\end{table*}

\subsection{Time-delay estimates}
\label{sec:timedelay}
Strong-lensing time-delay cosmography requires time-delay measurements and thus observational programs that monitor the lensed quasars. To facilitate the scheduling of the monitoring, estimations of the time delays of new quasar systems are useful. Furthermore, a comparison of the (blind) prediction of the time delays to the subsequently measured time delays provides a crash test of mass modeling procedures (\citealt{Treu.2016}).

When we obtain a good model for the lens potential $\psi(\vec{\theta})$ of the system from the previous subsections, we can use it to compute the relative time delays between the different images. Measured redshifts of the lens ($z_{\rm d}$) and of the source ($z_{\rm s}$) allow us to convert the dimensionless surface mass density into a physical quantity. We can then calculate the time-delay distance
\begin{equation}
        D_{\Delta t} = (1+z_{\rm d})\frac{D_{\rm d} D_{\rm s}}{D_{\rm ds}}
        \label{eq:tddistance}
\end{equation}
by assuming a cosmological model.  The distances on the right-hand side $D_{\rm d}$, $D_{\rm s}$ and $D_{\rm ds}$ are the angular diameter distance to the deflector or lens galaxy, to the source galaxy, and between the deflector and source galaxy, respectively. With the time-delay distance, \texttt{GLEE} can calculate the time delays between the quasar images \textit{i} and \textit{j},
\begin{equation}
        \Delta t_{ij} = \frac{1}{c}D_{\Delta t}\Delta \tau_{ij},
        \label{eq:td}
\end{equation}
where 
\begin{equation}
\label{eq:fermatpot}
        \Delta \tau_{ij}  = \left[\frac{1}{2}(\vec{\theta}_i-\vec{\beta})^2 - \psi(\vec{\theta}_i)\right] - \left[\frac{1}{2}(\vec{\theta}_j-\vec{\beta})^2 - \psi(\vec{\theta}_j)\right]
\end{equation}
is the Fermat potential difference between quasar images \textit{i} and \textit{j,} which is obtained from the final lens mass model. We have a separate code that calculates the time-delay distance from Eq.~(\ref{eq:tddistance}) and the relative time delays from Eq.~(\ref{eq:td}), including 1$\sigma$ uncertainties by using the final MCMC chain of the model parameters from the modeling code and the cosmological parameters and redshifts provided by the user.

\section{Observations}
\label{sec:observations}
\begin{figure*}[h]
        \subfigure{\includegraphics[width=0.33\textwidth]{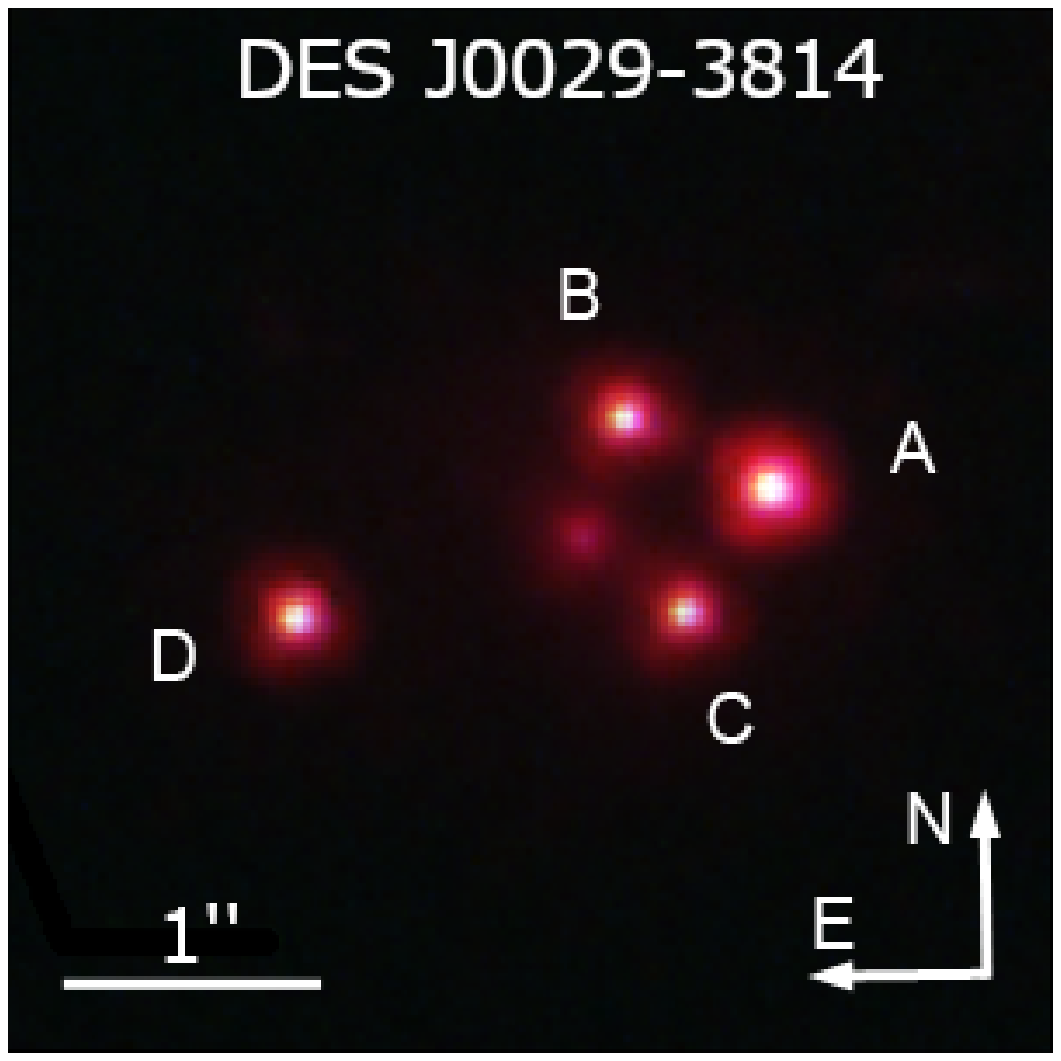}}
        \subfigure{\includegraphics[width=0.33\textwidth]{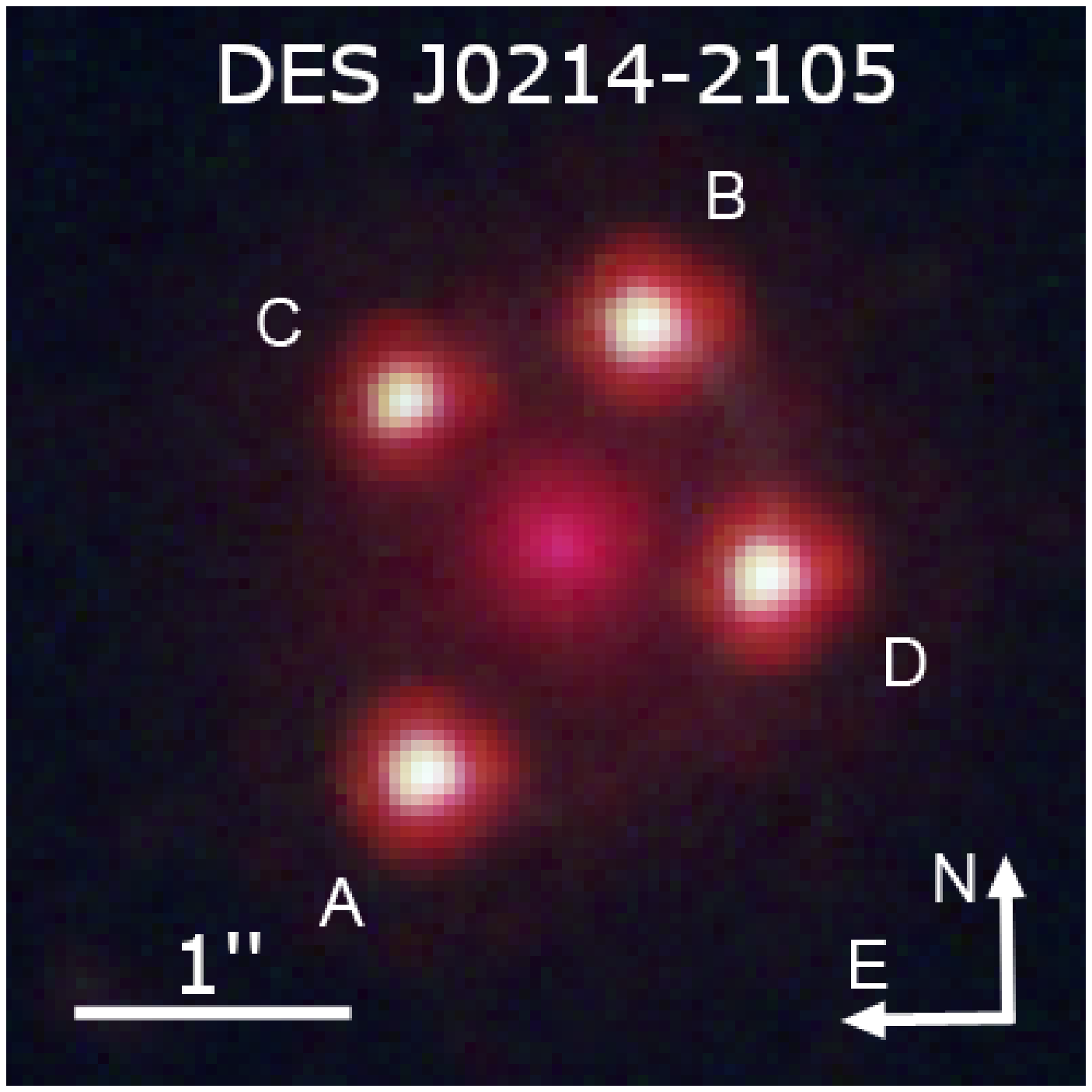}}
        \subfigure{\includegraphics[width=0.33\textwidth]{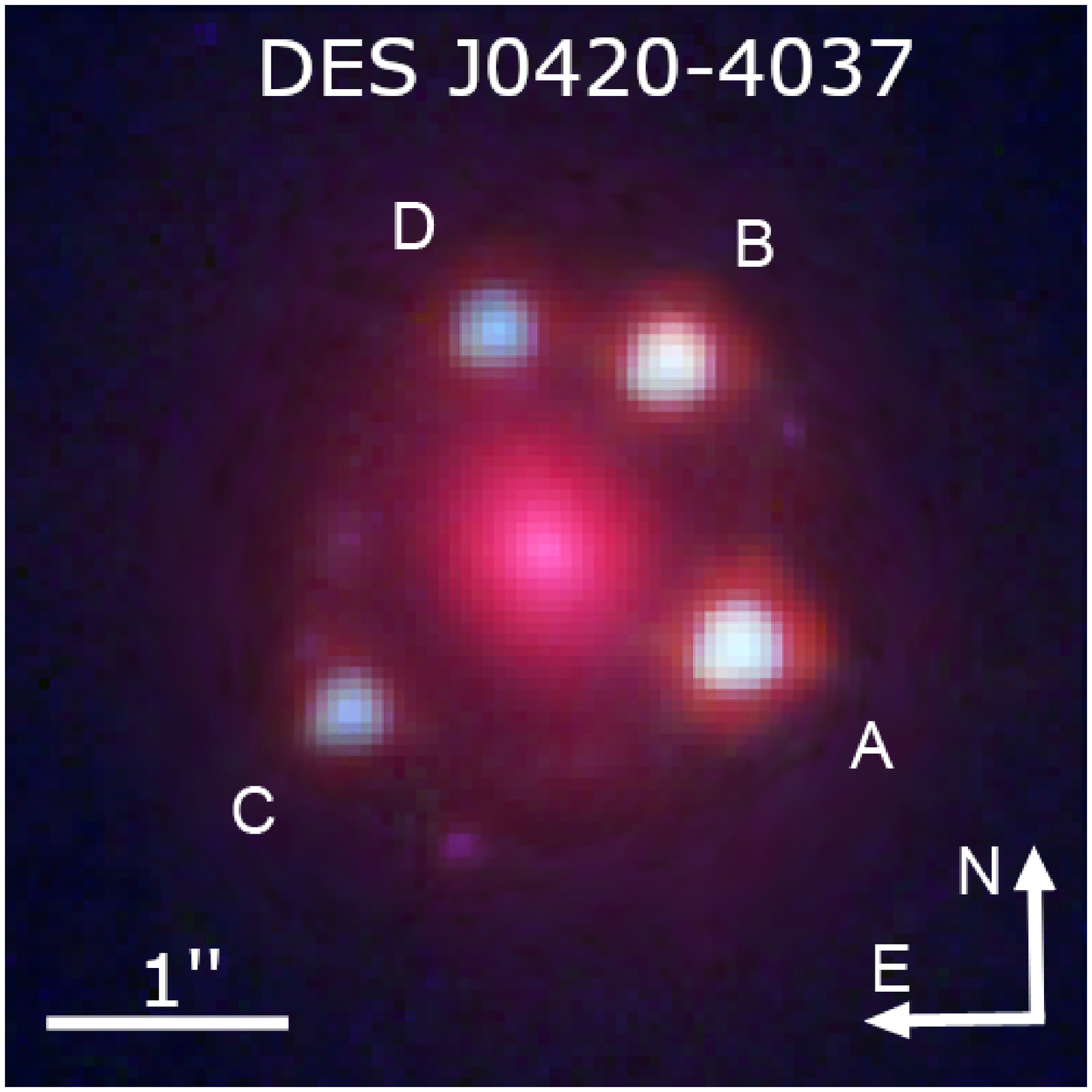}}
        \subfigure{\includegraphics[width=0.33\textwidth]{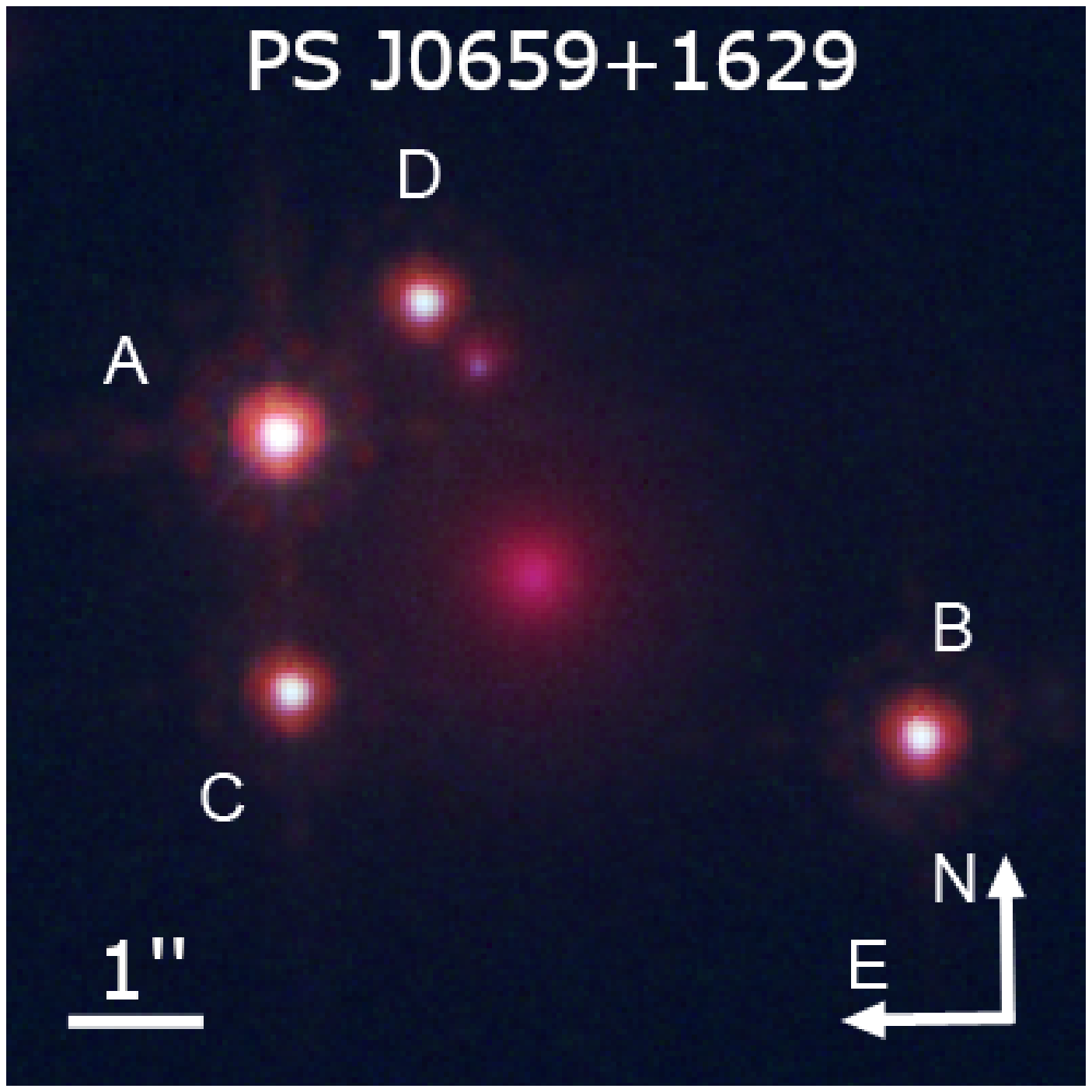}}
        \subfigure{\includegraphics[width=0.33\textwidth]{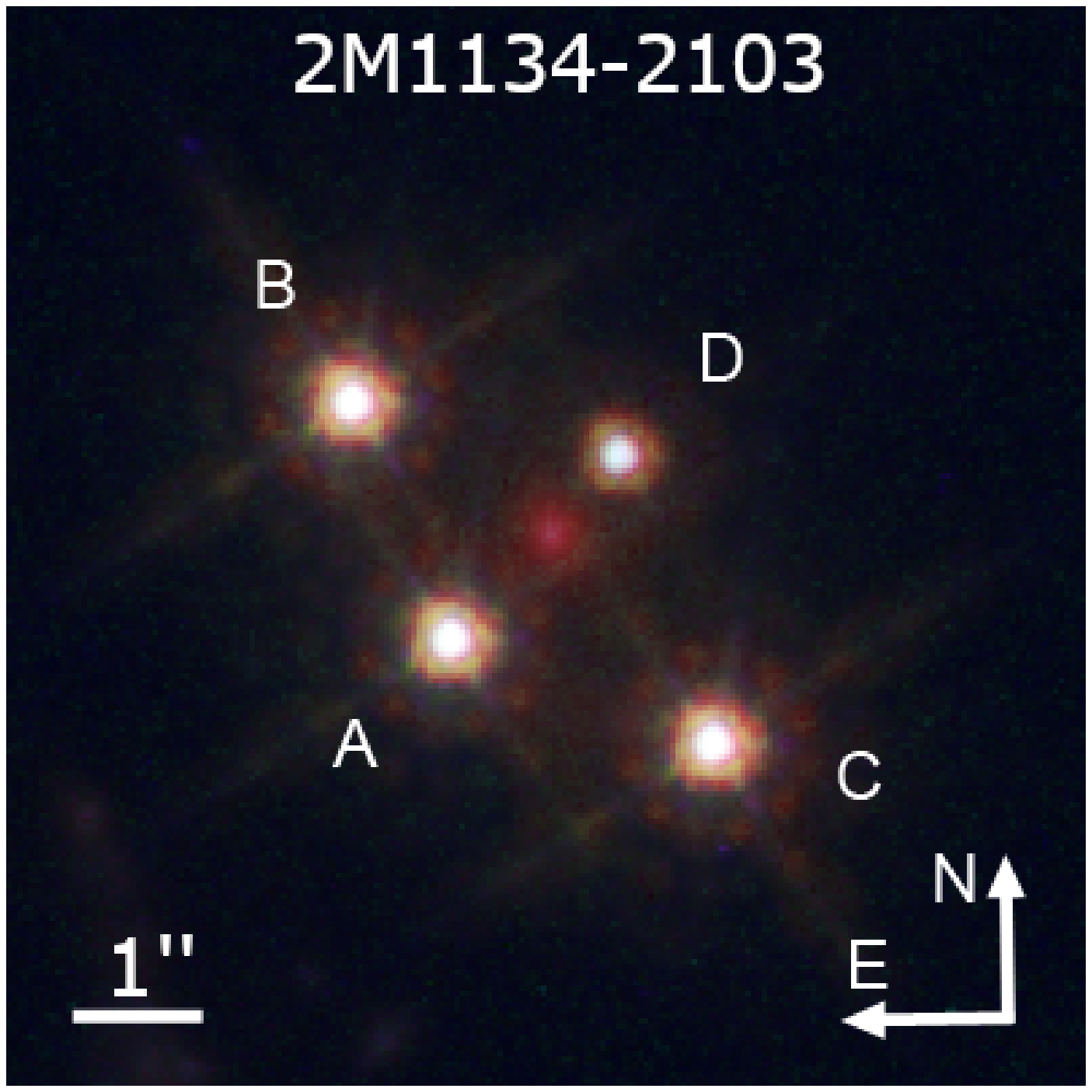}}
        \subfigure{\includegraphics[width=0.33\textwidth]{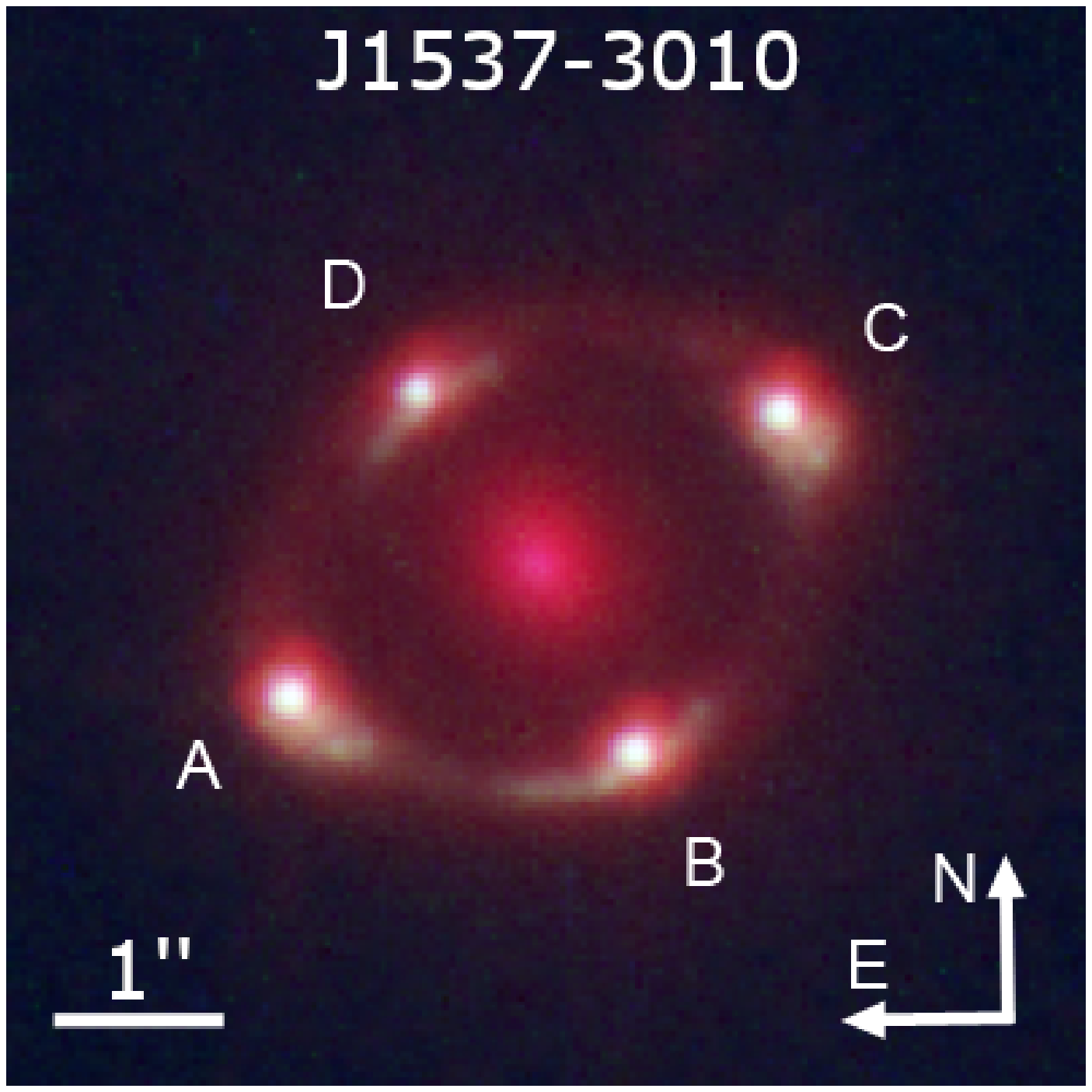}}
        \subfigure{\includegraphics[width=0.33\textwidth]{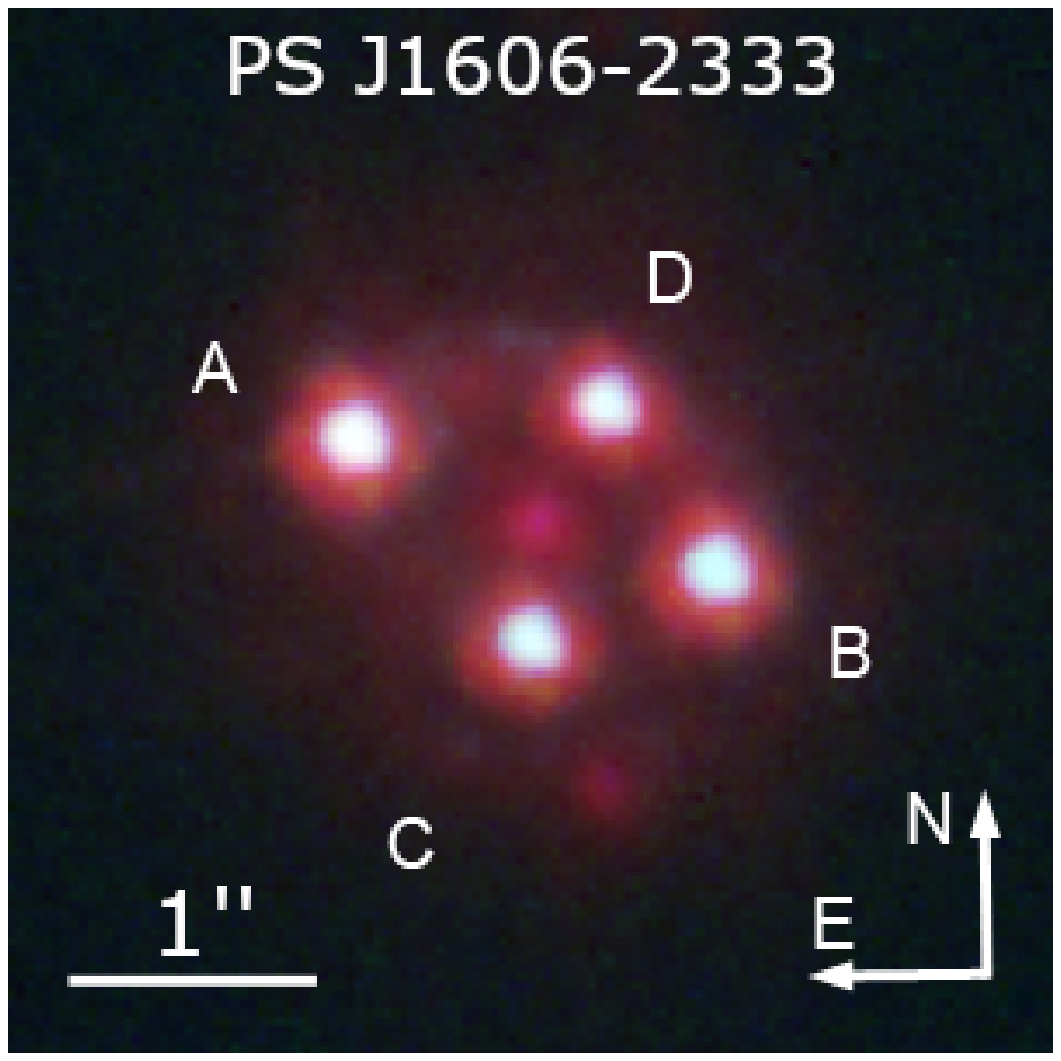}}
        \subfigure{\includegraphics[width=0.33\textwidth]{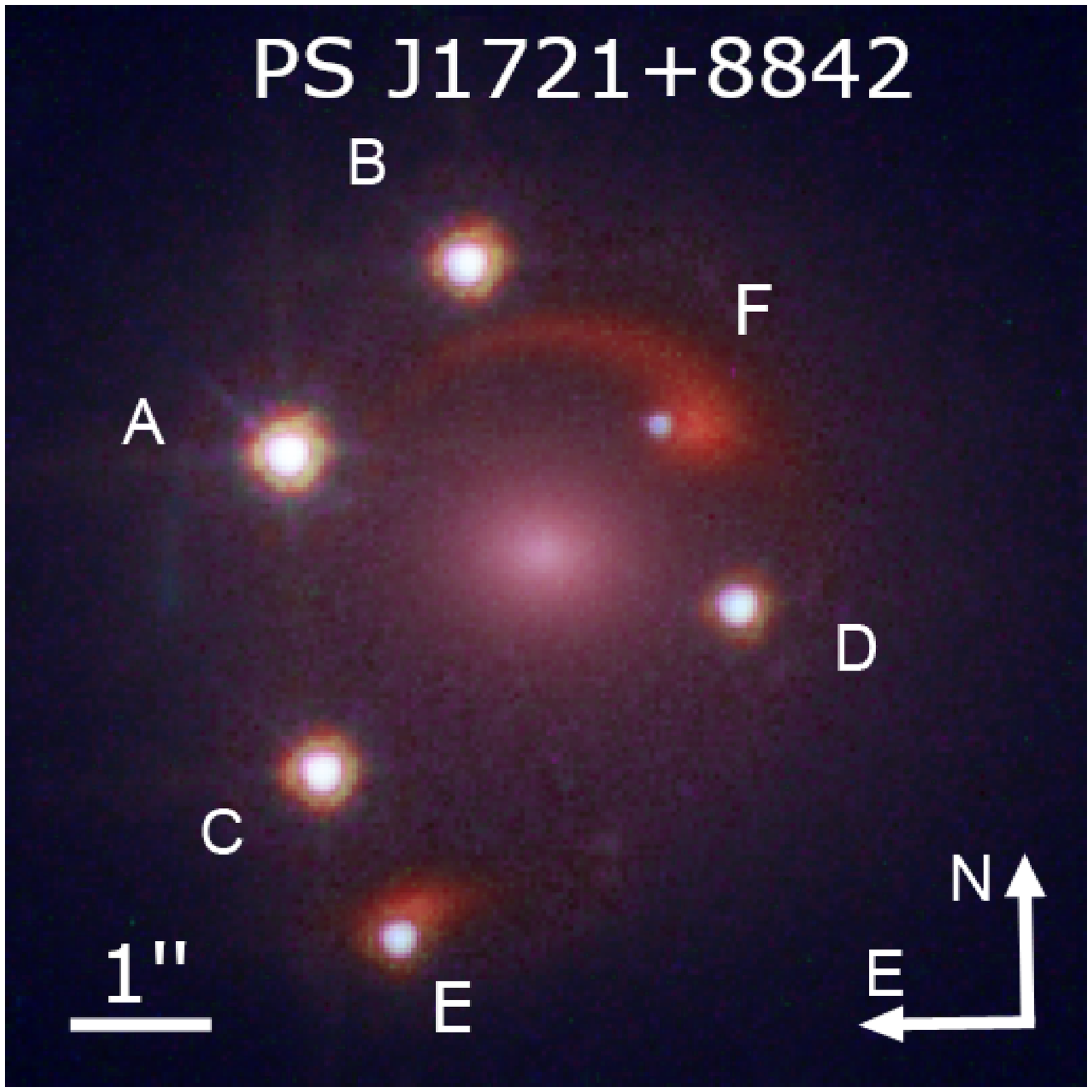}}
        \subfigure{\includegraphics[width=0.33\textwidth]{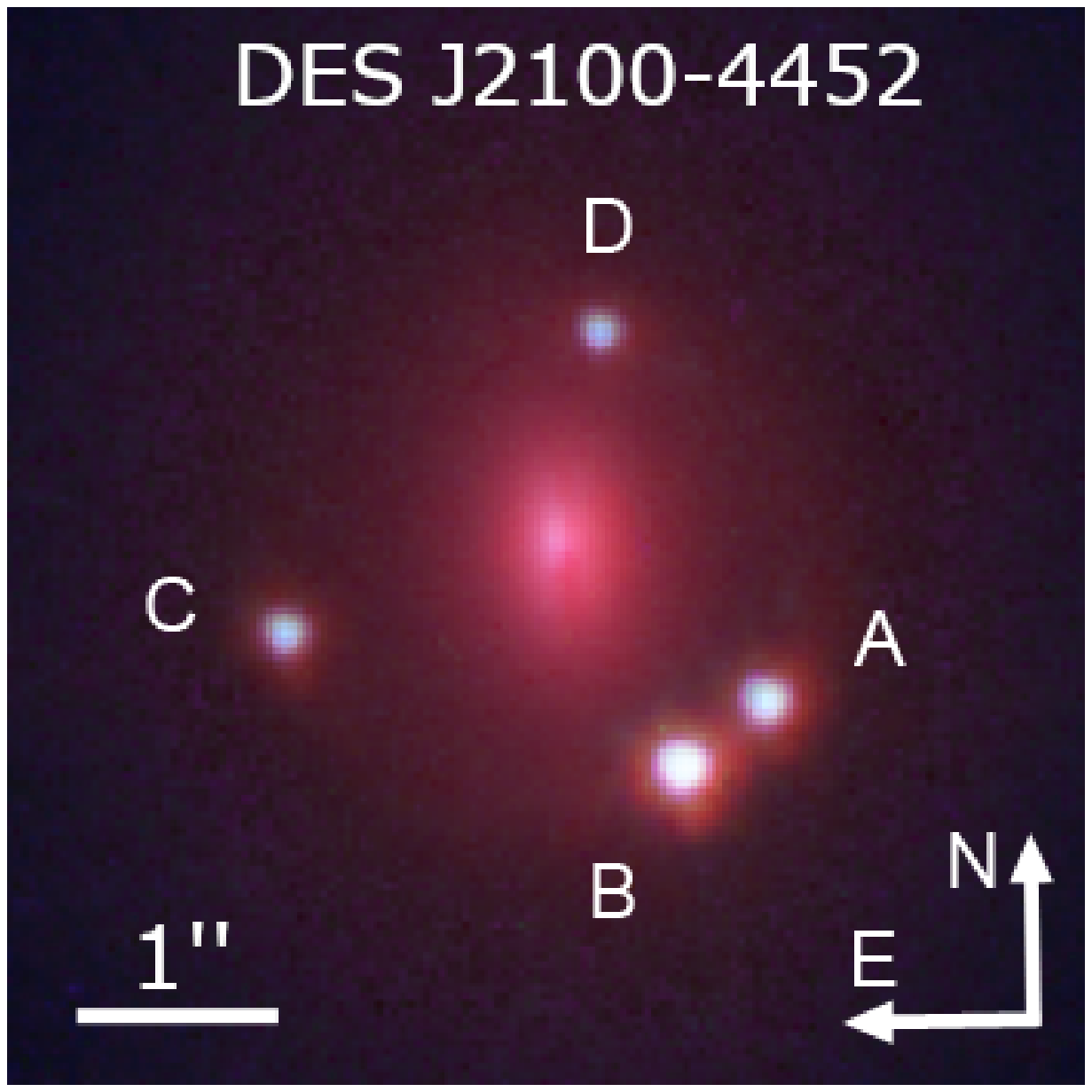}}
                                                        
        \caption{Color images for each of the nine lenses in our sample created with the three HST bands F160W (red), F475X (blue), and F814W (green).}
        \label{fig:colorimgs}
\end{figure*}

Our sample consisted of the nine strongly lensed quasars DES J0029$-$3814 (Schechter et al., in prep), DES J0214$-$2105 \citep{Agnello2019, Lee2019}, DES J0420$-$4037 (Ostrovski et al., in prep), PS J0659$+$1629 \citep{Delchambre2019}, 2M1134$-$2103 \citep{Lucey.2018,Rusu.2019}, J1537$-$3010 \citep{Lemon.2019,Delchambre2019,Stern2021}, PS J1606$-$2333 \citep{Lemon2018}, PS J1721$+$8842 \citep{Lemon2018}, and DES J2100$-$4452 \citep{Agnello2019}. 
A detailed description of the observations and the peculiarities of each lensing system are presented by \citetalias{Schmidt2022}.
In Fig.~\ref{fig:colorimgs} we show a color image that was created with the three HST bands F160W, F475X, and F814W, which were used as the red, blue, and green channels, respectively.

\section{Modeling results}
\label{sec:results}
In this section, we describe the modeling results of our sample of lensing systems obtained with the autonomous modeling pipeline that we presented in Sec.~\ref{sec:modeling}. 
The input files for each lensing system, that is, the PSF, error map, and masks, were obtained by following Sec.~\ref{sec:preparation}. Each system was modeled with a SPEMD profile and external shear. The lens light was modeled with two S\'{e}rsic profiles unless specified otherwise in the following subsections. For two systems (J0659$+$1629 and J1606$-$2333), we additionally modeled the light and mass distribution of a close-by satellite galaxy. We decided which bands to include by visual inspection of the residuals after modeling lens and quasar light, that is, the remaining light from the arcs, which are the lensed host galaxy of the quasar.
If there was significant arc light in one wavelength band, then we included that band in the final multiband model.

In Table \ref{tab:results_light} we present figures of the modeled light and reconstructed source for each of the nine lensing system in our sample. We show the observed image (third column), the modeled light and critical curves (fourth column), and the normalized residuals in a range between $-$5$\sigma$ and 5$\sigma$ (fifth column). For each pixel, the normalized residuals show the difference between data and model, normalized by the estimated standard deviation. We show cropped images instead of the full cutout that we used for modeling for better visibility and indicate 1\arcsec\ with a white line. The panels are oriented such that north is up and east is left. The sixth column shows the reconstructed SSB distribution of the quasar host galaxy on the image plane.  The seventh column shows the SSB distribution on the source plane with plotted caustic curves in red, and the mean weighted source position of the quasar as a blue star. We show rulers with length of 0.5\arcsec\ in $x$- and $y$-direction because the pixel size can be different in these directions.

In Table \ref{tab:results_param} we present the median parameter values of the lens mass distribution together with their 1$\sigma$ uncertainties from the final chain of each multiband model. We also show the best-fit lens and satellite light parameters in the F160W band (the reference filter). The centroid coordinates are given with respect to the bottom left corner of the image cutout. The light parameters in the F475X and F814W bands are presented in Tables \ref{tab:results_param_475} and \ref{tab:results_param_814}, respectively. 

In Table \ref{tab:chi2} we summarize the single-band $\chi_{\rm red}^2$ of each modeled band and a total $\chi_{\rm red, tot}^2$, which is the sum of the independent single-band $\chi^2$ divided by the DOF of the multiband model. We excluded pixels masked in the lens mask and the boosted quasar light regions in computing the DOF (see Sec.~\ref{sec:lenslight}). For two systems (2M1134$-$2103 and J1606$-$2333) the $\chi_{\rm red}^2$ is \textgreater 1 ($\chi_{\rm red}^2$ of 1.84 and 1.36, respectively). For these systems, the uncertainties could be underestimated.
In Table \ref{tab:kappa} we present the convergence $\kappa$, total shear strength $\gamma_{\rm tot}$, and magnification $\mu = 1/((1-\kappa)^2 - \gamma_{\rm tot}^2)$ at the (modeled) image positions.

In addition, we calculated the Fermat potential differences at the multiple modeled image positions and predict the relative time delays for each system. We used the redshifts mentioned in Sec. \ref{sec:observations}. If the lens redshift was not yet available, we assumed $z_{\rm d}\equiv0.5$, which is an approximate median value based on the sample of currently known lensed quasars with redshift measurements. The results are shown in Tables \ref{tab:fermat} and \ref{tab:timedelays}.

\begin{figure*}[h]
    \centering
        \subfigure{\includegraphics[width=0.7\textwidth]{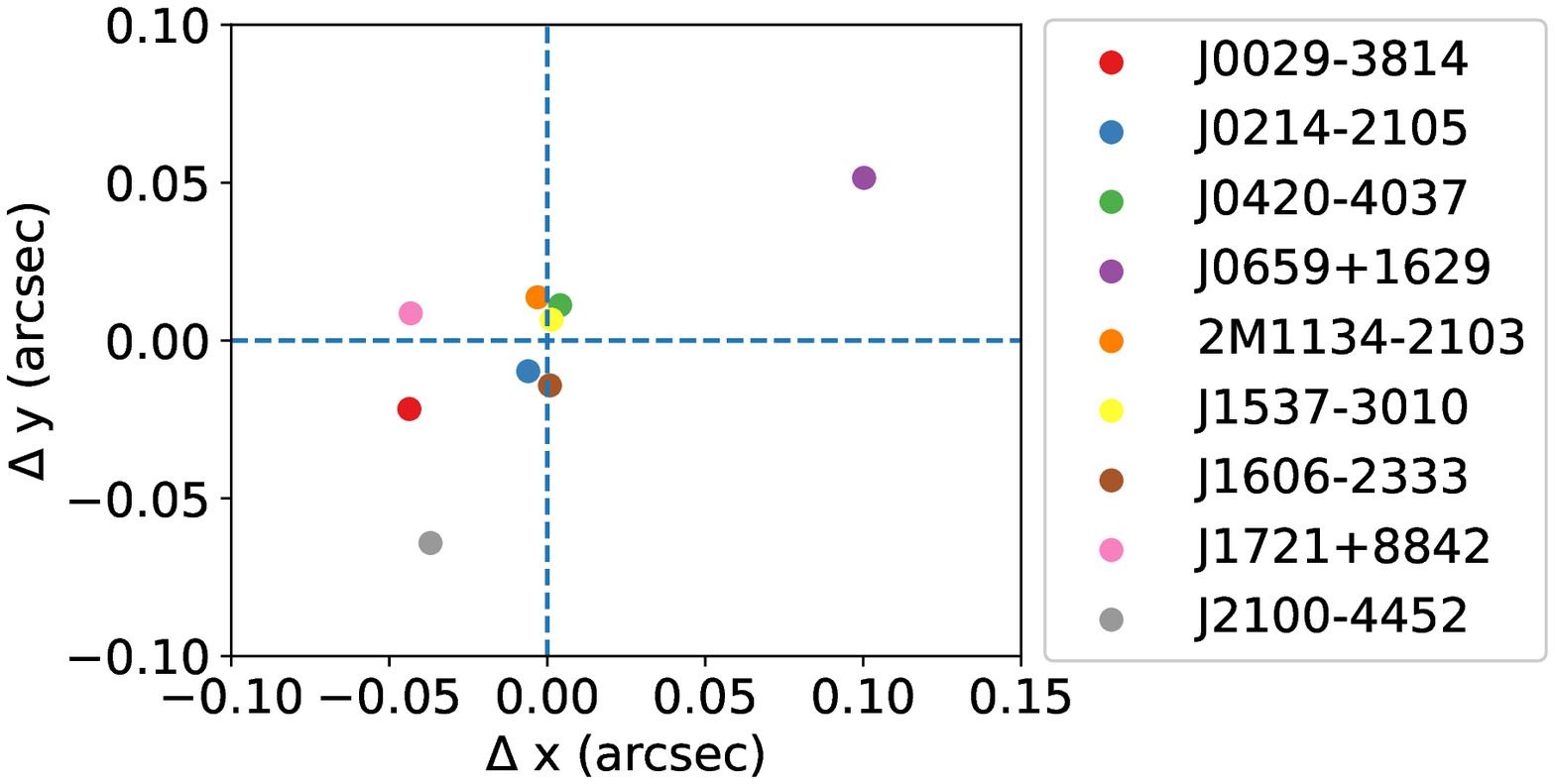}}
        \subfigure{\includegraphics[width=0.4\textwidth]{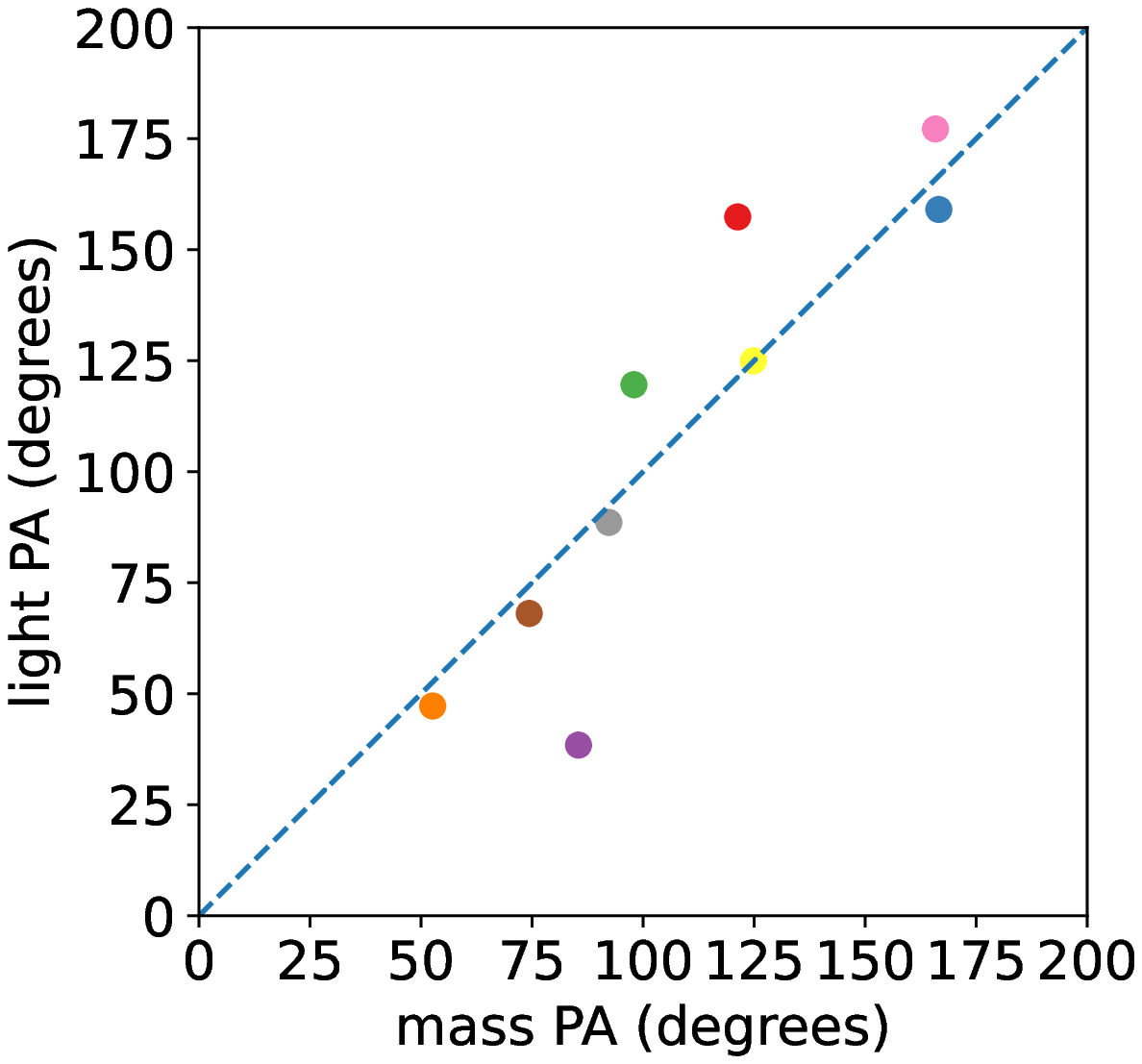}}
        \subfigure{\includegraphics[width=0.4\textwidth]{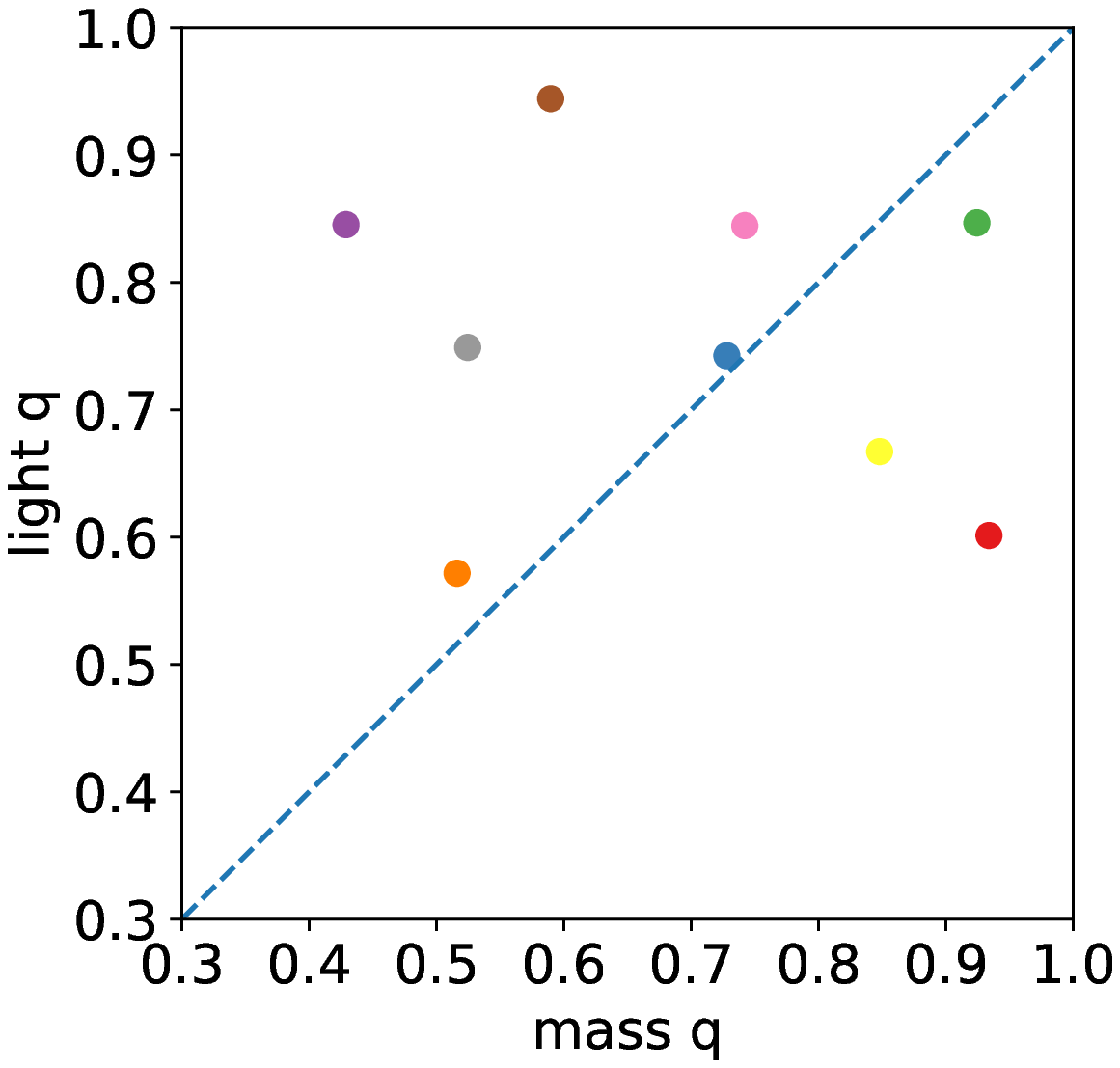}}
        \caption{Comparison between the position and structural parameters of the lens mass and light for eacht lensing system. \textit{Top:} Difference between the median mass and light centroid of our models, with $\Delta x = x_{\rm S} - x_{\rm m}$ and $\Delta y = y_{\rm S} - y_{\rm m}$. \textit{Bottom left:} Comparison between the median mass and light position angle. \textit{Bottom right:} Comparison between the median mass and light-axis ratio. The dashed lines show lines without a centroid offset (top) and the lines of equal position angle and axis ratio (bottom left and bottom right).}
        \label{fig:alignment}
\end{figure*}

Because the lens light is described by multiple \sersic \, components, we calculated the second brightness moments of the primary lens light model to compare the median centroid, axis ratio, and position angle to that of the mass distribution. The result is shown in Fig.~\ref{fig:alignment}.  The mass and light centroid align well within one pixel (0.08\arcsec) in most cases, with the exception of PS J0659$+$1629, which shows a large offset in the x-centroid. The position angle of mass and light agree mostly well within $\sim10\degree$ for five of the nine systems. The largest offset is also reached for J0659+1629 with $\sim$ 50$^\circ$. For the axis ratio, the models do not follow the 1:1 relation so closely. Five models tend to be more elliptical in mass compared to the light, where four of them either have a satellite or are highly sheared. This may explain the low mass-axis ratio.

Individual modeling details for each lensing system are discussed further in the following subsections.

\subsection{DES J0029-3814}
This lensing system was modeled solely in the IR F160W band. In the remaining bands, sufficient arc light from the host galaxy could not be identified by eye. We used two S\'{e}rsic profiles and a central point-source component for the light from the primary lens. The PSF was estimated by selecting two stars in the field.
The best-fit parameters in Table \ref{tab:results_param} show that the centers of the mass and light distribution are offset by $\sim$0.04\arcsec\ in the $x$-direction and by $\sim$0.02\arcsec\ in the $y$-direction. We obtain a high shear magnitude of $\sim$0.2 with a low position angle (14$^\circ$), which may be due to the overdense environment north of the lensing system.

\subsection{DES J0214-2105}
We modeled this system in all three considered bands (F160W, F475X, and F814W) and note that the quasar amplitudes in the F475X band are strongly underfit after step 5(a) (see Table \ref{tab:Modelingtable_multi}). The reason might be that the faint arc light and an imperfect PSF model result in higher residuals in regions of multiple quasar images. The underfitting is not directly obvious from the residuals because the missing quasar amplitude is compensated for by high source intensity values in a single pixel. To avoid negative impact of this band on the lens mass model, we present the F160W single-band model. The PSF was approximated with one star from the field in the F160W band and two stars in each of the F475X and F814 bands. The centers of the mass and light distribution are very well aligned, with an offset within $\sim$0.01\arcsec. 

\subsection{DES J0420-4037}
This lensing system was modeled in all three bands. As in the case of J0214$-$2105, the quasar amplitudes in the F475X band are strongly underfit. We present the F160W single-band model results.
We approximated the PSF in the F160W band by choosing two stars in the field; for the F475X and F814 bands, we used one star. Mass and light show a small offset of $<0.01$\arcsec\ in the $x$-direction and $\sim$0.01\arcsec\ in the $y$-directions. The modeling residuals indicate several light clumps near or in the arc. We confirmed two of them to be counter images, which indicates a second lensed background source.

\subsection{PS J0659+1629}
This lensing system was modeled in the IR F160W band because the arc light in the other bands was insufficient. We also included an SIS profile for the satellite mass and a S\'{e}rsic profile for the satellite light. The satellite is very close to the northern-most quasar image. We therefore boosted the error map in the satellite region in the same way as the quasar images in the last modeling step. The PSF was estimated with four stars from the field. We found two very faint light clumps in the residuals, which are at positions that suggest them to be counter images of a second, lensed background source at a similar redshift as the first source. The dark region in the source light reconstruction at the satellite position does not have an effect on the mass model because its counter image lies outside the reconstruction region (arc mask). For this system, the quasar amplitudes are also underfit after step 5(a) (see Table \ref{tab:Modelingtable_single}), which might have an effect on the lens mass model. This is difficult to overcome with automated modeling procedures and require individual tweaks that are deferred to future work for cosmographic analysis.
This system shows the largest difference between mass and light in our sample, with an offset of $\sim$0.1\arcsec\ in the the $x$-direction and $\sim$0.05\arcsec\ in $y$-direction. Furthermore, the modeled lens mass distribution is much more elliptical, with $\Delta q=q_{\rm S}-q_{\rm m} \sim -0.4,$ and the position angle is offset by $-45$\degree. The reason might be the satellite galaxy close to image D.

\subsection{2M1134-2103}
This system has a faint lens and very bright quasar images, which makes an exact modeling of the lens light and spectroscopic redshift of the lens difficult. We performed the modeling in the F160W and F814W bands. The PSF for the F160W band was estimated by using five stars. We chose three stars to build the PSF in the F814W band. 
The median mass and light centroid align very well with an offset smaller than $\sim$0.01\arcsec\ in the $x$-direction and $\sim$0.01\arcsec\ in the $y$-direction. This system has the highest external shear magnitude $\gamma_{\rm ext}$ of the sample. This is consistent with the elongated shape and the overdense environment in the field of view of this system \citep{Rusu.2019}.

\subsection{J1537-3010}
This system was modeled in all three bands. In each band, we chose five stars from the field to approximate the PSF.
The centers of the mass and light distribution align very well within $\sim$0.01\arcsec. In the model residuals shown in Table \ref{tab:results_light}, we find a small offset between the quasar images and the arc in bands F475X and F814W. A possible explanation for this offset is that these two bands are in the rest-frame UV of the quasar host galaxy, and thus are likely dominated by light from specific star-forming regions in the host galaxy. The power-law slope $\tilde{\gamma}_{\rm PL}\sim0.3$ hits the lower bound of the prior, which might be due to our imperfect PSF models in the F475X and F814W models. As we show in Sec.~\ref{sec:comparison}, the PSF has a significant effect on $\tilde{\gamma}_{\rm PL}$.
This system has one of the most prominent arcs in all three \textit{HST} filters in our sample, which is ideal for future cosmographic analysis.

\subsection{PS J1606-2333}
This lensing system was modeled in the F160W band with two S\'{e}rsic profiles and a central point source component for the primary lens light. Although we initially included the other bands because they show significant arc light, the source reconstruction shows no visible source (above the noise level) in these bands. We excluded these bands and optimized only the single-band model. For this lensing system, the PSF was stacked from five neighboring stars. The bright light clump in the south below image B that is embedded in a faint extended structure. We assumed this clump to be a satellite at the same redshift as the lens galaxy, and include an SIS profile for the satellite mass and a S\'{e}rsic profile for the satellite light. The error map was boosted at the satellite position in the last modeling step because of its proximity to the arc. Mass and light align very well within $\sim$0.01\arcsec. The power-law slope of this system is close to the lower bound of the prior with $\tilde{\gamma}_{\rm PL}\sim0.31$. Again, this might be due to an imperfect PSF model.

\subsection{PS J1721+8842}
\label{sec:1721}
This is an unusual and interesting lens system with six quasar images. It is composed of two quasar sources at the same redshift, with one quasar lensed into a quad (images A, B, C, and D) and the other one lensed into a double (images E and F) \citep{Lemon1721}. Most of the arc light in this system comes from the double system, therefore we only included these regions for the arc light modeling and SSB reconstruction. We modeled this system in all three considered bands with two S\'{e}rsic profiles and a central point-source component for the primary lens light. We picked three stars from the field to approximate the PSF in the F160W band, and one star in each of the F475X and F814W bands. The second source component was added manually because the current version of the automation code does not support an automated detection of multiple source components. It is represented as a green star in the source reconstruction plot in Table~\ref{tab:results_light}. Quasar image F is offset relative to the arc peak SB, which may be due to the quasar being kicked out of its host galaxy (\citealt{Mangat2021}). We used F475X as the reference band because image F is not clearly distinct from the arc in the F160W band where the arc is the brightest. The offset introduces a challenge to the quasar and arc light modeling. At the position of the second source component, the source intensity is negative. The mass and light centroids are offset by $\sim$0.04\arcsec\ in the $x$-direction, and align very well within $\sim$0.01\arcsec\ in the $y$-direction.

\subsection{DES J2100-4452}
The final model of this system includes only the F160W band. The F475X and F814W bands show small indications of arc light, but including these bands in the model led to no visible source reconstruction. The PSF was created by using five stars. A dust lane appears to cross the lens galaxy from north to south. Its influence on the model needs to be investigated in future work. The dust lane can be seen best in the bluest band (in our case, the F475X band), and given its rest frame wavelength of 1.3 microns in the F160W band, the effect on the model is probably very weak.
The mass is offset relative to the light by $\sim$0.04\arcsec\ in the $x$-direction and by $\sim$0.06\arcsec\ in the $y$-direction.

\begin{table}
        \caption{$\chi_{\rm red}^2$ for each individual band with total $\chi_{\rm red, tot}^2$.}
        \begin{tabular}{lccccc}
                \toprule System & $\chi_{\rm red, 160}^2$ & $\chi_{\rm red, 475}^2 $& $\chi_{\rm red, 814}^2 $ & $\chi_{\rm red, tot}^2 $\\ \toprule \vspace{2px}
                J0029$-$3814    & 0.80 & $-$ & $-$ & 0.80 \vspace{2px}\\
                J0214$-$2105    & 0.74 & $-$ & $-$ & 0.74 \vspace{2px}\\
                J0420$-$4037    & 0.56 & $-$ & $-$ & 0.56 \vspace{2px}\\
                J0659$+$1629    & 1.08 & $-$ & $-$ & 1.08 \vspace{2px}\\
                2M1134$-$2103   & 2.76 & $-$ & 1.62 & 1.84 \vspace{2px}\\
                J1537$-$3010    & 1.62 & 1.08 & 1.18 & 1.18 \vspace{2px}\\
                J1606$-$2333    & 1.36 & $-$ & $-$ & 1.36 \vspace{2px}\\
                J1721$+$8842    & 0.51 & 1.16 & 0.96 & 0.94 \vspace{2px}\\
                J2100$-$4452    & 0.94 & $-$ & $-$ & 0.94 \vspace{2px}\\ \bottomrule
                
        \end{tabular}
        \label{tab:chi2}
\end{table}

\section{Discussion}
\label{sec:discuss}
\subsection{Comparison of automated modeling to manual modeling with \texttt{GLEE}}
In this section, we compare the automated modeling with our automation code and the manual modeling of a lensing system with \texttt{GLEE} based on the total time actively spent by the user on modeling a lensing system. We report a significant reduction of preparation time for the input files, which are created in separate automated codes. The creation of the image cutout is sped up from $\sim$15 minutes to $\sim$5 minutes in the automated modeling. The time for creating the error map stays roughly the same because this was already automated prior to this work. The mask regions in both cases were obtained by eye and were marked manually, but the masks are now created fully automatically from the region files. This automation halves the user input time from roughly one hour in the manual case to $\sim$30 minutes when the automation code is used. The time for creating the PSF is significantly reduced from $\sim$2 hours to $\sim$15 minutes in the case of the automated modeling. When the PSF code is used, the only step the user has to conduct in order to obtain the PSF is to find suitable stars in the field and save their positions. The time-consuming technical steps such as saving the region file, stacking the files, subsampling if requested, and normalizing to create the PSF, are fully automated.

For the actual lens modeling process, we can also report a significant time reduction. The initial setup of the \texttt{GLEE} configuration file, which also includes estimations for the starting values of the parameters, is about one hour for manual modeling. In the automated case, the user only has to provide a region file, which can be created within 10 minutes. The setup of the configuration file and the estimations for the starting values are fully automated, so that it is not necessary for the user to learn how to use \texttt{GLEE} (e.g., how to set up the \texttt{GLEE} configuration file). Quality control and modifications to the configuration file between different modeling steps are also fully automated, which again saves several hours over the full modeling process. Moreover, there is no waiting time between the multiple MCMC chains, and the code obtains the optimal sampling parameters automatically. Together with short interaction between the user and the code and the inspection of the output, we estimate an average of one hour of total user time per lensing system for the automated modeling process. Table \ref{tab:usertime} gives a comparison between user input time in the automated and in the manual modeling case.
\begin{table*}
    \begin{center}
    \caption{User input time for each task in manual and automated modeling.}
                \begin{tabular}{lcc}\toprule
                        Task    & User time for manual modeling & User time for automated modeling   \\ \toprule
                        Creation of image cutout & $\sim$15 minutes & $\sim$5 minutes \\
                        Creation of error map & $\sim$10 minutes & $\sim$10 minutes \\
                        Creation of masks & $\sim$1 hour & $\sim$30 minutes \\
                        Creation of PSF & $\sim$2 hours & $\sim$15 minutes \\
                        Modeling process & $\sim$10 hours & $\sim$1 hour \\ \bottomrule
                \end{tabular}
        \ \\ \ \\
        \caption*{\textbf{Note: }This does not include computation time, which is summarized in Table \ref{tab:computationtime}.}
        \label{tab:usertime}
        \end{center}
\end{table*}

The computational time for the arc light modeling and SSB distribution reconstruction mainly depends on the arc mask and PSF size. The optimizing and sampling cycles with simulated annealing and MCMC chains use one core at a time. We used \texttt{EMCEE}, which can be highly parallelized, to obtain an initial sampling covariance matrix at the beginning of the last modeling steps (5(a) and 5(b) in single-band, and 4(a) and 4(b) in multi-band modeling) with a chain length of 400,000 on a cluster with Intel Xeon E5-2640 v4 CPU using 30 cores. The average computation time on the cluster for each \texttt{EMCEE} chain is $\sim$5 hours. Because of typical overhead in parallel computing, we estimate that the single core computation time for each \texttt{EMCEE} chain is $\sim$100 hours. \texttt{EMCEE} sampling is used twice in the automated modeling procedure, and thus we have a total \texttt{EMCEE} computation time of $\sim$200 hours per core. With this setup and modeling specifications, we estimated the average computation time per core per lens system for each modeling step described in Sec.~\ref{sec:modeling}, and summarize them in Table \ref{tab:computationtime}. The computationally expensive step of arc light modeling and SSB distribution reconstruction takes 15 to 20 days with a single core. Because of the parallelization of \texttt{EMCEE,} the effective computation time ranges between 7 to 14 days. When only approximate (and not full) convergence of a chain is sufficient, models can already be expected after 5 to 7 days. These models do not differ much from the final model that is fully converged and can be used for tentative analysis. We expect the computation time to be roughly the same in both the manual and the automated modeling. 
\begin{table*}[h]
\begin{center}
    \caption{Average computation time per core per lensing system for each modeling step.}
                \begin{tabular}{lcc}\toprule
                        Modeling step   &  Section & Average computation time   \\ \toprule
                        Lens light modeling & \ref{sec:lenslight} &$\sim$1 day \\
                        Quasar light modeling & \ref{sec:quasarlight} &$\sim$1 day \\
                        Lens mass modeling with source/image positions & \ref{sec:srcimgpos} & $\sim$1 minute  \\
                        Arc light modeling and SSB distribution reconstruction & \ref{sec:arclight} &$\sim15-20$ days \\\bottomrule
                \end{tabular}
                        \caption*{\textbf{Notes: }The last and most time expensive step is partly parallelized, and is thus effectively and notably shorter ($\sim 7-14$ days).}
                \label{tab:computationtime}
                \end{center}

\end{table*}

\subsection{Comparison of lens modeling results between \texttt{GLEE} and \texttt{Lenstronomy}}
\label{sec:comparison}
\subsubsection{Blind comparison}
All nine systems of our sample are part of a more extensive sample of 30 quads that were modeled independently by \citetalias{Schmidt2022} with a similar, automated pipeline based on the lens modeling software \texttt{Lenstronomy}. Both modeling frameworks are dedicated to high-resolution images of lensed quasars, and thus use similar modeling assumptions for the lenses: the lens mass distribution is modeled with a power-law profile with external shear, and the lens light distribution is modeled with \sersic \, profiles. 

There are important differences between the two procedures that need to be taken into account. First, the two softwares have different implementations for the SSB reconstruction: \texttt{GLEE} uses a pixelated grid of the source intensity distribution, whereas \texttt{Lenstronomy} uses circular \sersic\ profiles with additional optional shapelet components. 
Second, \citetalias{Schmidt2022} always modeled the three available bands of HST images, while in this work, we focused only on those in which the arcs are clearly visible by eye, typically F160W. Even though these are the most informative bands in our nine-system sample, they do not capture the full information, especially because the resolution is better in the UV/visible (UVIS) data.  Third, different priors have been chosen for some of the key parameters. \citetalias{Schmidt2022} adopted informative priors from the SLACS sample \citep[][]{Bolton2006,Bolton2008,Auger2010} on the radial slope of the mass density profile and the ellipticity of the mass distribution in order to avoid unphysical solutions when the parameters are poorly constrained by the data. In this study, we opted to adopt uniform priors within some bounds to constrain the parameters solely by the data at hand. In the best cases, the priors should not matter because the likelihood should dominate the posterior, but in practice, as we show below, many of the systems are poorly constrained by the data, and therefore the priors do matter.
Fourth, \citetalias{Schmidt2022} adopted an iterative scheme to improve the PSF estimate based on the multiple images of the quasar, sampled at the scale of the reduced data. In contrast, in this work, we used an subsampled PSF estimated from images of stars in the field, without additional corrections. As shown by \citet{Shajib2022}, the PSF is a crucial ingredient for cosmography-grade inference.

\begin{figure*}
        \subfigure{\includegraphics[width=0.33\textwidth]{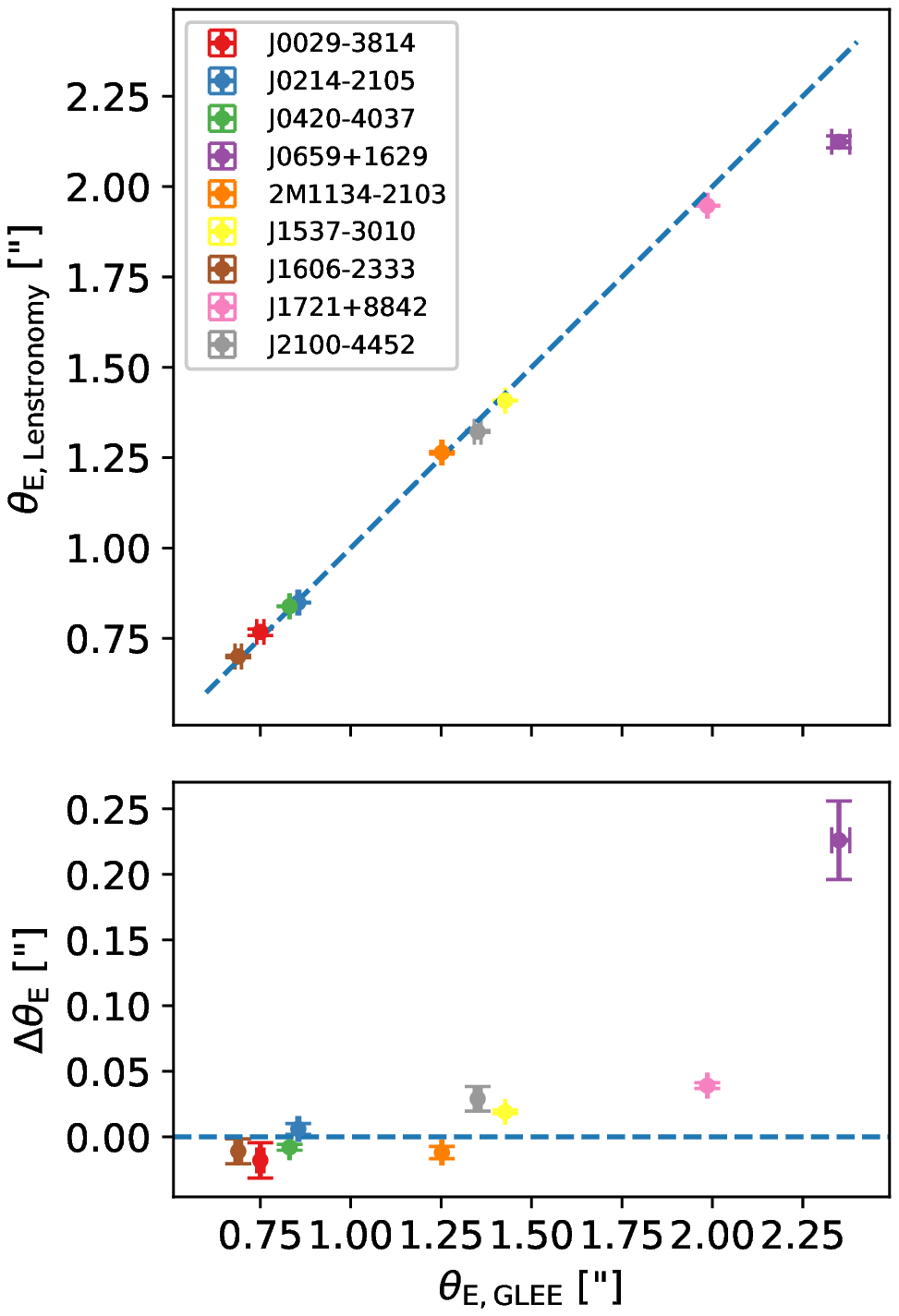}}
        \subfigure{\includegraphics[width=0.33\textwidth]{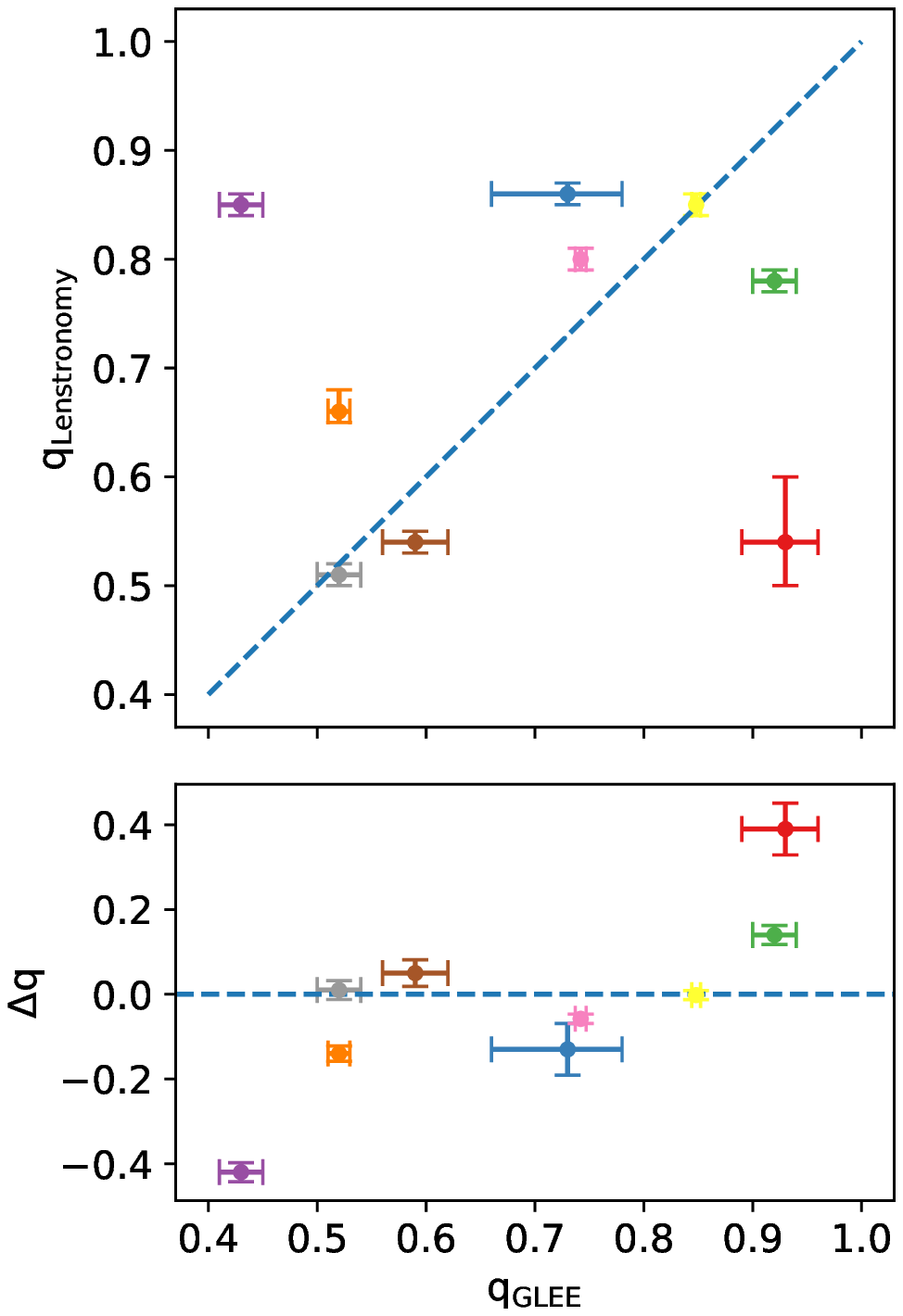}}
        \subfigure{\includegraphics[width=0.33\textwidth]{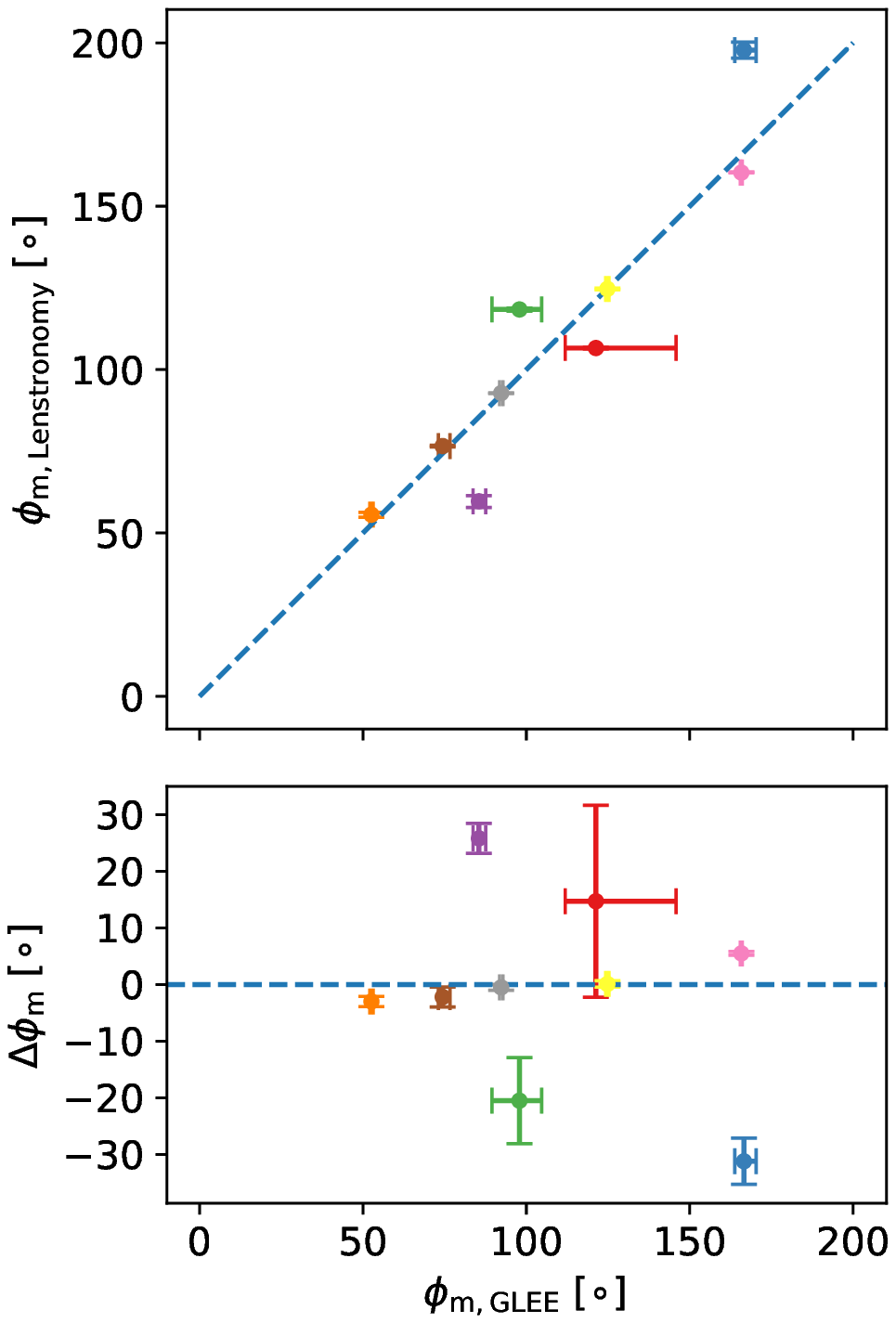}}
        \subfigure{\includegraphics[width=0.33\textwidth]{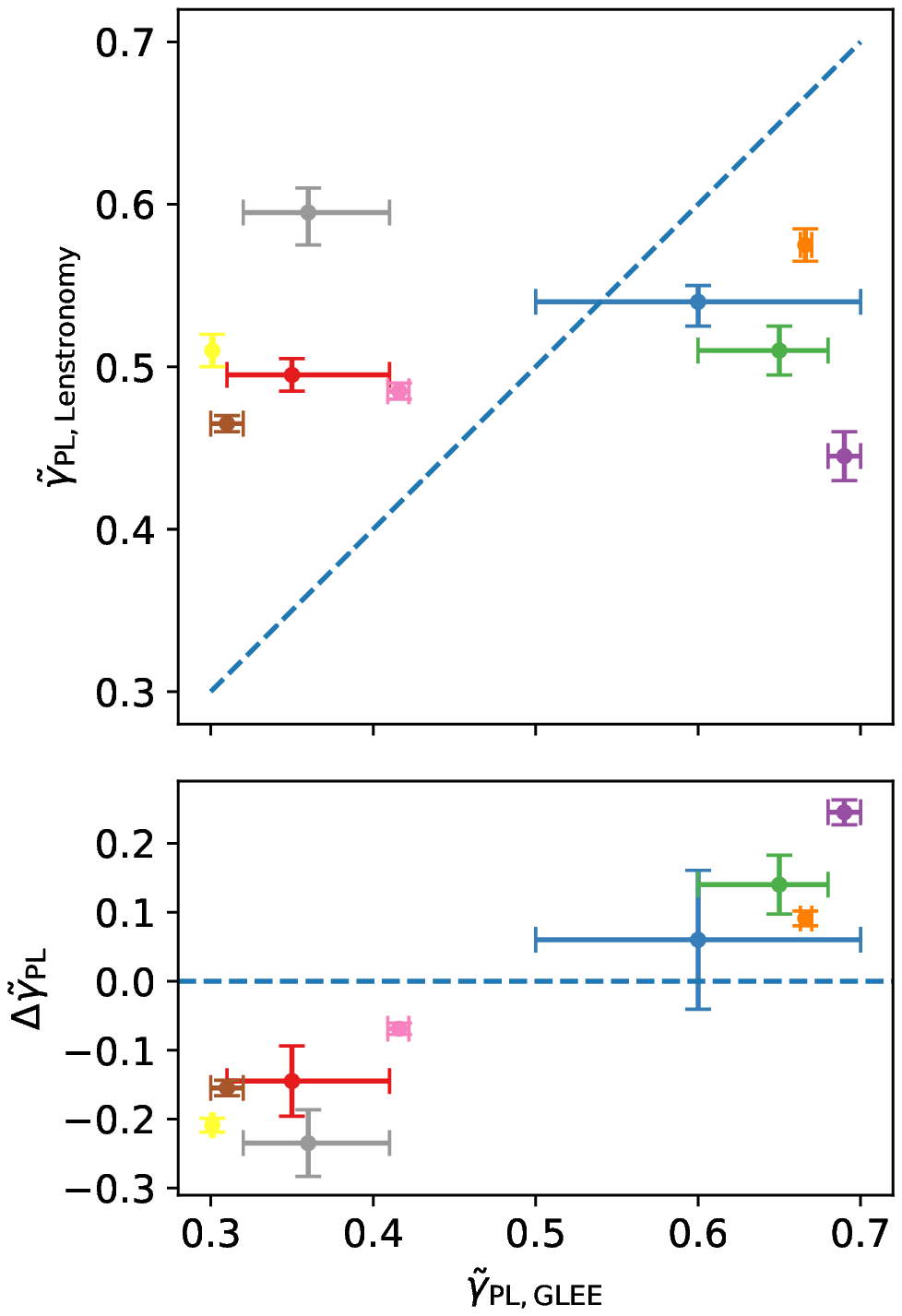}}
        \subfigure{\includegraphics[width=0.33\textwidth]{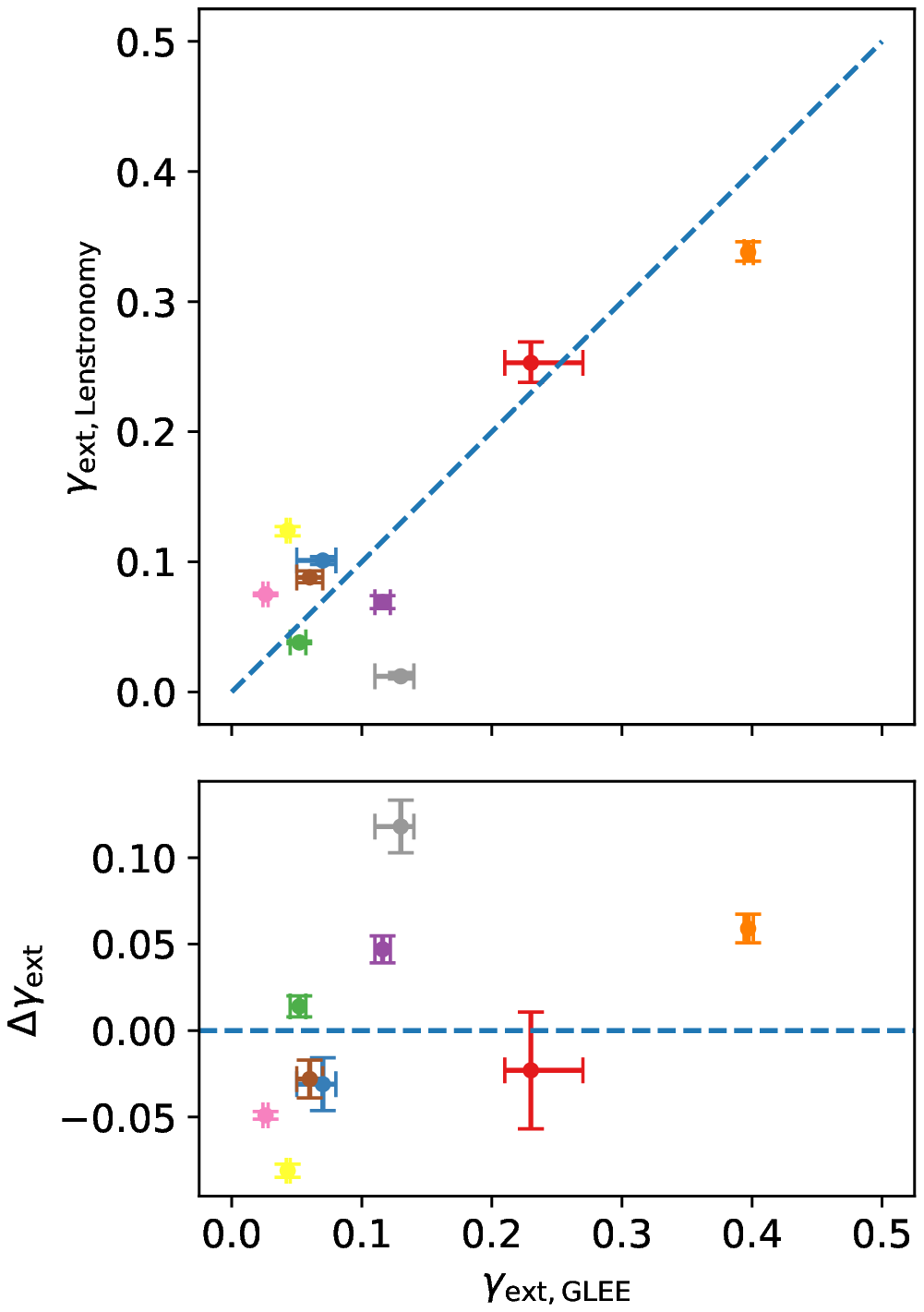}}
        \subfigure{\includegraphics[width=0.33\textwidth]{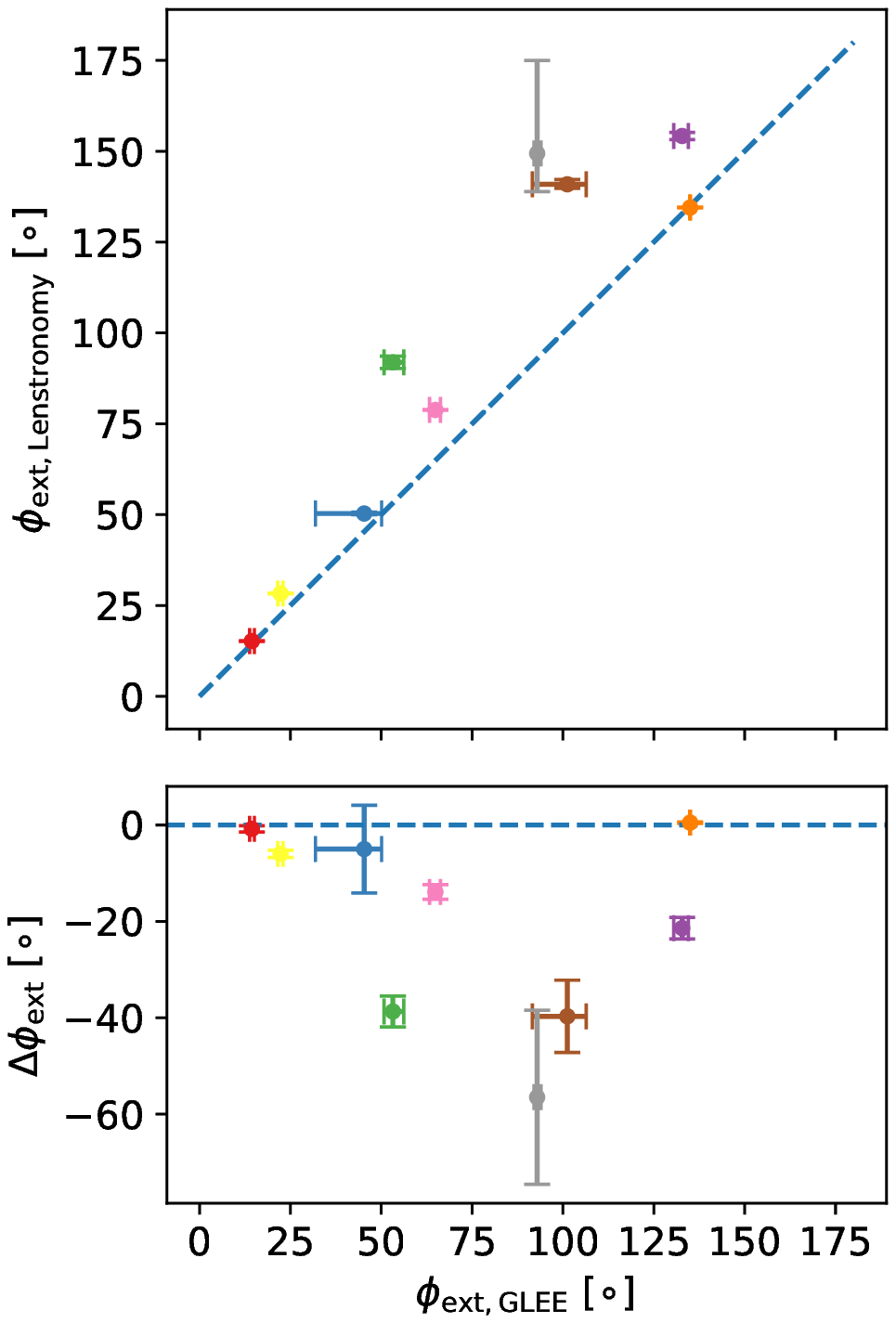}}
                                                        
        \caption{Direct comparison of lens mass parameter and external shear values between our models and \texttt{Lenstronomy} models from \citetalias{Schmidt2022}.}
        \label{fig:comp1}
\end{figure*}

Keeping the caveats in mind, we compared the lens mass parameters, external shear parameters, convergence and total shear at the quasar image positions, image magnifications, Fermat potential differences, and predicted time delays. The adopted cosmology is the same as in \citetalias{Schmidt2022}. We plot in Fig.~\ref{fig:comp1} the results of the two modeling codes together with a line showing the identity. In addition, we present the difference of the median values of both modeling codes against the \texttt{GLEE} parameter values to better illustrate the absolute differences between the final results.

\begin{figure*}
        \subfigure{\includegraphics[width=0.33\textwidth]{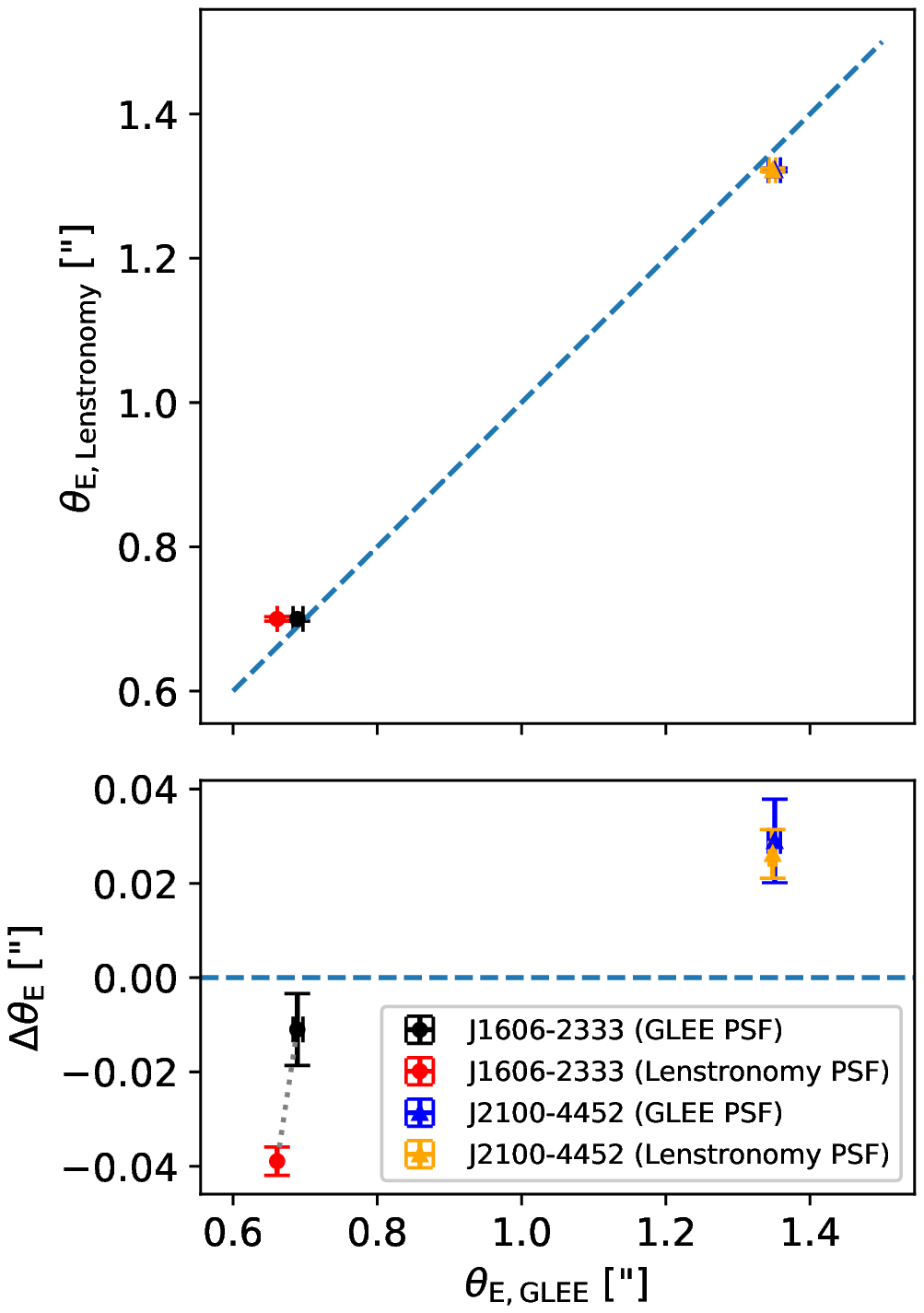}}
        \subfigure{\includegraphics[width=0.33\textwidth]{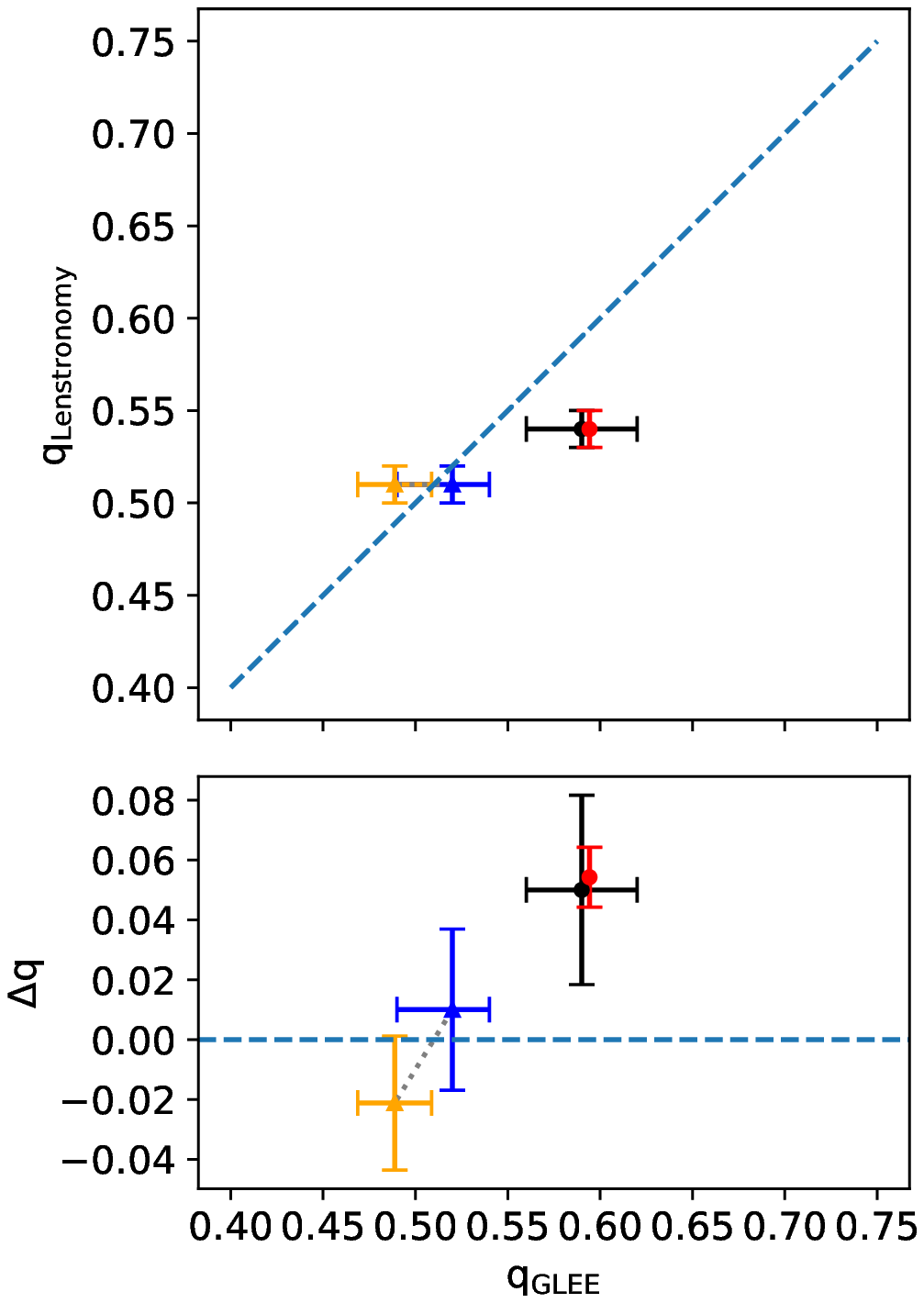}}
        \subfigure{\includegraphics[width=0.33\textwidth]{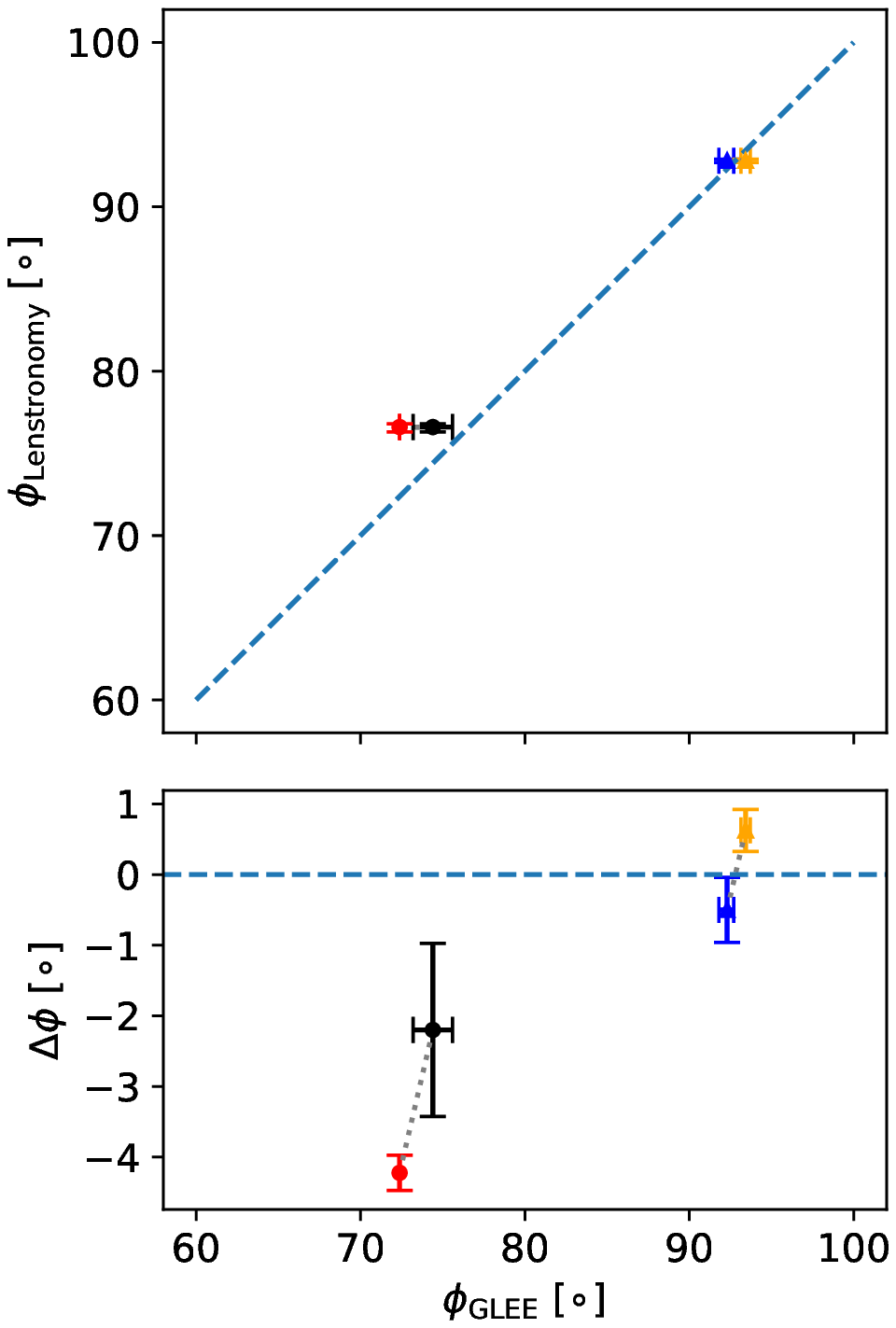}}
        \subfigure{\includegraphics[width=0.33\textwidth]{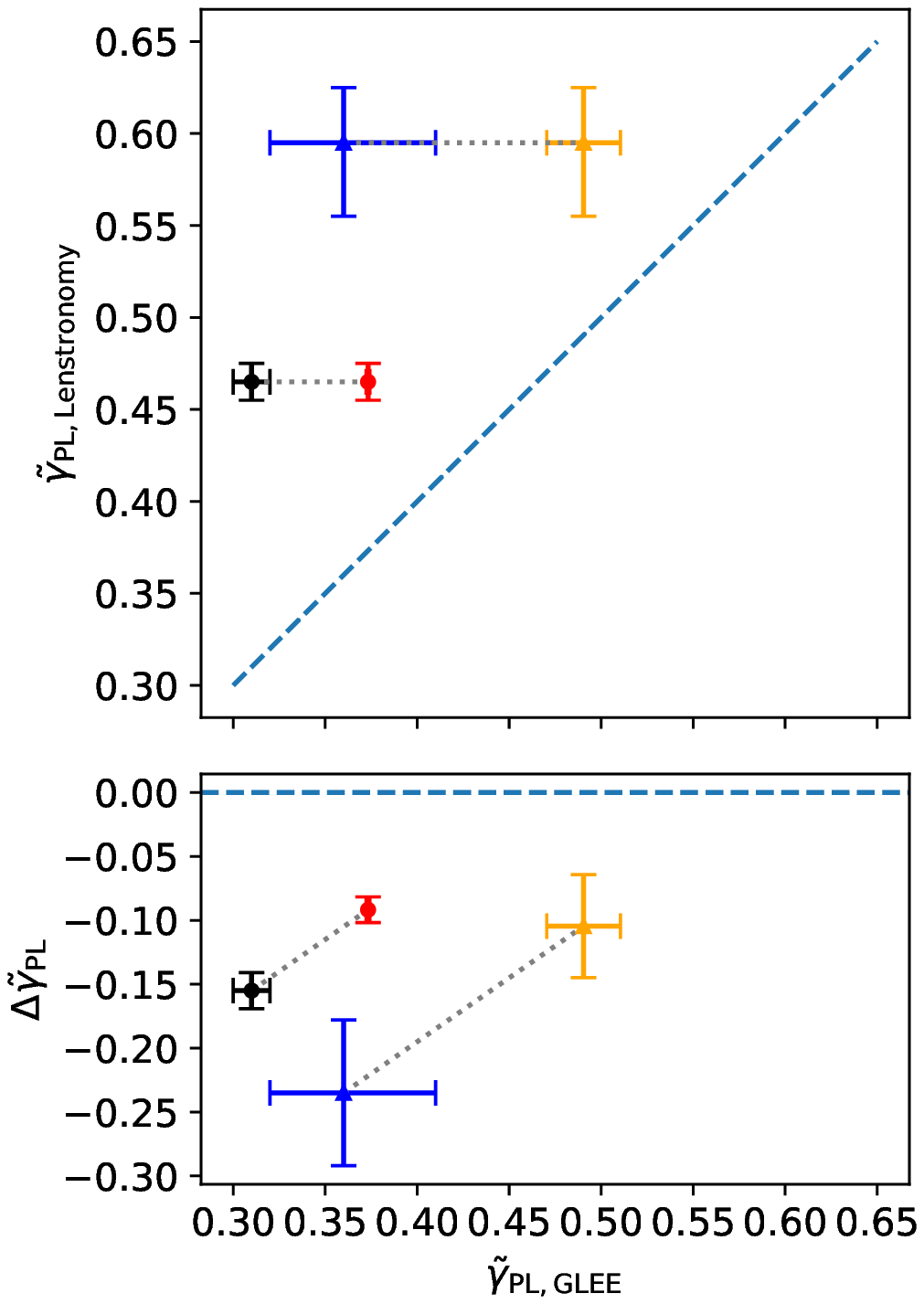}}
        \subfigure{\includegraphics[width=0.33\textwidth]{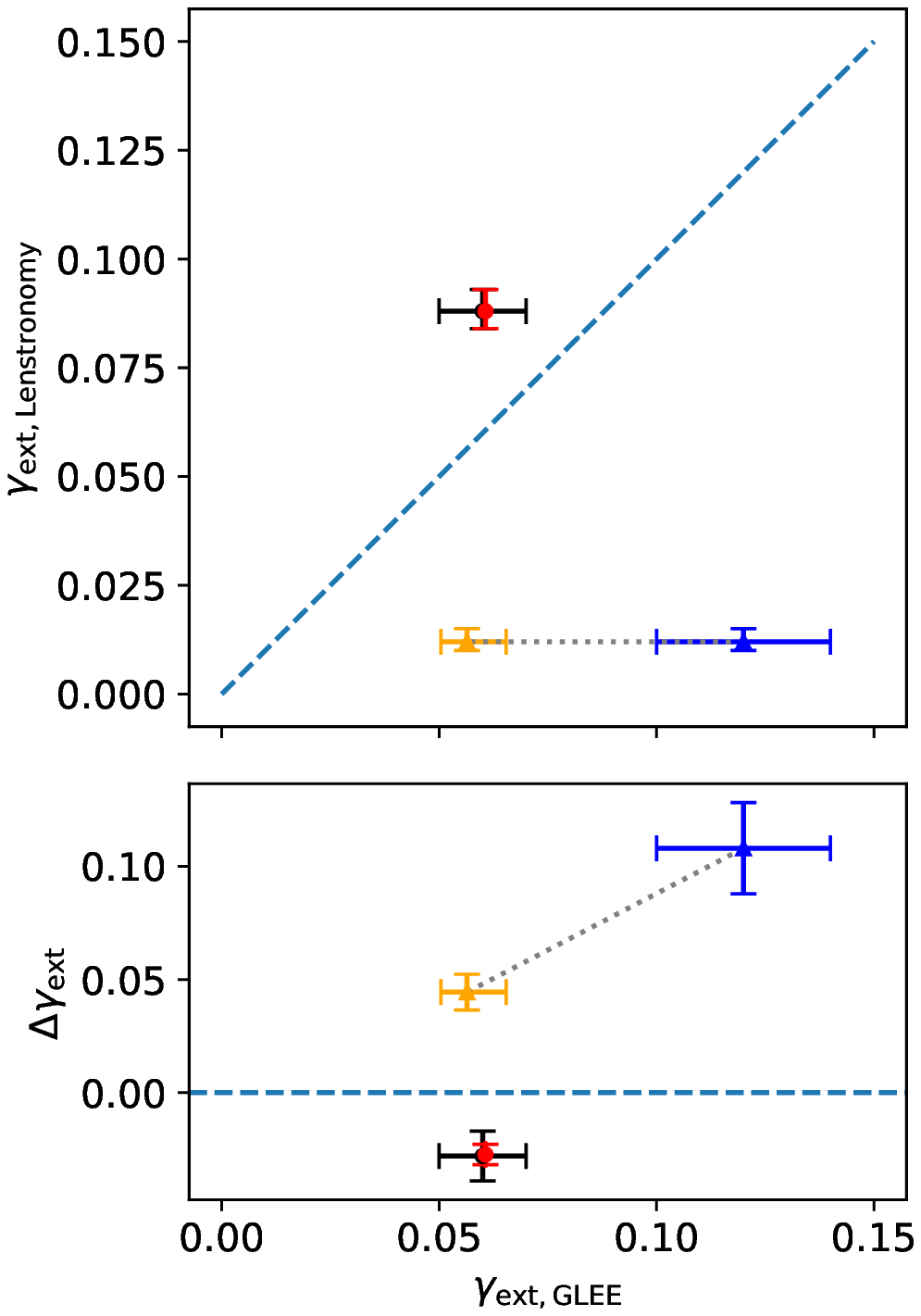}}
        \subfigure{\includegraphics[width=0.33\textwidth]{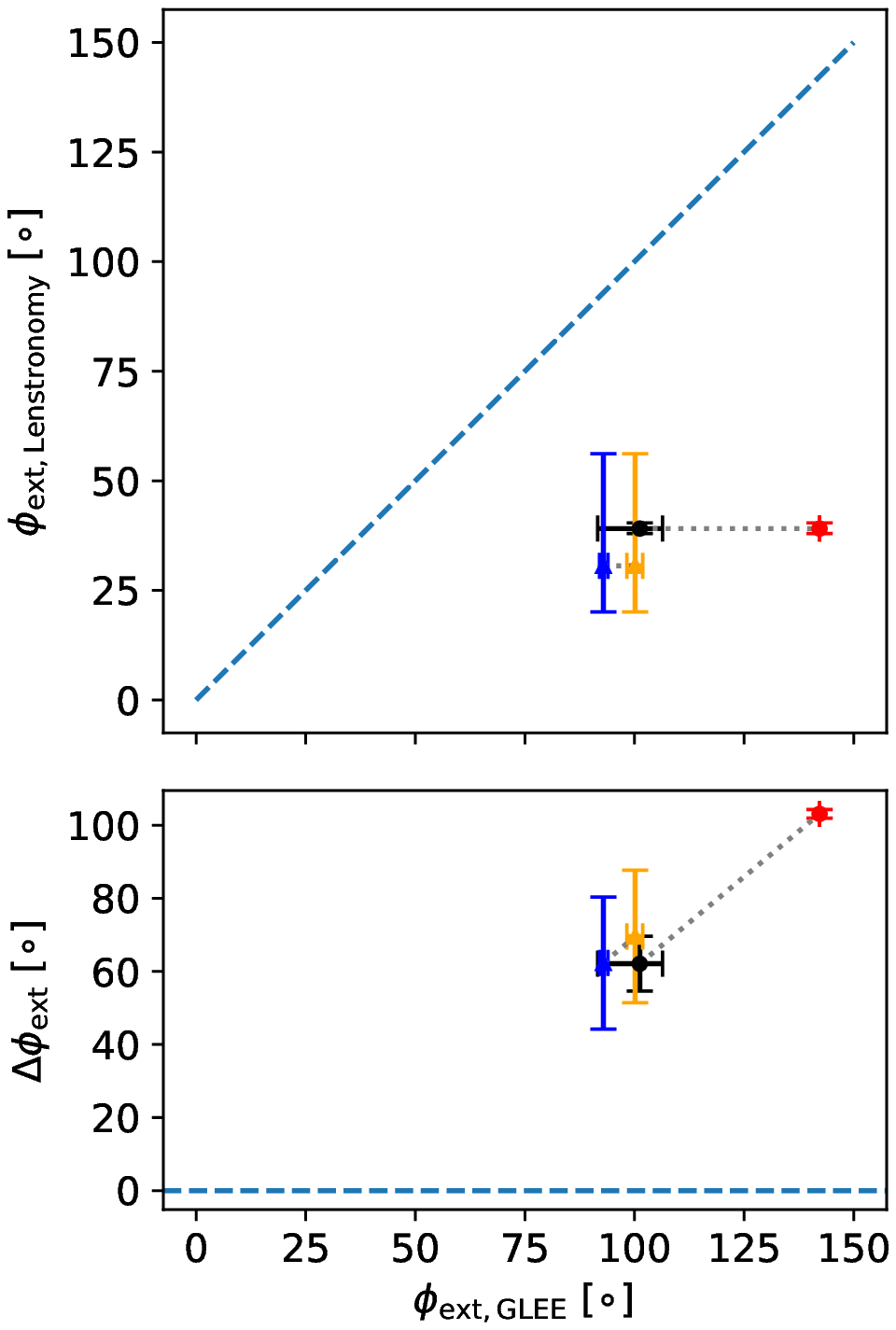}}
                                                        
        \caption{Change in lens mass parameter and external shear values when using the \texttt{Lenstronomy} PSF with our \texttt{GLEE} automation code for J1606$-$2333 and J2100$-$4452.}
        \label{fig:comp2}
\end{figure*}

We find excellent agreement in the Einstein radius of the primary lens galaxy, where the median values agree within 4\% for eight of the nine systems. This is expected, and it is reassuring that SL provides robust masses enclosed within the Einstein radius.  An apparent outlier is J0659$+$1629 with a difference of $\sim$0.2\arcsec, which is due to the presence of a satellite galaxy whose mass is degenerate with that of the primary lens. We calculated a value for the ``effective'' Einstein radius (the radius of the circle for which the enclosed mean $\kappa$ is 1) that included the satellite mass and compared it to the radius that was obtained by \citetalias{Schmidt2022}~for this purpose. Our value of $\theta_{\rm E,eff,GLEE}=2.368_{-0.02}^{+0.02}{\arcsec}$ matches very well $\theta_{\rm E,eff,Lenstronomy}=2.368 _{-0}^{+0.01}{\arcsec}$ for J0659$+$1629 (the distribution for $\theta_{\rm E,eff,Lenstronomy}$ is heavily skewed such that the median coincides with the 16th$^{\rm }$ percentile).
We thus conclude that for all systems, we recover the same Einstein radius to within a root-mean-square (r.m.s.) scatter of 1.6\% when the proper accounting for satellites is considered.

Flattening, position angle of the mass distribution, and external shear magnitude and direction are strongly degenerate with each other (and to some extent with the slope), and should therefore be looked at holistically. To illustrate this, we show the posterior distributions of the mass parameters and external shear of J0659$+$1629 in the appendix in Fig.~\ref{fig:degen}, where we highlight the strong degeneracy between the Einstein radius of the main deflector and the satellite.

For six out of the nine systems, the position angles of the lens mass distribution match within the errors (1 or 2 $\sigma$ at most). The outliers are J0659$+$1629, for which \texttt{GLEE} converges to a mass distribution that is highly flattened, more than the light distribution, and therefore likely not fully correct, while \citetalias{Schmidt2022} -- driven by the prior -- converges to a much rounder distribution, and J0214$-$2105 and J0420$-$4037, for which the discrepancy is due to a combination of differences in flattening, PA, and external shear. We note that for this system, the slope of the mass density profile is poorly constrained without a prior, suggesting that the information content of the data is not sufficient. 

The mass-flattening parameter agrees within $\sim0.1$ (larger than the formal uncertainty, which is therefore underestimated) except for two cases: i) J0659$+$1629 discussed above, and ii) J0029$+$3814, for which \citetalias{Schmidt2022} reported a stronger flattening than GLEE, perhaps driven by the prior on the axis ratio of the light distribution.
The inferred external shear magnitudes agree within 0.05 (again larger than the formal uncertainty); the two most discrepant systems are J1537$-$3010 and J2100$-$4452, for which the slope of the mass density profile is relatively poorly constrained without a prior, and which are therefore likely to have a poor information content of the data. Reassuringly, we find a good match of the shear position
angle for the two systems with the highest shear magnitude. 

As mentioned above, a main source of difference is the prior of the power-law slope $\tilde{\gamma}_{\rm PL}$, which reverberates through the other parameters. Not surprisingly, the inferred power-law slope $\tilde{\gamma}_{\rm PL}$ of the \texttt{Lenstronomy} models obtained by \citetalias{Schmidt2022} follows their imposed Gaussian prior, which is centered on a close-to-isothermal value of $\tilde{\gamma}_{\rm PL}=0.54$. In contrast, the slope values of the \texttt{GLEE} models span the uniform prior range of [0.3,0.7], with values closer to the bounds than to an isothermal value. As we discussed before, the difference arises from a combination of two factors: differences in the PSF, and information in the data that is insufficient to constrain the slope.

Not surprisingly, given the differences highlighted above, the inferred quantities $\kappa$, $\gamma_{\rm tot}$, and $\mu$ at the image positions show significant deviations (see Fig.~\ref{fig:comp_appendix1}). We calculated the average relative scatter of these quantities for each lensing system, and find that it ranges from $\sim$3\% to \textgreater 100\%, and the statistical uncertainties are underestimated in nearly all cases. 

The relative Fermat potential differences and predicted time delays (also shown in Fig.~\ref{fig:comp_appendix1}) follow the 1:1 line overall, although they often disagree within the statistical uncertainties. For these two quantities, we removed the outlier image C of J0659$+$1629 for better visibility in the plot, because its time delay is one order of magnitude longer than all other images.  Table \ref{tab:scatter} lists the relative deviation in Fermat potential differences $\delta(\Delta\tau) / |\Delta \tau_\text{GLEE}|$ with $\delta(\Delta\tau) = \Delta \tau_\text{GLEE} - \Delta \tau_\text{Lenstronomy}$. We estimated the statistical uncertainties on these relative deviations by symmetrizing the uncertainties on the GLEE and Lenstronomy parameter values (e.g., on $\Delta \tau_\text{GLEE}$ and $\Delta \tau_\text{Lenstronomy}$) and assuming that they follow Gaussian distributions.

\begin{table}
\caption{Relative deviation of Fermat potential differences.}
        \begin{tabular}{llp{1.8cm}p{2.2cm}}
                \toprule System & Image pair & $\delta(\Delta\tau)/|\Delta\tau_{\rm GLEE}|$ & $\delta(\Delta\tau_{\rm iso})/|\Delta\tau_{\rm GLEE,iso}|$  \\ \toprule  \ \\
                DES J0029$-$3814        & DA & $0.69$\textpm0.16 & $0.20$\textpm0.20 \\ 
                                    & DB & $0.74$\textpm0.18 & $0.23$\textpm0.21 \\
                                    & DC & $0.70$\textpm0.18 & $0.20$\textpm0.21 \vspace{5px}\\
                DES J0214$-$2105        & BA & $0.38$\textpm0.21 & $0.31$\textpm0.27 \\
                                    & CA & $0.34$\textpm0.20 & $0.27$\textpm0.26 \\
                                    & DA & $0.36$\textpm0.23 & $0.29$\textpm0.28 \vspace{5px}\\
                DES J0420$-$4037        & CA & $0.22$\textpm0.12 & $0.56$\textpm0.16 \\
                                    & CB & $0.34$\textpm0.12 & $0.71$\textpm0.17 \\
                                    & DC & $-0.33$\textpm0.12 & $-0.69$\textpm0.17 \vspace{5px}\\
                PS J0659$+$1629    & CA & $0.06$\textpm0.06 & $-0.46$\textpm0.10 \\
                & CB & $0.42$\textpm0.03 & $0.1$\textpm0.05 \\
                & DC & $0.45$\textpm0.17 & $-1.24$\textpm0.26  
                    \vspace{5px}\\
                2M1134$-$2103           & BA & $-0.19$\textpm0.02 & $-0.07$\textpm0.03 \\
                                    & CB & $-0.17$\textpm0.02 & $-0.04$\textpm0.03 \\
                                    & DB & $0.12$\textpm0.02 & $-0.02$\textpm0.02 \vspace{5px}\\
                J1537$-$3010        & BA & $-0.58$\textpm0.03 & $0.07$\textpm0.02\\
                                    & CA & $1.02$\textpm0.04 & $0.19$\textpm0.03 \\
                                    & DA & $-0.54$\textpm0.03 & $0.09$\textpm0.02 \vspace{5px}\\
                PS J1606$-$2333     & BA & $-0.45$\textpm0.08 & $0.03$\textpm0.08 \\
                                    & CA & $-0.46$\textpm0.05 & $0.03$\textpm0.05 \\
                                    & DA & $-0.46$\textpm0.05 & $0.03$\textpm0.06 \vspace{5px}\\
                PS J1721$+$8842     & DA & $-0.09$\textpm0.02 & $0.07$\textpm0.03 \\
                                    & DB & $-0.1$\textpm0.02 & $0.06$\textpm0.03 \\
                                    & DC & $-0.09$\textpm0.02 & $0.07$\textpm0.03 \vspace{5px}\\
                DES J2100$-$4452    & DA & $-0.62$\textpm0.11 & $0.05$\textpm0.16 \\
                                    & DB & $-0.66$\textpm0.12 & $0.03$\textpm0.17 \\
                                    & DC & $-0.54$\textpm0.14 & $0.1$\textpm0.18 \vspace{5px}\\ \bottomrule
                
        \end{tabular}
        \ \\
        \caption*{\textbf{Notes: }The third column shows the relative deviation of Fermat potential differences for each image pair. The fourth column shows the same comparison, but with the rescaled values retrieved when assuming an isothermal mass profile. The relative statistical uncertainty is underpredicted in most cases.}
        \label{tab:scatter}
\end{table}

This comparison shows us that for only one system (J1721$+$8842) are the relative Fermat potential differences and predicted time delays within 10\%. For seven of the nine systems, the deviation of the relative Fermat potential differences of the multiple images is 30\% or more on average. Therefore, for most systems, a more detailed modeling and/or better data are needed to bring all these models to cosmography-grade level. In some cases, the UVIS exposures may provide further constraints on the mass and light profile parameters, even though the source contribution is fainter than in the IR, especially because the resolution is twice as high in the UVIS.

\subsubsection{Post-blind analysis}
\label{subsec:postblind}
After completing the blind comparison between the results, we consider the efforts that have been made after unblinding to understand the origin of the discrepancies. To test the influence of the power-law slope $\tilde{\gamma}_{\rm PL}$ on the physical quantities $\kappa$ and $\Delta\tau$, we rescaled the \texttt{GLEE} and \texttt{Lenstronomy} values of $\Delta\tau$ with $\Delta\tau_{\rm iso} \simeq \Delta\tau/2\tilde{\gamma}_{\rm PL}$ \citep[e.g.,][]{Suyu2012b}, and calculated new $\kappa_{\rm iso}$ values with Eq. (\ref{eq:kappa_spemd}) and $\tilde{\gamma}_{\rm PL}=0.5$, thus assuming isothermal mass profiles. In this case, we obtained a much better agreement between the two modeling results overall, as shown in Fig.~\ref{fig:comp_appendix1}.
With these assumptions, all $\kappa$ values now agree within 20\%, and five of the nine systems have a scatter $\leq$10\%. The trend is similar for the relative Fermat potential differences at the multiple image positions, where six of the nine systems agree within $\sim$20\% (see the last column $\delta(\Delta\tau_{\rm iso})/|\Delta\tau_{\rm GLEE,iso}|$ in Table \ref{tab:scatter}). Two systems, J0420$-$4037 and J0659$+$1629, now have a higher relative deviation in the multiple images, but their absolute values of Fermat potential differences are small.  Therefore, most of the discrepancy in $\kappa$ and $\Delta\tau$ between the two independent models are due to the different power-law slope values.  The two modeling pipelines yield different slope values, as discussed above and as expected, because the priors and PSFs differ. 


To distinguish the two effects (priors and PSF), we used the final PSF of the \texttt{Lenstronomy} modeling by \citetalias{Schmidt2022} in our automated modeling procedure. The creation of the PSF of \citetalias{Schmidt2022} is different than in our pipeline because it includes iterative PSF correction, while our PSF is simply constructed from stars in the field (see Sec.~\ref{sec:preparation}). Moreover, the \texttt{GLEE} PSF is subsampled by a factor of 3 compared to the pixel scale of the data, while the \texttt{Lenstronomy} PSF uses the pixel scale of the original data. 

To ensure that the information content is sufficient (and thus the prior is less important), the two teams focused for this comparison on two systems with high signal-to-noise ratio (S/N) data and visible arcs: J1606$-$2333 and J2100$-$4452. We remodeled these two systems in the F160W (IR) band with our \texttt{GLEE}-based pipeline, but used the not subsampled \texttt{Lenstronomy} PSF. We compared the results with those based on the \texttt{GLEE} PSF. We show the comparison in Fig.~\ref{fig:comp2} and Fig.~\ref{fig:comp_appendix2}. In both figures, we present the \texttt{GLEE} blind results as presented in this work and the new post-blind results using the \texttt{Lenstronomy} PSF. In most cases, the values obtained with the \texttt{Lenstronomy} PSF are closer to the modeling results of \citetalias{Schmidt2022}. The Einstein radius of J1606$-$2333 is lower with the new PSF because it is degenerate with the Einstein radius of the satellite, which is now larger. The effective Einstein radius of this system remains unchanged. The influence on the power-law slope is evident, as the slope values move much closer to the values of \citetalias{Schmidt2022}, and it is closer to isothermal, without imposing a prior. This also leads to a better agreement of other quantities such as $\kappa$ or the Fermat potential differences to the \texttt{Lenstronomy} modeling results, as shown in \ref{fig:comp_appendix1}. We conclude from this that for data with a sufficiently high signal-to-noise ratio, the PSF is a crucial ingredient for accurately inferring the power-law slope, and thus the Fermat potential. Crucially, the PSF can be directly reconstructed from the data, thus improving the accuracy of cosmographic analysis. This result reinforces the necessity of PSF reconstruction, which has become best practice in recent years for cosmographic analyses using \texttt{GLEE} and \texttt{Lenstronomy} by members of our team and collaborators \citep[e.g.,][]{Wong2019,Birrer2018,Shajib2019,Chen2019}.
The counterpoint is that data of insufficient quality should not be used for cosmographic analysis, unless an accurate and informative prior is available. 

We conducted the same test with a subsampled version of the \texttt{Lenstronomy} PSF. It is subsampled in the same way as the \texttt{GLEE} PSF. The modeling results do not agree as well with the \texttt{Lenstronomy} values, and are closer to the \texttt{GLEE} values. This comparison shows that although subsampling in the IR band is important (\citealt{Suyu.2012}, \citealt{Shajib2022}), the modeling results depend on the way in which the PSF is subsampled. It should be done during the PSF reconstruction process, and not as a simple interpolation after the corrections of the PSF.

To conclude, we recall that the goal of this work was not to produce cosmography-ready models, but to automate the modeling procedure for a wide range of lens morphologies that can subsequently enable the construction of cosmography-grade models; thus differences in the results of the two modeling teams are expected. Our analysis shows that the origin of the differences can generally be understood, and that the two most significant factors are data quality and accuracy of PSF modeling.

\subsection{Comparison of astrometry}
\label{sec:astrometry}

To assess the precision of our astrometry, we compared the relative positions of the multiple quasar images from our model with the modeled positions from \citetalias{Schmidt2022} and \citet{Luhtaru2021} and the measured positions from the Gaia satellite data release 3 (\citealt{Brown2021}). Like in our pipeline, \citetalias{Schmidt2022} obtained positions by modeling the surface brightness distribution, while \citet{Luhtaru2021} used geometric properties in the image plane. The comparison with \citetalias{Schmidt2022} and  \citet{Luhtaru2021} was performed for all nine systems in our sample and the comparison with Gaia for seven of the nine systems. For the remaining two systems, no Gaia data are available. 

\begin{figure}[h]
        \centering
        \subfigure{\includegraphics[scale=0.52]{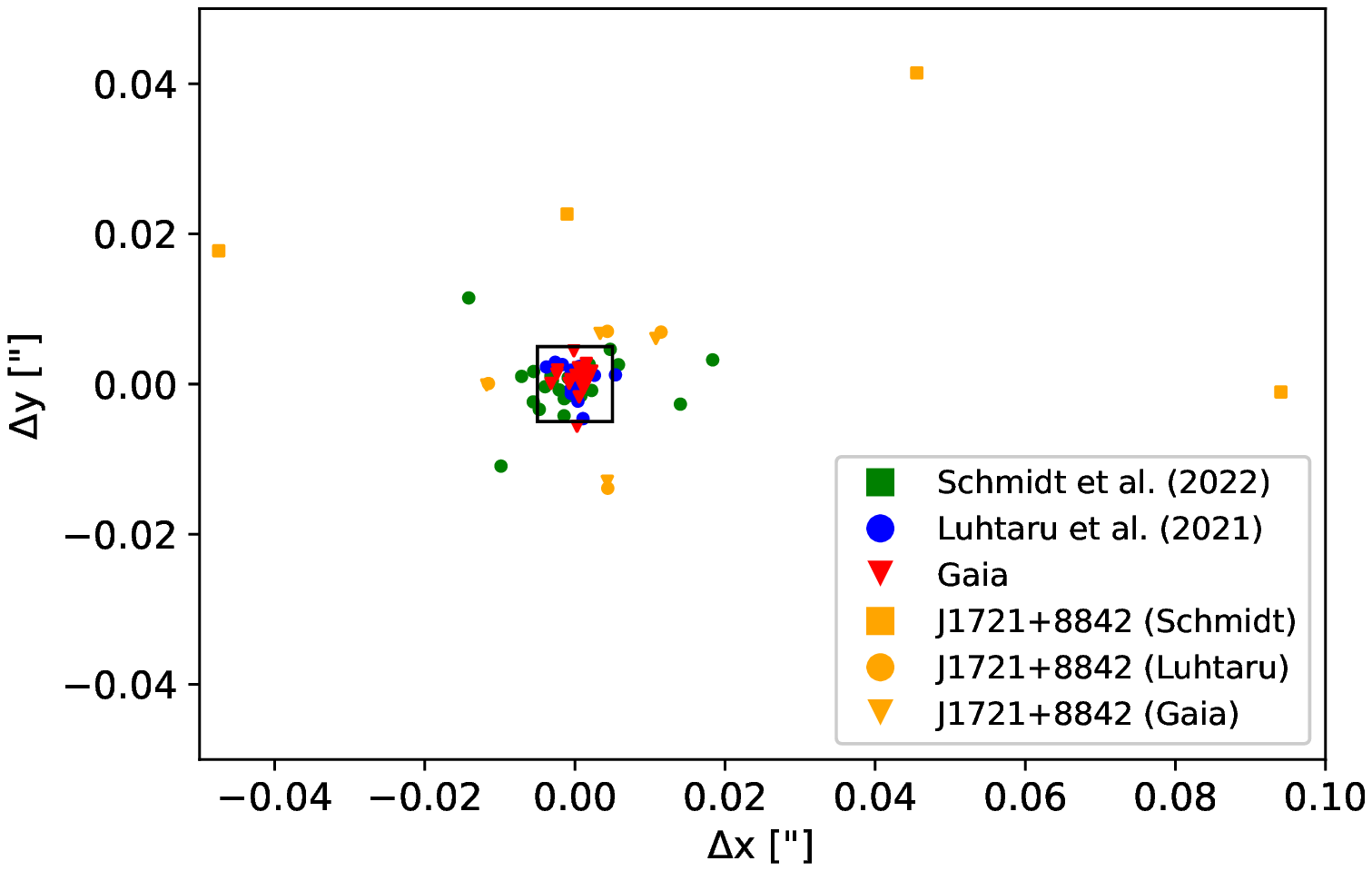}}
        \subfigure{\includegraphics[scale=0.7]{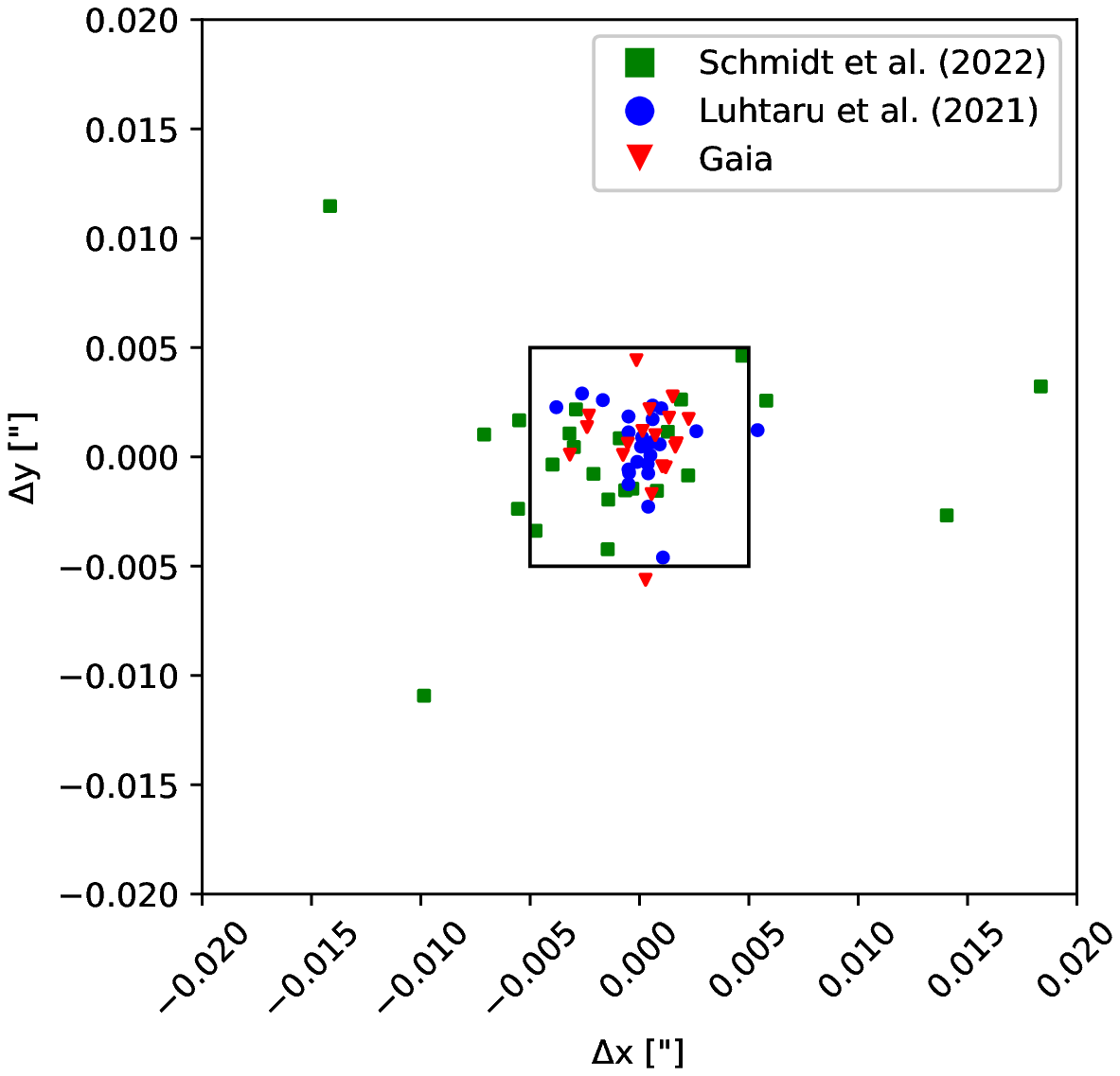}}
        \caption{Relative difference in quasar image positions between our results and \citetalias{Schmidt2022} (squares), \citet[][circles]{Luhtaru2021}, or Gaia (triangles) for systems that overlap in the multiple samples. The image positions of the outlier J1721$+$8842 are marked in orange (\textit{top}) or are excluded (\textit{bottom}).}
        \label{fig:astrometry}
\end{figure}

In all comparisons, J1721$+$8842 has the strongest offsets for all four images of the quad, which is expected because of its complexity (see Sec.~\ref{sec:1721}). When this system is excluded, we obtain the following results with respect to \citetalias{Schmidt2022}: We have an r.m.s.~scatter of $\sim$6 mas in the $x$-direction (RA) and $\sim$5 mas in the $y$-direction (dec). 
The comparison with \citet{Luhtaru2021} reveals an r.m.s.~scatter of $\sim$ 1.7 mas in the $x$- and $y$-directions. The r.m.s.~scatter for the Gaia comparison is 1.7 mas in the $x$- and 2.3 mas in the $y$-direction.
The top panel of Fig.~\ref{fig:astrometry} shows the difference in quasar image positions for the comparison with \citetalias{Schmidt2022}, \citet{Luhtaru2021}, and Gaia for all systems that are in all three samples (seven systems). The images of J1721$+$8842 are marked in orange as the mentioned outlier. The bottom panel shows the difference in quasar image positions for all systems that are in common with \citet{Luhtaru2021} and Gaia. To give an idea of how small the differences are, we include a box centered around zero offset with dimensions $\pm 5$mas. The vast majority of points lie within this box. Although we did not perform PSF reconstruction in our automated modeling pipeline, we have recovered the astrometry of the quasar images within $\sim$2 mas of the Gaia measurements (which are most precise or accurate because Gaia was designed to measure astrometry).

\citet{Birrer2019} provided an estimate of how the astrometric uncertainties translate into the uncertainty on the $H_0$ inference. The $\sim$ 2 mas rms offset to Gaia in our analysis translates into a $\sim$ 0.2 mas offset in the source plane, which leads to uncertainties on $H_0$ well below 5\%, which was chosen by \citet{Birrer2019} as an estimate of the total uncertainties (i.e., from modeling the lens mass potential and the contribution from the mass along the line of sight). This means that the astrometric uncertainties in this work do not contribute significantly to the cosmographic error budget, which shows that our automated pipeline can meet the astrometric requirements.

\subsection{Possible improvements of the automated lens modeling pipeline}
To further improve the automated uniform modeling in combination with detailed cosmographic analyses of lenses, we plan to implement several additions to the code. For the uniform modeling approach, we reconstruct the PSF from several stars in the field. This approach introduces inaccuracies that are negligible for uniform modeling, but need to be minimized for a detailed analysis, as discussed in Sec.~\ref{sec:comparison}. We will automate a method to iteratively update the initial PSF using the multiple lensed quasar images. This PSF correction might also resolve the tendency of the power-law slope parameter $\tilde{\gamma}_{\rm PL}$ to move to the boundaries of the prior. 

To speed up the modeling procedure and make it more convenient, we can automate the detection of extended structures in the cutout, for instance, the positions of multiple quasar images and objects that need to be masked. Other automated modeling codes already show that an accurate detection in the scope of uniform modeling is often possible , for example in \citealt{Savary2021} and \citealt{Rojas2021}, who also showed that objects are occasionally misidentified, and the automated modeling has to be stopped. Our current approach of manually identifying objects is thus a compromise to ensure very high accuracy. This works best for a medium-size sample (e.g., a few dozen), but is not applicable for samples of some hundred lenses or more.

The initial background subtraction of the data can be made more accurate by allowing for the selection of multiple sky regions. This is especially important for systems that show a gradient in background flux.  Another improvement can be using a $p$-value threshold for the lens light modeling (Sec.~\ref{sec:lenslight} and \ref{sec:multiband}). This would decrease the probability that the $\chi^2_{\rm red}$ criterion (Eq.~\ref{lenschi2red}) is met only due to specific noise realizations.

Other modeling teams have shown that optimizing methods that are based on gradient descent and can be heavily parallelized are superior to conventional methods in terms of computation time, for instance, \texttt{GIGA-Lens} (\citealt{Gu2022}). These directions are worth exploring for future developments of automated lens modeling.

\section{Conclusions}
\label{sec:conclude}

We presented a new automated modeling code for the uniform modeling of strongly lensed quasars in galaxy-scale systems with high-resolution data based on the well-tested software \texttt{GLEE}. We additionally developed several codes that create necessary input files (i.e., PSF, error map, and masks) that are used for the modeling. The lens mass distribution of the system is modeled in two steps. In the first step, the modeling code uses the predicted source position and the observed image positions to constrain the distribution of the lens mass parameters. In the second step, the light distributions of the lens, the multiple lensed quasar images, and their host galaxy are modeled. The latter is used to reconstruct the surface brightness distribution of the source. With the SSB distribution, the code models the light of the arc to further constrain the lens mass parameters. The modeling code and the creation of input files are nearly fully automated, requiring only minimum user input. Quality control and technical steps in the modeling process are fully automated. 

We tested the modeling code on a sample of nine strongly lensed quasars and obtained a good light fit and a good alignment of the mass and light distributions. The current automated pipeline is versatile enough to model a variety of lens systems in a uniform manner to obtain lens mass models.  
The models provide robust estimates of Einstein radii and astrometry of the quasar images. The accuracy of parameters such as the power-law slope depends crucially on the data quality and on the details of the modeling. This is highlighted by a blind comparison of our models with those of an  automated modeling framework from \citetalias{Schmidt2022}, based on the lens modeling code \texttt{Lenstronomy}. The two approaches are similar in the choice of parameterization and philosophy, but have a few crucial differences: the description of the lensed galaxy light (pixellated vs. \sersic+shapelets), the number of modeled bands (user choice vs. all available bands), priors (uniform vs. informative), and the PSF (initial guess based on stars vs. corrections based on the QSO images for \citetalias{Schmidt2022}). Despite the differences, Einstein radii and mass flattening agree well between the two studies, with a few exceptions that can be traced to inadequacy of the data or the PSF modeling.
Other quantities such as convergence and image magnifications show differences that are generally larger than the estimated formal statistical uncertainties, but they are still encouraging in amplitude, considering the automated approach.

The differences primarily arise from two effects, as shown by our post-blind analysis. First, when the data quality is insufficient to constrain the slope of the power-law mass density profile, the \citetalias{Schmidt2022} results are driven by the prior, while our current results tend to be more uncertain and constrained by the bounds of the uniform prior. We showed that if the same power-law slope is imposed, the agreement between the codes improves significantly.
Second, the reconstruction of the PSF is a key factor in determining the slope and other parameters for data of sufficient quality. We illustrated this point by modeling two high-quality systems with the same PSF as \citetalias{Schmidt2022}, and we found a much better agreement.
We conclude that for systems with high S/N, the PSF can be directly reconstructed from the data and the power-law slope can be stably inferred (this confirms the results obtained via detailed modeling by 
\citet{Wong2019,Birrer2018,Shajib2019,Chen2019,Shajib2022}). 

In terms of our automated modeling pipeline, we conclude that the models are a good starting point, but more work, particularly PSF corrections, and in some cases, better data, are needed to reach cosmography grade. 
Nonetheless, these automated modeling results provide important information about the lens systems, such as the approximate time delays, to help schedule the monitoring of the system. The lensing systems modeled with our automated pipeline are being followed up observationally to acquire the redshift and velocity dispersion of the lens, time delays, and environment properties, for example. We can use the modeling results that are presented here and models from lensing systems that are obtained with the automation code in the future, as input models for a more detailed modeling, in a fraction of the user time compared to conventional modeling.


\begin{acknowledgements}
We thank P. Schechter for data on the astrometric comparison and for helpful discussions. We further thank the anonymous referee for helpful comments.
This research is based on observations made with the NASA/ESA Hubble Space Telescope obtained from the Space Telescope Science Institute, which is operated by the Association of Universities for Research in Astronomy, Inc., under NASA contract NAS 5-26555. These observations are associated with programs HST-GO-15320 and HST-GO-15652. Support for the two programs was provided by NASA through a grant from the Space Telescope Science Institute, which is operated by the Association of Universities for Research in Astronomy, Inc., under NASA contract NAS 5-26555.

  SE, SS, and SHS thank the Max Planck Society for support through the Max Planck Research Group for SHS.

This project has received funding from the European Research Council (ERC)
under the European Union’s Horizon 2020 research and innovation
programme (LENSNOVA: grant agreement No 771776; COSMICLENS: grant
agreement No 787886).
This research is supported in part by the Excellence Cluster ORIGINS which is funded by the Deutsche Forschungsgemeinschaft (DFG, German Research Foundation) under Germany's Excellence Strategy -- EXC-2094 -- 390783311.

 TS and TT acknowledge support by the the National Science Foundation through grant NSF-AST-1906976 and NSF-AST-1907396 "Collaborative Research: Toward a 1\% measurement of the Hubble Constant with gravitational time delays". TT acknowledges support by the Packard Foundation through a Packard Research Fellowship. 

This work made use of \texttt{NumPy} \citep{Oliphant2015}, \texttt{SciPy} \citep{Jones2001}, \texttt{astropy} \citep{AstropyCollaboration2018}, \texttt{matplotlib} \citep{Hunter2007}, and \texttt{draw.io} at \href{https://www.draw.io}{https://www.draw.io}.
\end{acknowledgements}

\bibliographystyle{aa}
\bibliography{automation_paper}


\begin{appendix}

\begin{landscape}

\section{Final modeling results}

\begin{table}[h]
\caption{Lens modeling results.}
                \begin{tabularx}{\linewidth}{c|c|ccc|cc}\toprule \toprule
                        System & Filter & Observed & Model & \begin{tabular}{@{}c@{}}Normalized \\residuals\end{tabular}  & \begin{tabular}{@{}c@{}}Reconstructed \\arc\end{tabular} & \begin{tabular}{@{}c@{}}Reconstructed \\source\end{tabular} \\ \toprule \toprule
                        DES J0029$-$3814 & F160W & \raisebox{-.5\height}{\includegraphics[height=0.2\textwidth]{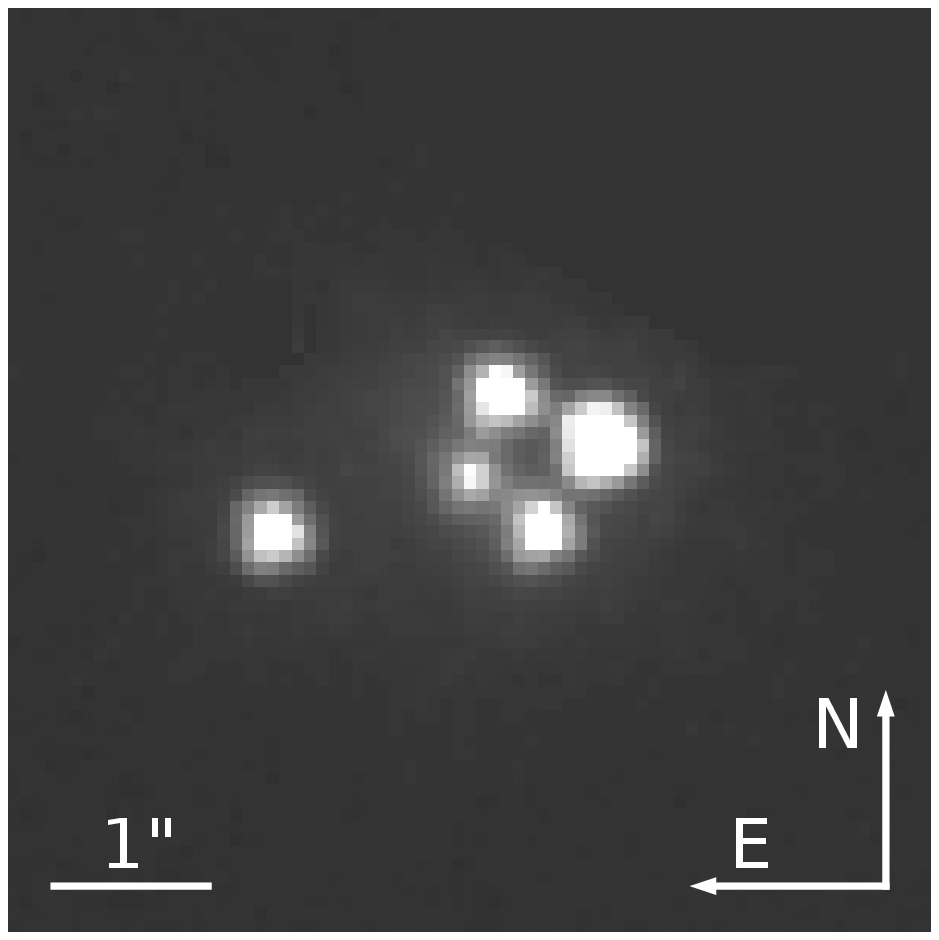}}  & \raisebox{-.5\height}{\includegraphics[height=0.2\textwidth]{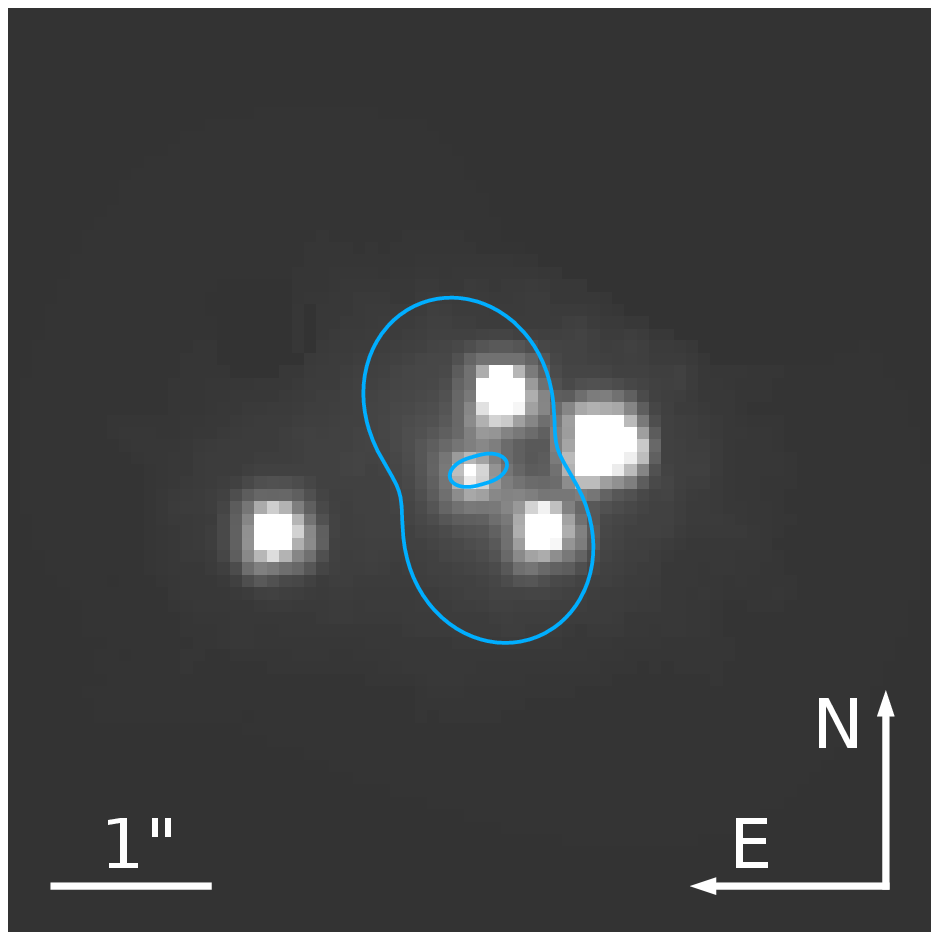}}  & \raisebox{-.5\height}{\includegraphics[height=0.2\textwidth]{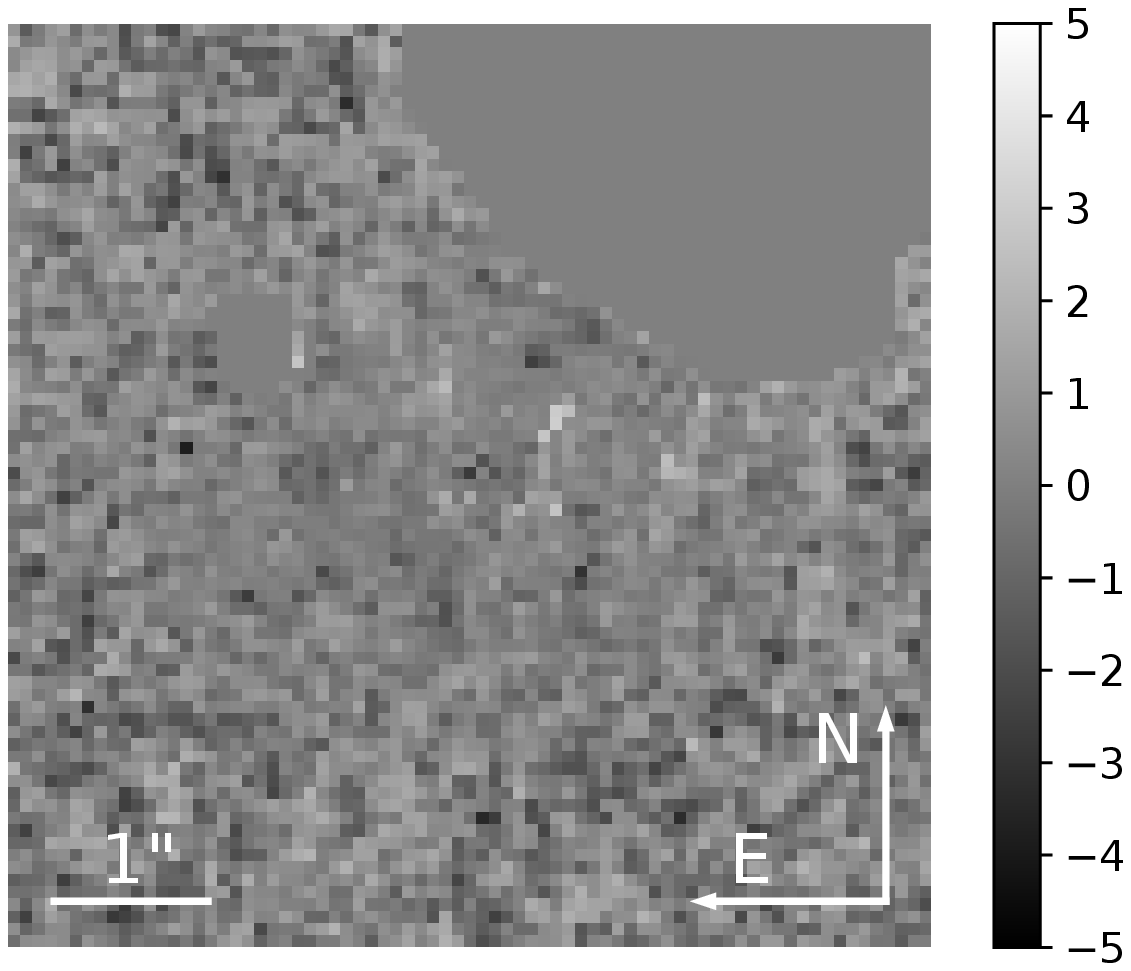}}  & 
                        \raisebox{-.5\height}{\includegraphics[height=0.2\textwidth]{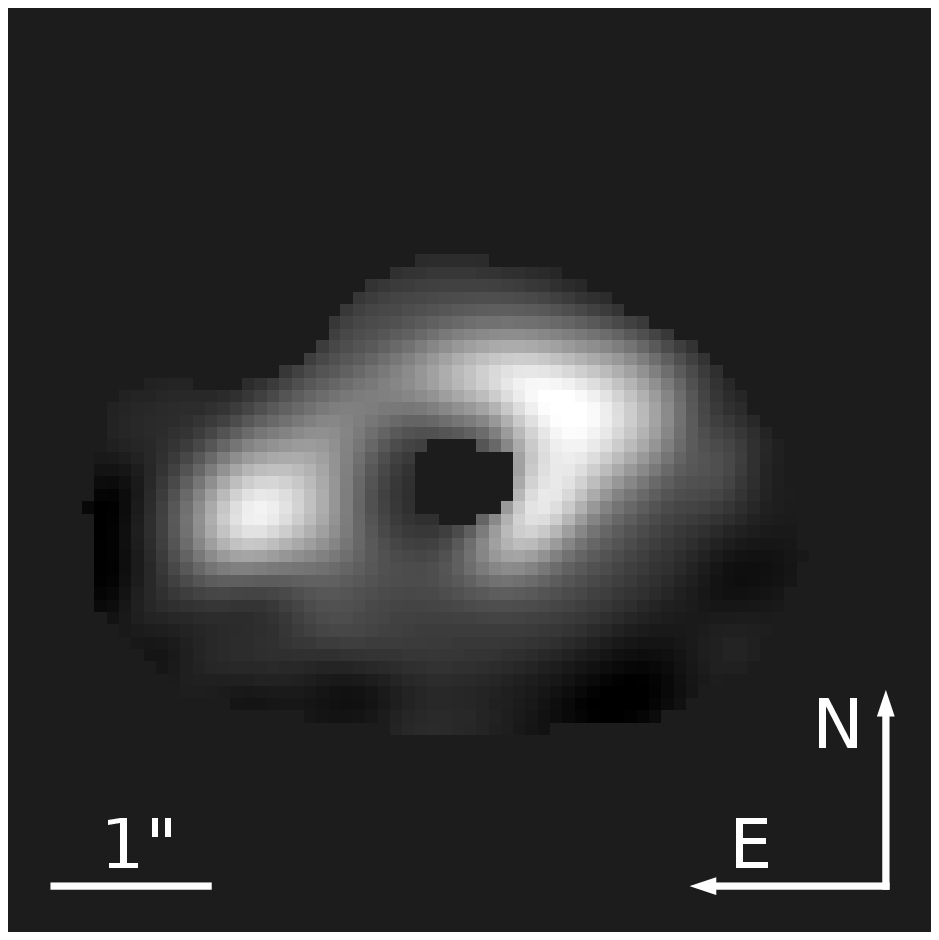}}  & 
                        \raisebox{-.5\height}{\includegraphics[width=0.2\textwidth]{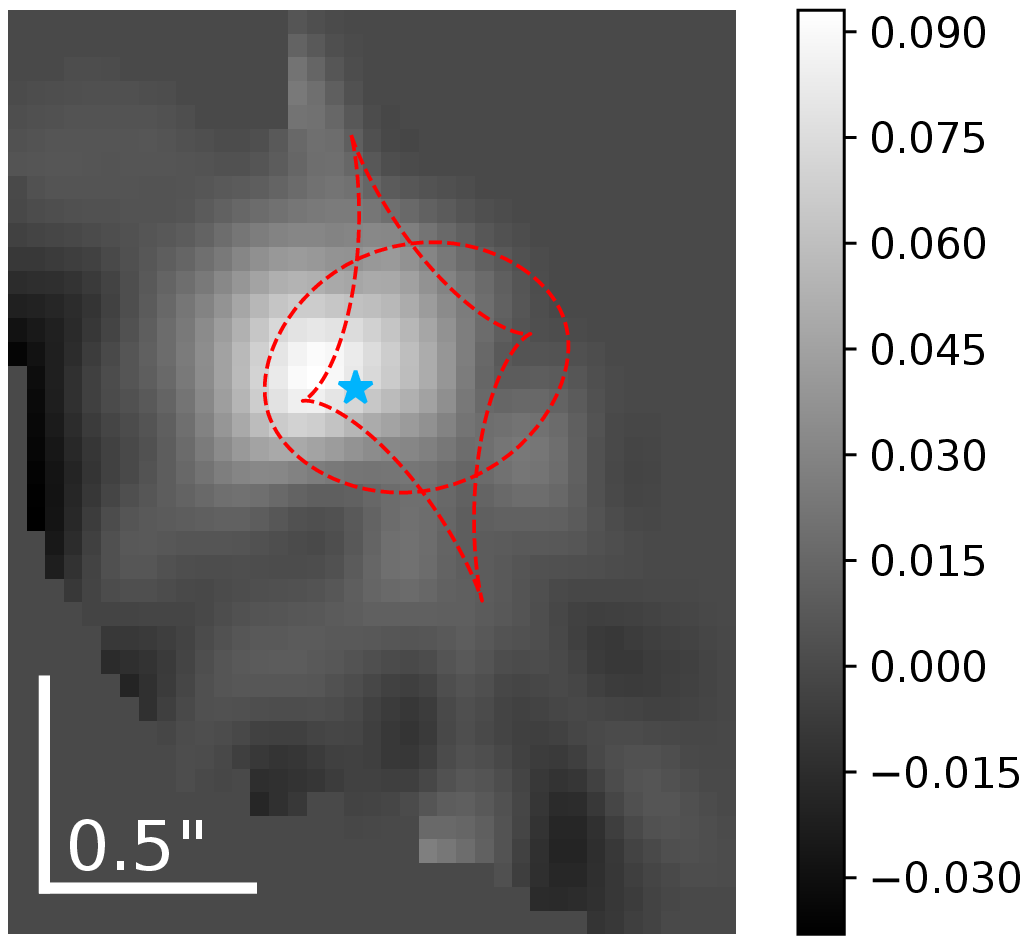}} \\ \midrule
                        DES J0214$-$2105 & F160W & \raisebox{-.5\height}{\includegraphics[height=0.2\textwidth]{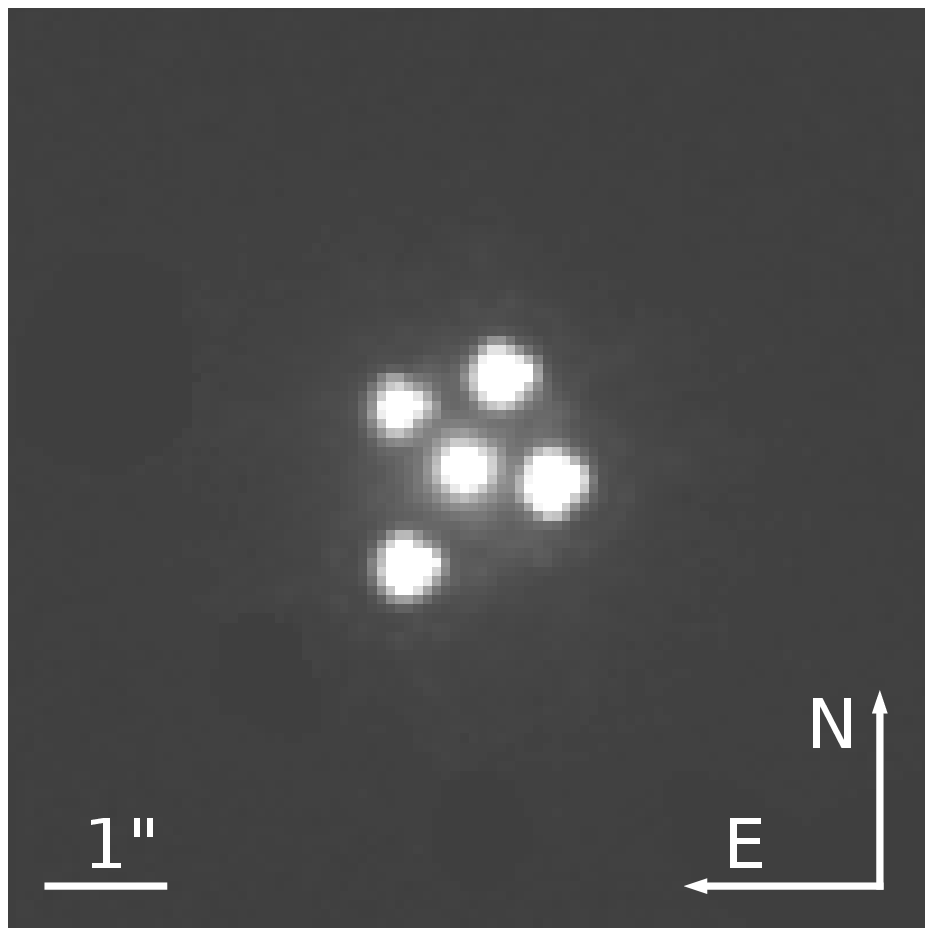}}  & \raisebox{-.5\height}{\includegraphics[height=0.2\textwidth]{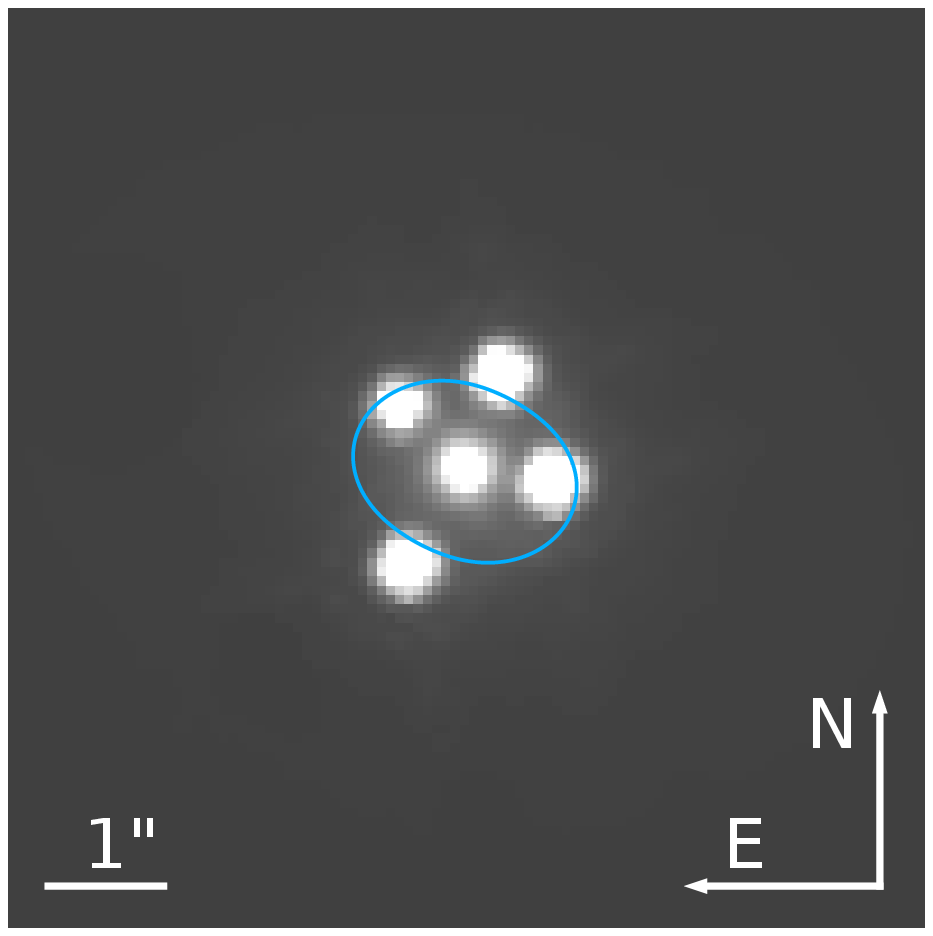}}  & \raisebox{-.5\height}{\includegraphics[height=0.2\textwidth]{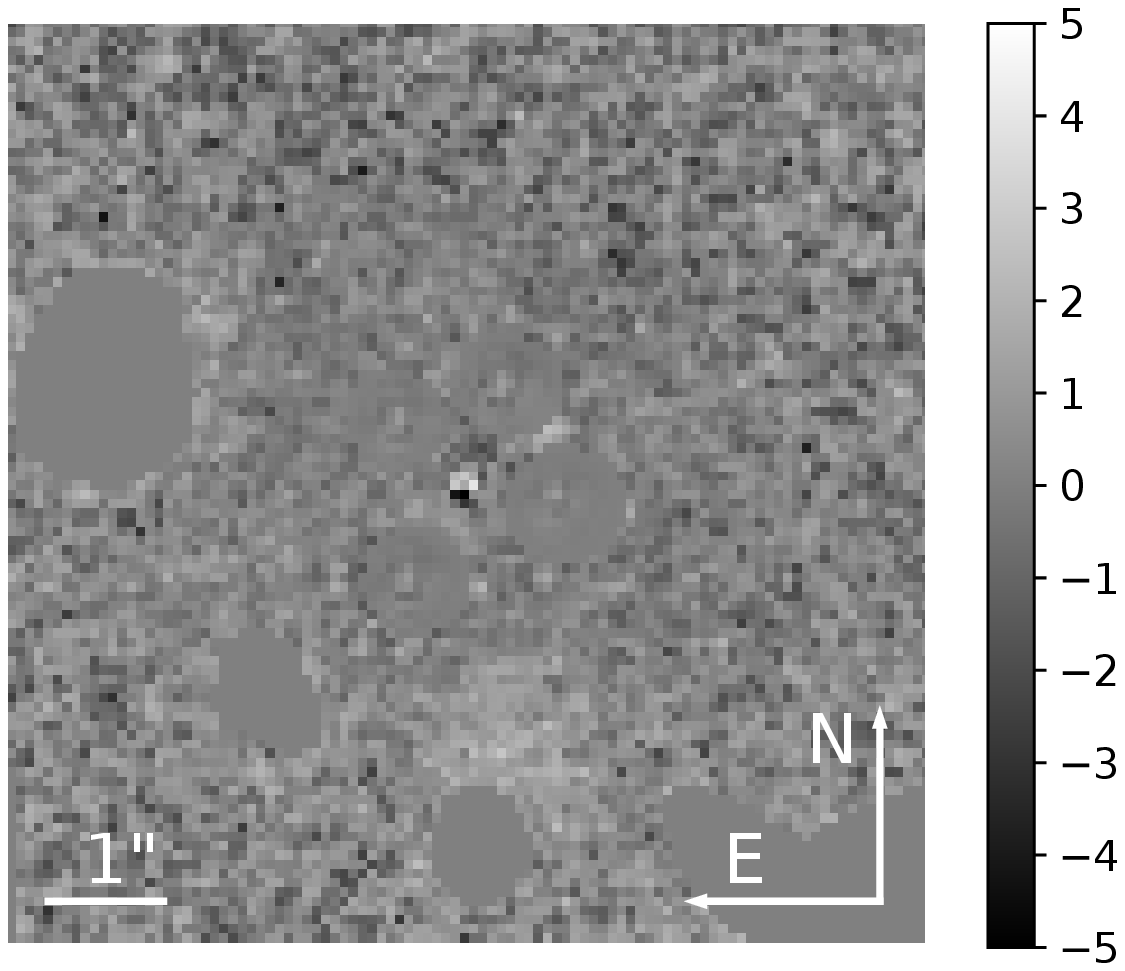}}  & 
                        \raisebox{-.5\height}{\includegraphics[height=0.2\textwidth]{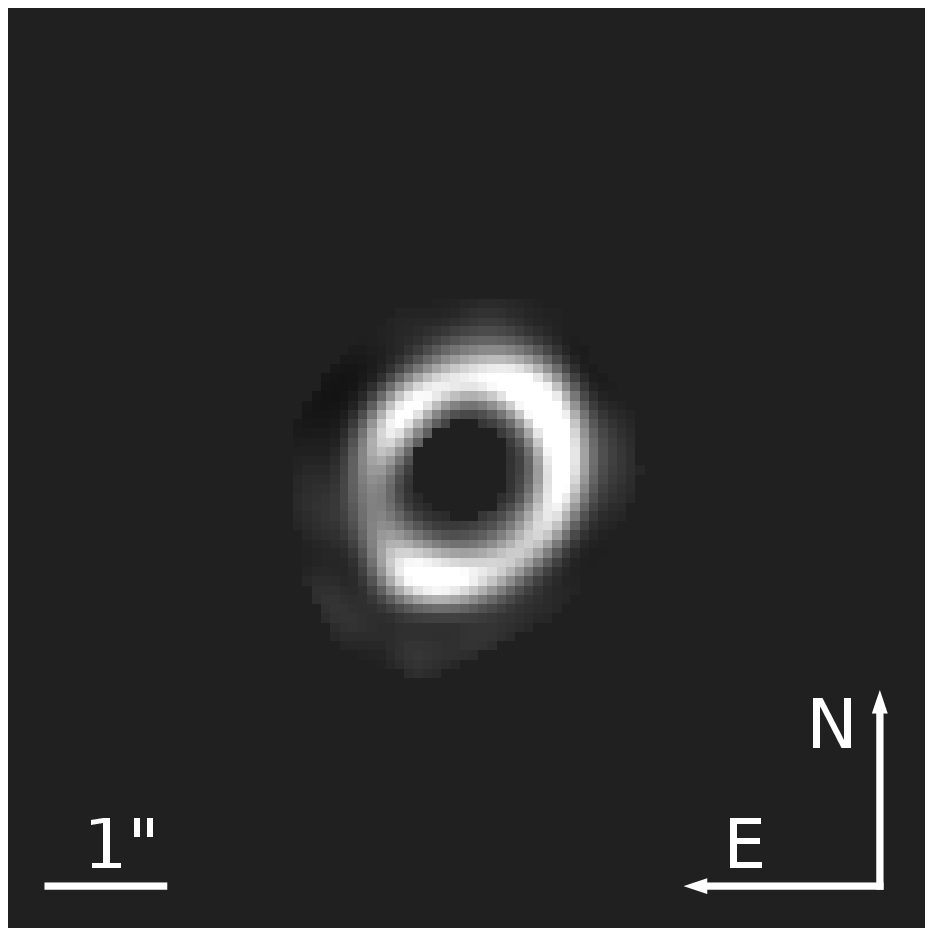}}  & 
                        \raisebox{-.5\height}{\includegraphics[width=0.2\textwidth]{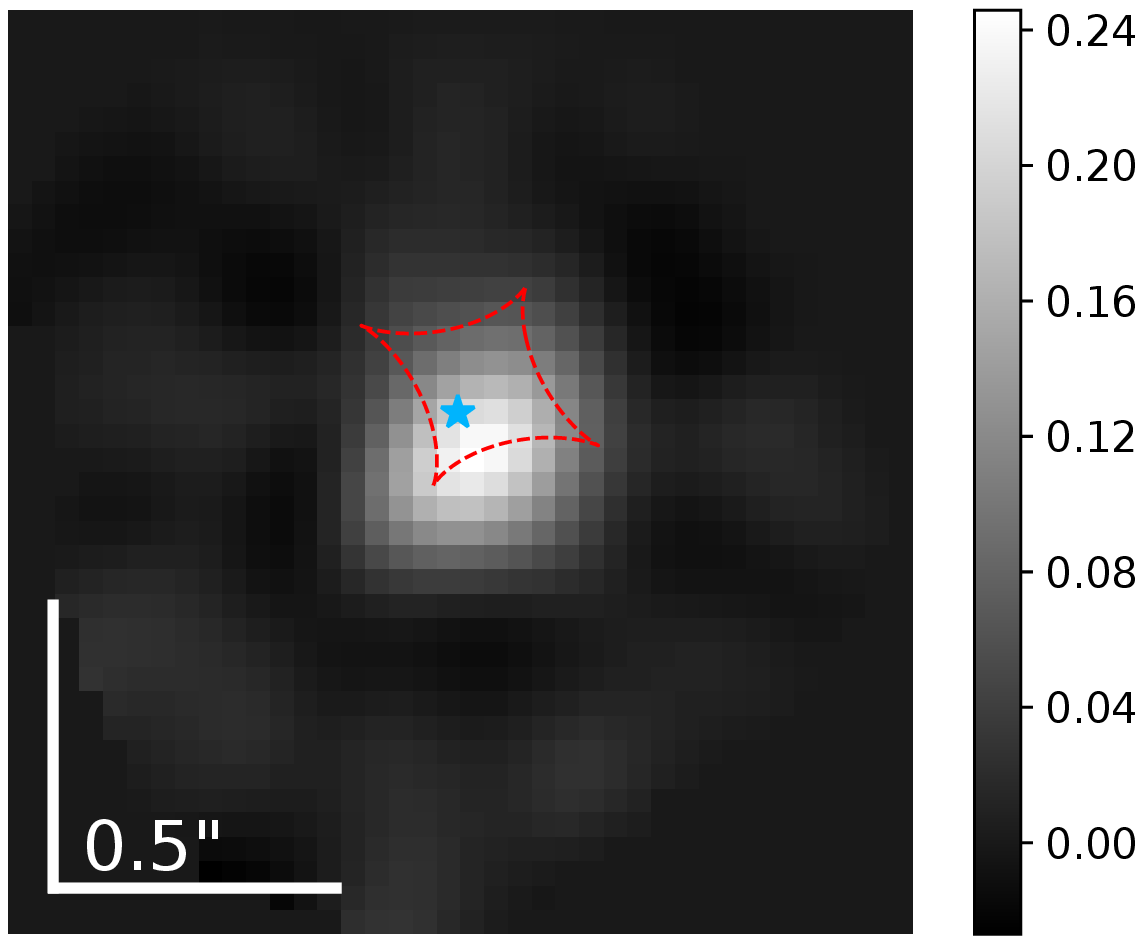}} \\ \midrule
                        DES J0420$-$4037 & F160W & \raisebox{-.5\height}{\includegraphics[height=0.2\textwidth]{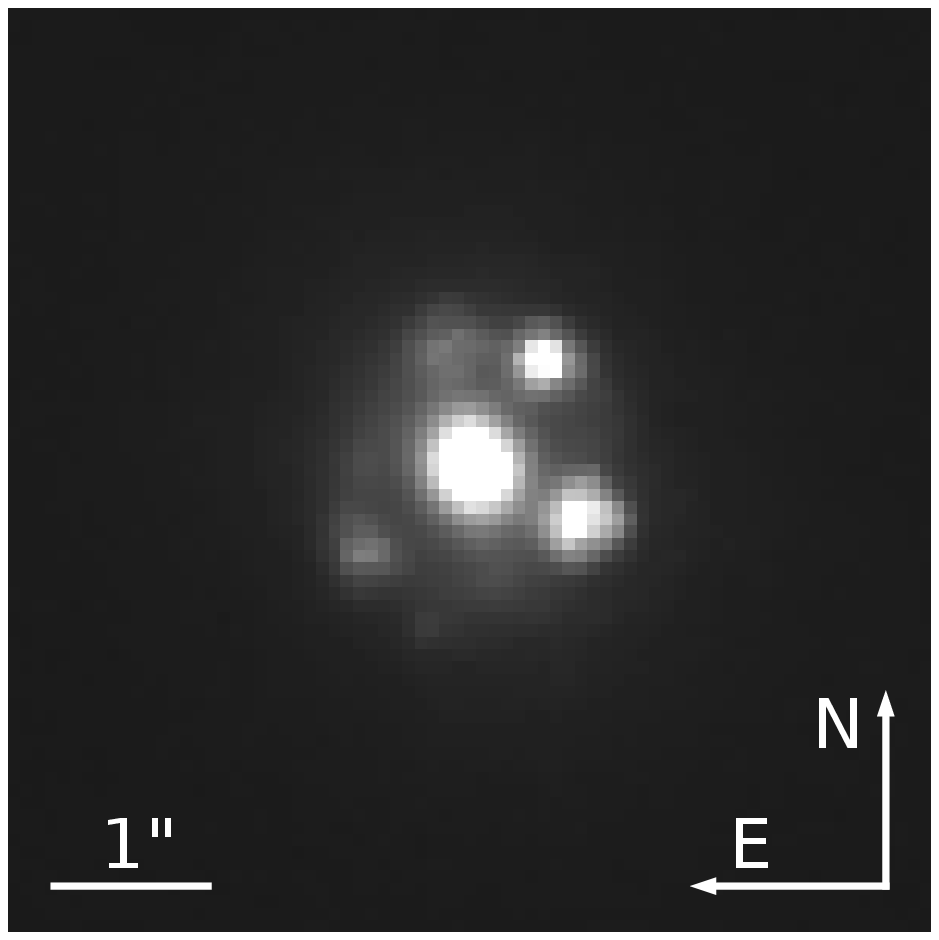}}  & \raisebox{-.5\height}{\includegraphics[height=0.2\textwidth]{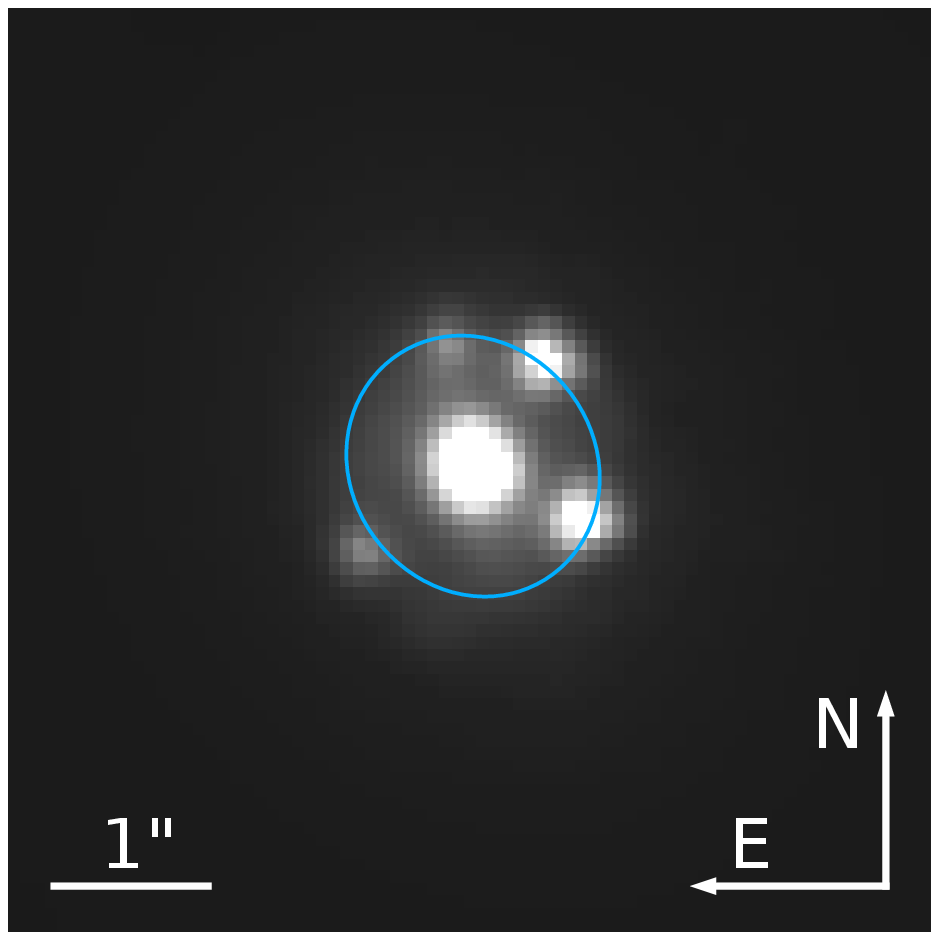}}  &  \raisebox{-.5\height}{\includegraphics[height=0.2\textwidth]{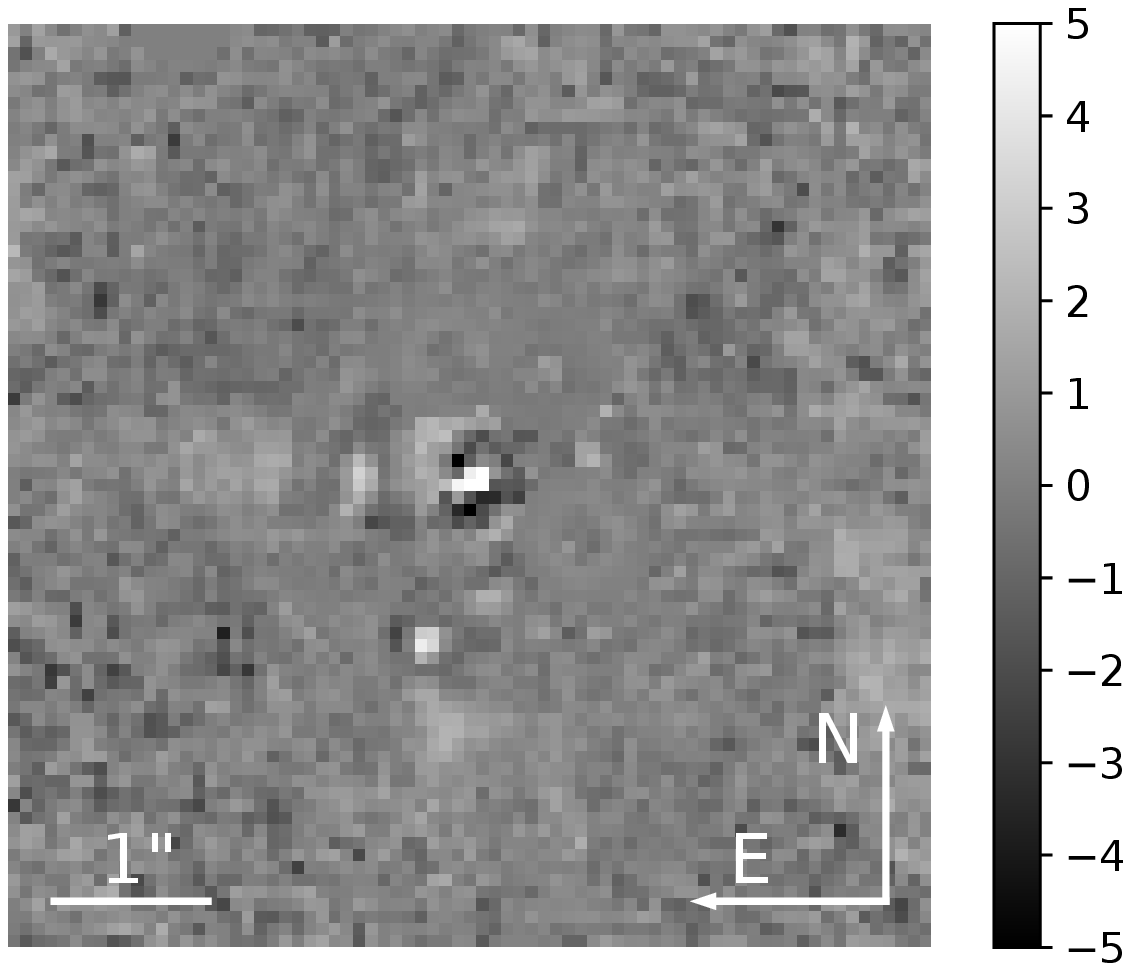}}  &
                        \raisebox{-.5\height}{\includegraphics[height=0.2\textwidth]{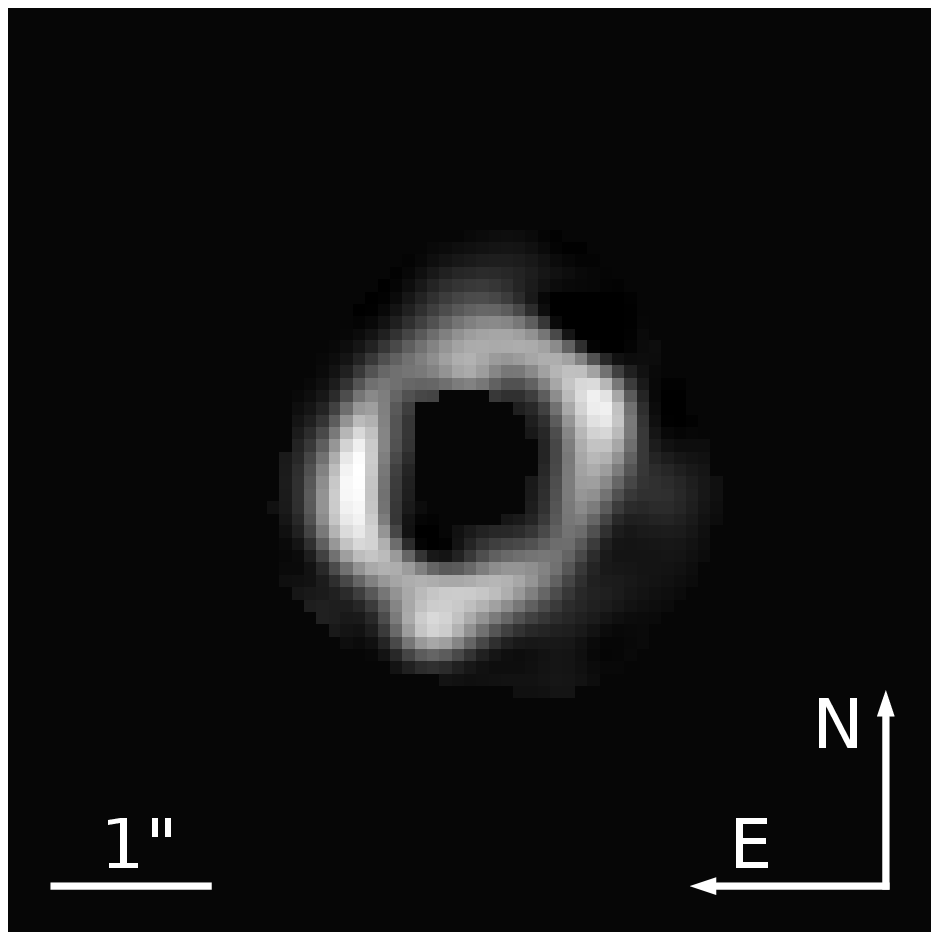}}  & 
                        \raisebox{-.5\height}{\includegraphics[width=0.2\textwidth]{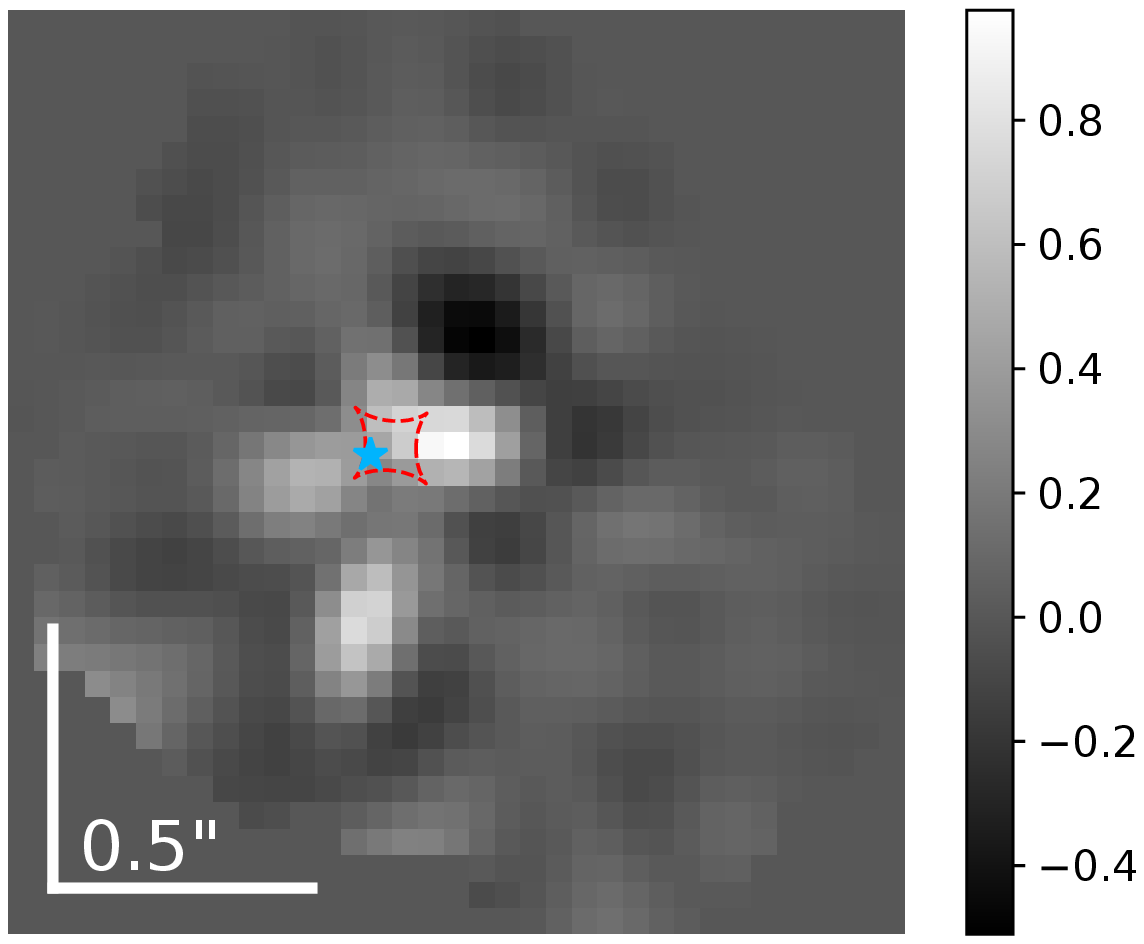}} \\ \bottomrule

                \end{tabularx}
        \ \\ \ \\
        \caption*{\textbf{Notes: }We modeled \textit{HST} filters where the arcs are visible, as indicated in the first two columns.   Column (3): Observed \textit{HST} image.  Column (4): Reconstructed image from our most probable model with critical curves (blue, solid lines).  Column (5): Normalized image residuals, i.e., the difference between (3) and (4), normalized by the estimated $1\sigma$ uncertainties.  Column (6): reconstructed intensity distribution of the lensed quasar host galaxy.  Column (7): Reconstructed quasar-host intensity distribution on a grid of pixels with caustic curves (red, dashed lines) and mean, weighted source position of the quad (blue star) and double (green star) quasar systems.  We show cropped images instead of the full cutout used for modeling for better visibility. The horizontal white line shows 1'' for each system and the figures are oriented such that north is up and east is left. The white lines in the source reconstruction indicate lengths of 0.5'' in the $x$- and $y$-direction.}
        \label{tab:results_light}
\end{table}
\clearpage
\begin{table}[]
                \captionsetup{labelformat=empty} 
            \caption*{\textbf{Table \ref{tab:results_light} continued.}}
            
                \begin{tabularx}{\linewidth}{c|c|ccc|cc}\toprule \toprule
        System & Filter & Observed & Model & \begin{tabular}{@{}c@{}}Normalized \\residuals\end{tabular}  &  \begin{tabular}{@{}c@{}}Reconstructed \\arc\end{tabular} & \begin{tabular}{@{}c@{}}Reconstructed \\source\end{tabular}\\ \toprule \toprule
                        PS J0659$+$1629 & F160W & \raisebox{-.5\height}{\includegraphics[height=0.2\textwidth]{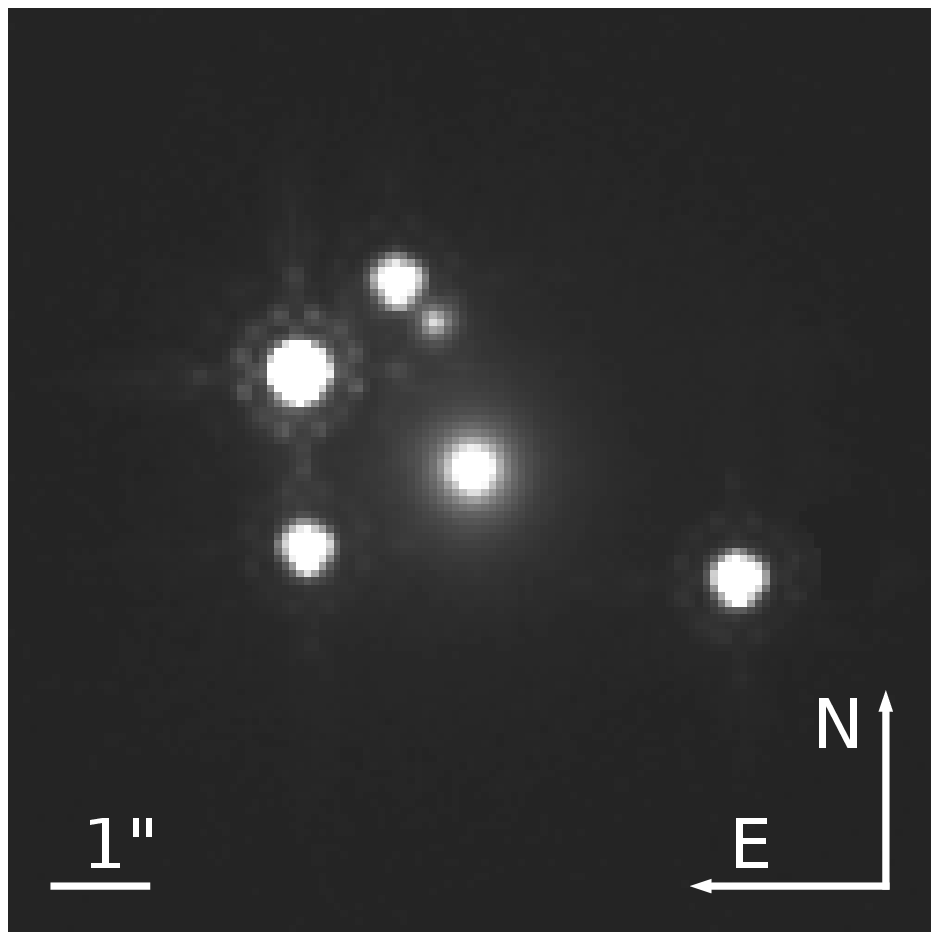}}  & \raisebox{-.5\height}{\includegraphics[height=0.2\textwidth]{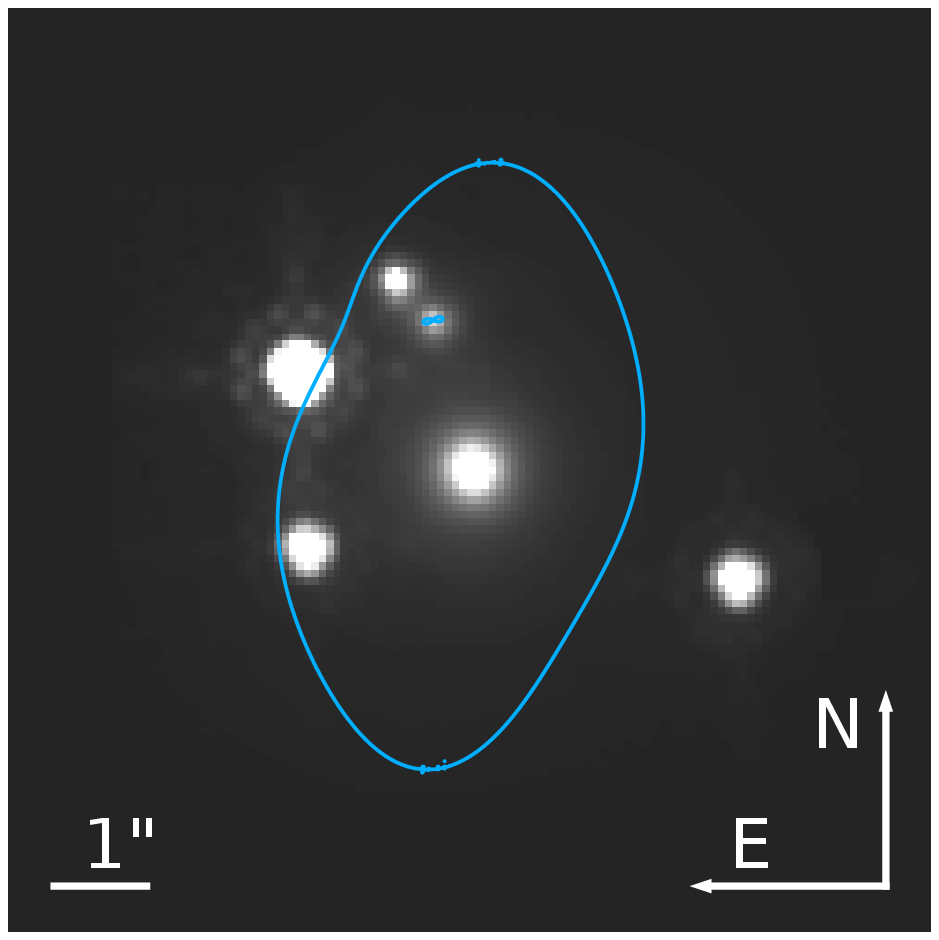}}  &  \raisebox{-.5\height}{\includegraphics[height=0.2\textwidth]{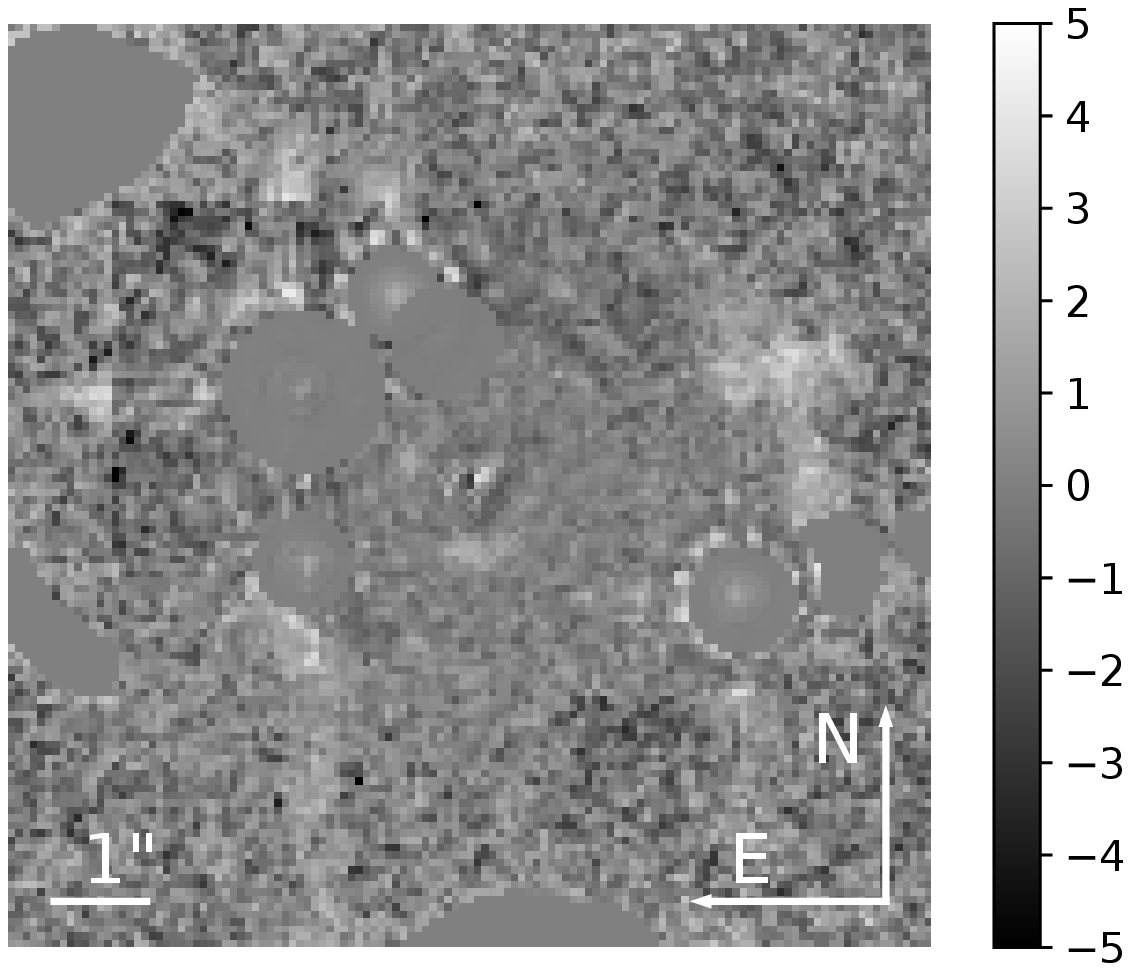}}  &
                        \raisebox{-.5\height}{\includegraphics[height=0.2\textwidth]{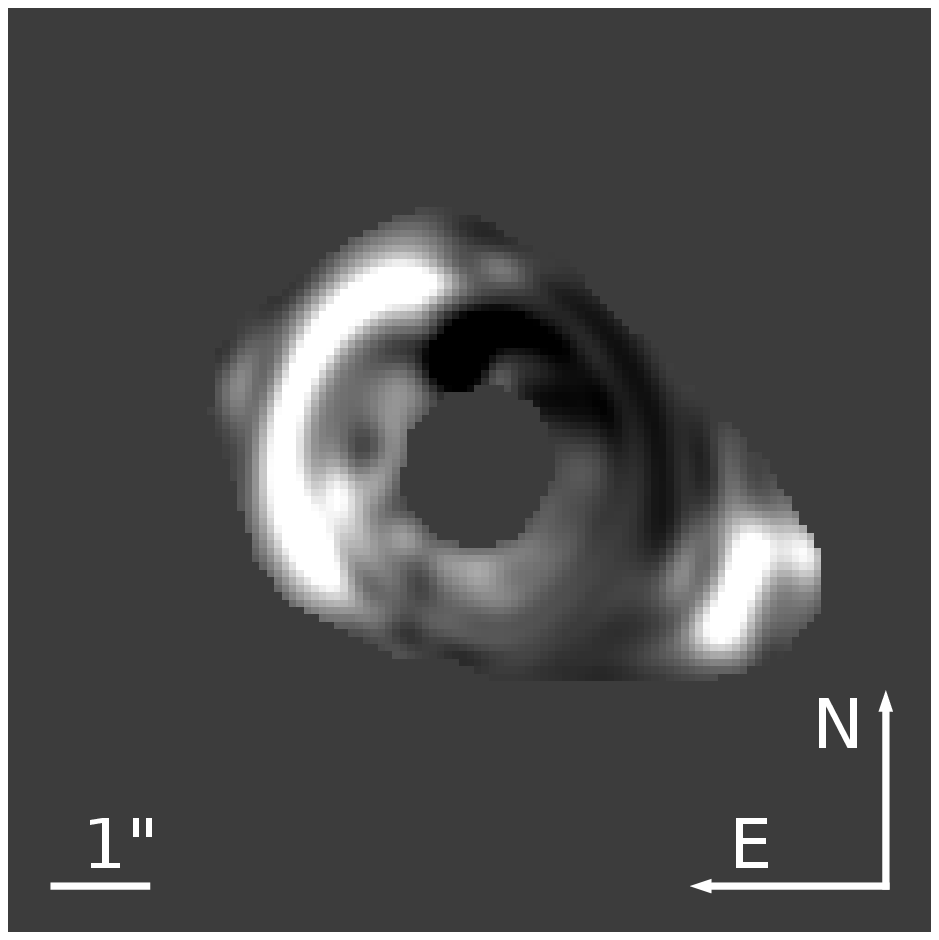}}  & 
                        \raisebox{-.5\height}{\includegraphics[width=0.2\textwidth]{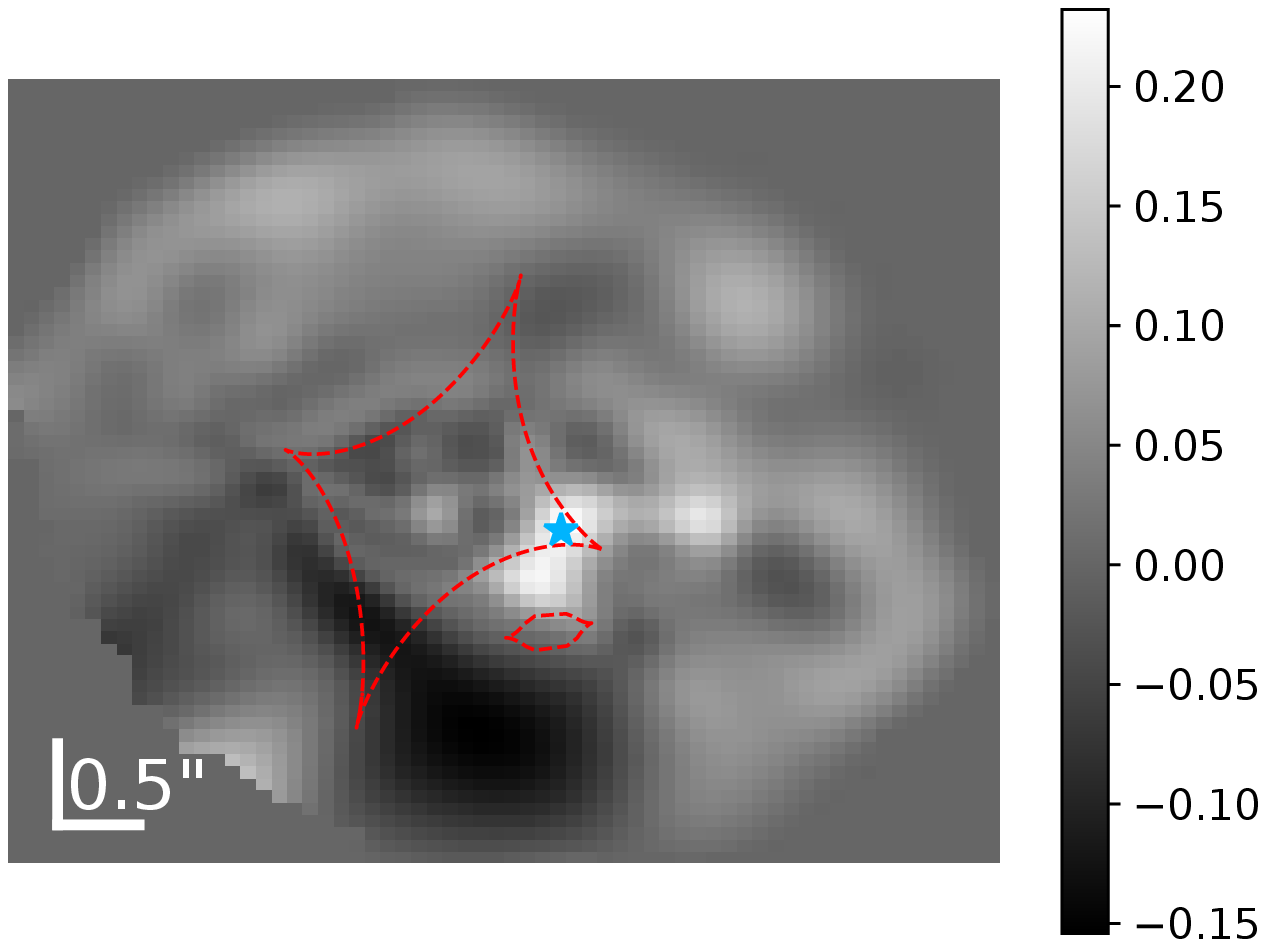}} \\ \midrule
                        2M1134$-$2103 & F160W & \raisebox{-.5\height}{\includegraphics[height=0.2\textwidth]{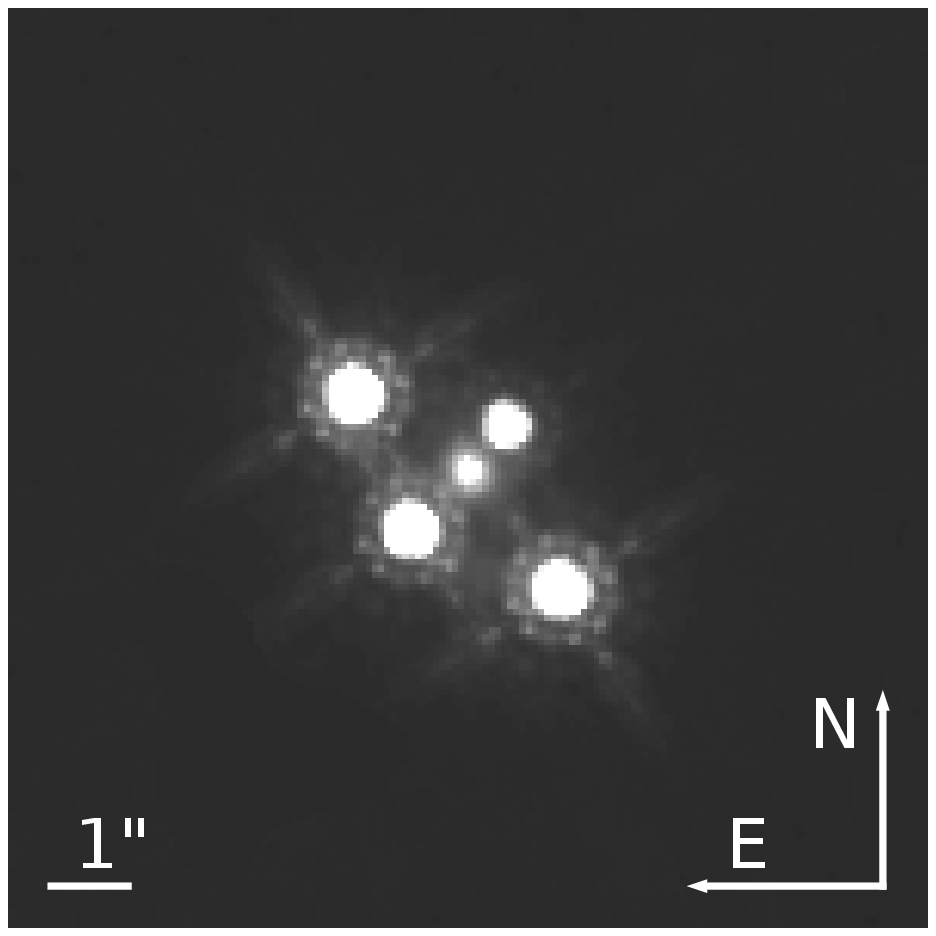}}  & \raisebox{-.5\height}{\includegraphics[height=0.2\textwidth]{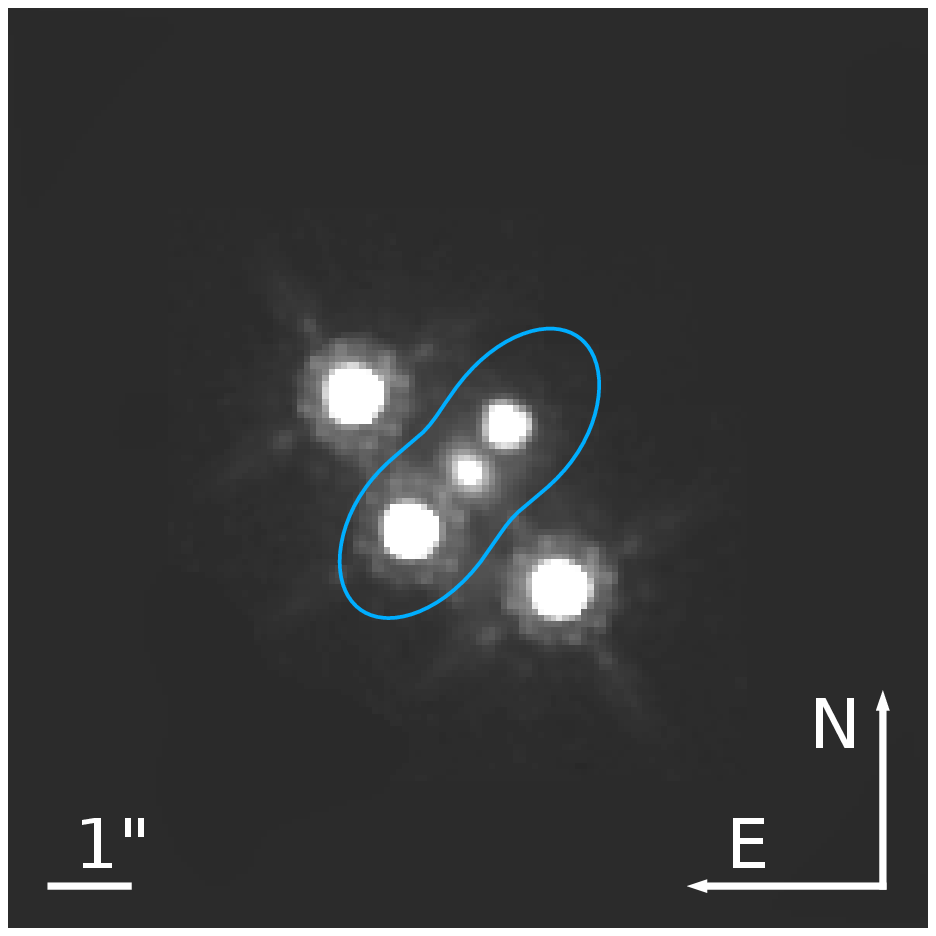}}  & \raisebox{-.5\height}{\includegraphics[height=0.2\textwidth]{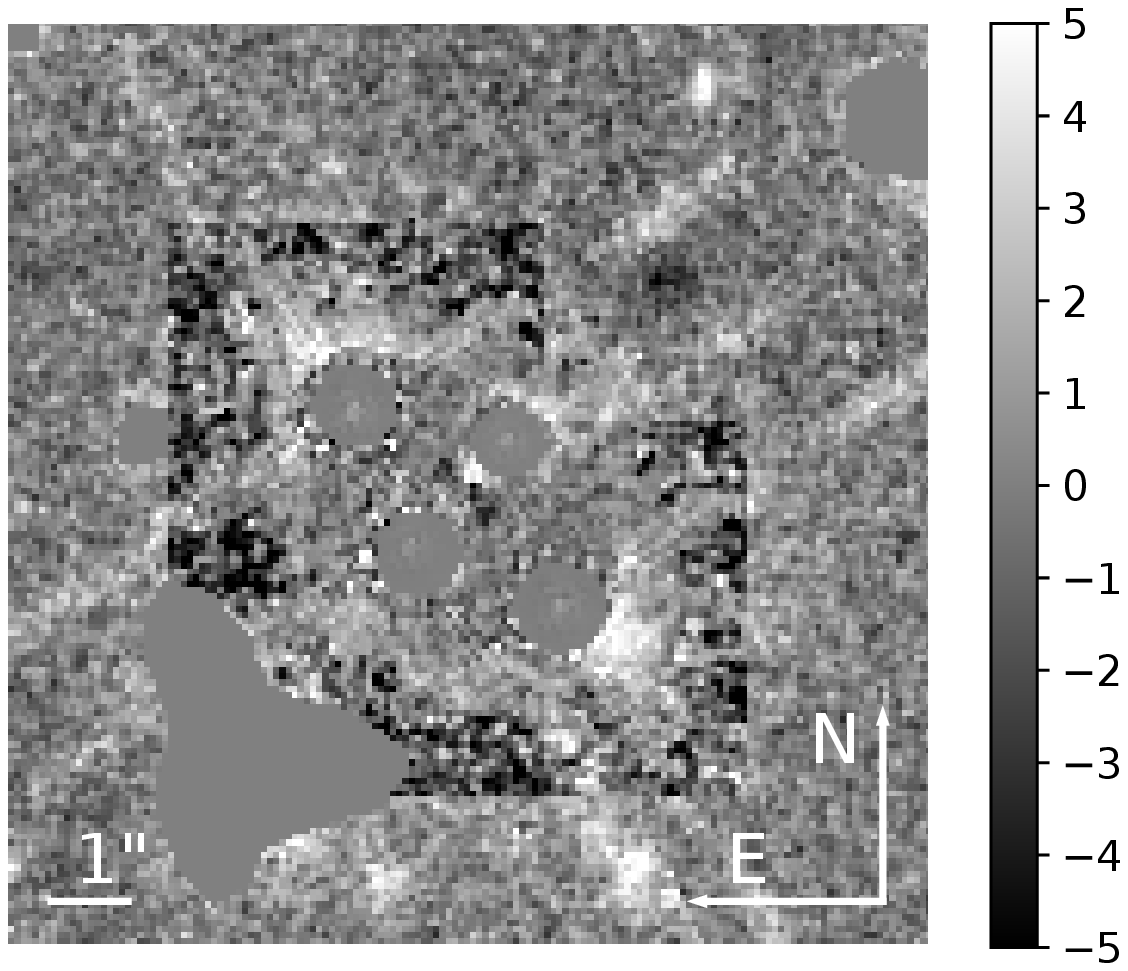}}  &
                        \raisebox{-.5\height}{\includegraphics[height=0.2\textwidth]{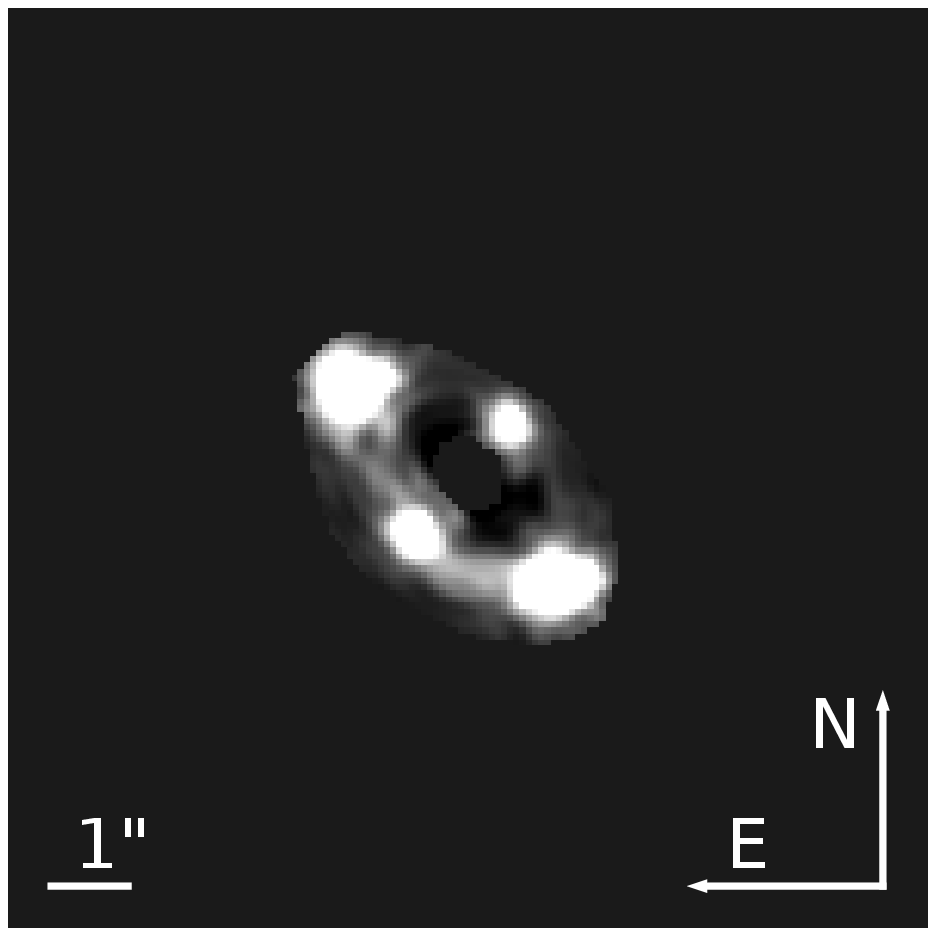}}  & 
                        \raisebox{-.5\height}{\includegraphics[width=0.2\textwidth]{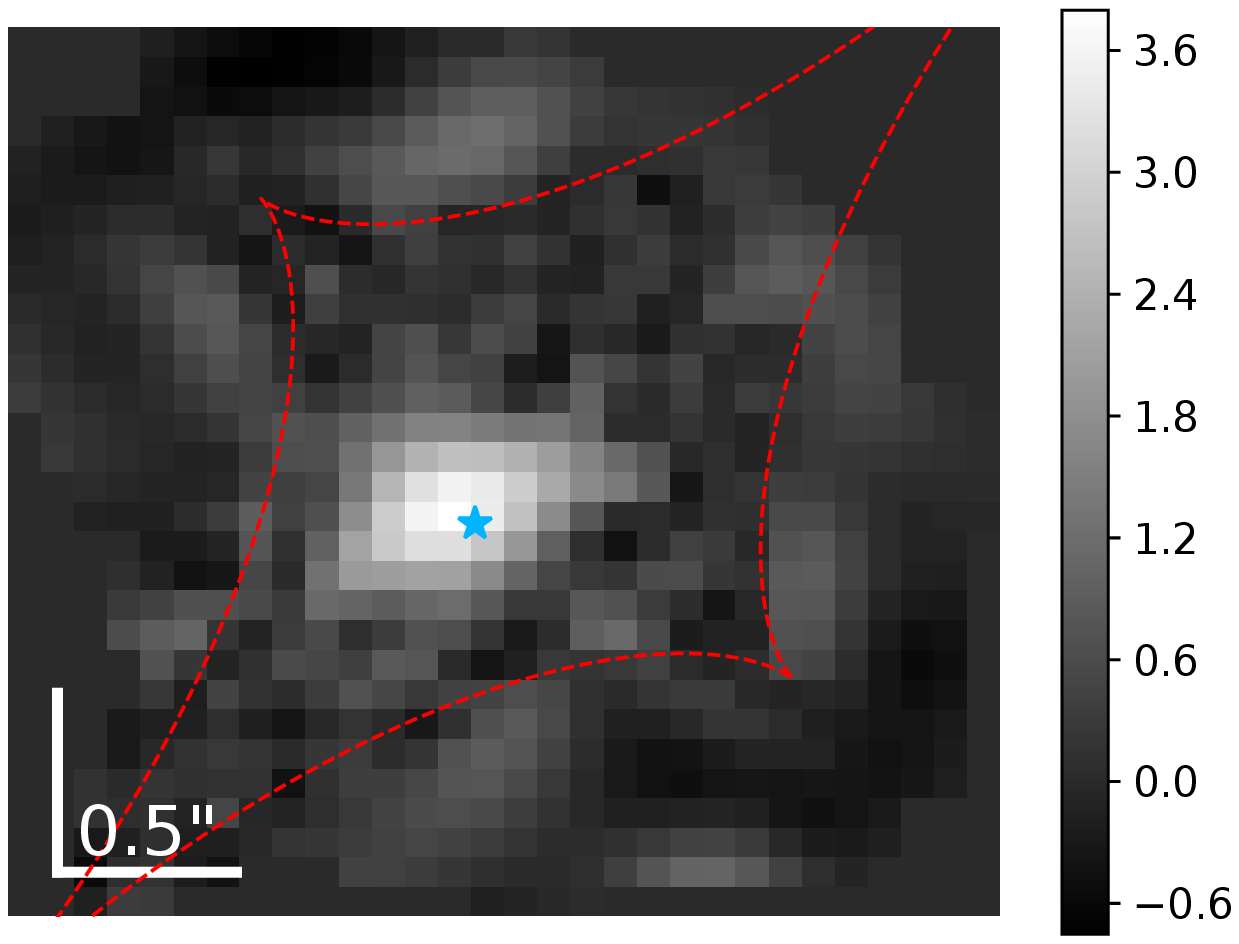}} \\ 
                        & F814W & \raisebox{-.5\height}{\includegraphics[height=0.2\textwidth]{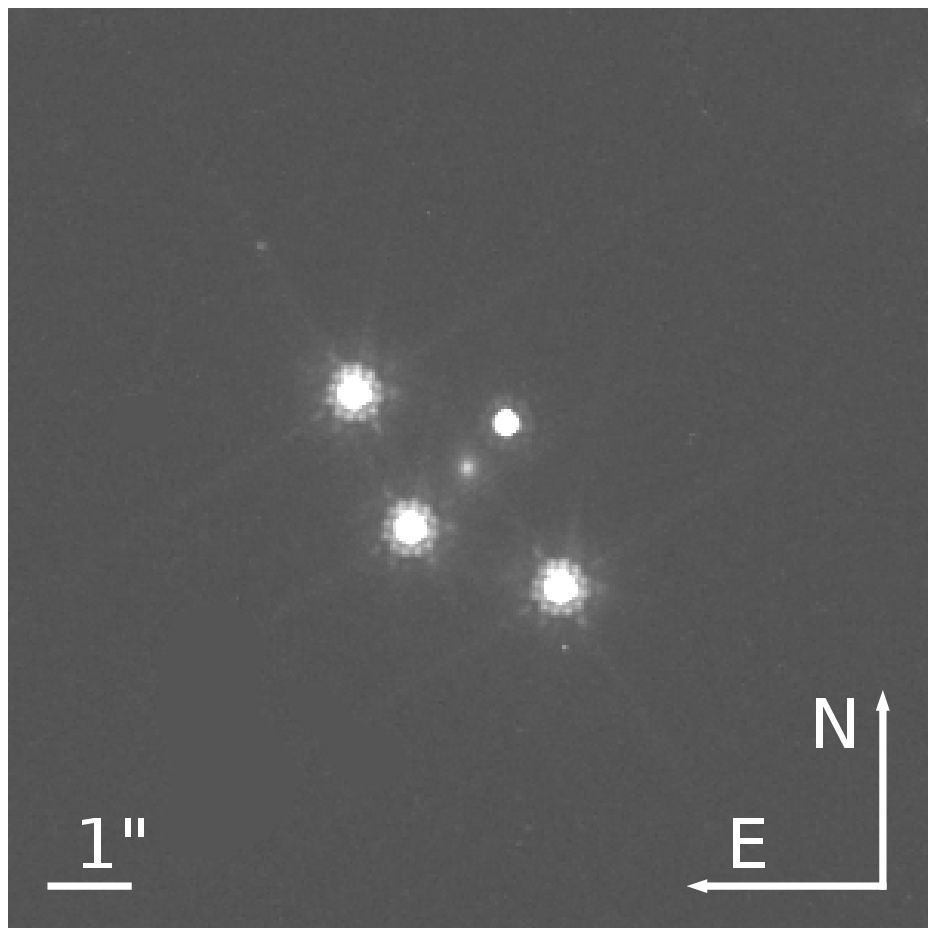}}  & \raisebox{-.5\height}{\includegraphics[height=0.2\textwidth]{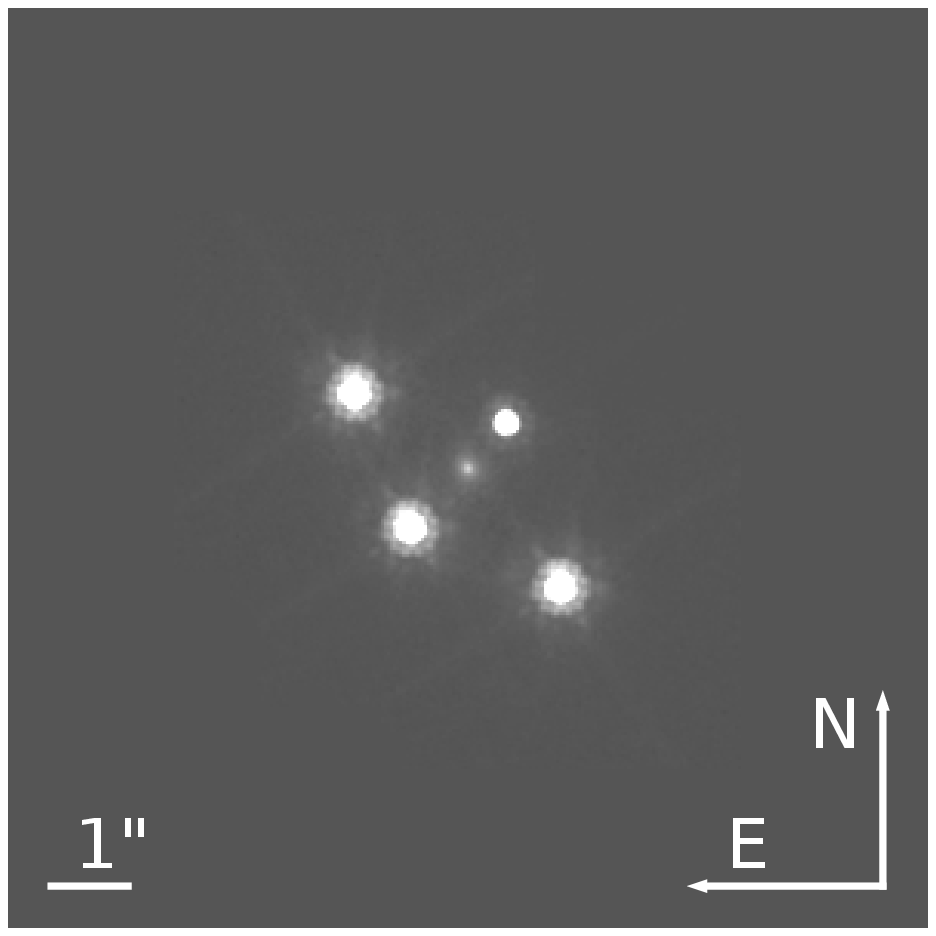}}  &  \raisebox{-.5\height}{\includegraphics[height=0.2\textwidth]{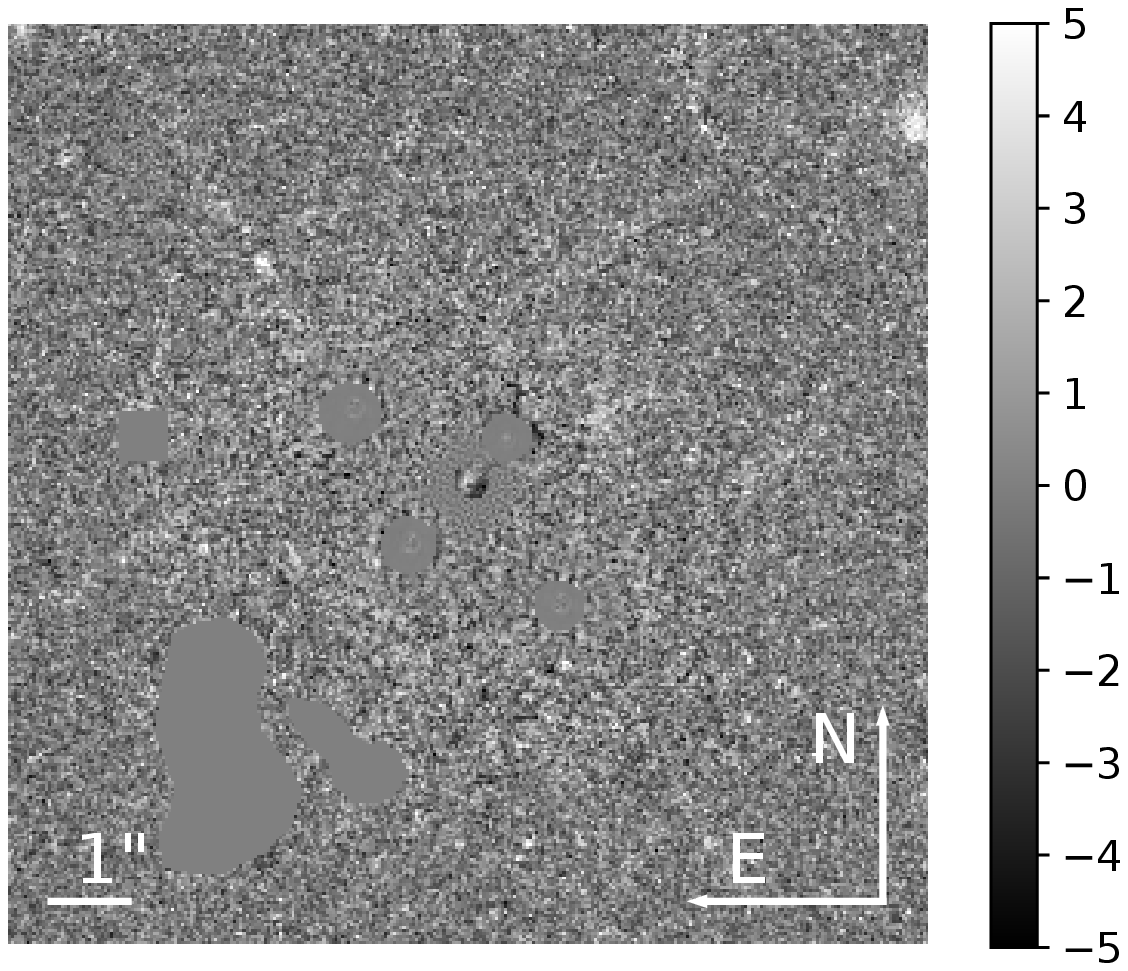}}  &
                        \raisebox{-.5\height}{\includegraphics[height=0.2\textwidth]{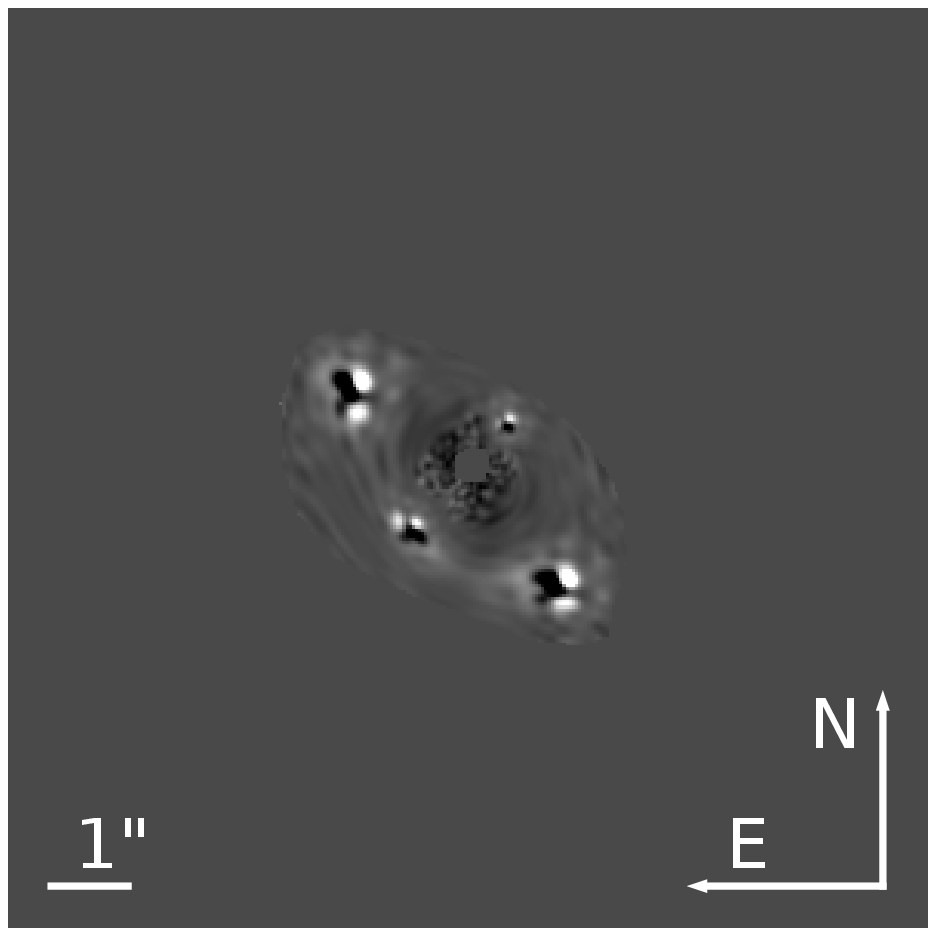}}  & 
                        \raisebox{-.5\height}{\includegraphics[width=0.2\textwidth]{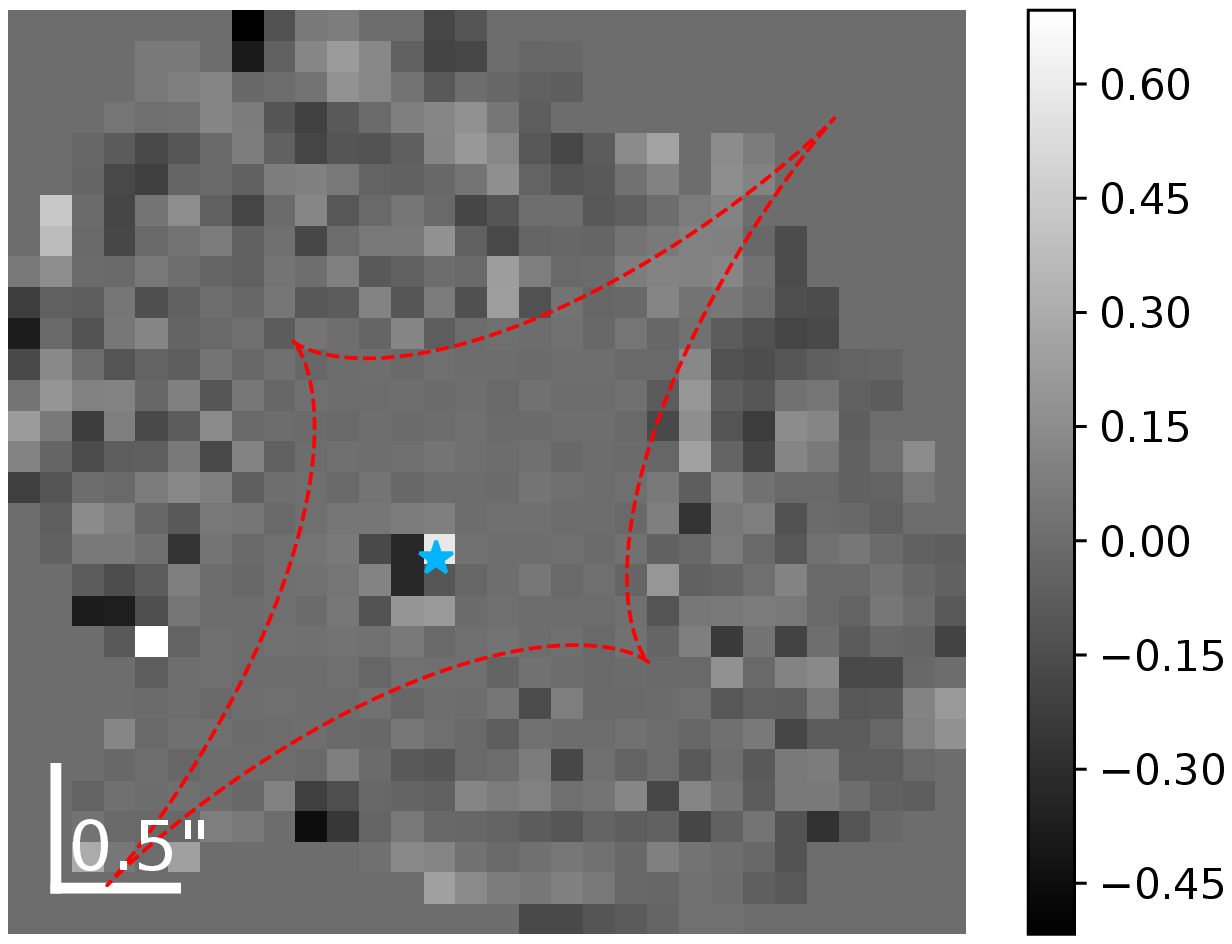}} \\ \midrule
                    J1537$-$3010 & F160W & \raisebox{-.5\height}{\includegraphics[height=0.2\textwidth]{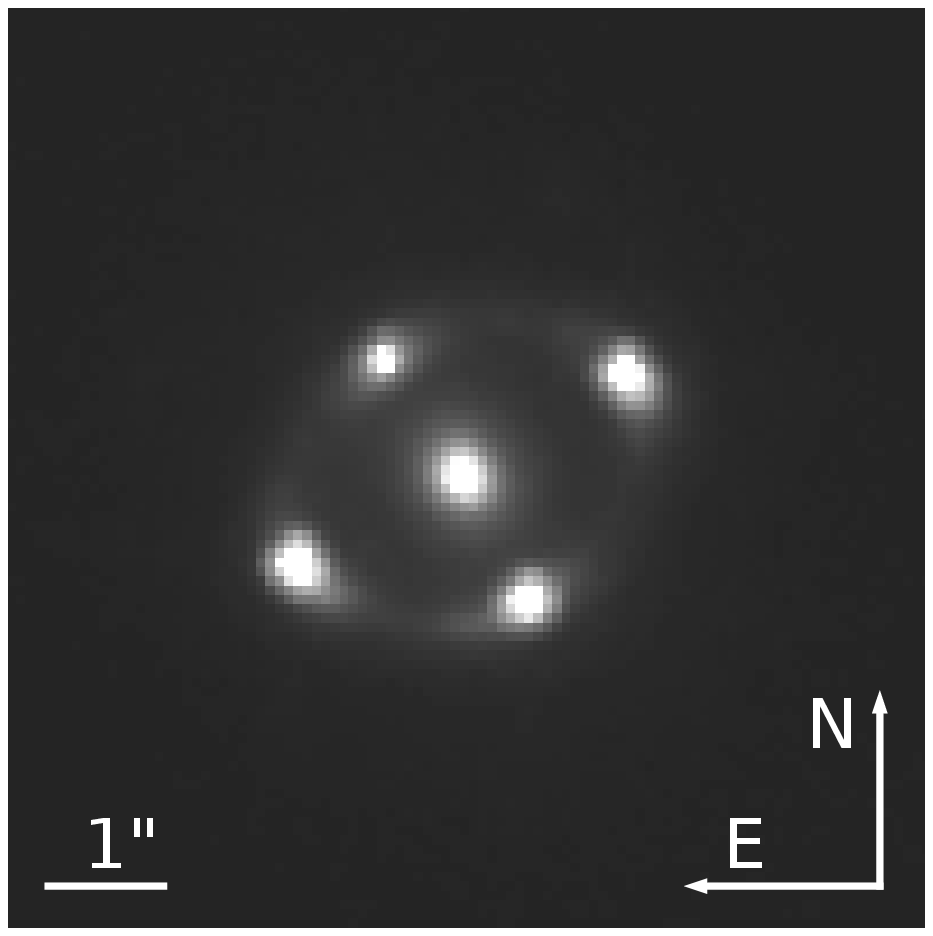}}  & \raisebox{-.5\height}{\includegraphics[height=0.2\textwidth]{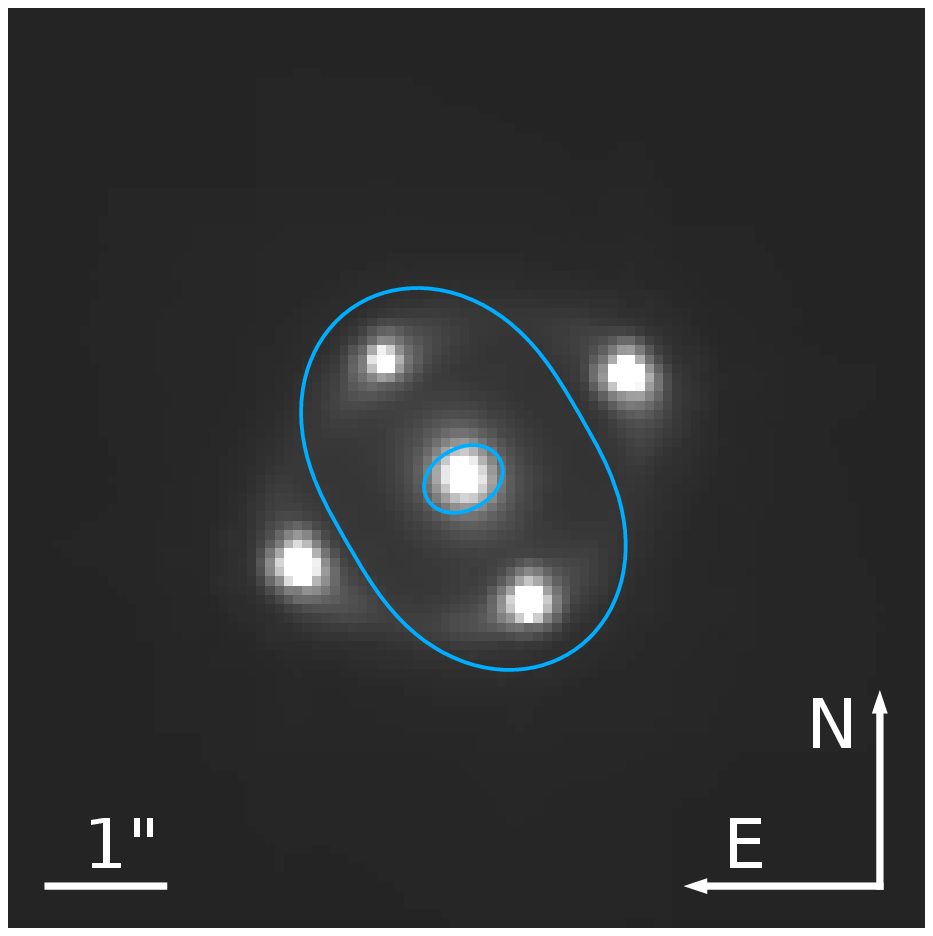}}  &  \raisebox{-.5\height}{\includegraphics[height=0.2\textwidth]{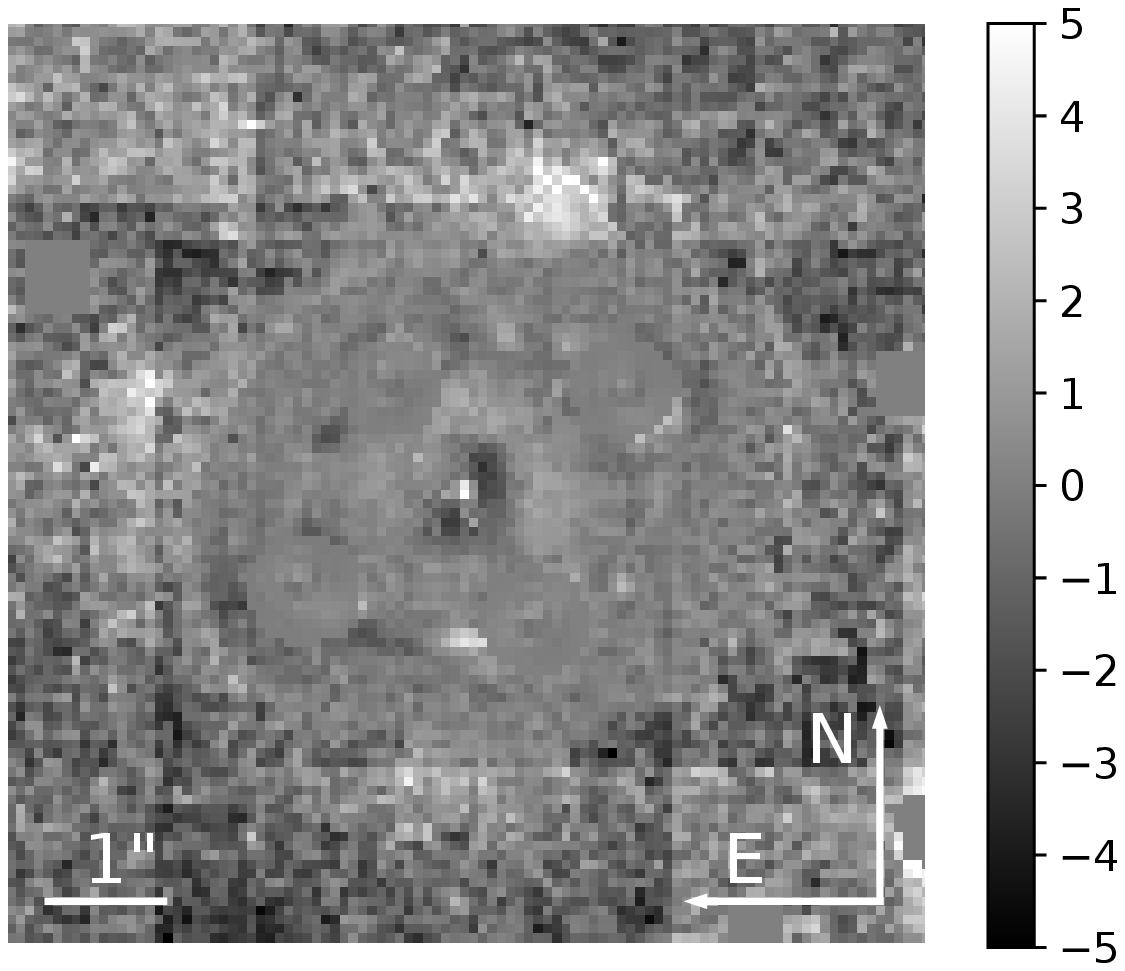}}  &
                        \raisebox{-.5\height}{\includegraphics[height=0.2\textwidth]{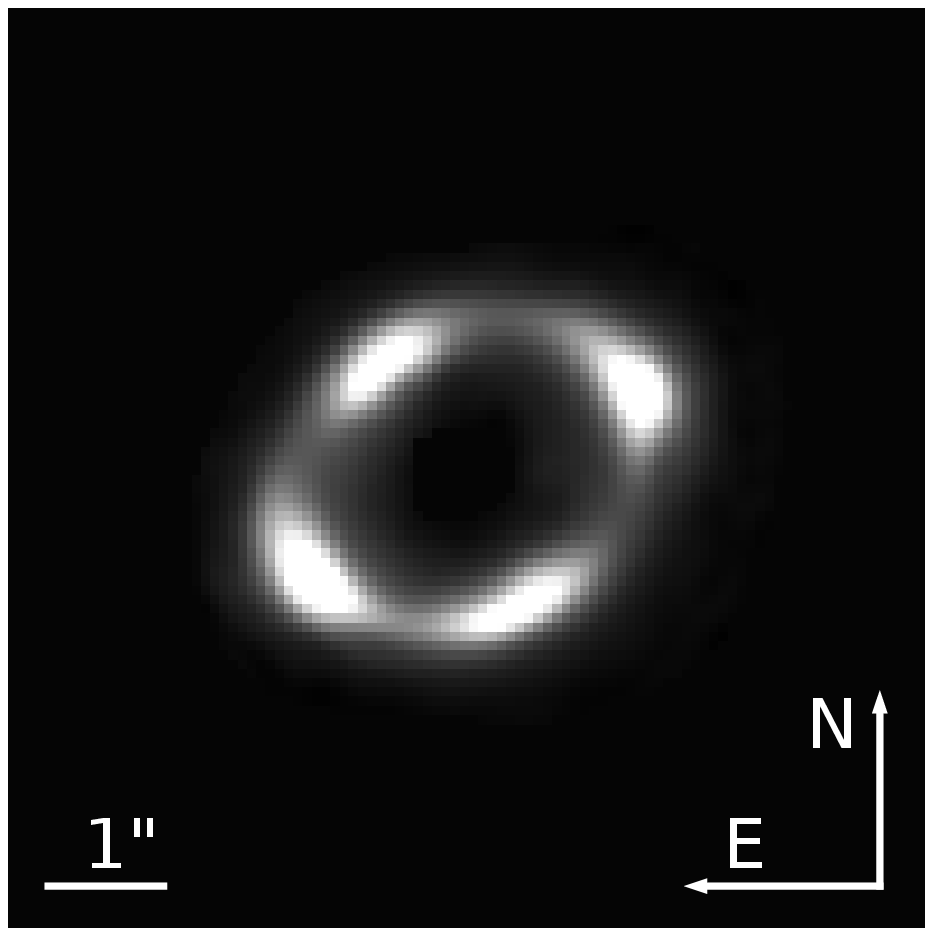}}  & 
                        \raisebox{-.5\height}{\includegraphics[width=0.2\textwidth]{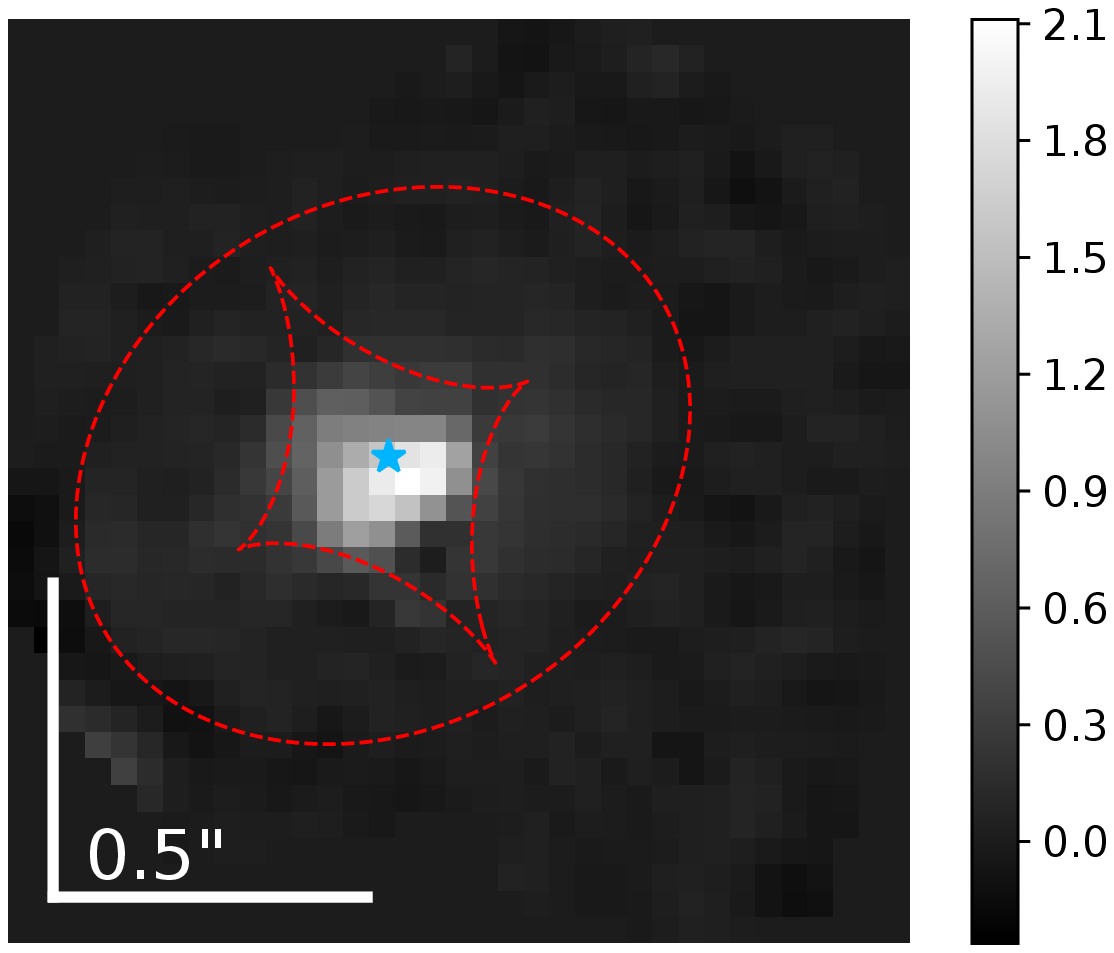}} \\

                \end{tabularx}
        \ \\
\end{table}

\clearpage
\begin{table}[]
                \captionsetup{labelformat=empty} 
            \caption*{\textbf{Table \ref{tab:results_light} continued.}}
            
                \begin{tabularx}{\linewidth}{c|c|ccc|cc}\toprule \toprule
            System & Filter & Observed & Model & \begin{tabular}{@{}c@{}}Normalized \\residuals\end{tabular}  &  \begin{tabular}{@{}c@{}}Reconstructed \\arc\end{tabular} & \begin{tabular}{@{}c@{}}Reconstructed \\source\end{tabular}\\ \toprule \toprule
            & F475X & \raisebox{-.5\height}{\includegraphics[height=0.2\textwidth]{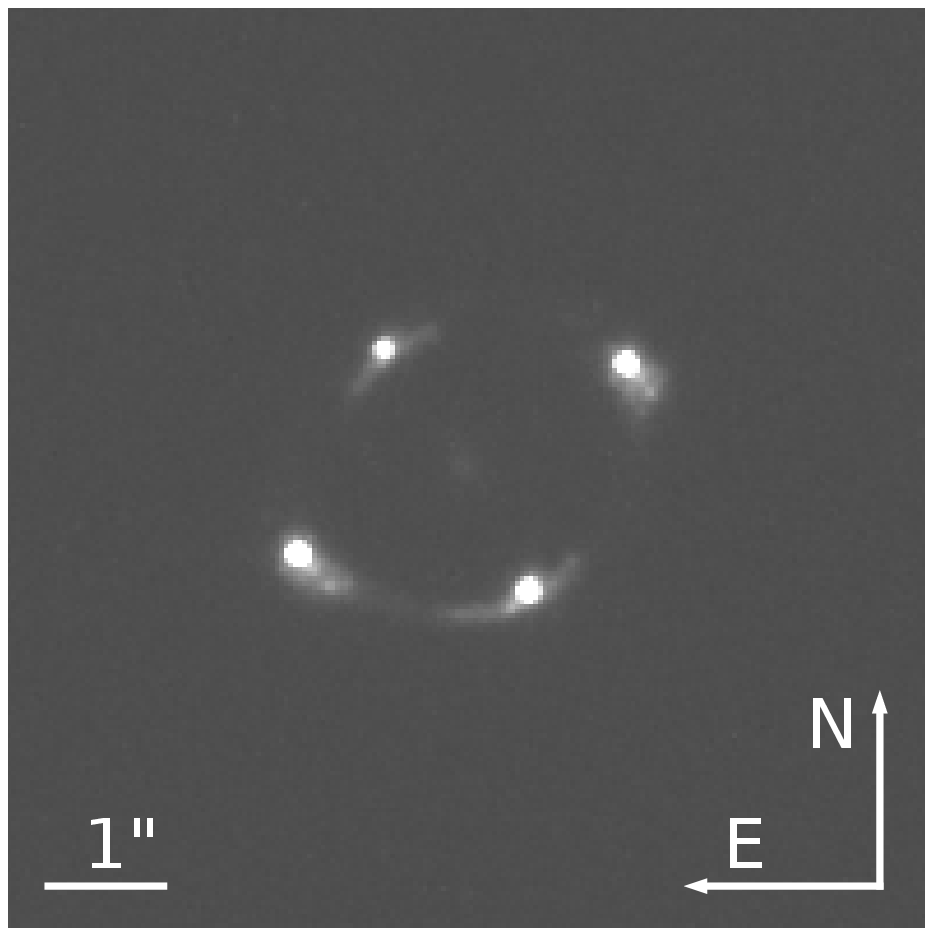}}  &  \raisebox{-.5\height}{\includegraphics[height=0.2\textwidth]{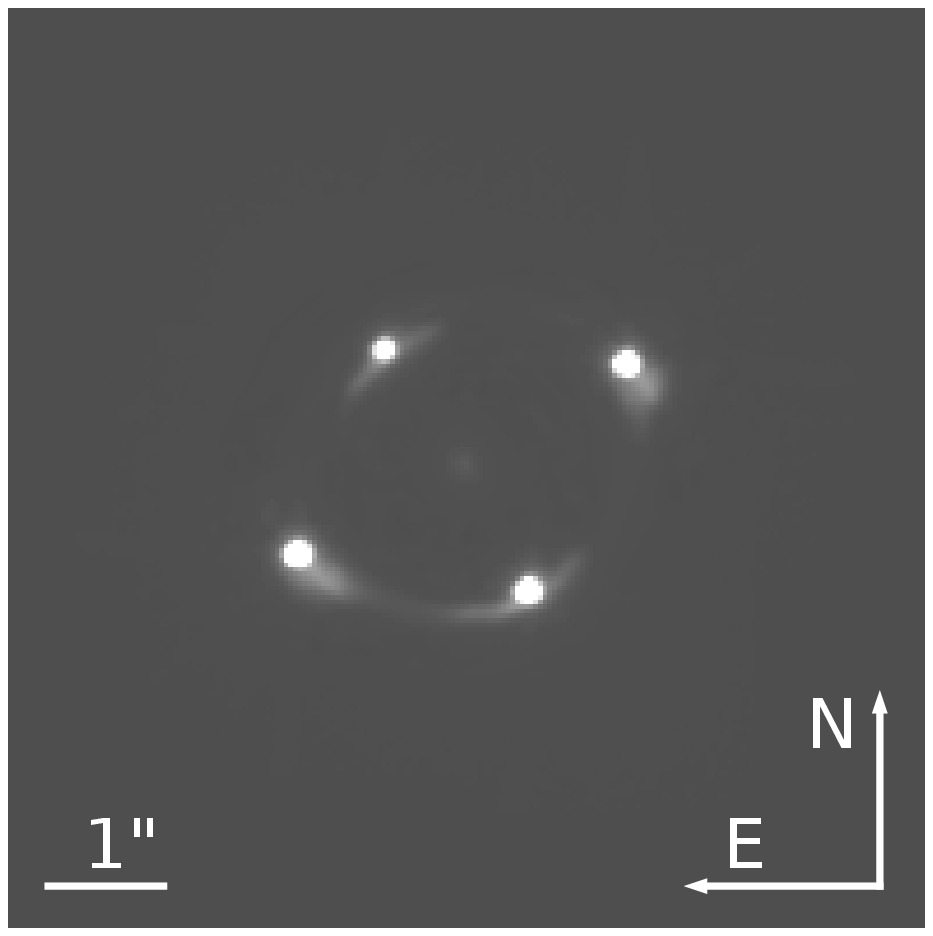}}  & \raisebox{-.5\height}{\includegraphics[height=0.2\textwidth]{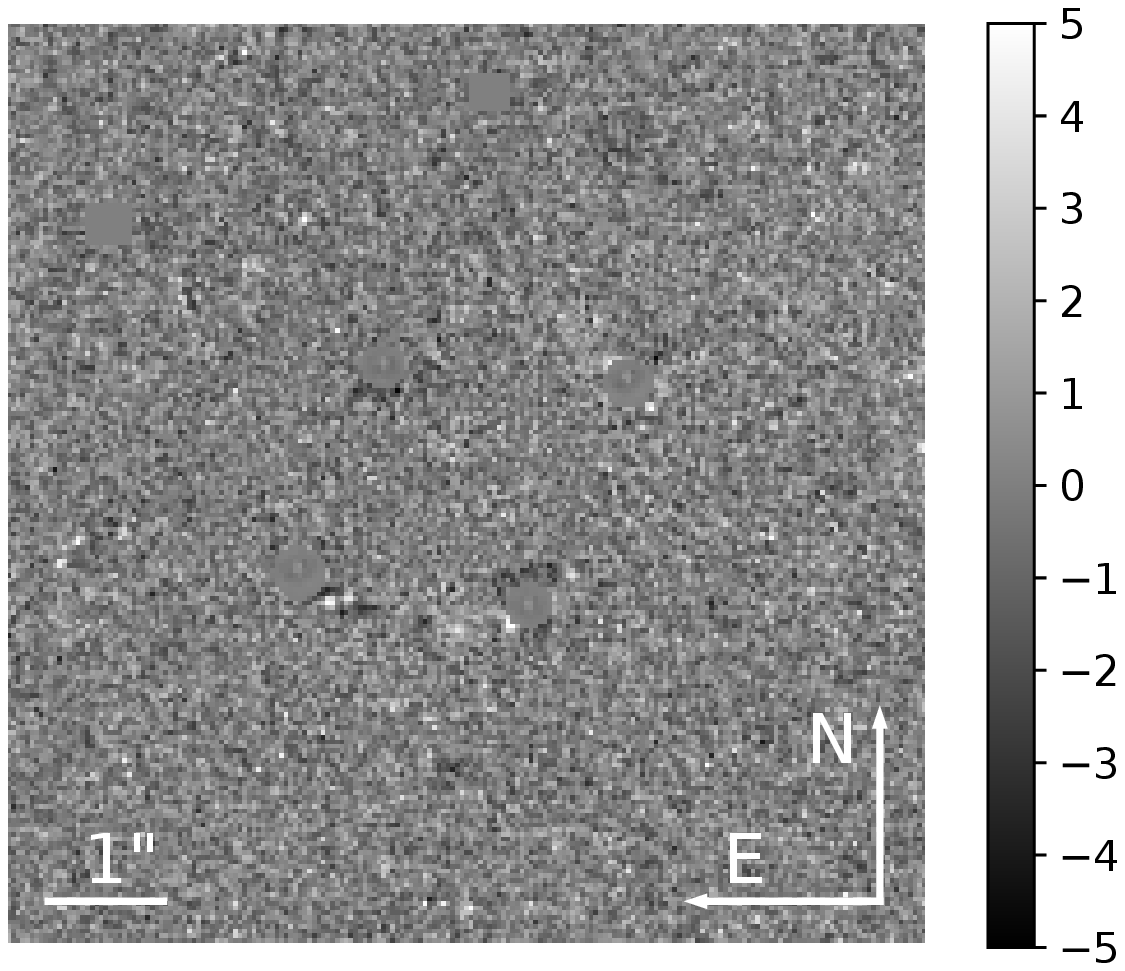}}  &
                         \raisebox{-.5\height}{\includegraphics[height=0.2\textwidth]{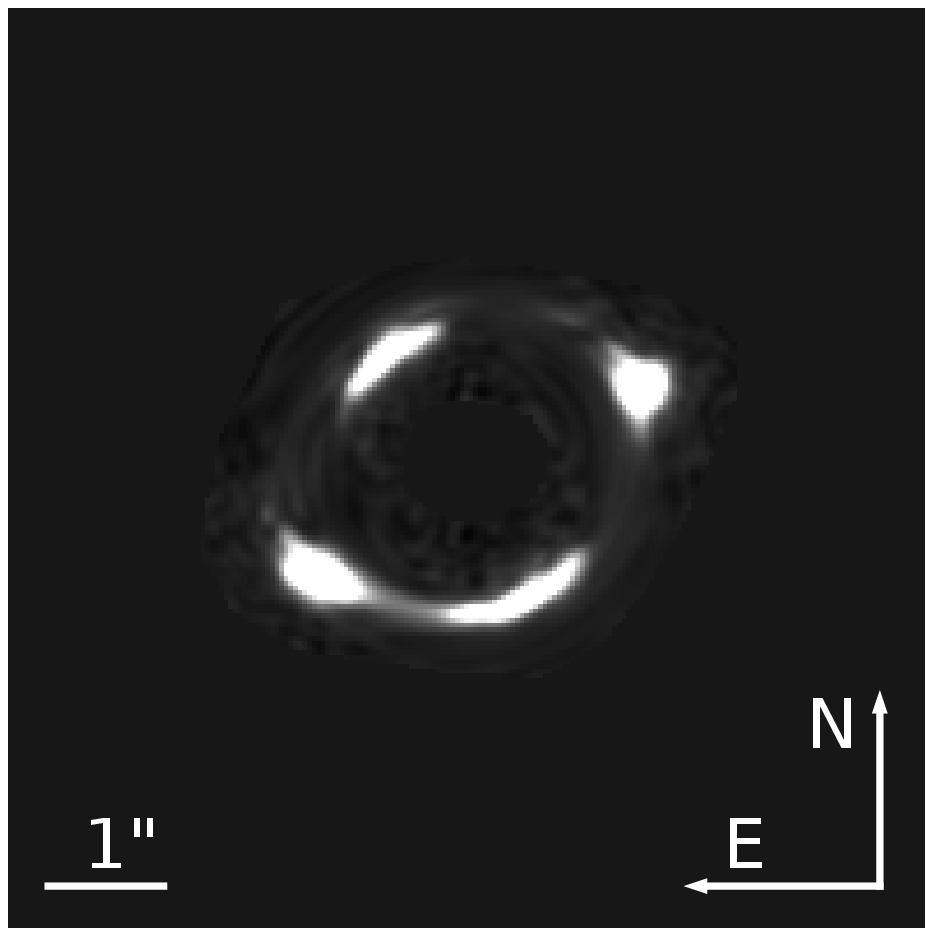}}  & 
                        \raisebox{-.5\height}{\includegraphics[width=0.2\textwidth]{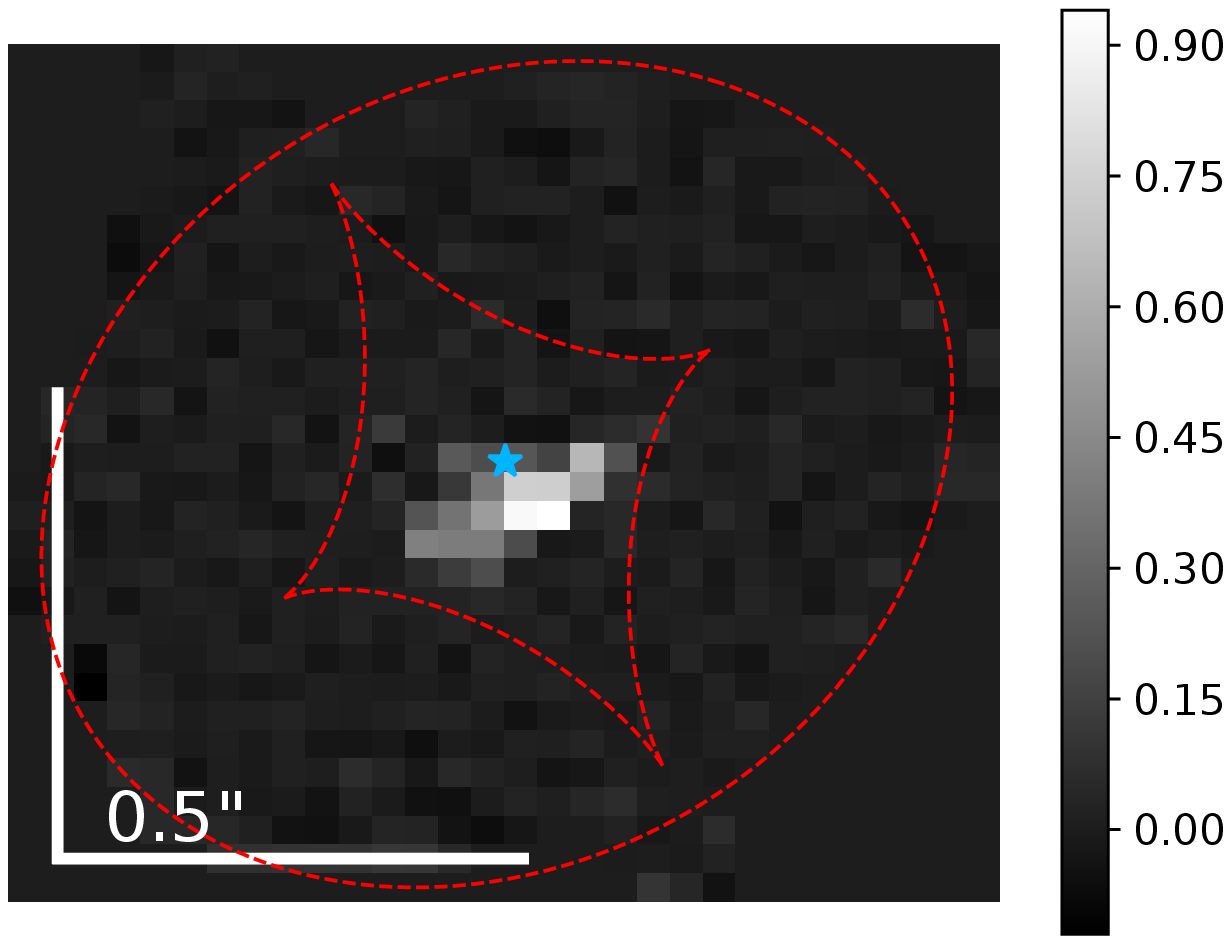}} \\ 
                        & F814W & \raisebox{-.5\height}{\includegraphics[height=0.2\textwidth]{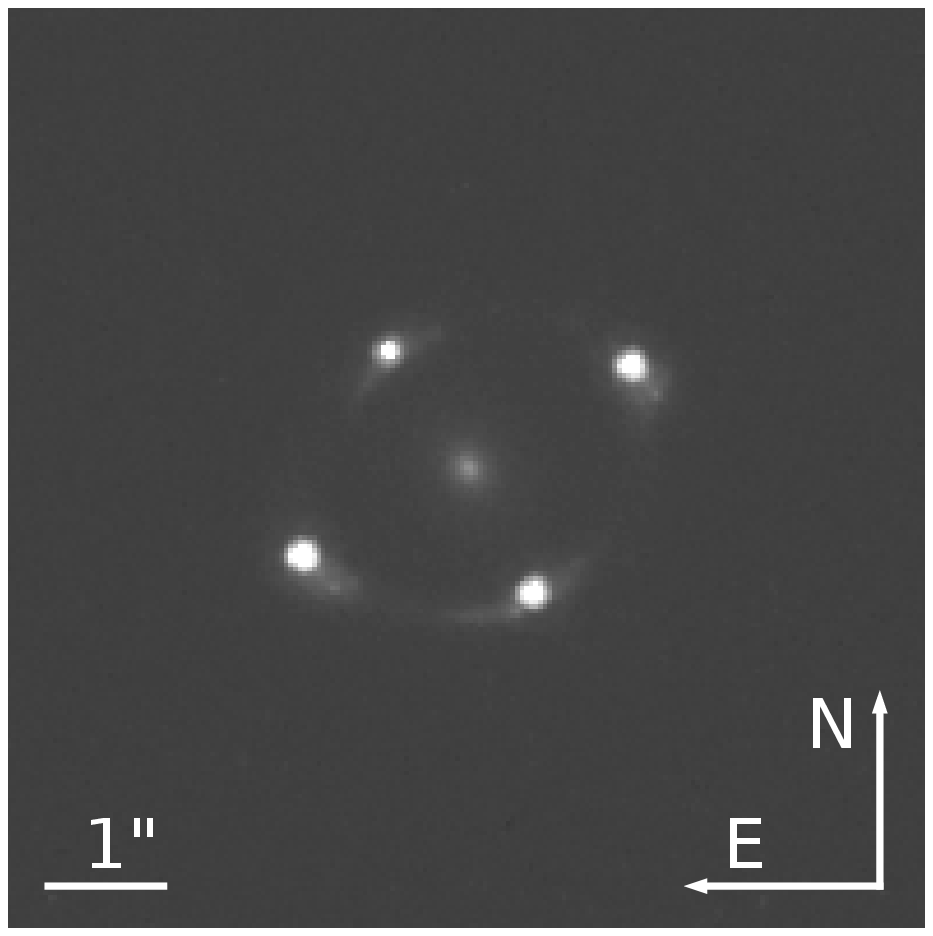}}  &  \raisebox{-.5\height}{\includegraphics[height=0.2\textwidth]{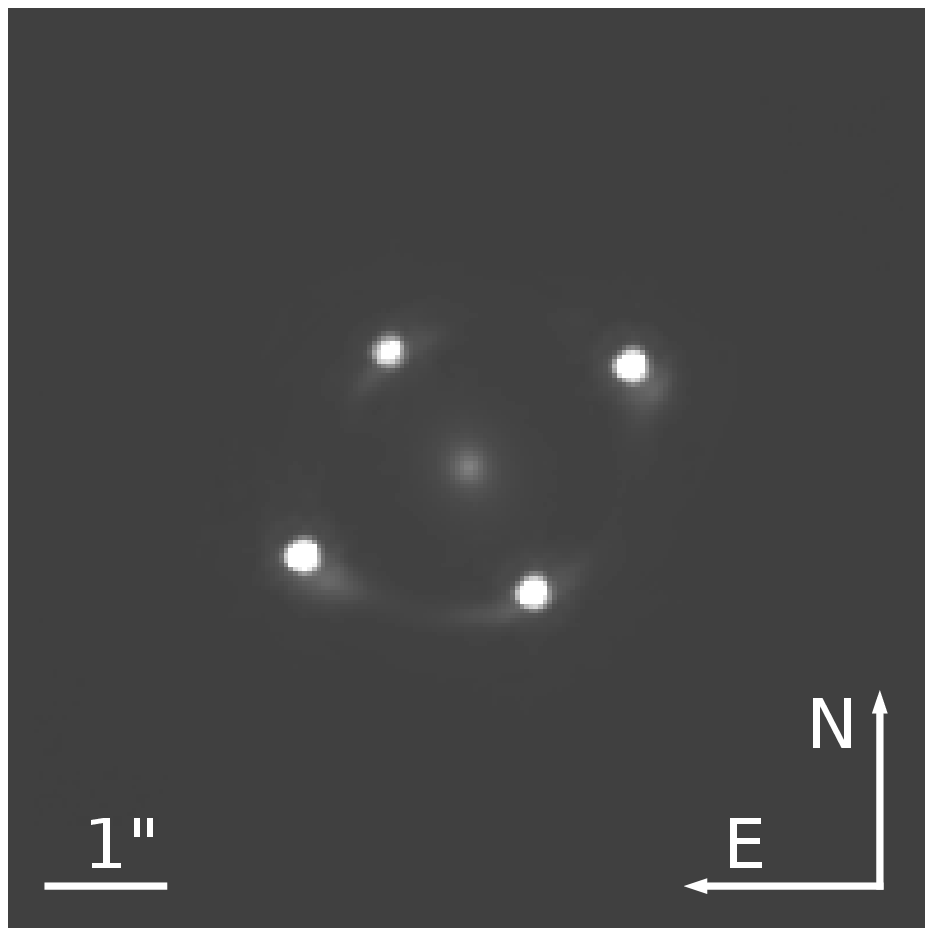}}  & \raisebox{-.5\height}{\includegraphics[height=0.2\textwidth]{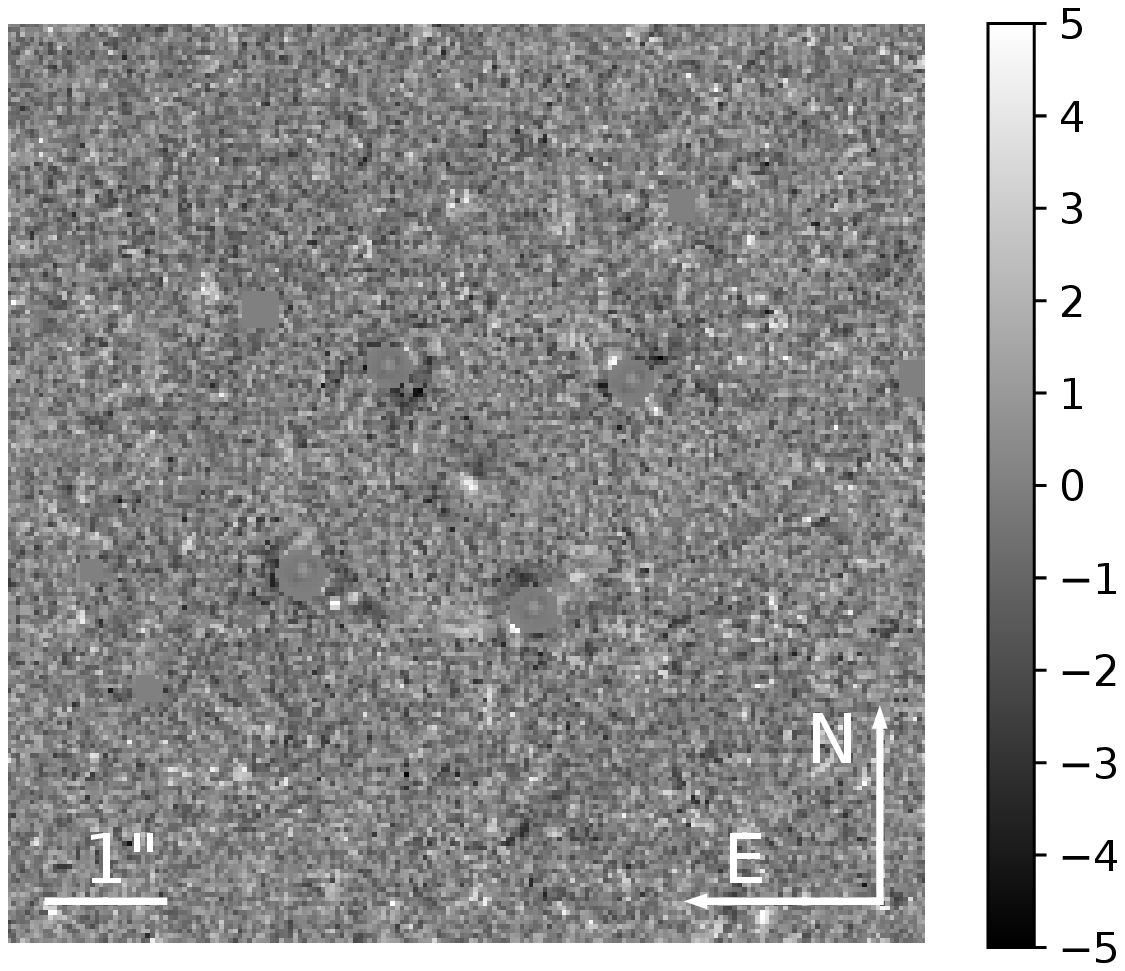}}  &
                        \raisebox{-.5\height}{\includegraphics[height=0.2\textwidth]{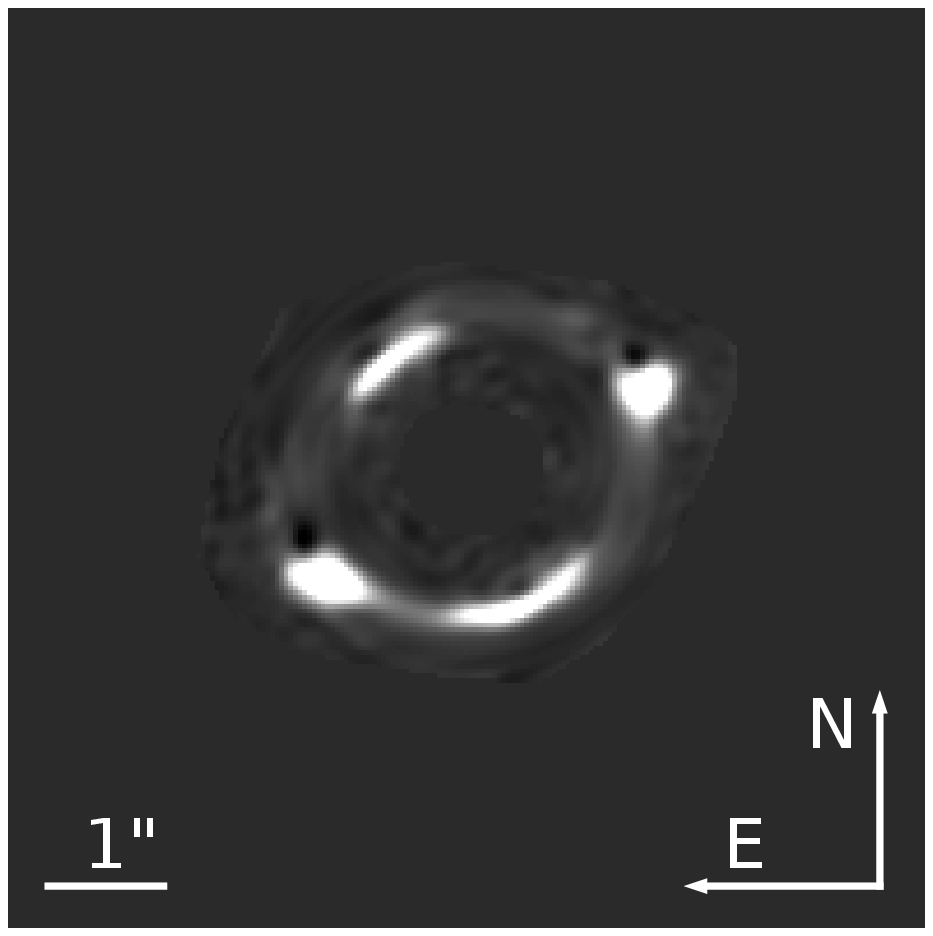}}  & 
                        \raisebox{-.5\height}{\includegraphics[width=0.2\textwidth]{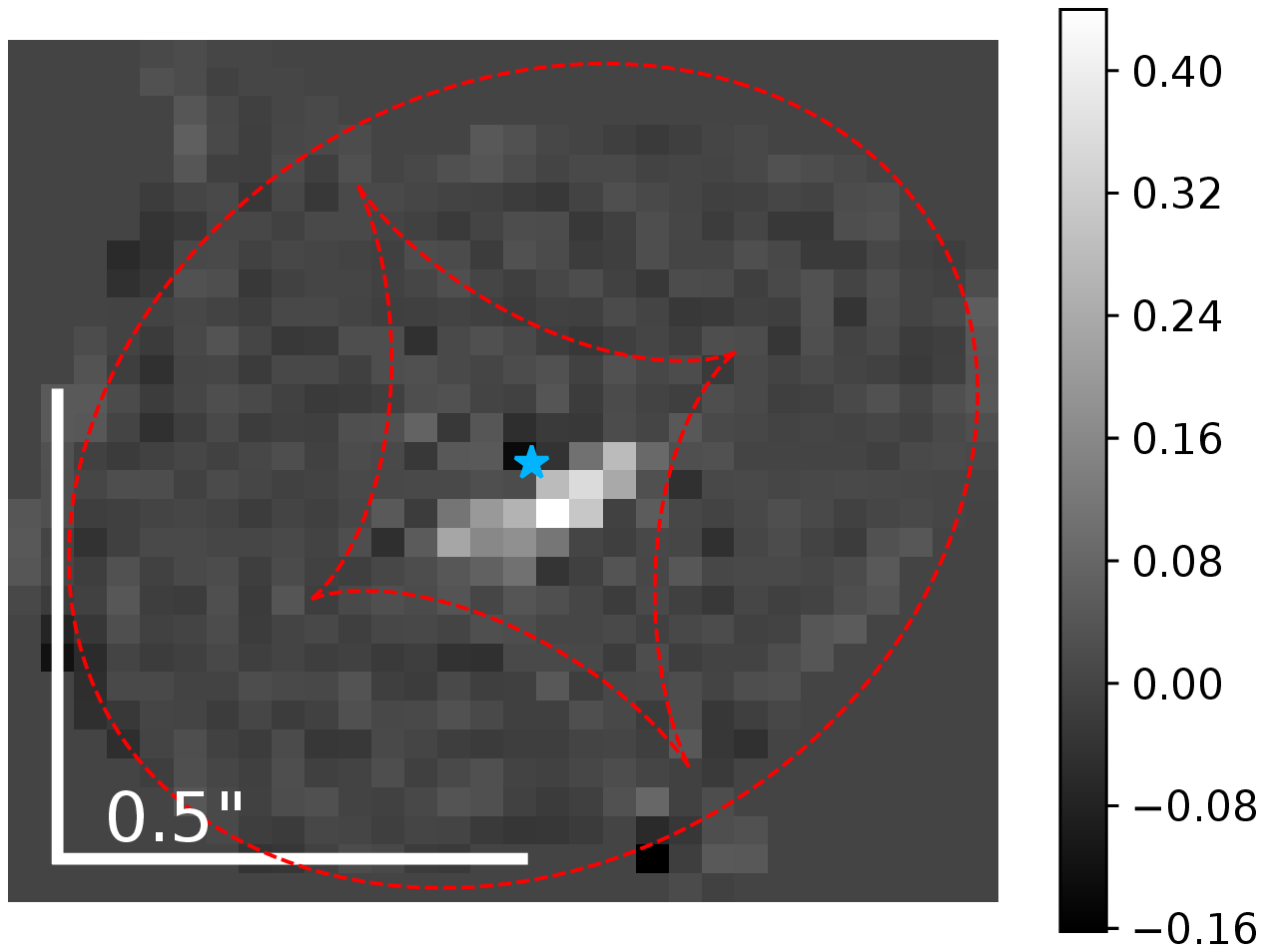}} \\ \midrule     
                        PS J1606$-$2333 & F160W & \raisebox{-.5\height}{\includegraphics[height=0.2\textwidth]{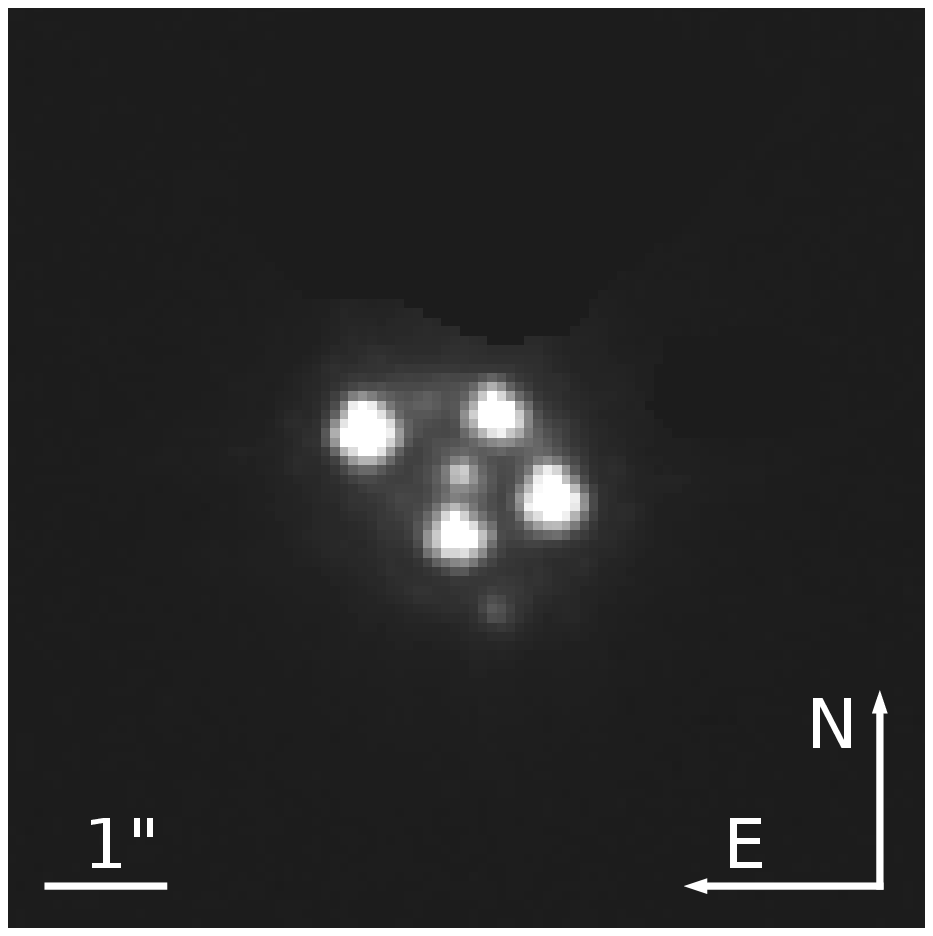}}  &  \raisebox{-.5\height}{\includegraphics[height=0.2\textwidth]{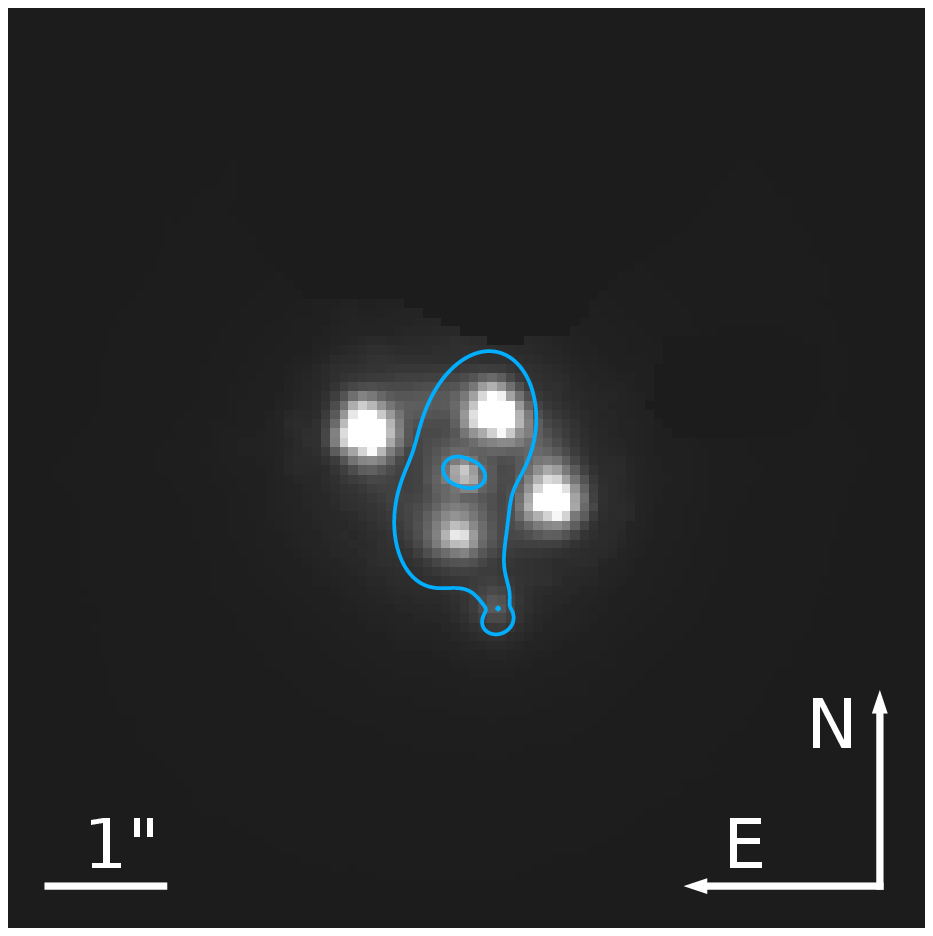}}  & \raisebox{-.5\height}{\includegraphics[height=0.2\textwidth]{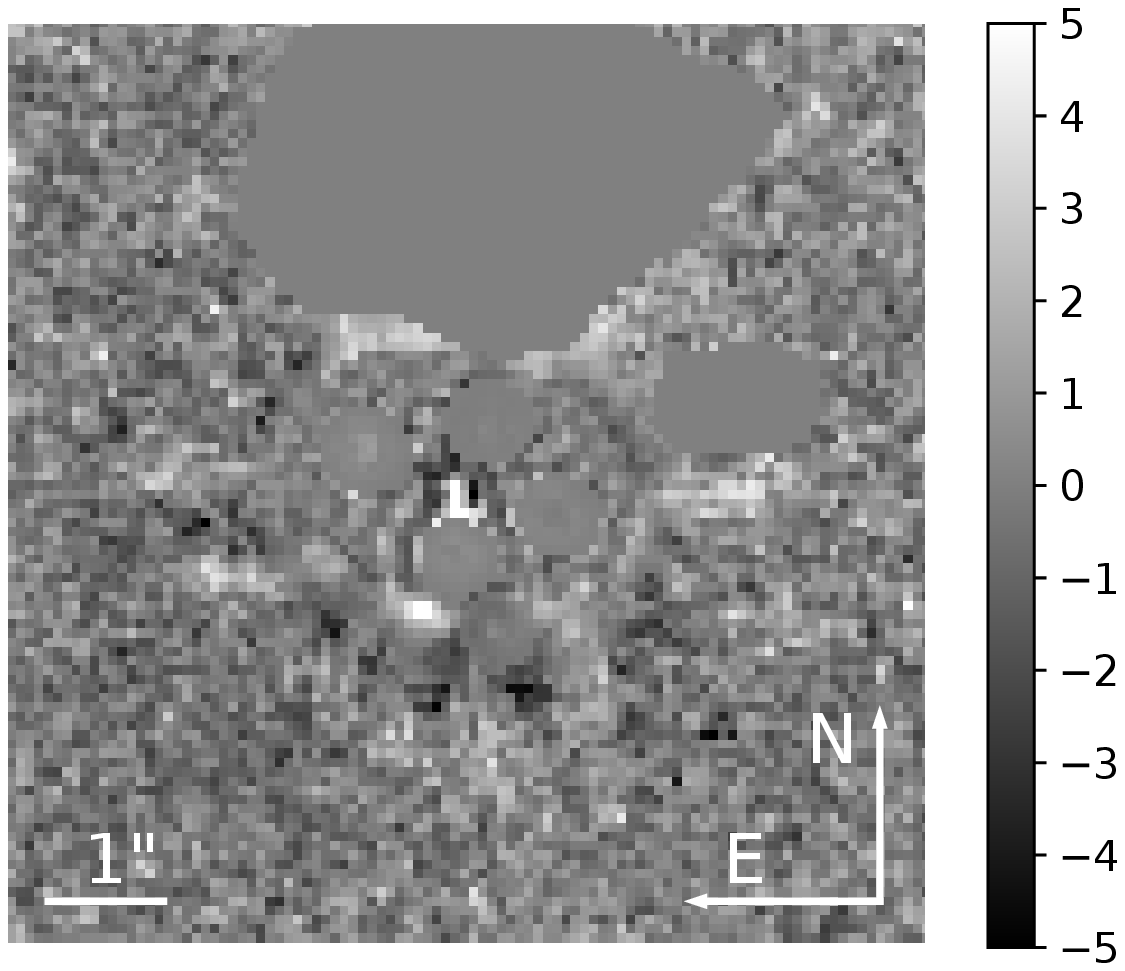}}  &
                    \raisebox{-.5\height}{\includegraphics[height=0.2\textwidth]{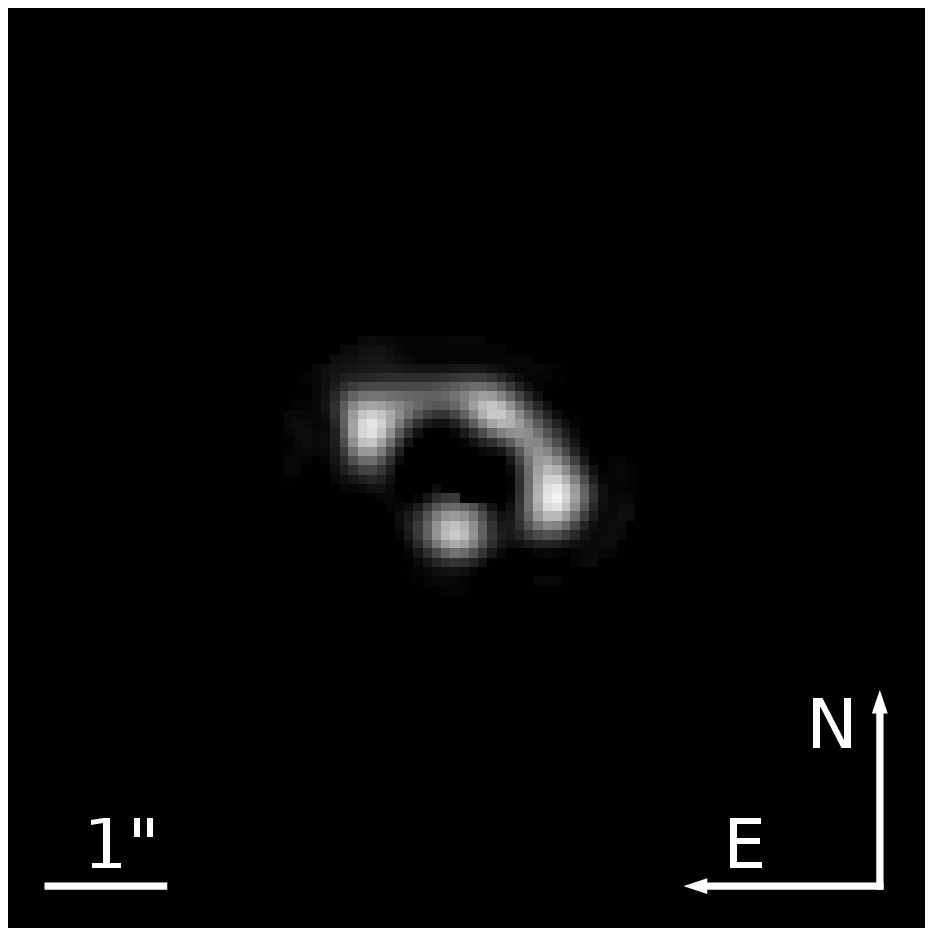}}  & 
                        \raisebox{-.5\height}{\includegraphics[width=0.2\textwidth]{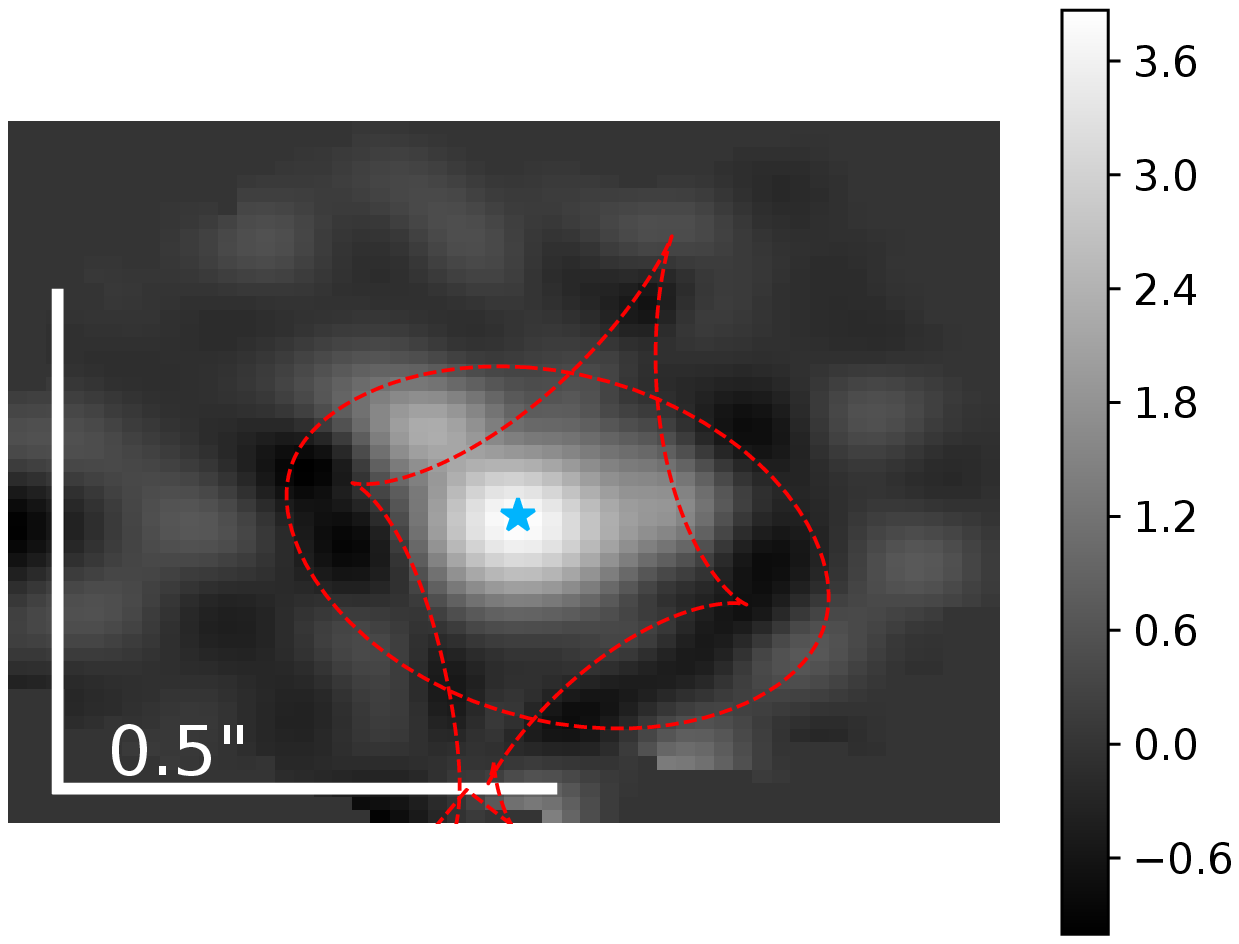}} \\ \midrule
            PS J1721$+$8842 & F160W & \raisebox{-.5\height}{\includegraphics[height=0.2\textwidth]{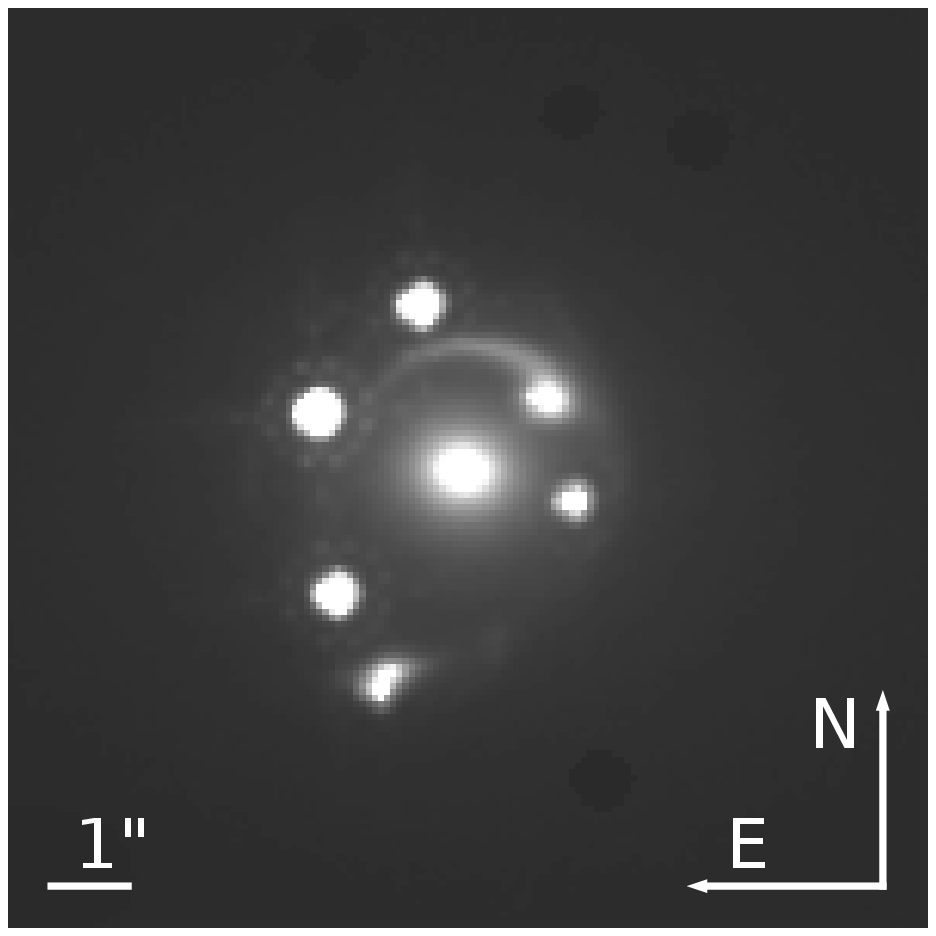}}  & \raisebox{-.5\height}{\includegraphics[height=0.2\textwidth]{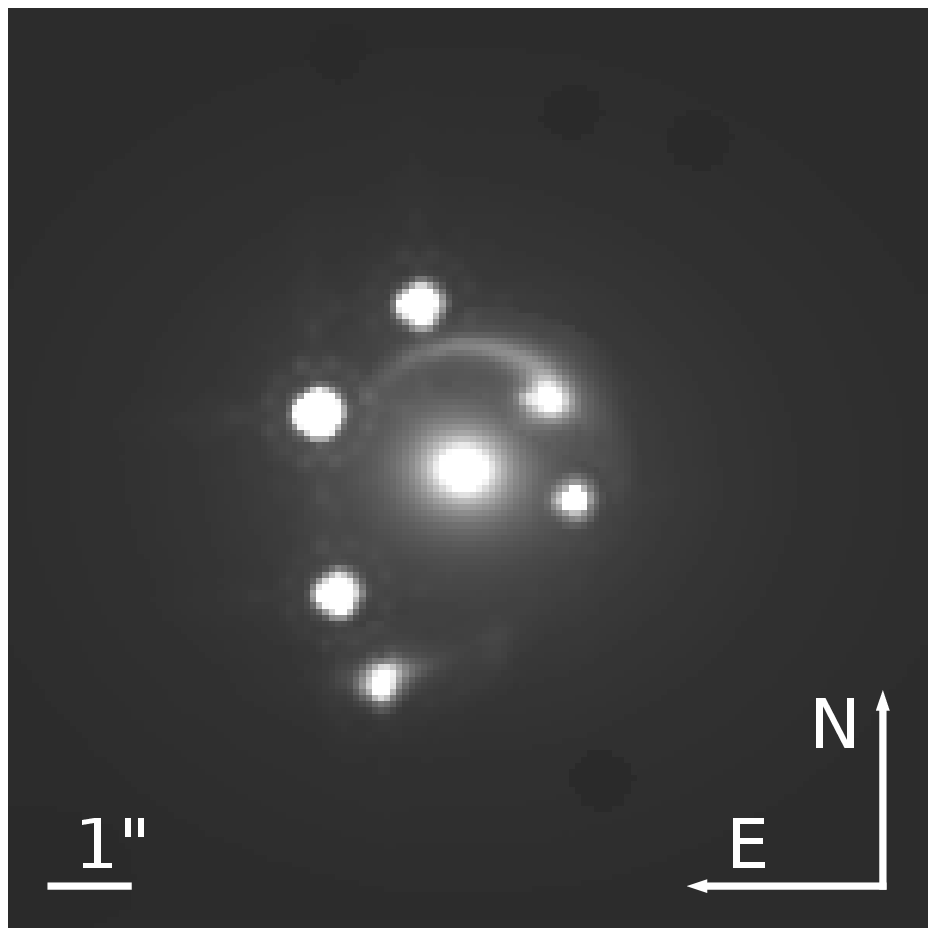}}  &  \raisebox{-.5\height}{\includegraphics[height=0.2\textwidth]{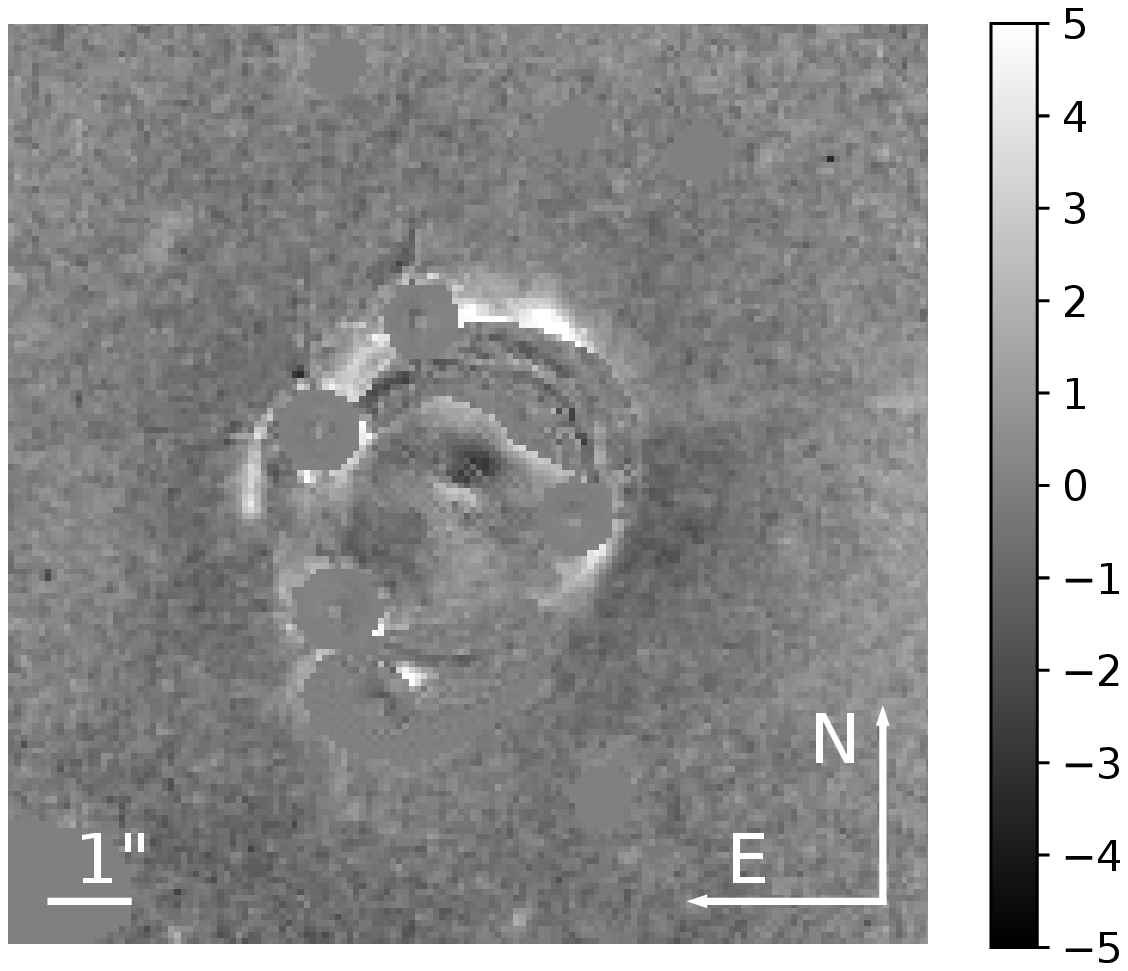}}  &
            \raisebox{-.5\height}{\includegraphics[height=0.2\textwidth]{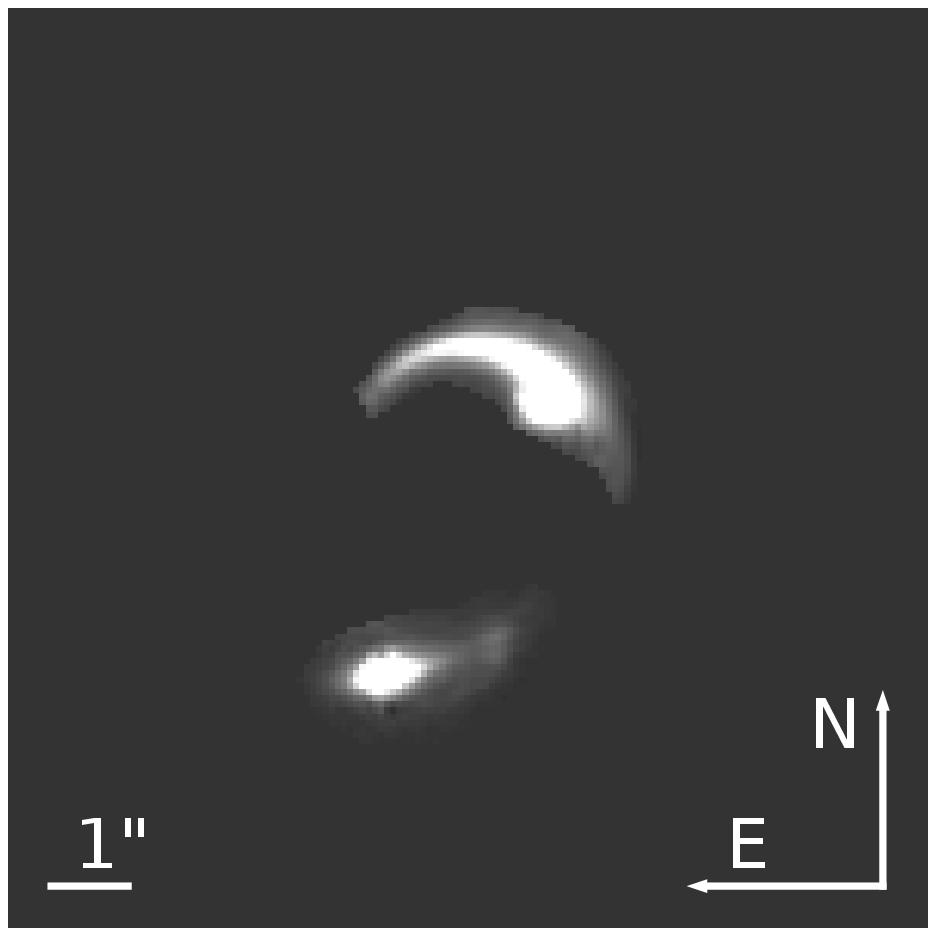}}  & 
                        \raisebox{-.5\height}{\includegraphics[width=0.2\textwidth]{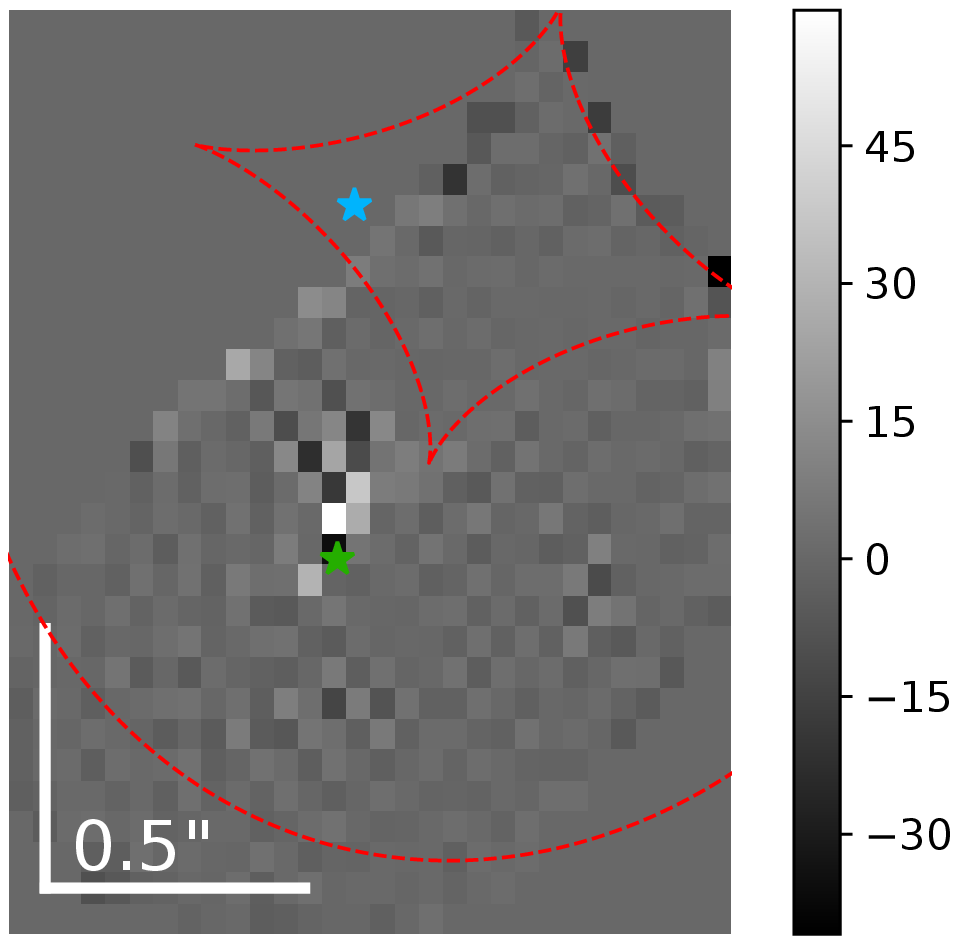}} \\ 

                \end{tabularx}
        \ \\
\end{table}
\clearpage
\begin{table}[]
                \captionsetup{labelformat=empty} 
            \caption*{\textbf{Table \ref{tab:results_light} continued.}}
            
                \begin{tabularx}{\linewidth}{c|c|ccc|cc}\toprule \toprule
            System & Filter & Observed & Model & \begin{tabular}{@{}c@{}}Normalized \\residuals\end{tabular}  &  \begin{tabular}{@{}c@{}}Reconstructed \\arc\end{tabular} & \begin{tabular}{@{}c@{}}Reconstructed \\source\end{tabular}\\ \toprule \toprule
            
            & F475X & \raisebox{-.5\height}{\includegraphics[height=0.2\textwidth]{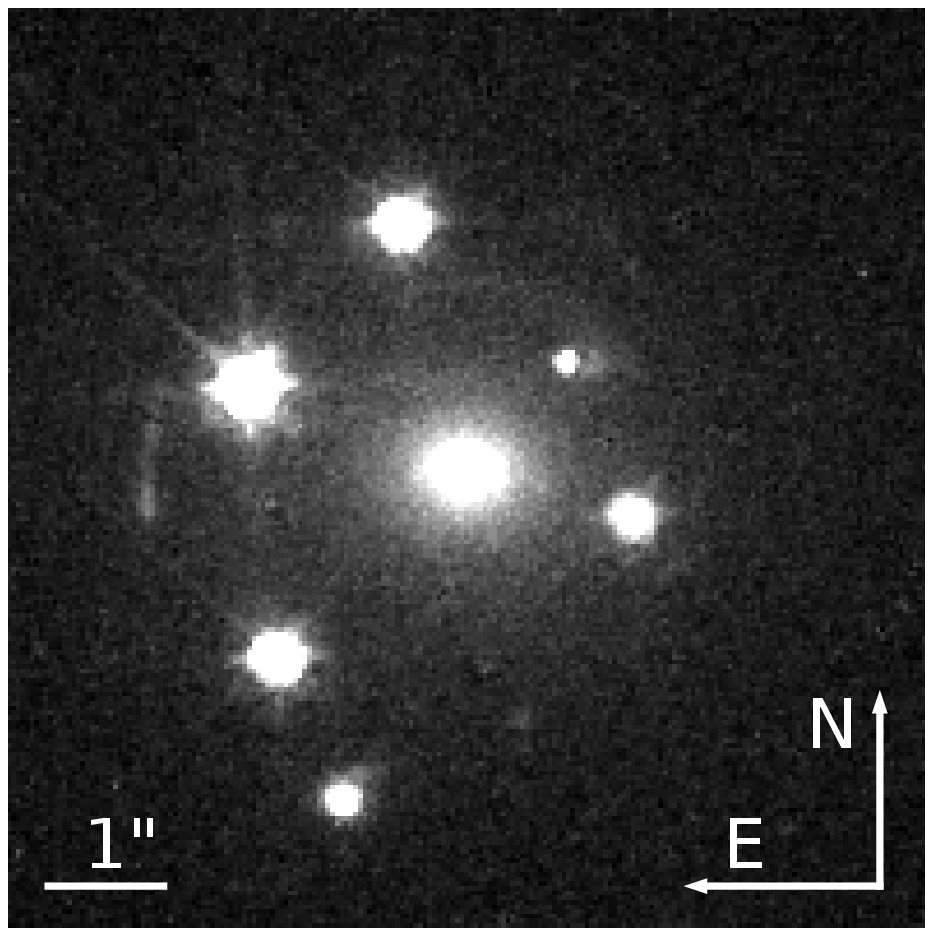}}  & \raisebox{-.5\height}{\includegraphics[height=0.2\textwidth]{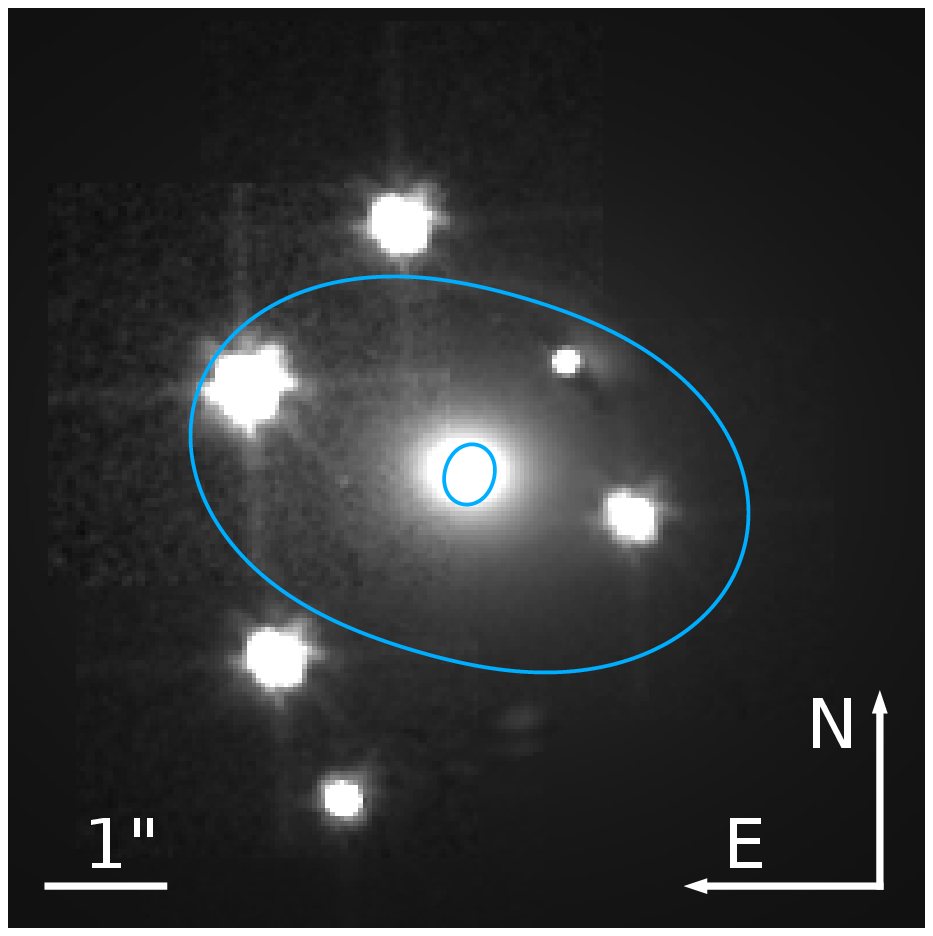}}  &  \raisebox{-.5\height}{\includegraphics[height=0.2\textwidth]{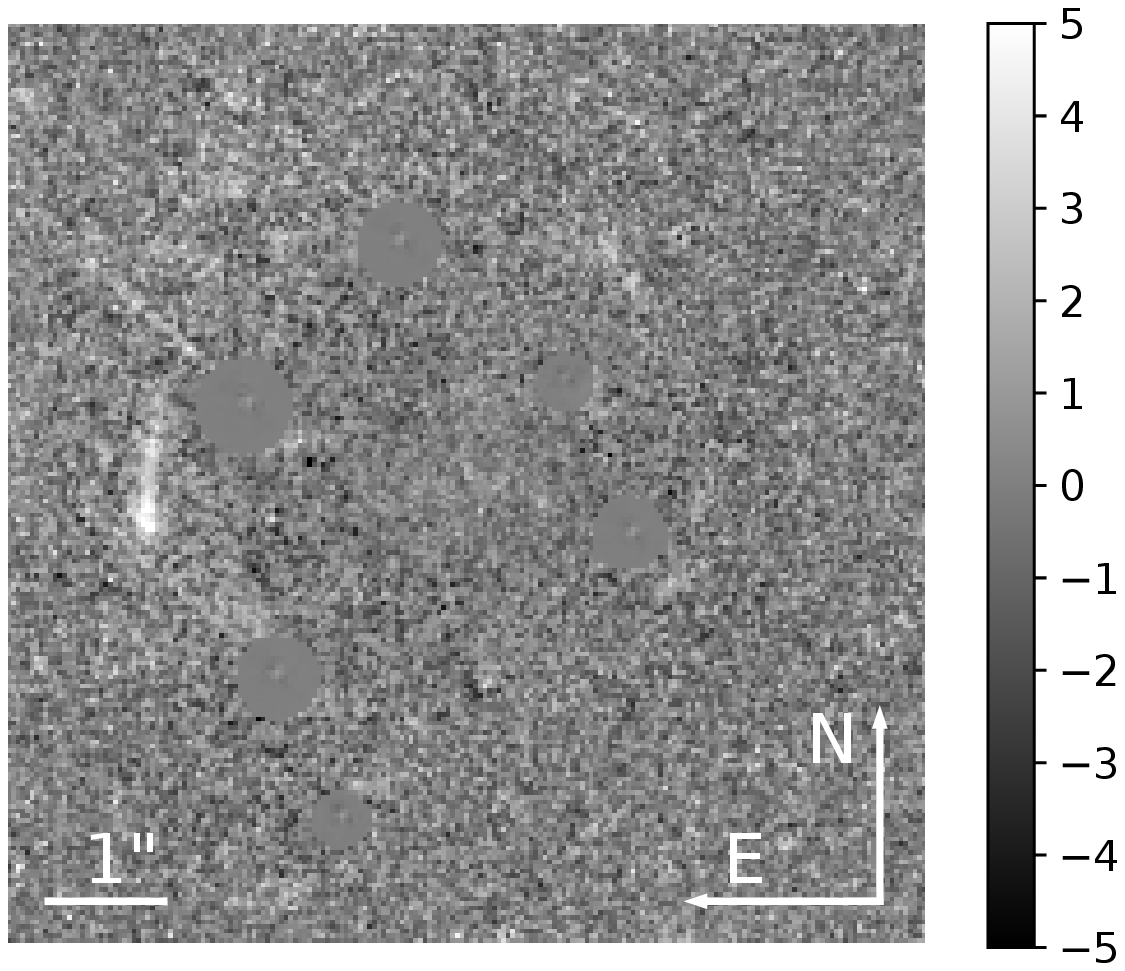}}  &
                         \raisebox{-.5\height}{\includegraphics[height=0.2\textwidth]{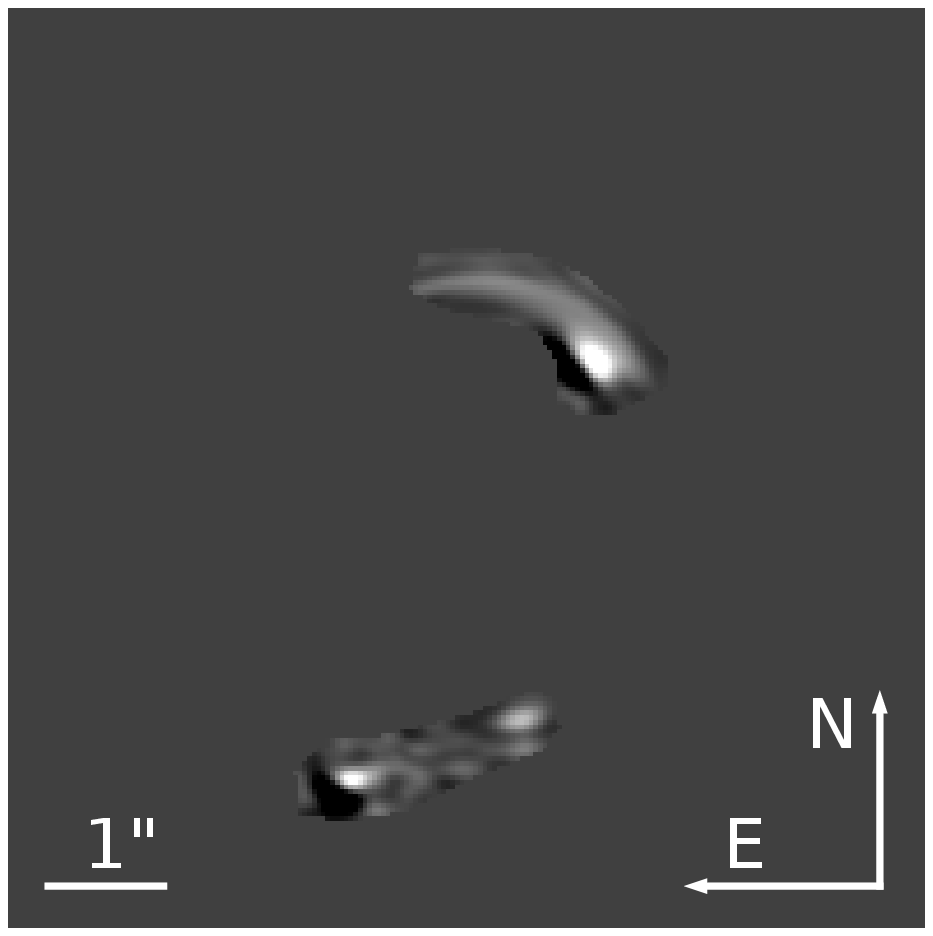}}  & 
                        \raisebox{-.5\height}{\includegraphics[width=0.2\textwidth]{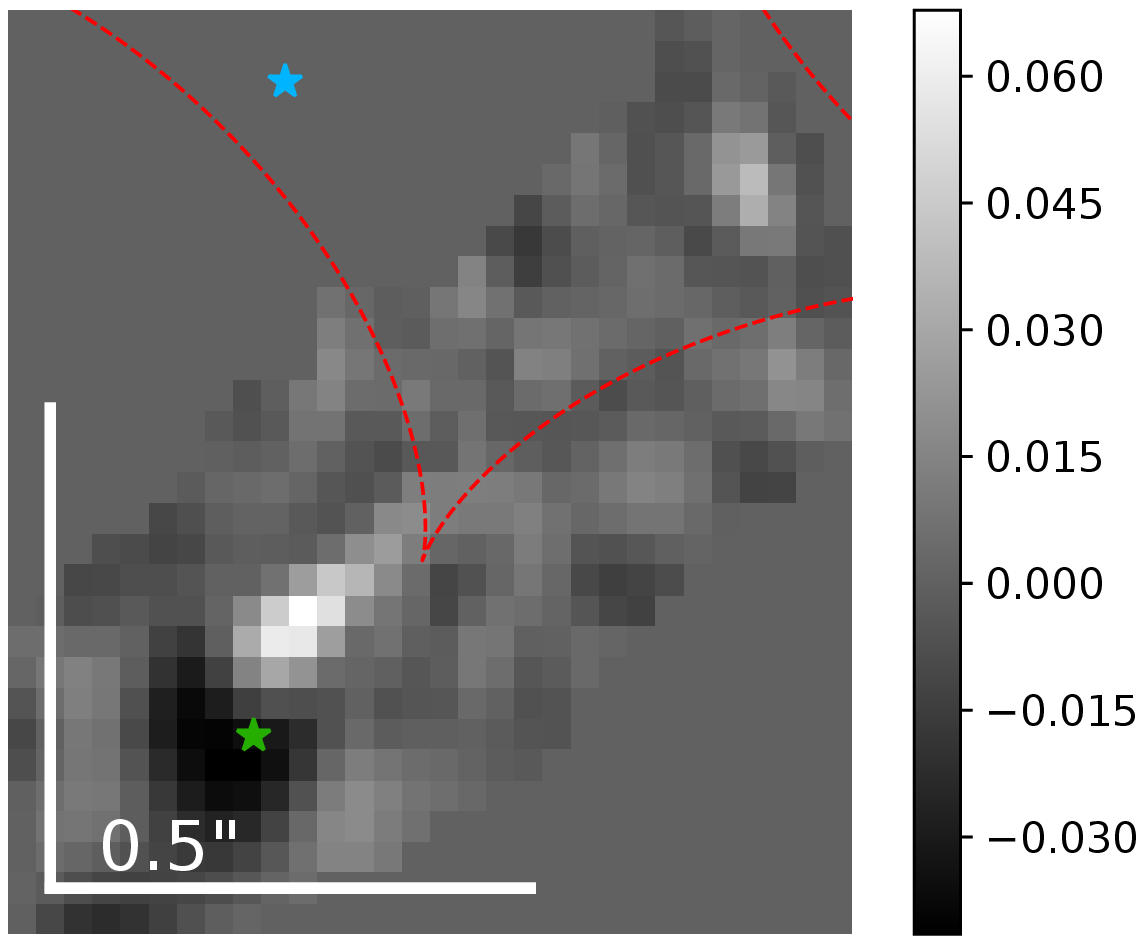}} \\ 
            
            & F814W & \raisebox{-.5\height}{\includegraphics[height=0.2\textwidth]{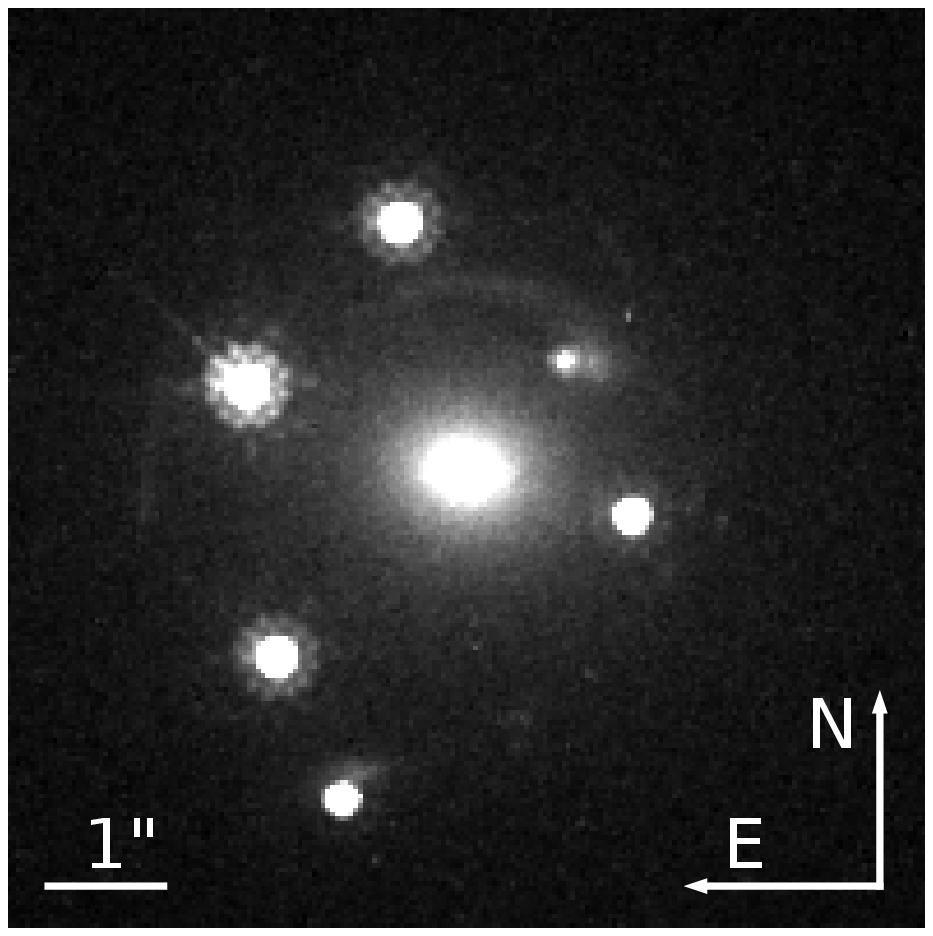}}  & \raisebox{-.5\height}{\includegraphics[height=0.2\textwidth]{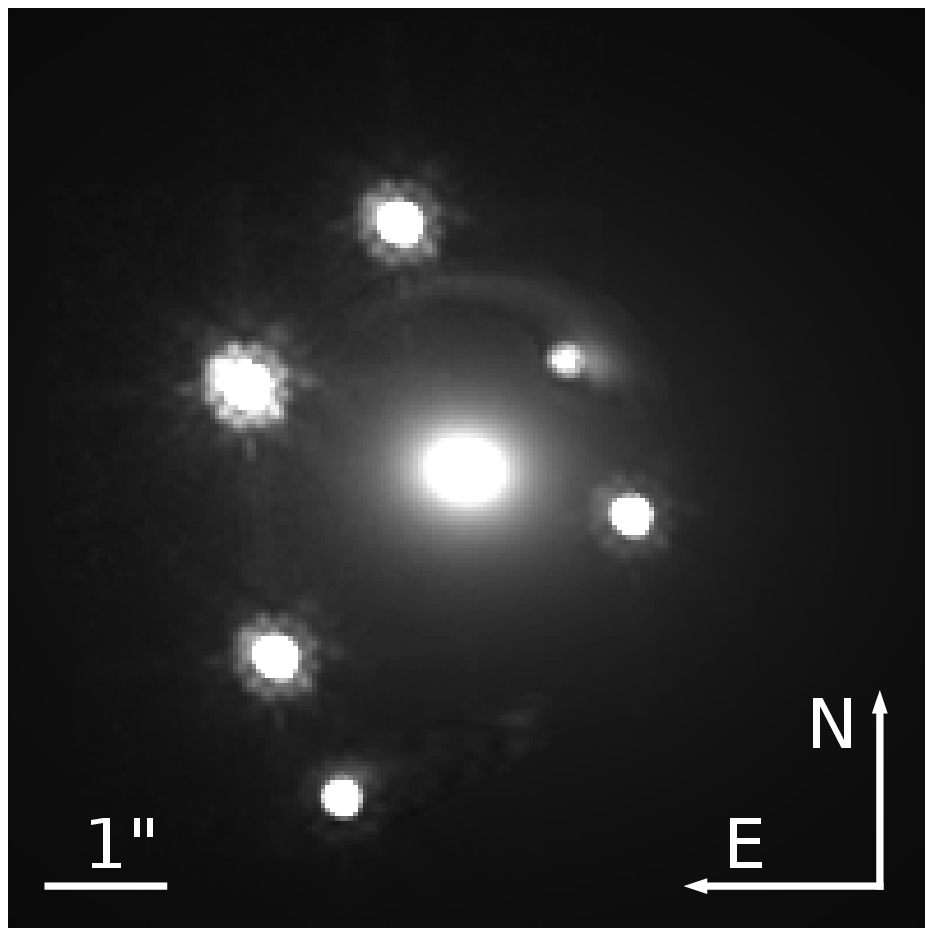}}  &  \raisebox{-.5\height}{\includegraphics[height=0.2\textwidth]{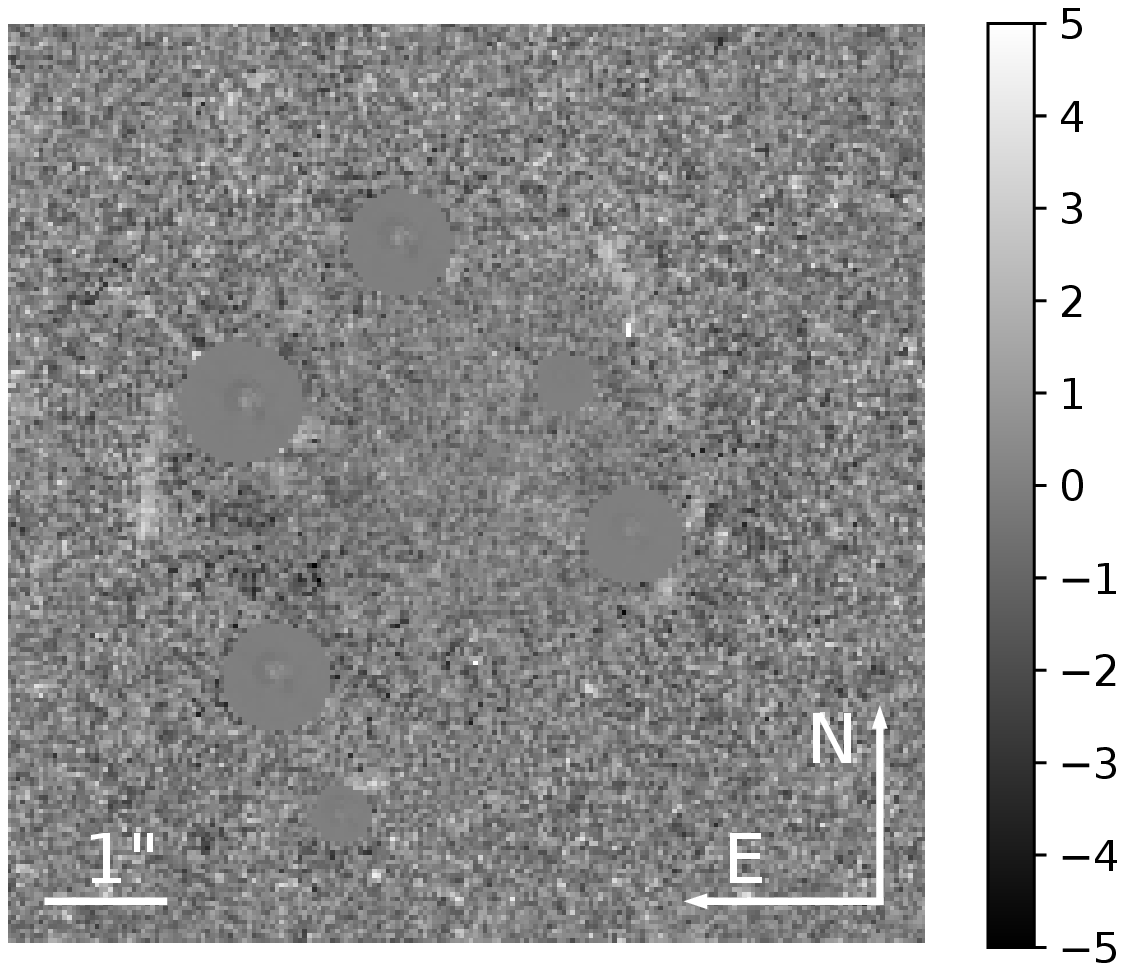}}  &
                        \raisebox{-.5\height}{\includegraphics[height=0.2\textwidth]{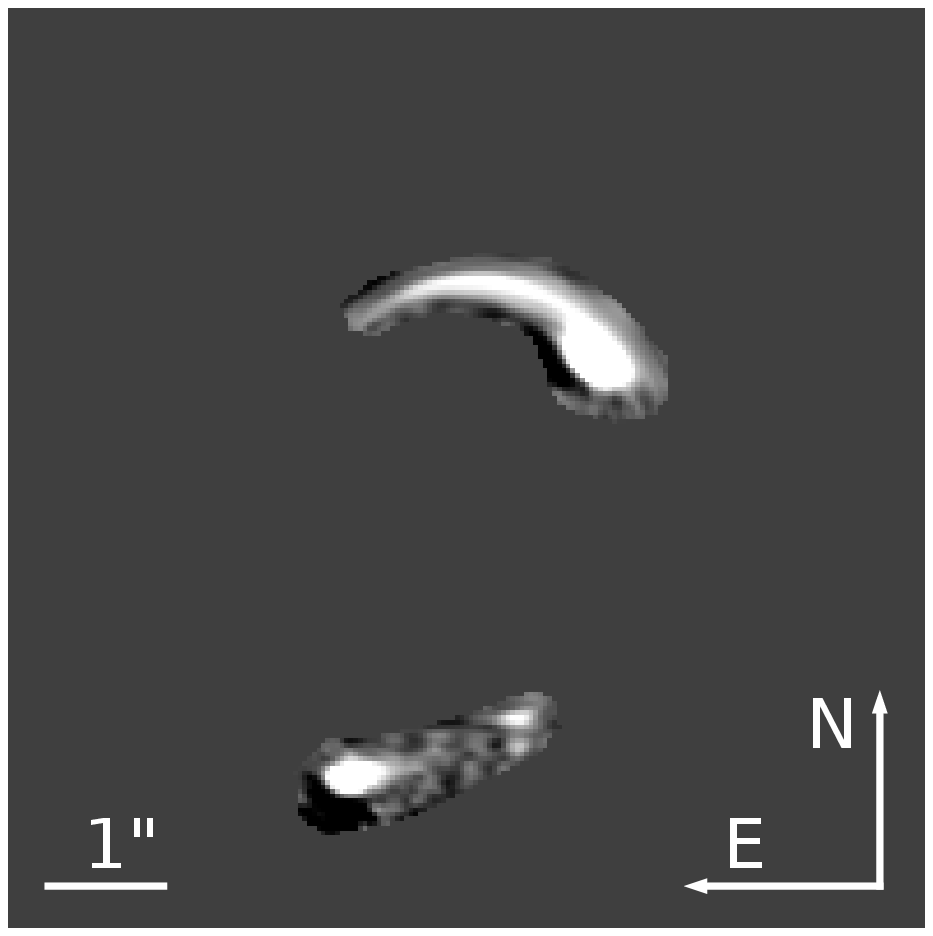}}  & 
                        \raisebox{-.5\height}{\includegraphics[width=0.2\textwidth]{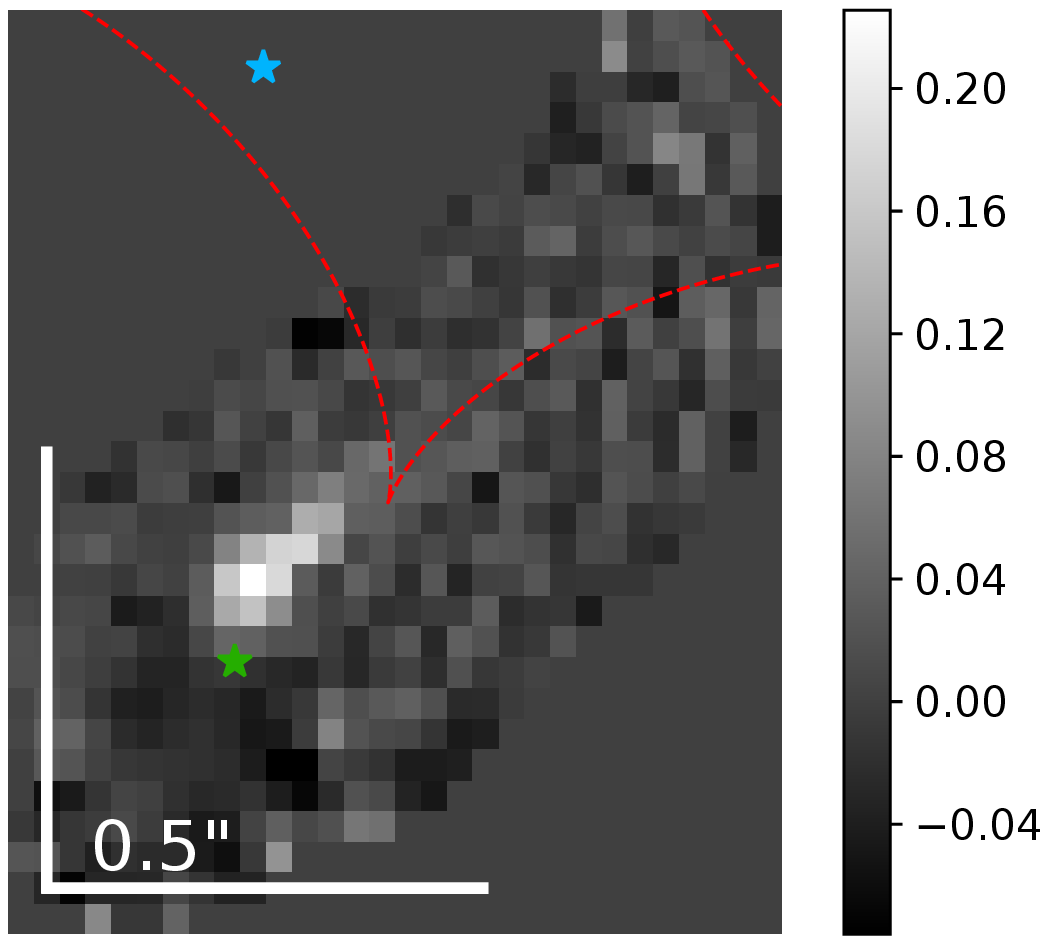}} \\ \midrule
                        
                        DES J2100$-$4452 & F160W & \raisebox{-.5\height}{\includegraphics[height=0.2\textwidth]{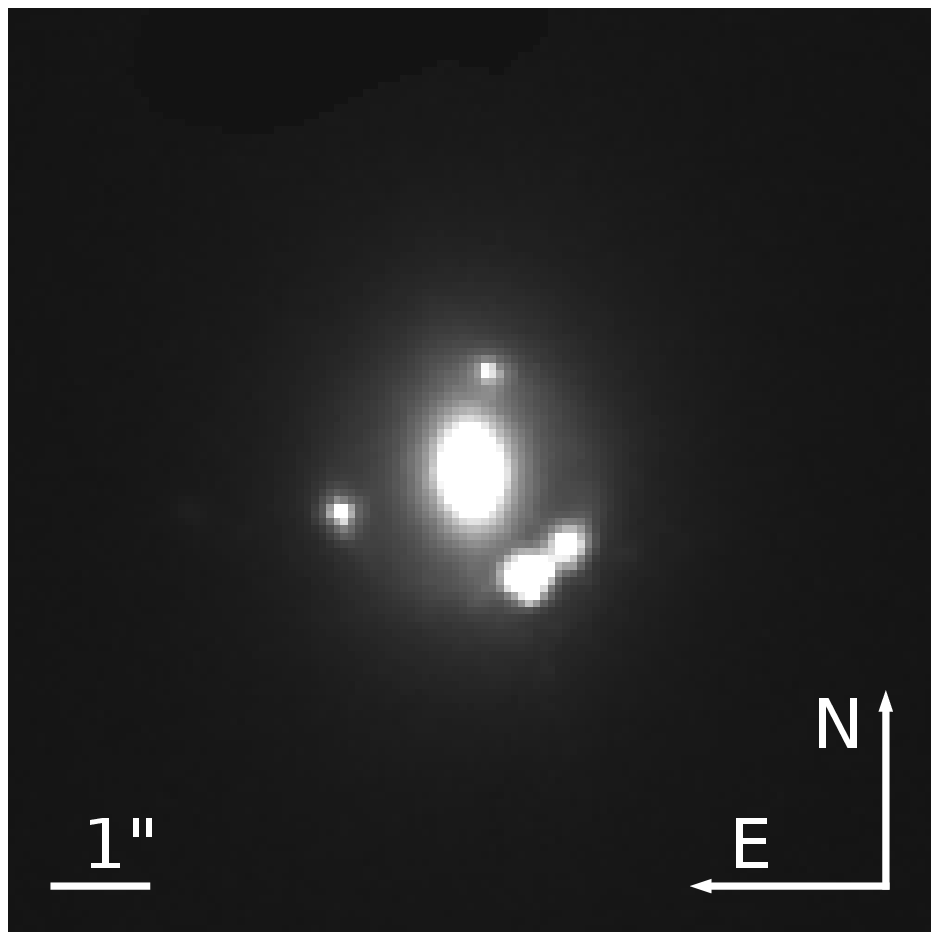}}  & \raisebox{-.5\height}{\includegraphics[height=0.2\textwidth]{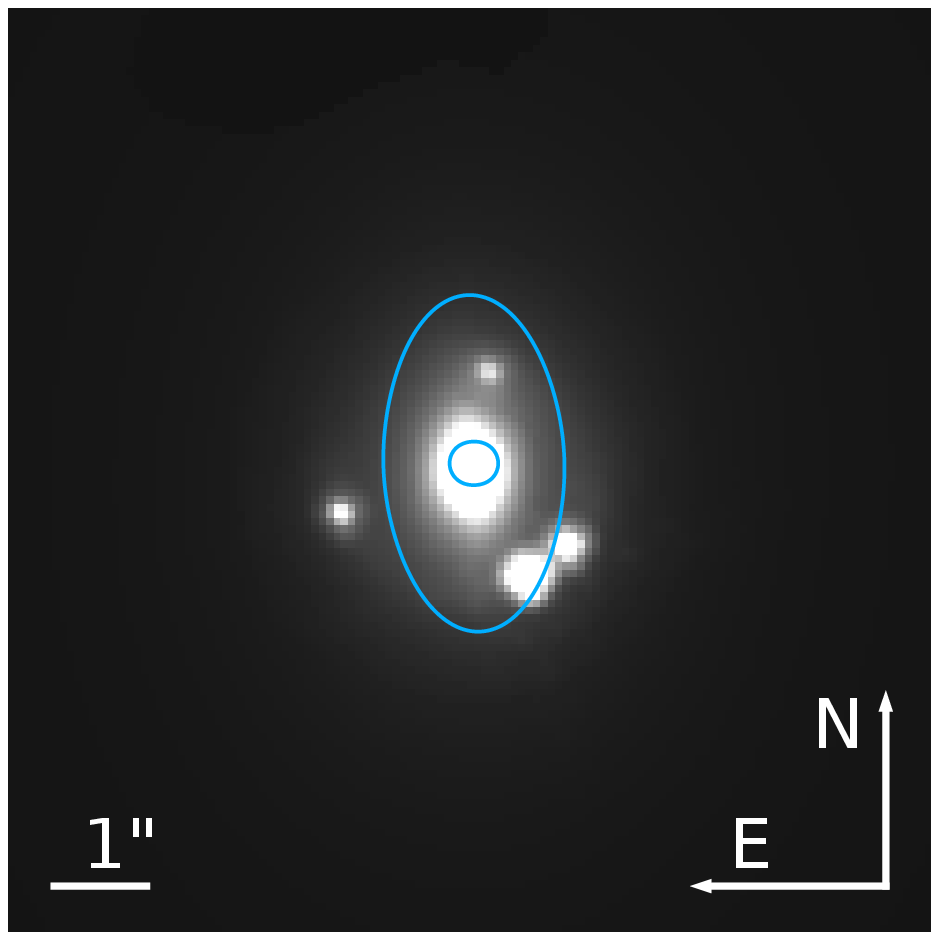}}  & \raisebox{-.5\height}{\includegraphics[height=0.2\textwidth]{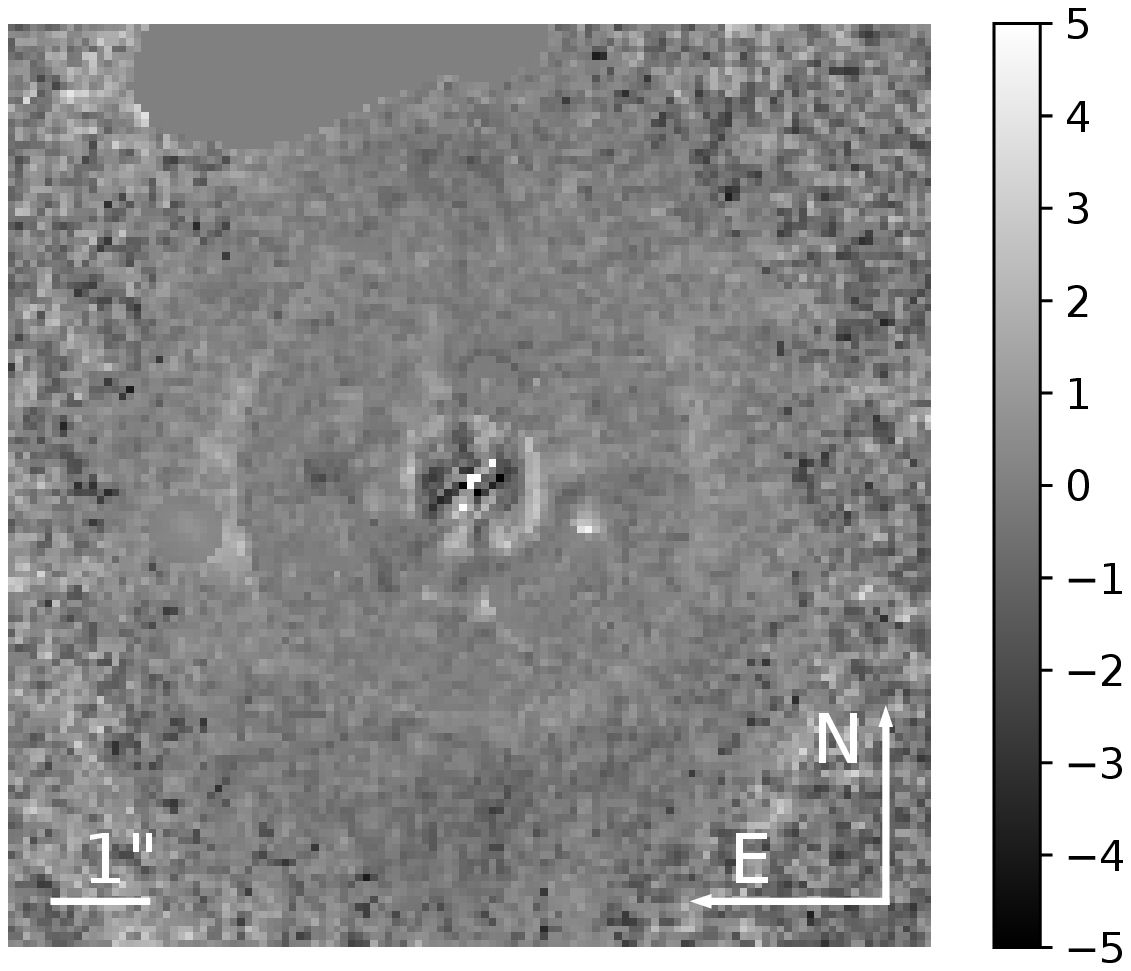}}  & 
                        \raisebox{-.5\height}{\includegraphics[height=0.2\textwidth]{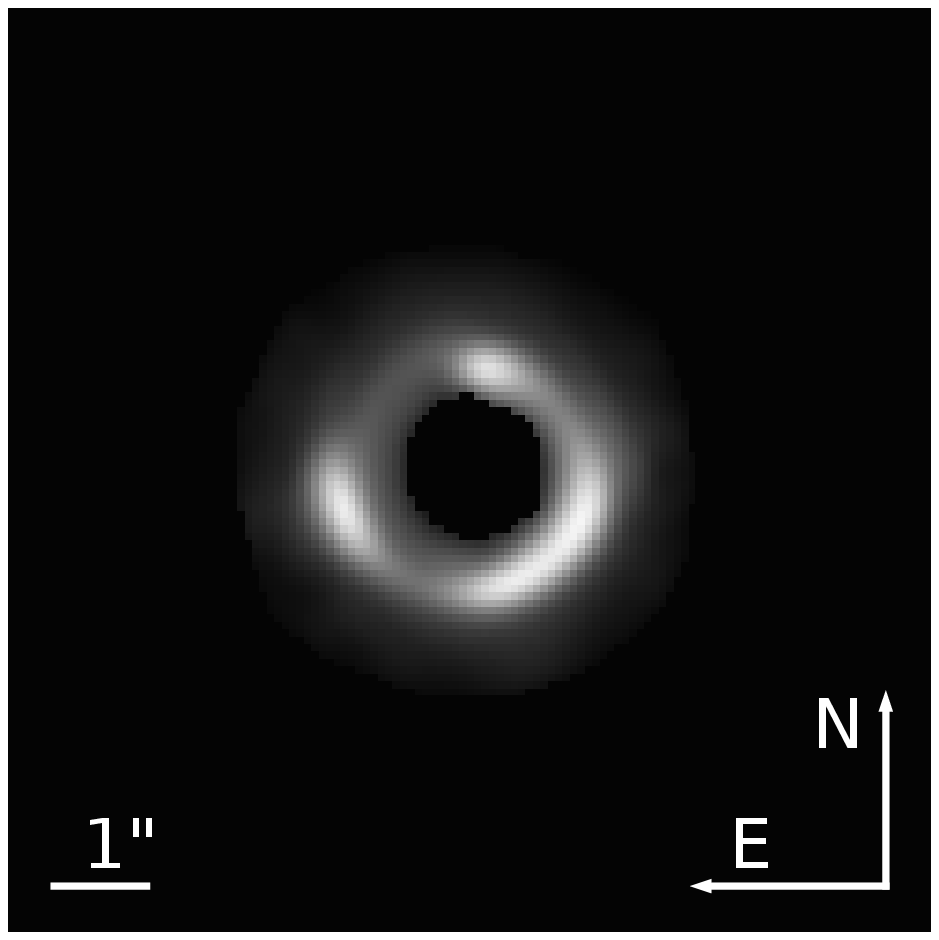}}  & 
                        \raisebox{-.5\height}{\includegraphics[width=0.2\textwidth]{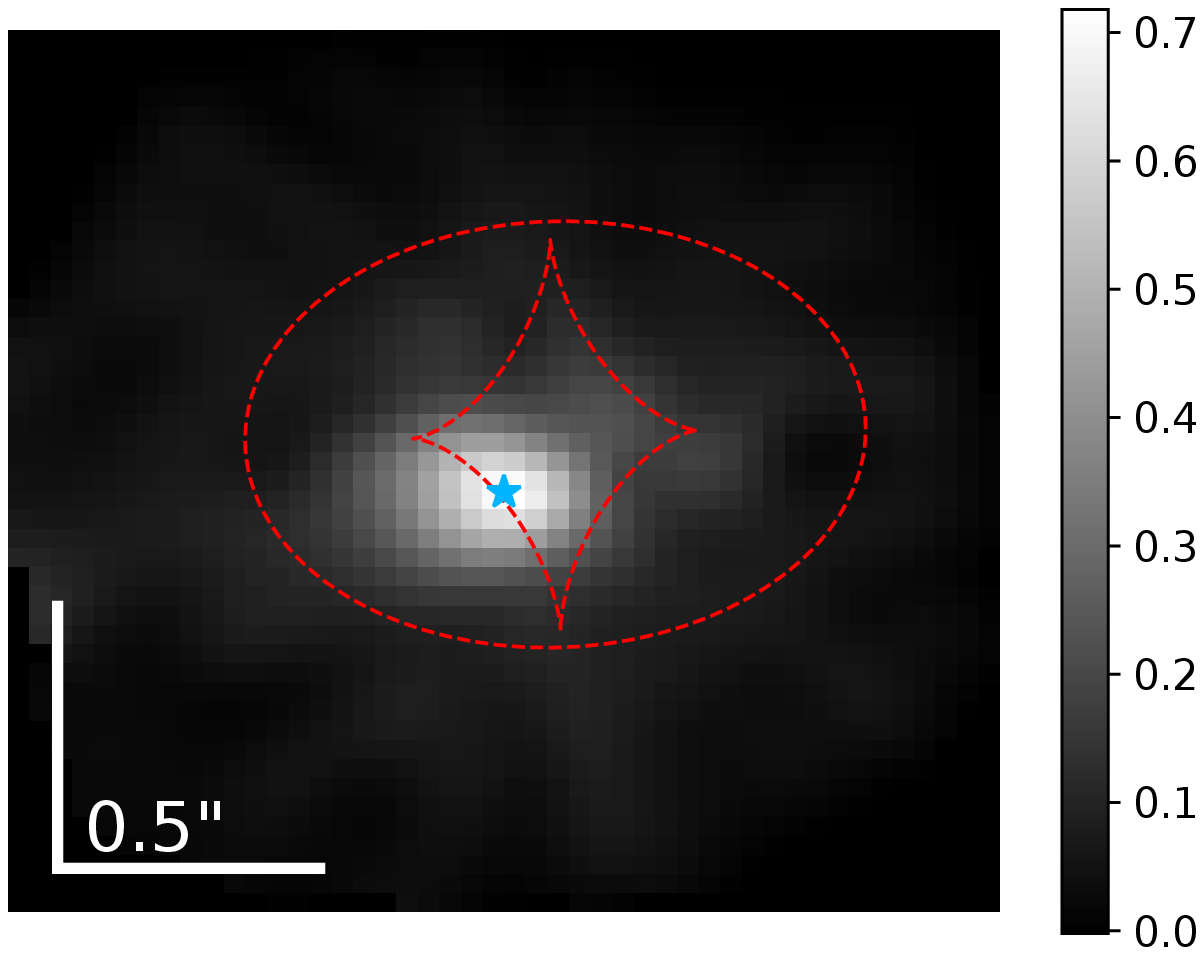}} \\ \bottomrule
                \end{tabularx}

        \ \\
\end{table}

\clearpage

\section{Final model parameters}

\begin{table}[h]
\caption{Median mass and light parameter values in the F160W band.}

\newcolumntype{L}{>{\raggedright\arraybackslash}X}
\fontsize{7}{8}\selectfont
\begin{tabularx}{\linewidth}{*{11}{L}}\toprule \toprule
        Parameter description & Parameter & \multicolumn{9}{c}{Lensing system} \\ \cmidrule(lr){3-11}
        & & DES J0029$-$3814& DES J0214$-$2105 & DES J0420$-$4037  & PS J0659$+$1629 & 2M1134$-$2103 & J1537$-$3010 & PS J1606$-$2333 & PS J1721$+$8842 & DES J2100$-$4452\\
        \toprule \toprule  
        \multicolumn{2}{l}{lens mass (SPEMD)} & & & & \\ \midrule
        \ \\
        $x$-centroid & $x_{\rm m}\ ['']$ & ${3.253}^{+0.009}_{-0.008} $ & ${4.012}^{+0.005}_{-0.004} $ & ${3.218}^{+0.003}_{-0.003} $ & ${5.12}^{+0.01}_{-0.02} $ & ${6.030}^{+0.002}_{-0.001} $ & ${3.9947}^{+0.0007}_{-0.0007} $ & ${3.994}^{+0.006}_{-0.006}$ & ${4.049}^{+0.002}_{-0.002} $ & ${8.042}^{+0.005}_{-0.005} $     \vspace{5px} \\ 
        
        $y$-centroid & $y_{\rm m}\ ['']$ & ${3.199}^{+0.004}_{-0.004} $ & ${3.994}^{+0.005}_{-0.005} $ & ${3.227}^{+0.002}_{-0.002} $ & ${5.178}^{+0.009}_{-0.008} $ & ${5.960}^{+0.002}_{-0.002} $ & ${3.925}^{+0.001}_{-0.001} $ & ${3.986}^{+0.005}_{-0.004} $ & ${3.964}^{+0.002}_{-0.002} $ & ${8.07}^{+0.01}_{-0.01} $   \vspace{5px} \\
        
        axis-ratio & $q_{\rm m}$  & 
        ${0.93}^{+0.03}_{-0.04} $ & ${0.73}^{+0.05}_{-0.07} $ & ${0.92}^{+0.02}_{-0.02} $ & ${0.43}^{+0.02}_{-0.02} $ & ${0.52}^{+0.01}_{-0.01} $ & ${0.848}^{+0.004}_{-0.004} $ & ${0.59}^{+0.03}_{-0.03} $ & ${0.742}^{+0.005}_{-0.005} $ & ${0.52}^{+0.02}_{-0.03} $    \vspace{5px} \\
        
        position angle & $\phi_{\rm m}$ [\degree]  & ${121.3}^{+24.5}_{-9.4} $ & ${166.6}^{+3.7}_{-2.8} $ & ${97.9}^{+6.8}_{-8.4} $ & ${85.5}^{+2.1}_{-1.8} $ & ${52.6}^{+0.5}_{-0.5} $ & ${124.8}^{+0.5}_{-0.5} $ & ${74.4}^{+2.2}_{-1.3} $ & ${165.8}^{+0.3}_{-0.3} $ & ${92.3}^{+0.5}_{-0.5} $   \vspace{5px} \\
        
        Einstein radius & $\theta_{\rm E}\ ['']$  & ${0.75}^{+0.01}_{-0.01}$   & ${0.855}^{+0.004}_{-0.004}$ & ${0.831}^{+0.002}_{-0.002}$  & ${2.35}^{+0.03}_{-0.02}$  & ${1.252}^{+0.003}_{-0.003}$  & ${1.427}^{+0.001}_{-0.001}$  & ${0.689}^{+0.009}_{-0.009}$ & 
        ${1.986}^{+0.002}_{-0.002}$ & 
        ${1.351}^{+0.009}_{-0.008}$   \vspace{5px} \\
        
        power-law index & $\tilde{\gamma}_{\rm PL}$  & 
        ${0.35}^{+0.06}_{-0.04} $ & ${0.6}^{+0.1}_{-0.1} $ & ${0.65}^{+0.03}_{-0.05} $ & ${0.69}^{+0.01}_{-0.01} $ & ${0.666}^{+0.004}_{-0.003} $ & ${0.301}^{+0.001}_{-0.001} $ & ${0.31}^{+0.01}_{-0.01} $ & ${0.416}^{+0.006}_{-0.007} $ & ${0.36}^{+0.05}_{-0.04} $ 
  \\
        \ \\ \midrule
                \multicolumn{2}{l}{satellite mass (SIS)} & & & & \\ \midrule
        \ \\
        $x$-centroid & $x_{\rm m, sat}\ ['']$  & $-$ & $-$ & $-$ &
        $4.80$ & $-$ & $-$ & 
        $4.29$ & $-$ & $-$  \vspace{5px} \\ 
        $y$-centroid & $y_{\rm m, sat}\ ['']$  & $-$ & $-$ & $-$ &
        $6.83$ & $-$ & $-$ & 
        $2.80$ & $-$ & $-$  \vspace{5px} \\ 
        Einstein radius & $\theta_{\rm E, sat}\ ['']$  & $-$ & $-$ & $-$ & ${0.14}^{+0.03}_{-0.04} $ & $-$ & $-$ & ${0.10}^{+0.02}_{-0.03} $ & $-$ & $-$\vspace{5px} \\
        
        \midrule
        \multicolumn{2}{l}{external shear} & & & & & \\ \midrule
        \ \\
        
        magnitude  & $\gamma_{\rm ext}$ &  ${0.23}^{+0.04}_{-0.02} $ & ${0.07}^{+0.01}_{-0.02} $ & ${0.052}^{+0.005}_{-0.007} $ & ${0.116}^{+0.006}_{-0.006} $ & ${0.397}^{+0.004}_{-0.003} $ & ${0.043}^{+0.002}_{-0.001} $ & ${0.06}^{+0.01}_{-0.01} $ & ${0.026}^{+0.002}_{-0.002} $ & ${0.12}^{+0.02}_{-0.02} $  \vspace{5px} \\ 
        position angle  & $\phi_{\rm ext}$ [\degree] & ${14.4}^{+0.7}_{-0.6} $ & ${45.3}^{+4.8}_{-13.4} $ & ${53.2}^{+3.0}_{-2.4} $ & ${132.8}^{+1.7}_{-2.3} $ & ${135.0}^{+0.1}_{-0.1} $ & ${22.3}^{+0.7}_{-0.8} $ & ${101.2}^{+5.2}_{-9.6} $ & ${64.9}^{+1.4}_{-1.6} $ & ${92.90}^{+1.05}_{-0.90} $    \\ 
        
        \ \\ \midrule
        \multicolumn{3}{l}{primary lens light (2 S\'{e}rsic \& point source)} & & & \\ \midrule
        \ \\
        
        $x$-centroid & $x_{\rm S}\ ['']$  & ${3.209}^{+0.002}_{-0.002} $ & ${4.006}^{+0.001}_{-0.001} $ & ${3.2221}^{+0.0006}_{-0.0006} $ & ${5.2210}^{+0.0007}_{-0.0007} $ & ${6.02666}^{+0.00002}_{-0.00002} $ & ${3.9961}^{+0.0007}_{-0.0007} $ & ${3.995}^{+0.001}_{-0.001} $ & ${4.0053}^{+0.0006}_{-0.0006} $ & ${8.0050}^{+0.0004}_{-0.0004} $  \vspace{5px} \\ 
        $y$-centroid & $y_{\rm S}\ ['']$  & ${3.177}^{+0.002}_{-0.002} $ & ${3.985}^{+0.001}_{-0.001} $ & ${3.2385}^{+0.0006}_{-0.0006} $ & ${5.2297}^{+0.0007}_{-0.0008} $ & 
        ${5.97334}^{+0.00002}_{-0.00002} $ & 
        ${3.9319}^{+0.0007}_{-0.0007} $ & ${3.972}^{+0.002}_{-0.002} $ & ${3.9728}^{+0.0005}_{-0.0005} $ & ${8.0035}^{+0.0005}_{-0.0005} $  \\ 
        \ \\
        \textit{S\'{e}rsic\#1} & & & & & \\
        axis ratio  & $q_{\rm S,1}$ & 
        ${0.59}^{+0.02}_{-0.02} $ & ${0.89}^{+0.02}_{-0.02} $ & ${0.51}^{+0.01}_{-0.01} $ & ${0.81}^{+0.01}_{-0.01} $ & ${0.18}^{+0.01}_{-0.01} $ & ${0.54}^{+0.01}_{-0.01} $ & ${0.893}^{+0.004}_{-0.004} $ & ${0.844}^{+0.001}_{-0.001} $ & ${0.749}^{+0.001}_{-0.001} $  \vspace{5px} \\ 
        position angle  & $\phi_{\rm S,1}$ [\degree] & ${135.5}^{+1.7}_{-1.6} $ & ${142.0}^{+4.8}_{-5.3} $ & ${113.0}^{+1.0}_{-1.0} $ & ${37.8}^{+0.9}_{-0.9} $ & ${42.4}^{+0.5}_{-0.5} $ & ${128.2}^{+0.5}_{-0.5} $ & ${146.9}^{+1.3}_{-1.3} $ & ${177.2}^{+0.2}_{-0.2} $ & ${88.4}^{+0.1}_{-0.1} $   \vspace{5px} \\ 
        amplitude & $A_{\rm S,1}$ & 
        ${0.016}^{+0.003}_{-0.002} $ & ${0.63}^{+0.09}_{-0.09} $ & ${9.8}^{+0.2}_{-0.4} $ & ${0.064}^{+0.006}_{-0.007} $ & ${0.00059}^{+0.00008}_{-0.00007} $ & 
        ${0.0207}^{+0.0008}_{-0.0007} $ & ${0.69}^{+0.01}_{-0.01} $ & 
        ${0.00618}^{+0.00007}_{-0.00007} $ & ${0.208}^{+0.003}_{-0.002} $   \vspace{5px} \\
        effective radius & $r_{\rm eff,1}\ ['']$ &  ${2.0}^{+0.2}_{-0.2} $ & ${0.37}^{+0.04}_{-0.03} $ & ${0.158}^{+0.004}_{-0.003} $ & ${2.9}^{+0.1}_{-0.1} $ & ${186.8}^{+9.4}_{-22.1} $ & ${3.14}^{+0.05}_{-0.06} $ & ${0.890}^{+0.003}_{-0.003} $ & ${4.92}^{+0.03}_{-0.03} $ & ${3.29}^{+0.02}_{-0.02} $   \vspace{5px} \\
        S\'{e}rsic index & $n_{\rm S,1}$ & ${5.6}^{+0.3}_{-0.4} $ & ${5.8}^{+0.2}_{-0.3} $ & ${5.9}^{+0.1}_{-0.1} $ & ${1.8}^{+0.1}_{-0.1} $ & ${2.3}^{+0.1}_{-0.1} $ & ${0.50}^{+0.001}_{-0.001} $ & ${0.501}^{+0.001}_{-0.001} $ & ${2.24}^{+0.04}_{-0.04} $ & ${4.76}^{+0.04}_{-0.04} $   \\
        \ \\
        \end{tabularx}

        \label{tab:results_param}
        \end{table}

        \clearpage
        
        \begin{table}\ContinuedFloat
        \captionsetup{labelformat=empty} 
        \caption{\textbf{Table \ref{tab:results_param} continued.}}
        
        \newcolumntype{L}{>{\raggedright\arraybackslash}X}
    \fontsize{7}{8}\selectfont
        \begin{tabularx}{\linewidth}{*{11}{L}}\toprule \toprule
        Parameter description & Parameter & \multicolumn{9}{c}{Lensing system} \\ \cmidrule(lr){3-11}
        & & DES J0029$-$3814 & DES J0214$-$2105 & DES J0420$-$4037  & PS J0659$+$1629 & 2M1134$-$2103 & J1537$-$3010 & PS J1606$-$2333 & PS J1721$+$8842 & DES J2100$-$4452\\
        \toprule \toprule  
        \multicolumn{3}{l}{primary lens light (2 S\'{e}rsic \& point source)} & & & \\ \midrule
        \textit{S\'{e}rsic\#2} & & & & & \\
        axis ratio  & $q_{\rm S,2}$ &
        ${0.60}^{+0.03}_{-0.03} $ & ${0.70}^{+0.01}_{-0.02} $ & ${0.854}^{+0.005}_{-0.005} $ & ${0.94}^{+0.01}_{-0.01} $ & ${0.74}^{+0.01}_{-0.01} $ & ${0.803}^{+0.006}_{-0.005} $ & ${0.871}^{+0.008}_{-0.008} $ & ${0.793}^{+0.004}_{-0.004} $ & ${0.302}^{+0.003}_{-0.001} $   \vspace{5px} \\ 
        position angle  & $\phi_{\rm S,2}$ [\degree] & ${0.6}^{+2.9}_{-1.6} $ & ${161.0}^{+1.0}_{-1.1} $ & ${123.2}^{+1.1}_{-1.1} $ & ${117.6}^{+3.3}_{-3.4} $ & ${120.6}^{+0.8}_{-0.8} $ & ${115.5}^{+0.8}_{-0.9} $ & ${63.6}^{+1.9}_{-1.9} $ & ${171.5}^{+0.6}_{-0.6} $ & ${97.3}^{+0.3}_{-0.3} $ \vspace{5px} \\ 
        amplitude & $A_{\rm S,2}$ &
        ${0.022}^{+0.003}_{-0.003} $ & ${0.021}^{+0.001}_{-0.001} $ & ${0.31}^{+0.01}_{-0.01} $ & ${1.5}^{+0.2}_{-0.2} $ & ${0.148}^{+0.006}_{-0.005} $ & ${1.08}^{+0.03}_{-0.03} $ & ${0.046}^{+0.002}_{-0.002} $ & ${0.054}^{+0.001}_{-0.001} $ & ${7.9}^{+0.3}_{-0.3} $   \vspace{5px} \\
        effective radius & $r_{\rm eff,2}\ ['']$ & ${1.60}^{+0.08}_{-0.06} $ & ${3.1}^{+0.1}_{-0.1} $ & ${1.13}^{+0.01}_{-0.01} $ & ${0.41}^{+0.04}_{-0.03} $ & ${1.26}^{+0.02}_{-0.03} $ & ${0.495}^{+0.007}_{-0.007} $ & ${2.34}^{+0.04}_{-0.05} $ & ${0.455}^{+0.007}_{-0.007} $ & ${0.288}^{+0.005}_{-0.005} $  \vspace{5px} \\
        S\'{e}rsic index & $n_{\rm S,2}$ &  ${0.53}^{+0.05}_{-0.02} $ & ${1.5}^{+0.2}_{-0.2} $ & ${1.71}^{+0.04}_{-0.04} $ & ${3.3}^{+0.2}_{-0.1} $ & ${4.7}^{+0.1}_{-0.1} $ & ${3.13}^{+0.04}_{-0.04} $ & ${0.86}^{+0.06}_{-0.05} $ & ${1.24}^{+0.02}_{-0.03} $ & ${3.02}^{+0.08}_{-0.08} $  \\       
        \ \\
        \textit{Point source} & & & & & \\
        amplitude & $A_{\rm ps}$ &
        ${10.5}^{+0.9}_{-1.0} $ & $-$ & $-$ & $-$ & $-$ & $-$ & ${73.5}^{+0.2}_{-0.3} $ & ${0.2}^{+0.1}_{-0.1} $ & $-$    \\
        \midrule
\multicolumn{2}{l}{satellite light (S\'{e}rsic)} & & & & \\ \midrule
        \ \\
        
        $x$-centroid & $x_{\rm sat}\ ['']$  &  $-$ & $-$ & $-$ &
        $4.80$ & $-$ & $-$ & 
        $4.29$ & $-$ & $-$  \vspace{5px} \\ 
        $y$-centroid & $y_{\rm sat}\ ['']$  &  $-$ & $-$ & $-$ &
        $6.83$ & $-$ & $-$ & 
        $2.80$ & $-$ & -  \vspace{5px} \\ 
        axis ratio  & $q_{\rm sat}$ & $-$ & $-$ & $-$ &
        $0.76$ & $-$ & $-$ & 
        $0.34$ & $-$ & $-$  \vspace{5px} \\
        position angle  & $\phi_{\rm sat}$ [\degree] & $-$ & $-$ & $-$ &
        $160.08$ & $-$ & $-$ & 
        $83.54$ & $-$ & $-$  \vspace{5px} \\
        amplitude & $A_{\rm sat}$ & $-$ & $-$ & $-$ &
        $0.08$ & $-$ & $-$ & 
        $0.36$ & $-$ & $-$  \vspace{5px} \\
        effective radius & $r_{\rm eff,sat}\ ['']$ & $-$ & $-$ & $-$ &
        $1.11$ & $-$ & $-$ & 
        $0.20$ & $-$ & $-$  \vspace{5px} \\
        S\'{e}rsic index & $n_{\rm sat}$ & $-$ & $-$ & $-$ &
        $4.77$ & $-$ & $-$ & 
        $5.87$ & $-$ & $-$  \vspace{5px} \\
         \midrule
         
        quasar light & & & & & \\ \midrule
        \ \\
        \textit{Quasar\#A} & & & & & \\
        $x$-centroid \hspace{85px} & $x_{\rm QSO,A}\ ['']$ &
        $4.0486$   &
        $3.5069$  &
        $3.9264$ &
        $3.3407$ &
        $5.2792$ & 
        $2.5612$&
        $3.1513$ &
        $2.1149$ &
        $9.0466$ \vspace{5px} \\ 
        
        $y$-centroid & $y_{\rm QSO,A}\ ['']$  &
        $3.3999$ &
        $3.1362$ &
        $2.8945$ &
        $6.2785$ &
        $5.1995$ &
        $3.1630$ &
        $4.3453$ &
        $4.7010$ &
        $7.1921$ \vspace{5px} \\  
        
        amplitude & $A_{\rm QSO,A}$ &
        $156.11$ &
        $272.95$ &
        $172.25$ &
        $2272.19$ & 
        $2421.62$  &
        $341.53$  &
        $323.23$ &
        $774.05$ &
        $184.10$ \\
        \ \\
        \textit{Quasar\#B} & & & & & \\
        $x$-centroid & $x_{\rm QSO,B}\ ['']$  &
        $3.3976$ &
        $4.3237$ &
        $3.6861$&
        $8.0907$ &
        $4.5511$ &
        $4.5556$ &
        $4.7760$ &
        $3.4383$ &
        $8.6241$ \vspace{5px} \\ 
        
        $y$-centroid & $y_{\rm QSO,B}\ ['']$  &
        $3.7123$ &
        $4.7914$ &
        $3.9210$ &
        $4.0381$ &
        $6.9546$ &
        $2.8461$ &
        $3.7531$ &
        $6.1091$ &
        $6.8608$ \vspace{5px} \\  
        
        amplitude & $A_{\rm QSO,B}$ &
        $74.42$ &
        $270.04$ &
        $138.73$ &
        $694.04$ & 
        $2874.75$  &
        $284.55$  &
        $215.42$  &
        $354.28$ &
        $480.00$ \\
        \ \\
        \textit{Quasar\#C} & & & & & \\
        $x$-centroid & $x_{\rm QSO,C}\ ['']$  &
        $3.6689$ &
        $3.4361$ &
        $2.5129$ &
        $3.4261$ &
        $7.2261$ &
        $5.4041$ &
        $3.9512$ &
        $2.3668$ &
        $6.6032$ \vspace{5px} \\ 
        
        $y$-centroid & $y_{\rm QSO,C}\ ['']$  & 
        $2.8328$ & 
        $4.5200$ & 
        $2.6768$  &
        $4.3756$ &
        $4.4272$ &
        $4.8098$ &
        $3.4377$ &
        $2.3534$ &
        $7.5404$ \vspace{5px} \\  
        
        amplitude & $A_{\rm QSO,C}$ &
        $56.14$ &
        $156.48$ & 
        $41.67$ &
        $559.82$ &
        $2942.03$ &
        $300.08$&
        $67.05$ &
        $340.03$ &
        $82.30$ \\
        \ \\
        \textit{Quasar\#D} & & & & & \\
        $x$-centroid & $x_{\rm QSO,D}\ ['']$  &
        $1.9218$ &
        $4.7708$ &
        $3.0603$ &
        $4.4029$ &
        $6.5268$ &
        $3.3031$  &
        $4.2779$ &
        $5.4424$ &
        $8.2033$ \vspace{5px} \\ 
        
        $y$-centroid & $y_{\rm QSO,D}\ ['']$  &
        $2.8143$ & 
        $3.8688$ &
        $4.0253$ &
        $7.2629$ &
        $6.5686$ &
        $4.9344$  &
        $4.4968$ &
        $3.5779$ &
        $9.0707$ \vspace{5px} \\  
        
        amplitude & $A_{\rm QSO,D}$ &
        $72.44$ & 
        $290.99$ &
        $39.08$ &
        $455.58$ &
        $559.82$  &
        $142.76$&
        $286.37$&
        $158.33$ &
        $47.55$ \\
        
        \bottomrule
        \end{tabularx}

        \end{table}
        
        \clearpage
        
        \begin{table}\ContinuedFloat
    \captionsetup{labelformat=empty} 
        \caption{\textbf{Table \ref{tab:results_param} continued.}}
        
        \newcolumntype{L}{>{\raggedright\arraybackslash}X}
    \fontsize{7}{8}\selectfont
        \begin{tabularx}{\linewidth}{*{11}{L}}\toprule \toprule
        Parameter description & Parameter & \multicolumn{9}{c}{Lensing system} \\ \cmidrule(lr){3-11}
        & & DES J0029$-$3814 & DES J0214$-$2105 & DES J0420$-$4037  & PS J0659$+$1629 & 2M1134$-$2103 & J1537$-$3010 & PS J1606$-$2333 & PS J1721$+$8842 & DES J2100$-$4452\\
        \toprule \toprule  
        quasar light & & & & & \\ \midrule
\ \\
        \textit{Quasar\#E} & & & & & \\
        $x$-centroid & $x_{\rm QSO,E}\ ['']$  & $-$ & $-$ & $-$ & $-$ & $-$ & $-$ & $-$ &
        $2.9345$ &
        $-$ \vspace{5px} \\ 
        $y$-centroid & $y_{\rm QSO,E}\ ['']$ & $-$ & $-$ & $-$ & $-$ & $-$ & $-$ & $-$ &
        $1.1298$ &
        $-$ \vspace{5px} \\  
        amplitude & $A_{\rm QSO,E}$ & $-$ & $-$ & $-$ & $-$ & $-$ & $-$ & $-$ &
        $87.26$ & 
        $-$ \vspace{5px}
        \\
        \ \\
        \textit{Quasar\#F} & & & & & \\
        $x$-centroid & $x_{\rm QSO,F}\ ['']$  & $-$ & $-$ & $-$ & $-$ & $-$ & $-$ & $-$ &
        $4.8644$ &
        $-$ \vspace{5px} \\ 
        $y$-centroid & $y_{\rm QSO,F}\ ['']$  & $-$ & $-$ & $-$ & $-$ & $-$ & $-$ & $-$ &
        $4.9188$ &
        $-$ \vspace{5px} \\  
        amplitude & $A_{\rm QSO,F}$ & $-$ & $-$ & $-$ & $-$ & $-$ & $-$ & $-$ & 
        $14.61$ &
        $-$ \vspace{5px}
        \\
        \bottomrule
        \end{tabularx}

        \end{table}
        
        \clearpage
        
        \begin{table}[h]
\caption{Median light parameter values in the F475X band}

\newcolumntype{L}{>{\raggedright\arraybackslash}X}
\fontsize{7}{8}\selectfont
\begin{tabularx}{\linewidth}{*{11}{L}}\toprule \toprule
        Parameter description & Parameter & \multicolumn{9}{c}{Lensing system} \\ \cmidrule(lr){3-11}
        & & DES J0029$-$3814 & DES J0214$-$2105 & DES J0420$-$4037  & PS J0659$+$1629 & 2M1134$-$2103 & J1537$-$3010 & PS J1606$-$2333 & PS J1721$+$8842 & DES J2100$-$4452\\
        \toprule \toprule  
        \multicolumn{3}{l}{primary lens light (2 S\'{e}rsic \& point source)} & & & \\ \midrule
        \textit{S\'{e}rsic\#1} & & & & & \\
        amplitude & $A_{\rm S,1 (475)}$&
        $-$ &
        $-$ &
        $-$ &
        $-$ &
        $-$ & 
        $0.002_{-0.0002}^{+0.0002}$ & 
        $-$ & 
        $0.0137_{-0.0002}^{+0.0001}$ & 
        $-$ \vspace{3px} \\
        \textit{S\'{e}rsic\#2} & & & & & \\
        amplitude & $A_{\rm S,2 (475)}$ &
        $-$ &
        $-$ &
        $-$ &
        $-$ &
        $-$ & 
        $0.0073_{-0.0005}^{+0.0005}$ & 
        $-$ & 
        $0.145_{-0.003}^{+0.003}$ & 
        $-$ \vspace{3px} \\
        \textit{Point source} & & & & & \\
        amplitude & $A_{\rm ps (475)}$ &
        $-$ &
        $-$ &
        $-$ &
        $-$ &
        $-$ & 
        $-$ & 
        $-$ & 
        $4.25_{-0.39}^{+0.55}$ & 
        $-$ \vspace{3px} \\
        \midrule
        \multicolumn{2}{l}{satellite light (S\'{e}rsic)} & & & & \\ \midrule
        amplitude & $A_{\rm sat (475)}$ &
        $-$ &
        $-$ &
        $-$ &
        $-$ &
        $-$ & 
        $-$ & 
        $-$ & 
        $-$ & 
        $-$ \vspace{3px}\\
        \midrule
        quasar light & & & & & \\ \midrule
        \textit{Quasar\#A} & & & & & \\
        amplitude & $A_{\rm QSO,A (475)}$ &
        $-$ &
        $-$ &
        $-$ &
        $-$ &
        $-$ & 
        $233.32$ & 
        $-$ & 
        $669.25$ & 
        $-$ 
        \vspace{3px} \\
        \textit{Quasar\#B} & & & & & \\
        amplitude & $A_{\rm QSO,B (475)}$ &
        $-$ &
        $-$ &
        $-$ &
        $-$ &
        $-$ & 
        $192.35$ & 
        $-$ & 
        $362.34$ & 
        $-$ \vspace{3px} \\
        \textit{Quasar\#C} & & & & & \\
        amplitude & $A_{\rm QSO,C (475)}$ &
        $-$ &
        $-$ &
        $-$ &
        $-$ &
        $-$ & 
        $202.20$ & 
        $-$ & 
        $345.94$ & 
        $-$ \vspace{3px} \\
        \textit{Quasar\#D} & & & & & \\
        amplitude & $A_{\rm QSO,D (475)}$ &
        $-$ &
        $-$ &
        $-$ &
        $-$ &
        $-$ & 
        $75.88$ & 
        $-$ & 
        $139.11$ & 
        $-$  \vspace{3px} \\
        \textit{Quasar\#E} & & & & & \\
        amplitude & $A_{\rm QSO,E (475)}$ &
        $-$ & $-$ & $-$ & $-$ & $-$ & $-$ & $-$ & $91.63$ & $-$ \vspace{3px} \\
        \textit{Quasar\#F} & & & & & \\
        amplitude & $A_{\rm QSO,F (475)}$ &
        $-$ & $-$ & $-$ & $-$ & $-$ & $-$ & $-$ & $22.76$ & $-$ \vspace{3px} \\ \bottomrule
        
        \end{tabularx}
        \ \\
        \caption*{\textbf{Notes: }The parameter values and uncertainties are taken from the final (fully) converged chain.}
        \label{tab:results_param_475}
        \end{table}

        \clearpage
        
\begin{table}
\caption{Median light parameter values in the F814W band}

\newcolumntype{L}{>{\raggedright\arraybackslash}X}
\fontsize{7}{8}\selectfont
\begin{tabularx}{\linewidth}{*{11}{L}}\toprule \toprule
        Parameter description & Parameter & \multicolumn{9}{c}{Lensing system} \\ \cmidrule(lr){3-11}
        & & DES J0029$-$3814 & DES J0214$-$2105 & DES J0420$-$4037  & PS J0659$+$1629 & 2M1134$-$2103 & J1537$-$3010 & PS J1606$-$2333 & PS J1721$+$8842 & DES J2100$-$4452\\
        \toprule \toprule  
        \multicolumn{3}{l}{primary lens light (2 S\'{e}rsic \& point source)} & & & \\ \midrule
        \textit{S\'{e}rsic\#1} & & & & & \\
        amplitude & $A_{\rm S,1 (814)}$&
        $-$ &
        $-$ &
        $-$ &
        $-$ &
        $0.000026_{-0.000006}^{+0.000020}$ & 
        $0.0031_{-0.0002}^{+0.0002}$ & 
        $-$ & 
        $0.328_{-0.004}^{+0.003}$ & 
        $-$ \vspace{3px} \\
        \textit{S\'{e}rsic\#2} & & & & & \\
        amplitude & $A_{\rm S,2 (814)}$ &
        $-$ &
        $-$ &
        $-$ &
        $-$ &
        $0.0071_{-0.0006}^{+0.0008}$ & 
        $0.0363_{-0.002}^{+0.002}$ & 
        $-$ & 
        $3.60_{-0.06}^{+0.06}$ & 
        $-$ \vspace{3px} \\
        \textit{Point source} & & & & & \\
        amplitude & $A_{\rm ps (814)}$ &
        $-$ &
        $-$ &
        $-$ &
        $-$ &
        $-$ & 
        $-$ & 
        $-$ & 
        $7.5_{-0.6}^{+0.9}$ & 
        $-$ \\ \vspace{3px} \\
        \midrule
        \multicolumn{2}{l}{satellite light (S\'{e}rsic)} & & & & \\ \midrule
        amplitude & $A_{\rm sat (814)}$ &
        $-$ &
        $-$ &
        $-$ &
        $-$ &
        $-$ & 
        $-$ & 
        $-$ & 
        $-$ & 
        $-$ \vspace{3px}\\
        \midrule
        quasar light & & & & & \\ \midrule
        \textit{Quasar\#A} & & & & & \\
        amplitude & $A_{\rm QSO,A (814)}$ &
        $-$ &
        $-$ &
        $-$ &
        $-$ &
        $1114.33$ & 
        $151.38$ & 
        $-$ & 
        $1946.88$ & 
        $-$ \vspace{3px} \\
        \textit{Quasar\#B} & & & & & \\
        amplitude & $A_{\rm QSO,B (814)}$ &
        $-$ &
        $-$ &
        $-$ &
        $-$ &
        $1206.04$ & 
        $119.95$ & 
        $-$ & 
        $1129.38$ & 
        $-$ \vspace{3px} \\
        \textit{Quasar\#C} & & & & & \\
        amplitude & $A_{\rm QSO,C (814)}$ &
        $-$ &
        $-$ &
        $-$ &
        $-$ &
        $1170.66$ & 
        $131.65$ & 
        $-$ & 
        $1079.34$ & 
        $-$ \vspace{3px} \\
        \textit{Quasar\#D} & & & & & \\
        amplitude & $A_{\rm QSO,D (814)}$ &
        $-$ &
        $-$ &
        $-$ &
        $-$ &
        $274.30$ & 
        $69.48$ & 
        $-$ & 
        $413.29$ & 
        $-$ \vspace{3px} \\
        \textit{Quasar\#E} & & & & & \\
        amplitude & $A_{\rm QSO,E (814)}$ &
        $-$ & $-$ & $-$ & $-$ & $-$ & $-$ & $-$ & $254.68$ & $-$ \vspace{3px} \\
        \textit{Quasar\#F} & & & & & \\
        amplitude & $A_{\rm QSO,F (814)}$ &
        $-$ & $-$ & $-$ & $-$ & $-$ & $-$ & $-$ & $44.33$ & $-$ \vspace{3px} \\ \bottomrule
        
        \end{tabularx}
        \ \\
        \caption*{\textbf{Notes: }The parameter values and uncertainties are taken from the final (fully) converged chain.}
        \label{tab:results_param_814}
        \end{table}

\end{landscape}
\clearpage

\section{Results and predictions for other lensing quantities}

\begin{table}[h]
        \caption{Convergence $\kappa$, total shear strength $\gamma_{\rm tot}$, position angle $\phi_{\gamma_{\rm tot}}$ of the total shear ($\gamma_{\rm tot}$), and magnification $\mu$ at the (modeled) image positions.}

        \begin{tabular}{lccccc}
                \toprule System & Image & Convergence & Total shear & Position angle of total shear & Image mag.  \\ & & $\kappa$ & $\gamma_{\rm tot}$ & $\phi_{\gamma_{\rm tot}}$ & $\mu$ \\ \toprule
                J0029$-$3814    & A & $0.60_{-0.07}^{+0.04}$ & $0.08_{-0.01}^{+0.01}$ & $102.5_{-2.2}^{+2.1}$ & $6.4_{-1.7}^{+1.6}$ \vspace{2px}\\
                & B & $0.82_{-0.05}^{+0.03}$ & $0.6_{-0.1}^{+0.1}$ & $174.6_{-0.2}^{+0.3}$ & $-3.0_{-0.9}^{+0.9}$ \vspace{2px}\\
        & C & $0.82_{-0.05}^{+0.02}$ & $0.59_{-0.07}^{+0.1}$ & $37.6_{-0.2}^{+0.2}$ & $-3.2_{-0.9}^{+1}$ \vspace{2px}\\
                & D & $0.41_{-0.07}^{+0.05}$ & $0.02_{-0.01}^{+0.02}$ & $171.8_{-14.7}^{+5.9}$ & $2.9_{-0.6}^{+0.5}$ \vspace{2px}\\ \midrule
                J0214$-$2105    & A & $0.3_{-0.1}^{+0.1}$ & $0.36_{-0.05}^{+0.04}$ &  $152.5_{-0.3}^{+0.3}$ & $3.2_{-0.8}^{+1.3}$  \vspace{2px}\\
                                        & B & $0.4_{-0.1}^{+0.1}$ & $0.44_{-0.07}^{+0.07}$ & $162.1_{-0.5}^{+0.5}$ &  $5.4_{-1.4}^{+2.4}$ \vspace{2px}\\
                                & C & $0.5_{-0.1}^{+0.1}$ & $0.7_{-0.1}^{+0.1}$ & $47.0_{-0.3}^{+0.3}$ & $-3.7_{-1.8}^{+1.2}$ \vspace{2px}\\
                                & D & $0.6_{-0.1}^{+0.1}$ & $0.7_{-0.1}^{+0.1}$ &  $78.2_{-0.2}^{+0.2}$ & $-2.8_{-1.4}^{+0.9}$\vspace{2px}\\ \midrule
                J0420$-$4037    & A & $0.37_{-0.03}^{+0.05}$ & $0.74_{-0.06}^{+0.04}$ & $64.3_{-0.3}^{+0.3}$ & $-6.8_{-1.3}^{+0.9}$ \vspace{2px}\\
                                        & B & $0.35_{-0.03}^{+0.04}$ & $0.59_{-0.04}^{+0.03}$ & $146.0_{-0.4}^{+0.4}$ & $14.3_{-1.9}^{+2.6}$ \vspace{2px}\\
                                        & C & $0.31_{-0.03}^{+0.04}$ & $0.54_{-0.03}^{+0.02}$ &  $126.4_{-0.2}^{+0.2}$ & $5.5_{-0.7}^{+0.9}$ \vspace{2px}\\
                                    & D &  $0.38_{-0.03}^{+0.05}$ & $0.69_{-0.05}^{+0.04}$ & $13.4_{-0.4}^{+0.4}$ & $-11.6_{-2.2}^{+1.6}$ \vspace{2px}\\ \midrule
                J0659$+$1629      & A & $0.278_{-0.009}^{+0.009}$ & $0.588_{-0.008}^{+0.008}$ & $66.1_{-0.3}^{+0.4}$ & $5.7_{-0.2}^{+0.2}$ \vspace{2px}\\ 
                                        & B & $0.136_{-0.006}^{+0.007}$ & $0.33_{-0.01}^{+0.01}$ & $78.0_{-0.9}^{+1.1}$ & $1.57_{-0.03}^{+0.04}$ \vspace{2px}\\
                                                        & C& $0.30_{-0.02}^{+0.01}$ & $0.90_{-0.02}^{+0.01}$ & $116.0_{-0.3}^{+0.3}$ & $-3.2_{-0.2}^{+0.2}$ \vspace{2px}\\
                                        & D & $0.541_{-0.008}^{+0.009}$ & $0.81_{-0.02}^{+0.01}$ & $21.4_{-0.4}^{+0.4}$ & $-2.2_{-0.1}^{+0.1}$ \vspace{2px}\\ \midrule
                2M1134$-$2103           & A & $0.627_{-0.009}^{+0.009}$ & $1.40_{-0.01}^{+0.02}$ & $134.84_{-0.08}^{+0.09}$ & $-0.55_{-0.01}^{+0.01}$ \vspace{2px}\\
                & B & $0.134_{-0.004}^{+0.003}$ & $0.145_{-0.003}^{+0.003}$ & $105.5_{-0.6}^{+0.6}$ & $1.37_{-0.01}^{+0.01}$ \vspace{2px}\\
                                        & C & $0.123_{-0.004}^{+0.003}$ & $0.113_{-0.003}^{+0.004}$ & $152.3_{-0.6}^{+0.7}$ & $1.32_{-0.01}^{+0.01}$ \vspace{2px}\\
                                    & D & $0.98_{-0.02}^{+0.02}$ & $1.95_{-0.03}^{+0.03}$ & $139.35_{-0.08}^{+0.09}$ & $-0.263_{-0.007}^{+0.008}$ \vspace{2px}\\ \midrule
                J1537$-$3010            & A & $0.617_{-0.001}^{+0.001}$ & $0.1892_{-0.0004}^{+0.0006}$ & $118.9_{-0.1}^{+0.1}$ & $8.99_{-0.06}^{+0.05}$ \vspace{2px}\\
                                        & B & $0.808_{-0.002}^{+0.001}$ & $0.432_{-0.001}^{+0.002}$ & $27.46_{-0.04}^{+0.04}$ & $-6.68_{-0.05}^{+0.05}$ \vspace{2px}\\
                                        & C & $0.605_{-0.001}^{+0.001}$ & $0.1855_{-0.0004}^{+0.0005}$ & $122.2_{-0.1}^{+0.1}$ & $8.22_{-0.06}^{+0.05}$ \vspace{2px}\\
                                    & D & $0.800_{-0.002}^{+0.001}$ & $0.427_{-0.001}^{+0.002}$ & $32.33_{-0.04}^{+0.05}$ & $-7.01_{-0.05}^{+0.06}$ \vspace{2px}\\ \midrule
                J1606$-$2333        & A & $0.52_{-0.02}^{+0.01}$ & $0.178_{-0.009}^{+0.009}$ & $70.2_{-0.7}^{+1.0}$ & $5.0_{-0.4}^{+0.2}$ \vspace{2px}\\
                                            & B & $0.58_{-0.02}^{+0.01}$ & $0.151_{-0.005}^{+0.007}$ & $85.4_{-1.2}^{+1.4}$ & $6.3_{-0.5}^{+0.3}$ \vspace{2px}\\
                                        & C & $0.99_{-0.02}^{+0.02}$ & $0.53_{-0.02}^{+0.03}$ & $175.1_{-1.5}^{+1.0}$ & $-3.6_{-0.2}^{+0.3}$ \vspace{2px}\\

                                    & D & $0.90_{-0.01}^{+0.02}$ & $0.51_{-0.01}^{+0.02}$ & $151.3_{-0.4}^{+0.5}$ & $-3.9_{-0.3}^{+0.3}$ \vspace{2px}\\ \midrule
                J1721$+$8842                                            & A & $0.632_{-0.005}^{+0.006}$ & $0.508_{-0.008}^{+0.007}$ & $69.13_{-0.08}^{+0.08}$ & $-8.2_{-0.3}^{+0.3}$ \vspace{2px}\\
                                        & B & $0.498_{-0.007}^{+0.007}$ & $0.341_{-0.005}^{+0.004}$ & $20.5_{-0.1}^{+0.1}$  & $7.4_{-0.2}^{+0.3}$ \vspace{2px}\\

                                        & C & $0.485_{-0.007}^{+0.008}$ & $0.314_{-0.004}^{+0.004}$ & $131.4^{+0.2}_{-0.2}$ & $6.0_{-0.2}^{+0.2}$ \vspace{2px}\\
                                        & D & $0.868_{-0.005}^{+0.005}$ & $0.69_{-0.01}^{+0.01}$ & $75.5_{-0.05}^{+0.05}$ & $-2.20_{-0.08}^{+0.08}$ \vspace{2px}\\
                                        & E & $0.364_{-0.007}^{+0.008}$ & $0.213_{-0.002}^{+0.002}$ & $158.1^{+0.1}_{-0.1}$ & $2.79_{-0.07}^{+0.07}$ \vspace{2px}\\
                                    & F & $0.786_{-0.007}^{+0.007}$ & $0.51_{-0.01}^{+0.01}$ & $139.2_{-0.1}^{+0.1}$ & $-4.7_{-0.2}^{+0.2}$ \vspace{2px}\\
                         \midrule
                J2100$-$4452 & A & $0.60_{-0.05}^{+0.04}$ & $0.32_{-0.03}^{+0.04}$ & $53.8_{-0.6}^{+0.6}$ & $18.8_{-4.3}^{+4.4}$ \vspace{2px}\\
                                    & B & $0.72_{-0.04}^{+0.03}$ & $0.38_{-0.04}^{+0.06}$ & $28.2_{-0.3}^{+0.3}$ & $-16.0_{-4.1}^{+4.0}$ \vspace{2px}\\
                                    & C & $0.48_{-0.06}^{+0.05}$ & $0.29_{-0.02}^{+0.03}$ &  $106.7_{-0.4}^{+0.4}$ & $5.5_{-1.2}^{+1.2}$ \vspace{2px}\\
                                    & D & $0.96_{-0.03}^{+0.02}$ & $0.55_{-0.06}^{+0.09}$ & $170.3_{-0.2}^{+0.2}$ & $-3.3_{-0.9}^{+0.9}$ \vspace{2px}\\
                                                                        \bottomrule

        \end{tabular}

        \label{tab:kappa}
\end{table}

\begin{landscape}

\begin{table}
\caption{Fermat potential differences}

        \begin{tabular}{lcccccccccc}
                \toprule System & $\Delta \tau_{\rm BA}$ &$\Delta \tau_{\rm CA}$ & $\Delta \tau_{\rm DA}$ & $\Delta \tau_{\rm FE}$ & $\Delta \tau_{\rm CB}$ & $\Delta \tau_{\rm DB}$ & $\Delta \tau_{\rm DC}$ \\ \toprule  \ \\
                DES J0029$-$3814        & $0.038_{-0.004}^{+0.007}$ & $0.032_{-0.004}^{+0.006}$  & $-0.34_{-0.05}^{+0.03}$ &
                $-$ & $-0.006_{-0.002}^{+0.002}$ & $-0.37_{-0.06}^{+0.04}$ & $-0.37_{-0.06}^{+0.04}$   \vspace{5px}\\
                DES J0214$-$2105        & 
                $0.13_{-0.02}^{+0.03}$ &
                $0.16_{-0.03}^{+0.03}$  & 
                $0.19_{-0.04}^{+0.04}$ &
                $-$ & $0.04_{-0.006}^{+0.007}$  & $0.06_{-0.01}^{+0.01}$ & $0.026_{-0.006}^{+0.006}$  \vspace{5px}\\
                DES J0420$-$4037        & 
                $-0.021_{-0.003}^{+0.002}$ &
                $-0.09_{-0.01}^{+0.01}$ & 
                $-0.017_{-0.002}^{+0.002}$ & 
                $-$& $-0.065_{-0.007}^{+0.007}$ & $0.0046_{-0.0005}^{+0.0006}$ & $0.070_{-0.007}^{+0.008}$   \vspace{5px}\\
                PS J0659$+$1629 & 
                $-3.9_{-0.1}^{+0.1}$ &
                $0.113_{-0.005}^{+0.005}$ &
                $0.086_{-0.003}^{+0.003}$ & 
                $-$ & $4.0_{-0.1}^{+0.1}$ & $4.0_{-0.1}^{+0.1}$ & $-0.028_{-0.004}^{+0.004}$  \vspace{5px}\\
                2M1134$-$2103   & 
        $-0.584_{-0.006}^{+0.006}$ & 
        $-0.803_{-0.008}^{+0.008}$  &
        $0.497_{-0.006}^{+0.008}$ &
        $-$ & $-0.219_{-0.002}^{+0.002}$ & $1.08_{-0.01}^{+0.01}$ & $1.30_{-0.01}^{+0.01}$  \vspace{5px}\\
                J1537$-$3010    & 
                $0.207_{-0.001}^{+0.001}$  &
                $-0.0483_{-0.0006}^{+0.0005}$  &
                $0.190_{-0.001}^{+0.001}$ &
                $-$  & $-0.255_{-0.001}^{+0.001}$  & $-0.0174_{-0.0008}^{+0.0009}$ & $0.238_{-0.002}^{+0.002}$  \vspace{5px}\\
                PS J1606$-$2333 & 
                $0.076_{-0.005}^{+0.005}$ &
                $0.148_{-0.005}^{+0.005}$ &
                $0.128_{-0.005}^{+0.005}$ & 
                $-$ & $0.072_{-0.002}^{+0.003}$ & $0.053_{-0.002}^{+0.002}$ & $-0.02_{-0.002}^{+0.002}$  \vspace{5px}\\
                PS J1721$+$8842 & 
            $-0.103_{-0.002}^{+0.002}$ & 
            $-0.202_{-0.004}^{+0.004}$  &
            $0.98_{-0.02}^{+0.01}$  & 
            $3.11_{-0.04}^{+0.04}$ & $-0.099_{-0.003}^{+0.002}$ & $1.08_{-0.02}^{+0.02}$ & $1.18_{-0.02}^{+0.02}$ \vspace{5px}\\
                DES J2100$-$4452        & 
                $0.0042_{-0.0005}^{+0.0008}$ &
                $-0.27_{-0.05}^{+0.03}$ &
                $0.26_{-0.02}^{+0.03}$ &
                $-$ & $-0.27_{-0.05}^{+0.03}$ & $0.25_{-0.02}^{+0.03}$ & $0.53_{-0.08}^{+0.05}$  \vspace{5px}\\ \bottomrule
                
        \end{tabular}
        \ \\
        \caption*{\textbf{Notes: } Computed based on Eq.~(\ref{eq:fermatpot}). We give the median value and include the corresponding 1$\sigma$ uncertainties.}
        \label{tab:fermat}
\end{table}

\begin{table}
\caption{Time-delay predictions}

        \begin{tabular}{lcccccccccccc}
                \toprule System & $z_{\rm d}$ & $z_{\rm s}$ & $\Delta t_{\rm BA}$ (days) &$\Delta t_{\rm CA}$ (days) & $\Delta t_{\rm DA}$ (days)& $\Delta t_{\rm FE}$ (days) & $\Delta t_{\rm CB}$ (days) & $\Delta t_{\rm DB}$ (days) & $\Delta t_{\rm DC}$ (days) \\ \toprule  \ \\
                DES J0029$-$3814        & 0.863 & 2.821 & $6.0_{-0.7}^{+1.0}$ & $5.1_{-0.6}^{+0.9}$ & $-53.5_{-8.1}^{+5.3}$  & 
                $-$ &
                $-0.9_{-0.3}^{+0.3}$ &
                $-59.4_{-9.2}^{+6.0}$  &
                $-58.6_{-9.1}^{+5.9}$ \vspace{5px}\\
                DES J0214$-$2105        & 0.22 & 3.229 & 
                $3.6_{-0.7}^{+0.8}$  &
                $4.7_{-0.9}^{+1.0}$ & 
                $5.4_{-1.0}^{+1.2}$ &
                $-$ &
                $1.0_{-0.2}^{+0.2}$ &
                $1.8_{-0.3}^{+0.4}$  &
                $0.8_{-0.2}^{+0.2}$ \vspace{5px}\\
                DES J0420$-$4037        & 0.358 & 2.4 &
                $-1.1_{-0.1}^{+0.1}$ &
                $-4.5_{-0.5}^{+0.5}$ & 
                $-0.9_{-0.1}^{+0.1}$ & 
                $-$ &
                $-3.4_{-0.4}^{+0.3}$ & 
                $0.24_{-0.03}^{+0.03}$  &
                $3.7_{-0.4}^{+0.4}$  \vspace{5px}\\
                PS J0659$+$1629 & 0.766 & 3.083 &
                $-509.4_{-14.1}^{+15.3}$ &
                $14.7_{-0.6}^{+0.6}$ &
                $11.1_{-0.4}^{+0.4}$  & 
                $-$ &
                $524.4_{-14.3}^{+16.0}$ & 
                $520.5_{-15.6}^{+14.5}$ &
                $-3.6_{-0.5}^{+0.5}$  \vspace{5px}\\
                2M1134$-$2103   & $\equiv$ 0.5 & 2.77 &
        $-44.7_{-0.4}^{+0.5}$ & 
        $-61.4_{-0.6}^{+0.6}$  &
        $38.0_{-0.5}^{+0.6}$ &
        $-$ &
        $-16.7_{-0.2}^{+0.2}$ &
        $82.7_{-0.7}^{+0.8}$ &
        $99.4_{-0.8}^{+0.9}$  \vspace{5px}\\
                J1537$-$3010    & 0.592 & 1.721 &
                $23.4_{-0.1}^{+0.1}$ &
                $-5.46_{-0.06}^{+0.06}$  &
                $21.4_{-0.1}^{+0.1}$  &
                $-$ &
                $-28.9_{-0.2}^{+0.2}$ &
                $-2.0_{-0.1}^{+0.1}$ &
                $26.9_{-0.2}^{+0.2}$  \vspace{5px}\\
                PS J1606$-$2333 & $\equiv$ 0.5 & 1.69 &
                $6.7_{-0.3}^{+0.3}$ &
                $13.1_{-0.5}^{+0.5}$ &
                $11.3_{-0.5}^{+0.5}$ &
                $-$ &
                $6.4_{-0.1}^{+0.1}$ &
                $4.7_{-0.2}^{+0.2}$ & 
                $-1.8_{-0.2}^{+0.2}$ \vspace{5px}\\
                PS J1721$+$8842 & 0.184 & 2.37 & 
            $-2.50_{-0.05}^{+0.05}$ & 
            $-4.9_{-0.1}^{+0.1}$ &
            $23.9_{-0.4}^{+0.4}$ & 
            $73.6_{-0.9}^{+1.0}$ & 
            $-2.42_{-0.06}^{+0.06}$  & 
            $26.4_{-0.4}^{+0.4}$ &
            $28.8_{-0.5}^{+0.5}$ \vspace{5px}\\
                DES J2100$-$4452        & 0.203 & 0.92 & 
                $0.13_{-0.02}^{+0.02}$ &
                $-8.5_{-1.6}^{+0.9}$ &
                $8.2_{-0.7}^{+1.0}$ &
                $-$ &
                $-8.6_{-1.6}^{+0.9}$ &
                $8.1_{-0.7}^{+1.0}$ &
                $16.7_{-2.5}^{+1.6}$  \vspace{5px}\\ \bottomrule
                
        \end{tabular}
        \ \\
        \caption*{\textbf{Notes: }We present the measured or assumed ($\equiv$) redshifts, which are fixed for the prediction. The predicted time delays are with respect to the reference image A for source 1 and E for source 2. In addition we present all other combinations of time delays. We give the median value, and include the corresponding 1$\sigma$ uncertainties.}
        \label{tab:timedelays}
\end{table}

\end{landscape}
\section{Degeneracies between the mass parameters}

\begin{figure}[h]
    \includegraphics[width=0.9\textwidth]{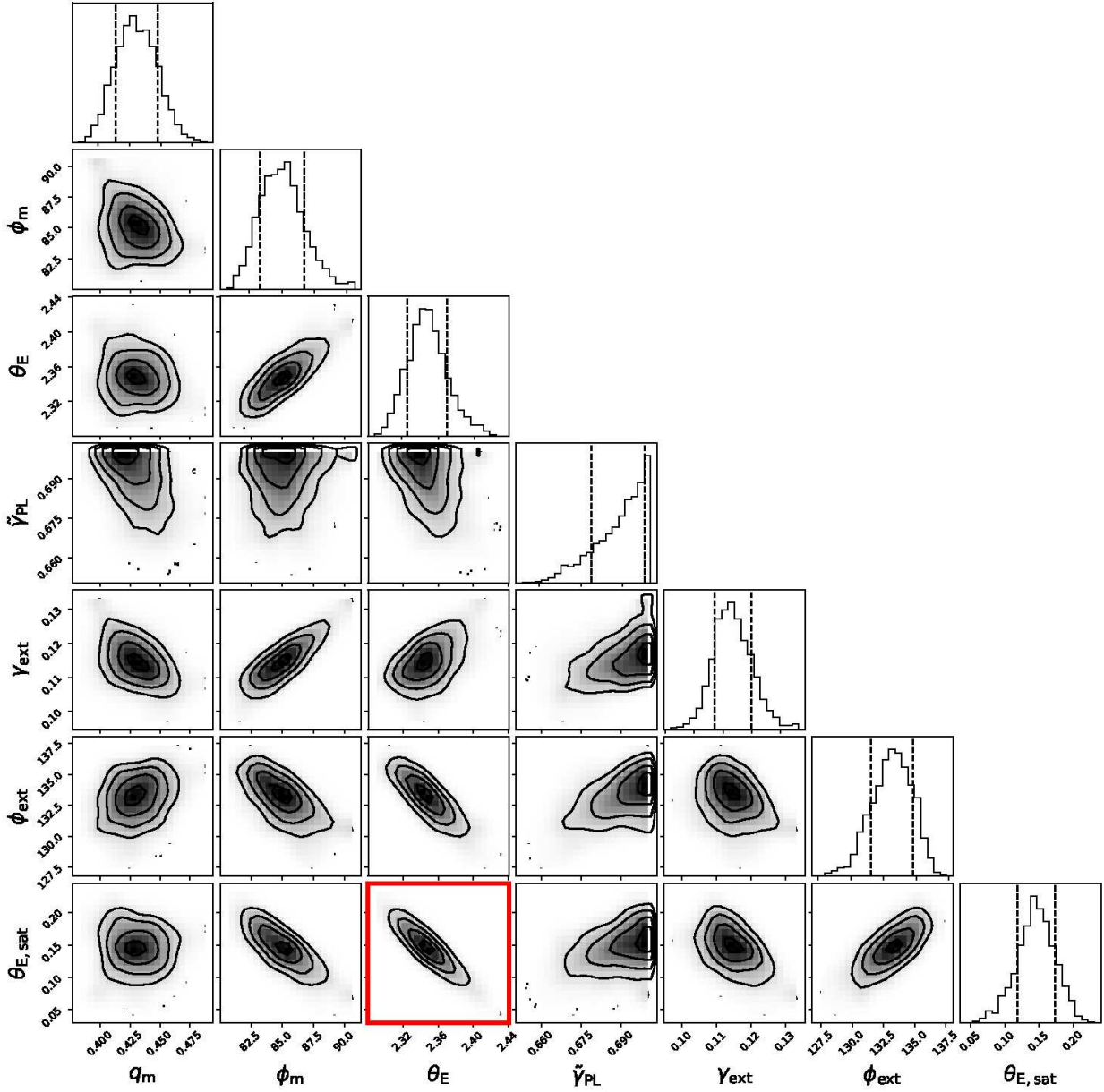}
        \caption{\label{fig:degen}Posterior distributions for the main mass parameters and external shear of J0659$+$1629. The correlation between the Einstein radius of the lens and the satellite is highlighted in red. The three shaded areas show the 1, 2, and 3$\sigma$ credible regions. The one-dimensional histograms show the marginalized posterior distribution for the selected mass parameters, and the vertical lines mark the 1$\sigma$ confidence intervals.}
\end{figure}

\clearpage

\onecolumn
\section{Plots for the comparison with \texttt{Lenstronomy} models}
\begin{figure*}[h]
        \subfigure{\includegraphics[height=0.37\textwidth]{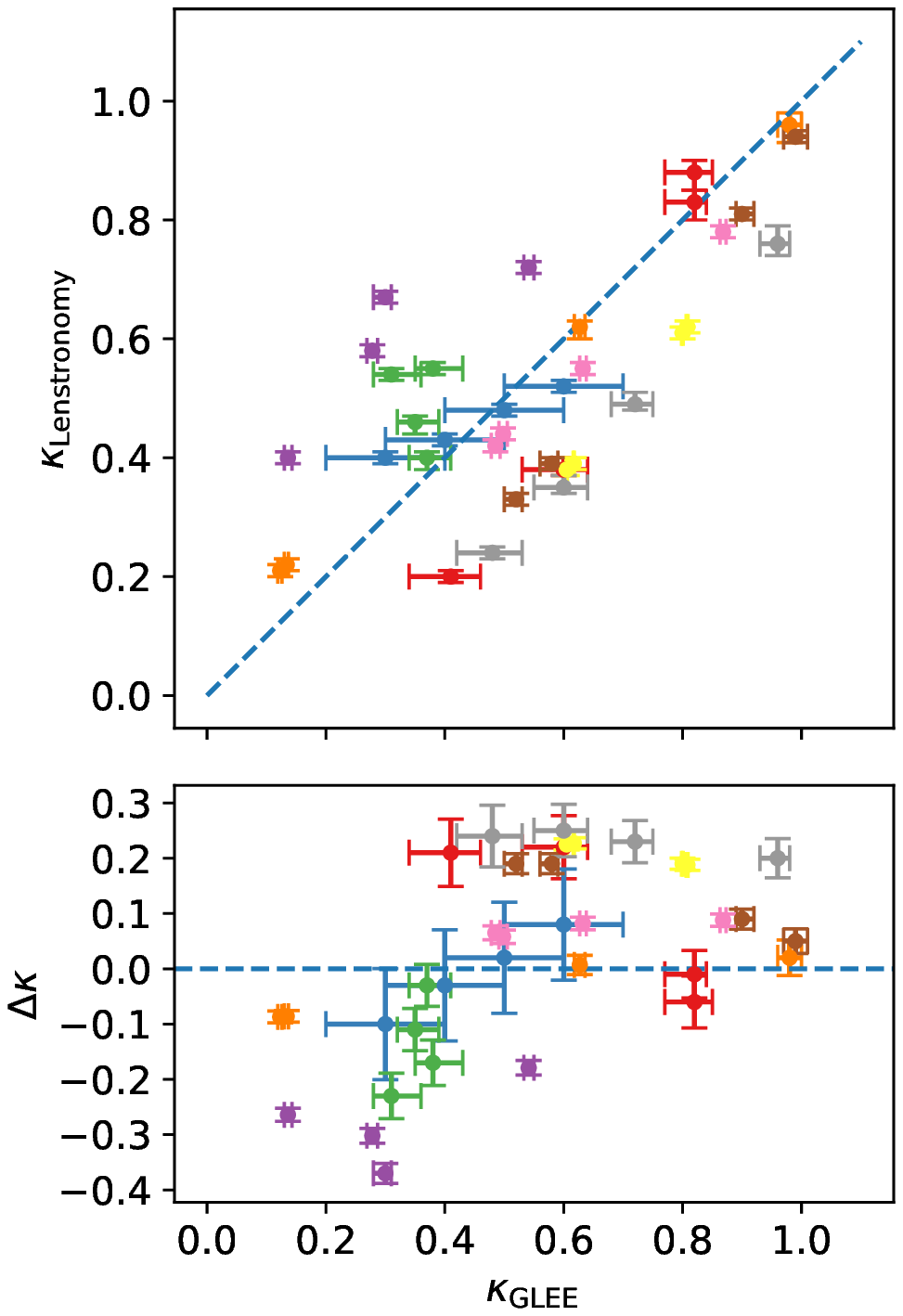}}
        \subfigure{\includegraphics[height=0.37\textwidth]{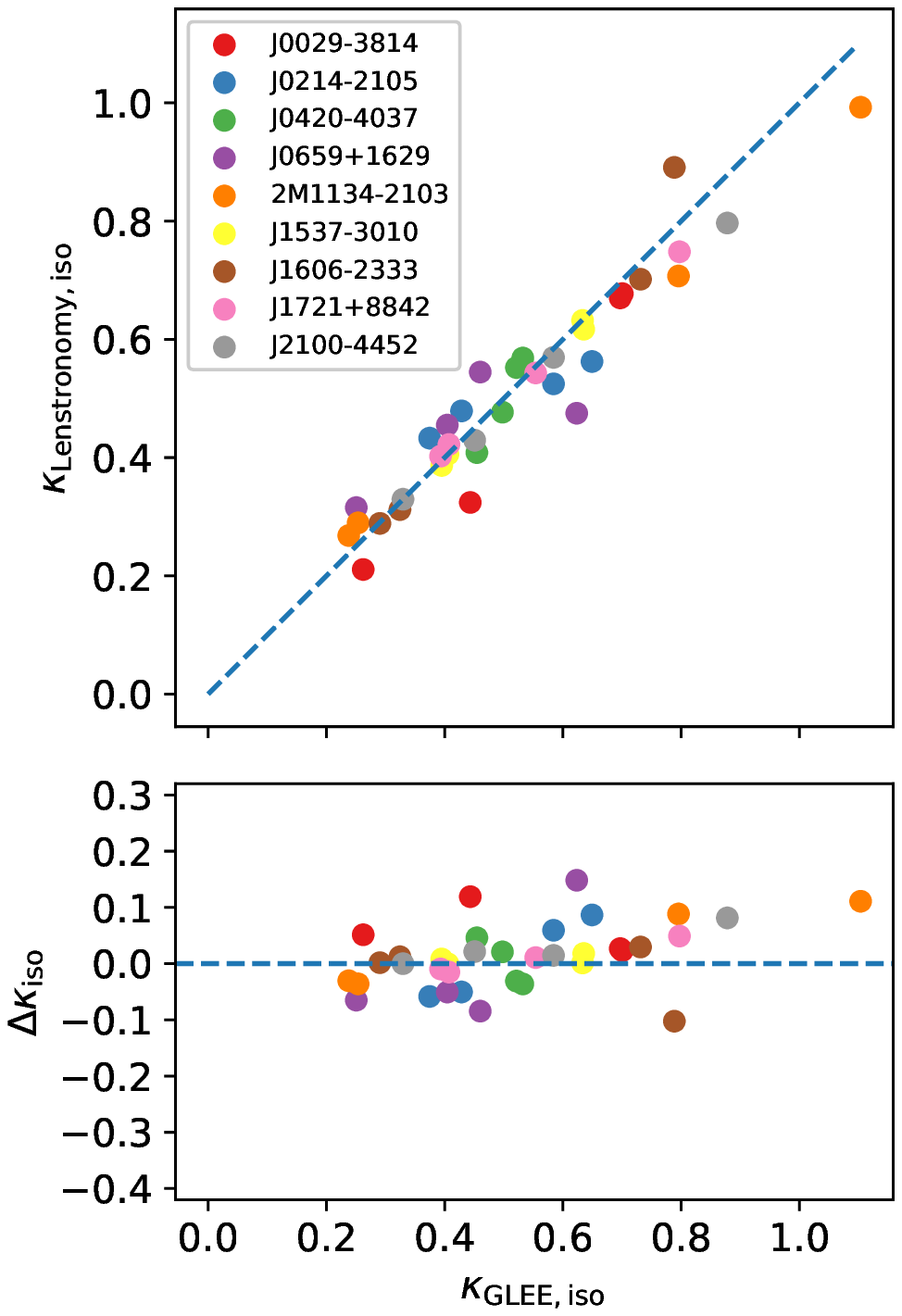}}
        \subfigure{\includegraphics[height=0.37\textwidth]{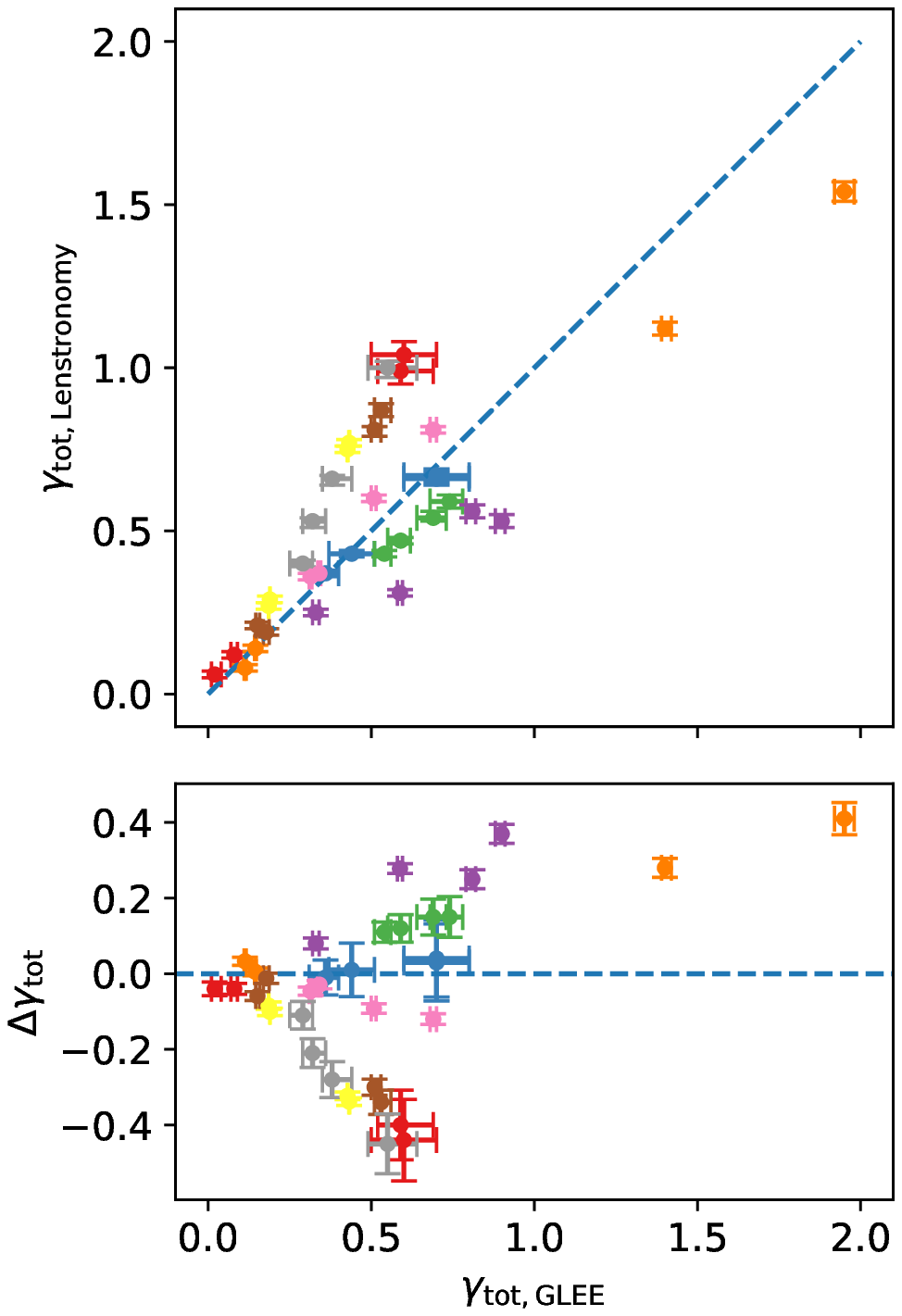}}
        \subfigure{\includegraphics[height=0.37\textwidth]{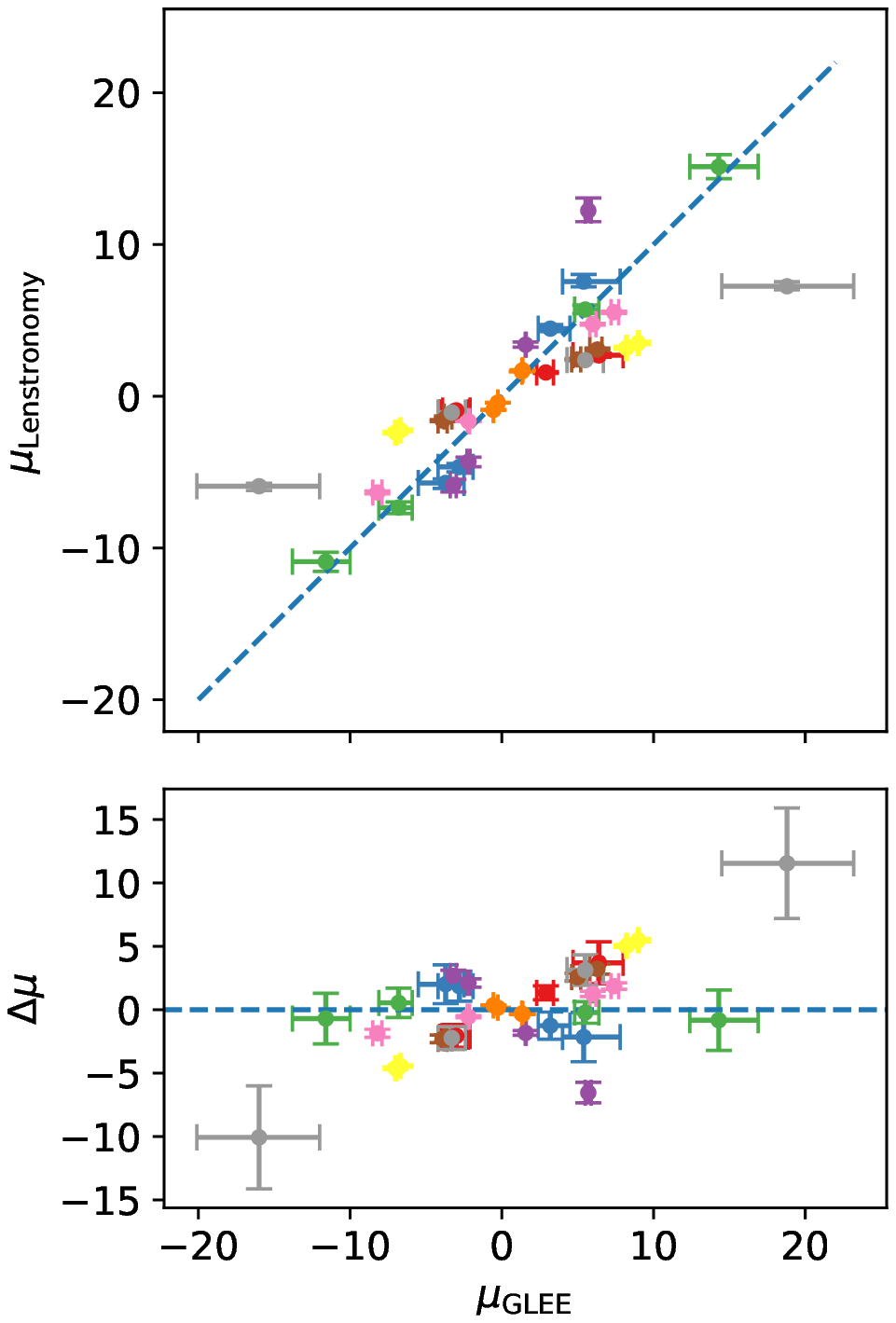}}
        \subfigure{\includegraphics[height=0.37\textwidth]{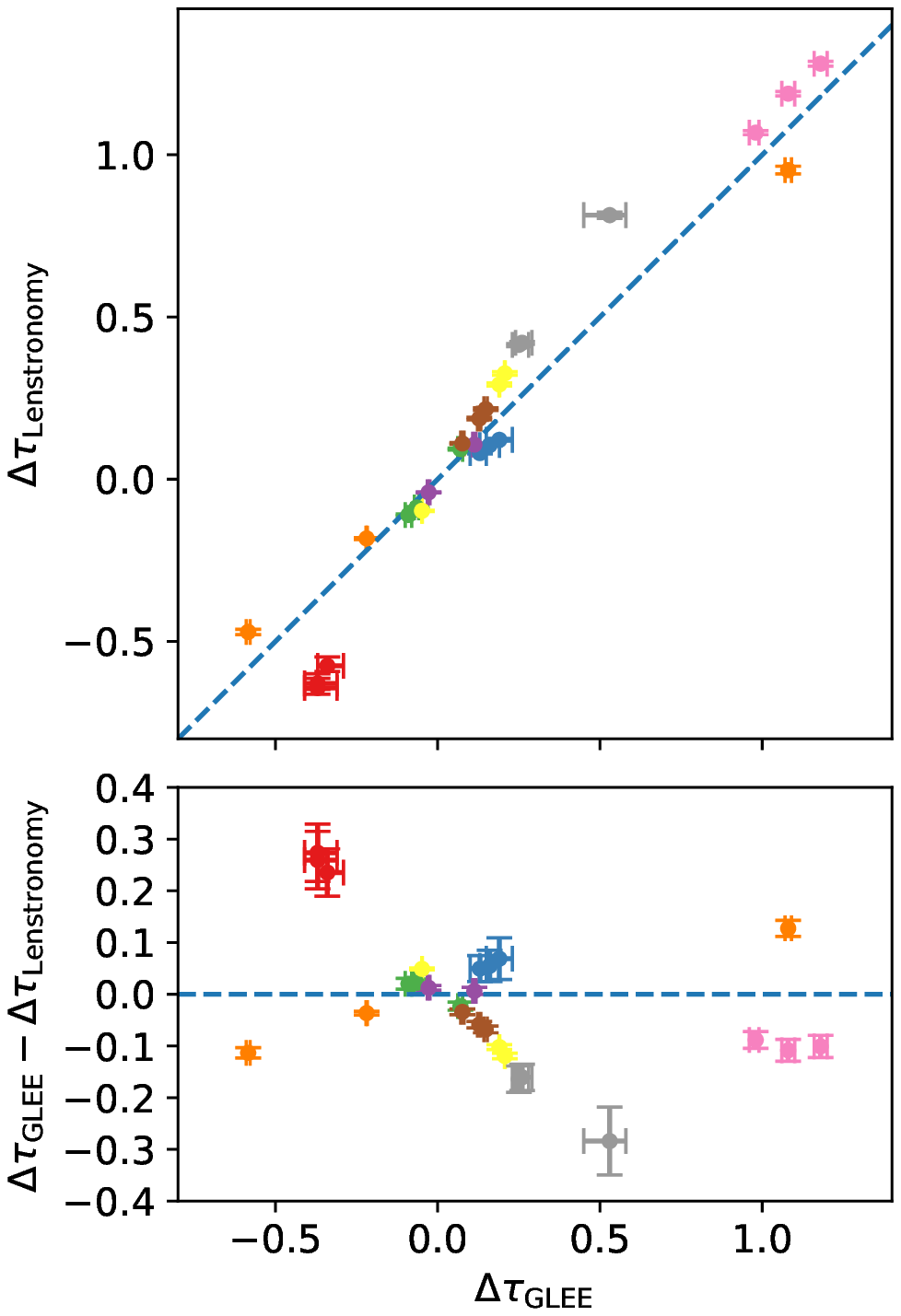}}
        \subfigure{\includegraphics[height=0.37\textwidth]{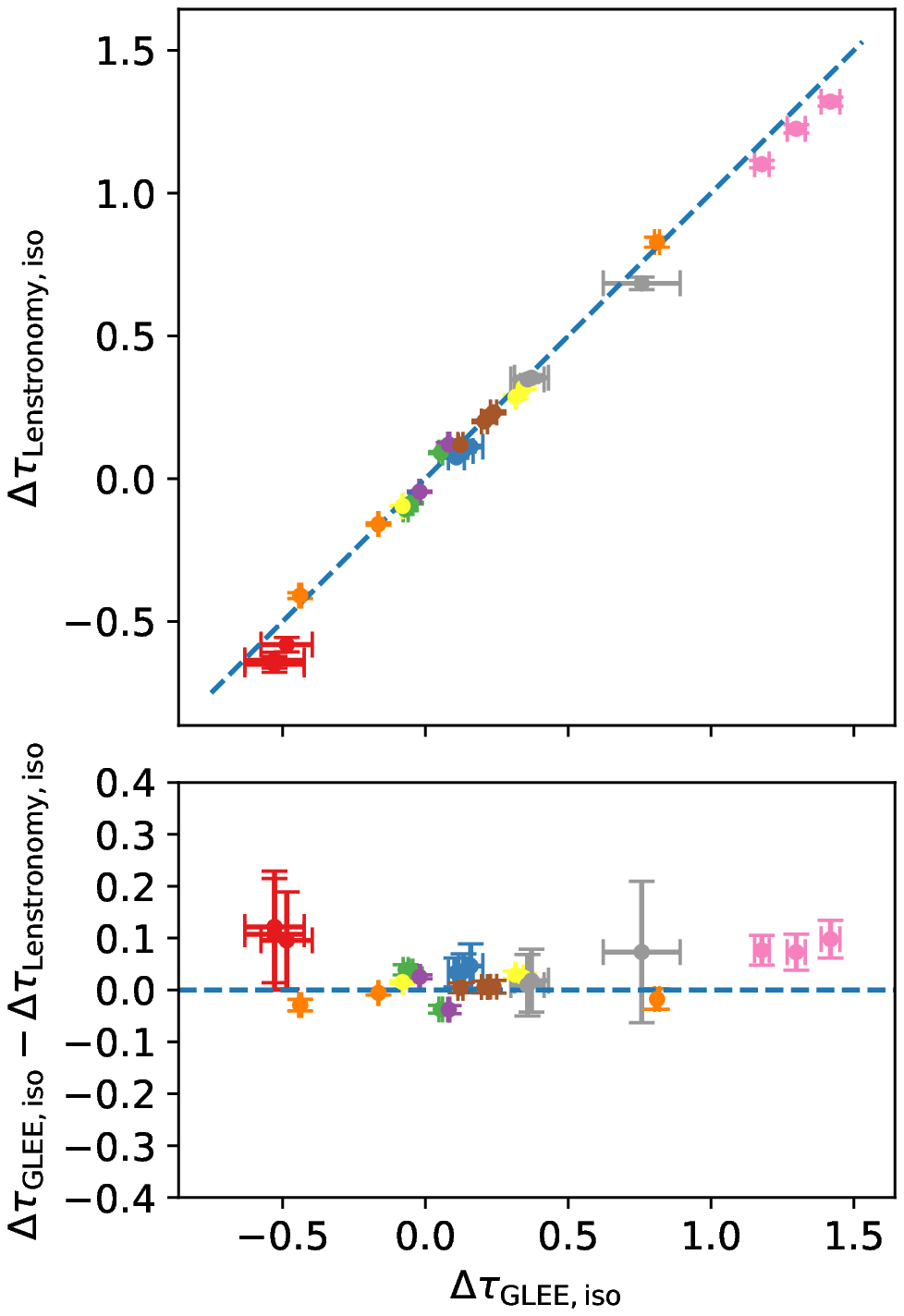}}
        \subfigure{\includegraphics[height=0.37\textwidth]{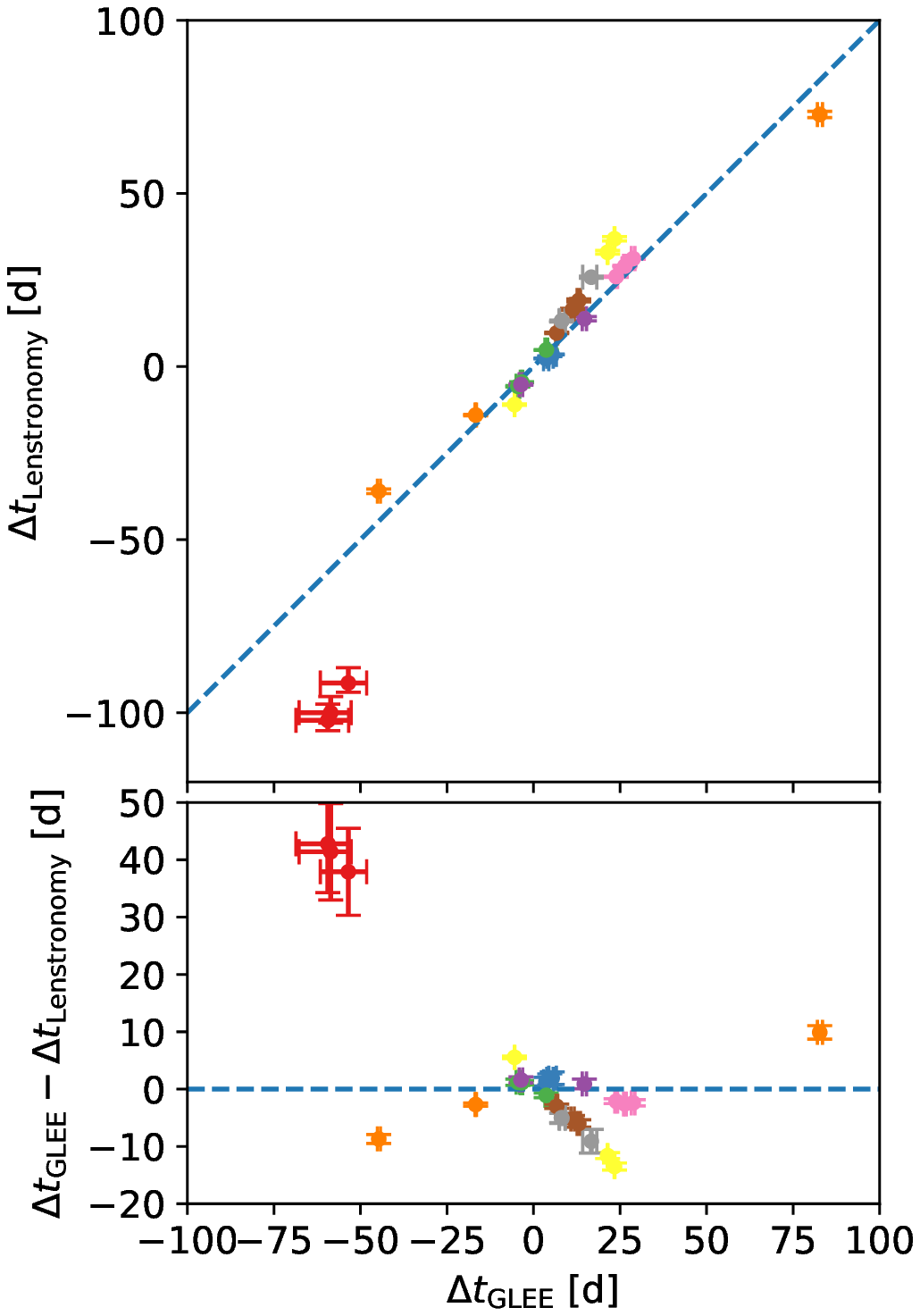}}
                                                        
        \caption{Comparison of $\kappa$, $\gamma_{\rm tot}$, $\mu$, Fermat potential difference $\Delta\tau$, and predicted time delay $\Delta t$ between the \texttt{GLEE} and \texttt{Lenstronomy} models. Additionally, we compare $\kappa_{\rm iso}$ and $\Delta\tau_{\rm iso}$, where we assume isothermal instead of power-law mass profiles.}
        \label{fig:comp_appendix1}
\end{figure*}

\begin{figure*}
        \subfigure{\includegraphics[height=0.4\textwidth]{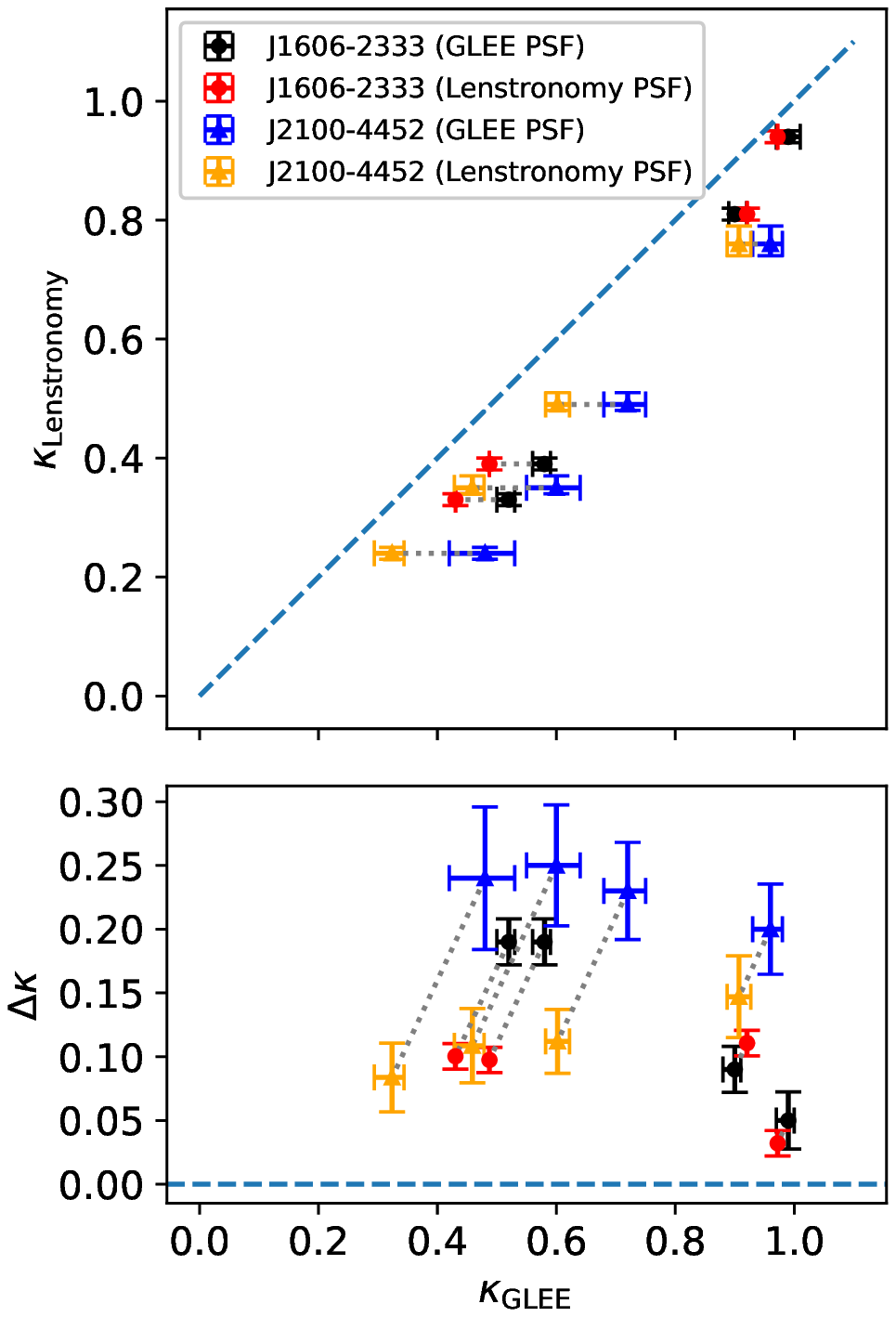}}
        \subfigure{\includegraphics[height=0.4\textwidth]{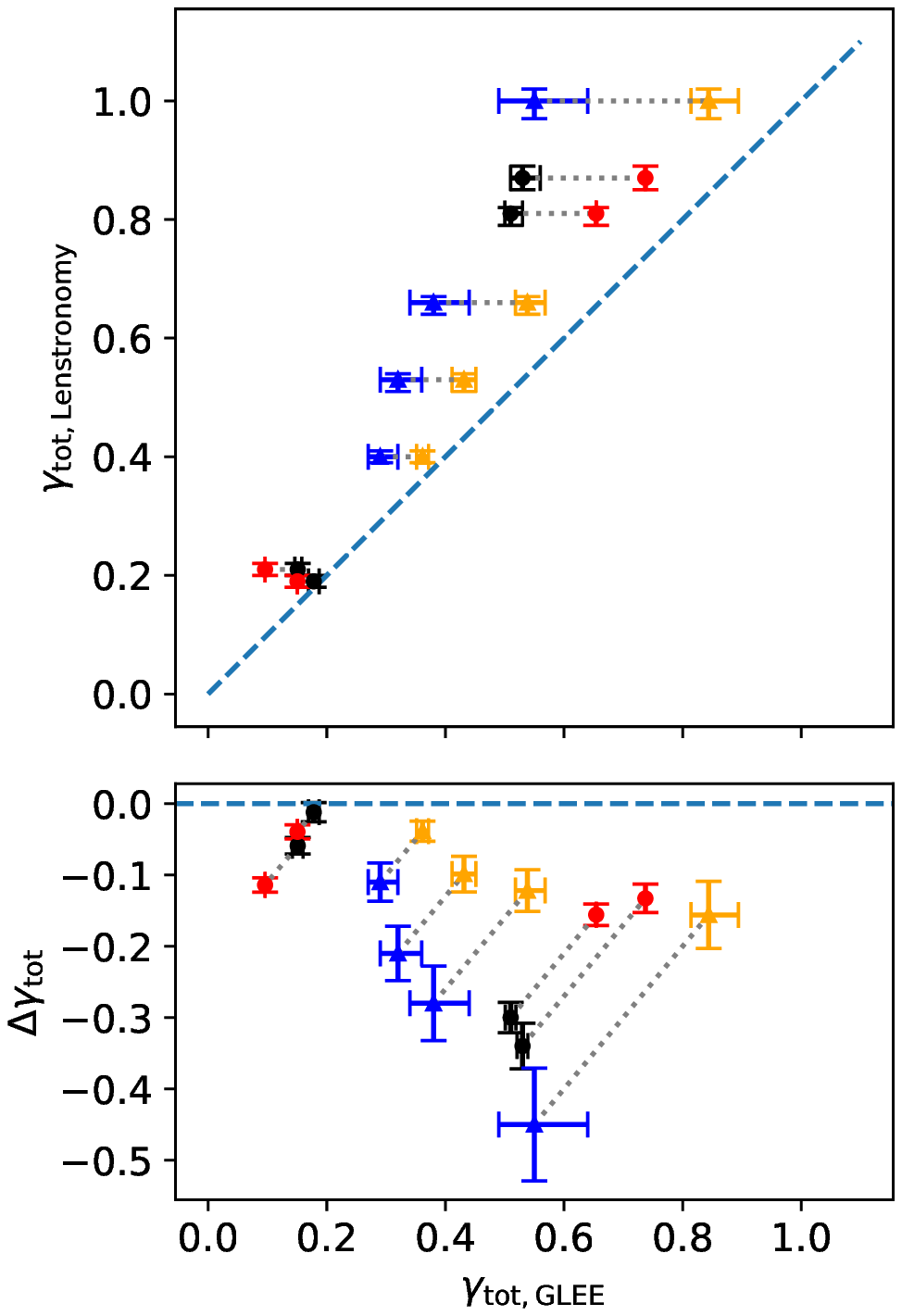}}
        \subfigure{\includegraphics[height=0.4\textwidth]{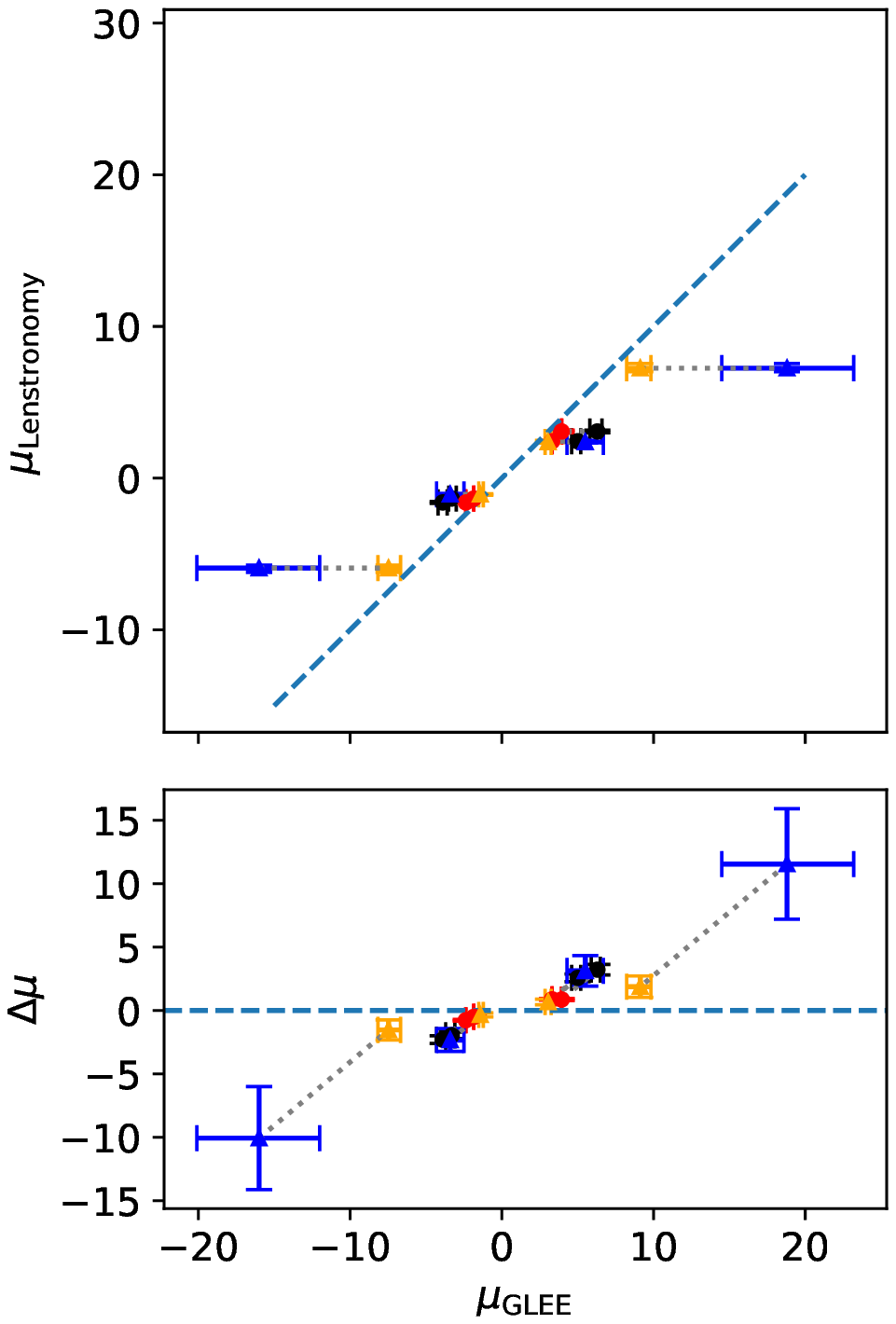}}
        \subfigure{\includegraphics[height=0.4\textwidth]{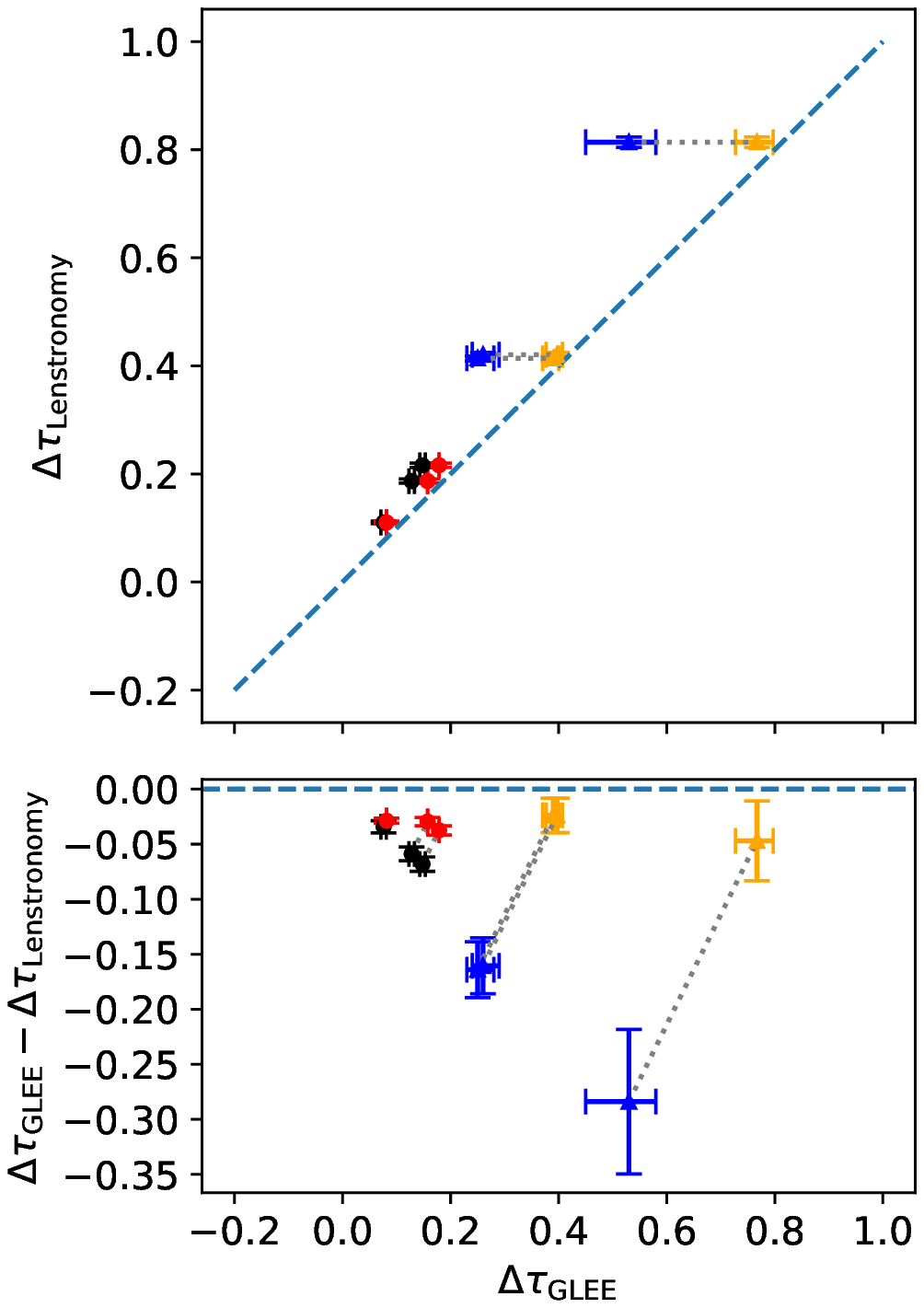}}
        \subfigure{\includegraphics[height=0.4\textwidth]{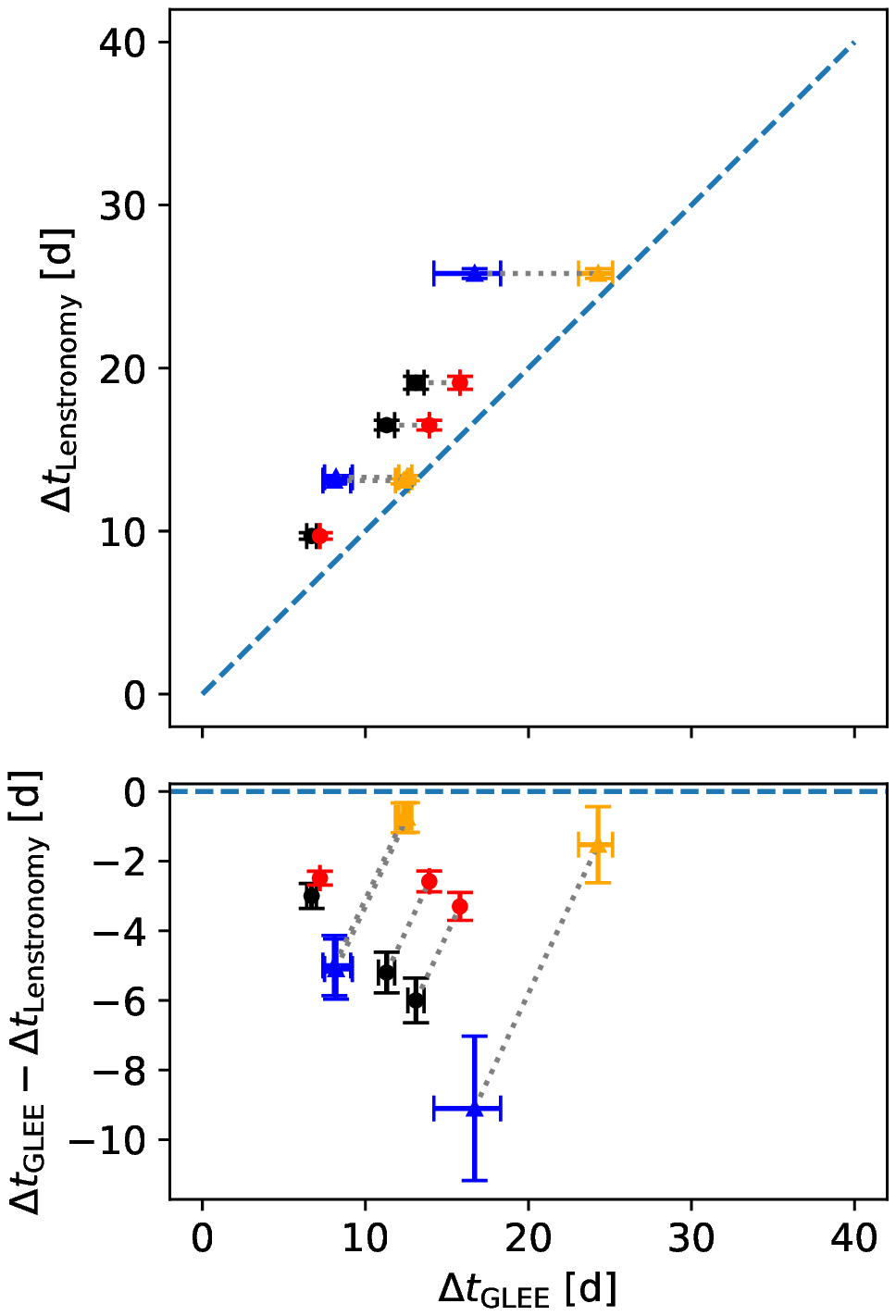}}

        \caption{Change in $\kappa$, $\gamma_{\rm tot}$, $\mu$, Fermat potential difference $\Delta\tau$, and predicted time delay $\Delta t$ when using the \texttt{Lenstronomy} PSF with our \texttt{GLEE} automation code for J1606$-$2333 and J2100$-$4452 in the F160W (IR) band.}
        \label{fig:comp_appendix2}
\end{figure*}

\end{appendix}

\end{document}